\documentclass[prb,showpacs,preprintnumbers,amsmath,amssymb]{revtex4}

\usepackage{epsfig}
\usepackage{graphicx}
\usepackage{dcolumn}
\usepackage{bm}

\newcommand{\EQ}{\begin{equation}}
\newcommand{\EN}{\end{equation}}
\newcommand{\ea}{\end{eqnarray}}
\newcommand{\ba}{\begin{eqnarray}}

\newcommand{\bear}{\begin{eqnarray}}
\newcommand{\ear}{\end{eqnarray}}

\begin{document}

\title{The square-lattice quantum liquid of charge $c$ fermions and spin-neutral two-spinon $s1$ fermions}
\author{J. M. P. Carmelo} 
\affiliation{GCEP-Centre of Physics, University of Minho, Campus Gualtar, P-4710-057 Braga, Portugal}

\date{7 July 2009}


\begin{abstract}
The momentum bands, energy dispersions, and velocities of the
charge $c$ fermions and spin-neutral two-spinon $s1$ fermions
of a square-lattice quantum liquid referring to the
Hubbard model on such a lattice of edge length $L$ in the one- and 
two-electron subspace are studied. The model involves the effective
nearest-neighbor integral $t$ and on-site repulsion $U$ and can be experimentally realized in 
systems of correlated ultra-cold fermionic atoms on an optical lattice and 
thus our results are of interest for such systems. Our investigations 
profit from a general rotated-electron description, 
which is consistent with the model global $SO(3)\times SO(3)\times U(1)$ symmetry.
For the model in the one- and two-electron subspace the discrete 
momentum values of the $c$ and $s1$ fermions are good quantum numbers so that in contrast to the original 
strongly-correlated electronic problem their interactions 
are residual. The use of our description renders
an involved many-electron problem into a quantum liquid with
some similarities with a Fermi liquid. For the Hubbard model on a square lattice 
in the one- and two-electron subspace a composite $s1$ fermion consists of 
a spin-singlet spinon pair plus an infinitely thin flux tube attached to it.
In the $U/4t\rightarrow\infty$ limit of infinite on-site interaction the $c$
fermions become non-interacting spinless fermions and the $s1$ fermion 
occupancy configurations that generate the spin degrees of freedom of spin-density
$m=0$ ground states become within a suitable mean-field approximation for the fictitious magnetic field 
$B_{s1}\,{\vec{e}}_{x_3}$ brought about by the correlations of the original $N$ electron problem
those of a full lowest Landau level with $N/2$ degenerate one-$s1$ fermion states of the two-dimensional 
quantum Hall effect. In turn, for $U/4t$ finite the degeneracy of the $N/2$ one-$s1$-fermion states is removed by the emergence
of a finite-energy-bandwidth $s1$ fermion dispersion yet the number of $s1$
band discrete momentum values remains being given by $B_{s1}\,L^2/\Phi_0$
and the $s1$ effective lattice spacing by $a_{s1}=l_{s1}/\sqrt{2\pi}$
where $l_{s1}$ is the fictitious-magnetic-field length and in our units the fictitious-magnetic-field 
flux quantum reads $\Phi_0=1$. Elsewhere it is found that the use of the square-lattice quantum liquid
of charge $c$ fermions and spin-neutral two-spinon $s1$ fermions investigated here contributes 
to the further understanding of the role of electronic correlations in the unusual properties of the
hole-doped cuprate superconductors. This indicates that quantum-Hall-type behavior 
with or without magnetic field may be ubiquitous in nature. 
\end{abstract}

\maketitle

\section{Introduction}

The Hubbard model on a square lattice is the simplest 
realistic toy model for description of the electronic correlation effects  
in general many-electron problems with short-range interaction
on such a lattice and therefore is the obvious starting point 
for the study of the role of such effects in the exotic physics of the hole-doped cuprates
\cite{two-gaps,k-r-spaces,duality,2D-MIT,Basov,ARPES-review,Tsuei,pseudogap-review}.
The model involves two effective parameters: the in-plane nearest-neighbor
transfer integral $t$ and the effective on-site repulsion $U$.
Despite that it is among the mostly studied models in condensed 
matter physics, there is no exact solution and few controlled 
approximations exist for finite $U/4t$ values.

In this paper we study the momentum bands, energy dispersions, and velocities of the
charge $c$ fermions and spin-neutral two-spinon $s1$ fermions
introduced for the model on the square lattice in Ref. \cite{companion}. The $s1$ fermions emerge 
from the spin-neutral two-spinon $s1$ bond particles \cite{companion,s1-bonds} 
through an extended Jordan-Wigner transformation 
\cite{J-W,Fradkin,Wang,Feng}. Our study has as starting point the 
properties of the Hubbard model on the square 
lattice in the one- and two-electron subspace defined
in Ref. \cite{companion} and profits from the general rotated-electron description 
introduced in that reference. 

There is consensus about the scientific interest of the Hubbard 
model on the square lattice as simplest toy model for describing the 
effects of electronic correlations in the 
high-$T_c$ superconductors \cite{two-gaps,k-r-spaces,duality,2D-MIT,Basov,ARPES-review,Tsuei,pseudogap-review} and
their Mott-Hubbard insulators parent compounds
\cite{Cuprates-insulating-phase,LCO-neutr-scatt}. However,
many open questions about its properties remain 
unsolved. Interestingly, the model can be experimentally 
realized with unprecedented precision in systems of correlated ultra-cold fermionic 
atoms on an optical lattice \cite{Zoller} and one may expect very detailed 
experimental results over a wide range of parameters to be available.

The square-lattice quantum liquid of charge $c$ fermions
and spin-neutral two-spinon $s1$ fermions and the related 
general rotated-electron description
of Ref. \cite{companion} are consistent with the global 
$SO(3)\times SO(3)\times U(1)=[SO(4)\times U(1)]/Z_2$ symmetry
found recently in Ref. \cite{bipartite} for the model
on any bipartite lattice. Such a global symmetry is an extension 
of the $SO(4)$ symmetry known to occur for
the model on such lattices \cite{HL,Zhang}. 
The extended global symmetry is related to the rotated 
electrons, which for on-site repulsion $U>0$ emerge
from the the electrons through a unitary transformation
of the type considered in Ref. \cite{Stein}, and to
the local symmetries and unitary transformations
studied in Ref. \cite{U(1)-NL}. 

The building blocks of the general description
introduced in Ref. \cite{companion} are the
$\eta$-spin-$1/2$ $\eta$-spinons, spin-$1/2$ spinons, 
and spinless and $\eta$-spinless charge $c$ fermions
whose occupancy configurations
generate the state representations of the
$\eta$-spin $SU(2)$ symmetry, spin $SU(2)$ symmetry,
and $U(1)$ symmetry, respectively,
associated with the model global 
$SO(3)\times SO(3)\times U(1)=[SU(2)\times SU (2)\times U(1)]/Z_2^2$ symmetry. 
Such three basic objects are well defined for $U>0$. 
These state representations are found in Ref. \cite{companion} to correspond to a
complete set of momentum eigenstates. The $\eta$-spin-$1/2$ $\eta$-spinons
describe the $\eta$-spin degrees of freedom of
the rotated-electron occupancy configurations that generate such states involving doubly occupied 
and unoccupied sites, the spin-$1/2$ spinons the
spin degrees of freedom of the rotated-electron configurations of the singly occupy sites,
and the $c$ fermions the charge excitations associated with
rotated-electron motion that conserves the numbers
of singly occupied, doubly occupied, and unoccupied sites. 

For $U/4t\gg 1$ the $c$ fermion, spinon,
and $\eta$-spinon operators become the
quasicharge, spin, and pseudospin operators,
respectively, obtained from the transformation 
considered in Ref. \cite{Ostlund-06}, which 
does not introduce Hilbert-space constraints.
Such quasicharge, spin, and pseudospin operators
have for $U/4t\gg 1$ the same expressions 
in terms of creation and annihilation electron operators as the $c$ fermion, spinon,
and $\eta$-spinon operators for $U>0$ in terms of 
rotated-electron creation and annihilation operators. The unitary character of the 
transformation between electrons and rotated electrons
then assures that the transformation that generates
the $c$ fermion, spinon, and $\eta$-spinon operators
from the electronic operators also does not introduce 
Hilbert-space constraints. The vacuum of the theory is 
given in Eq. (\ref{23}) of Appendix A, where some basic
information on the square-lattice quantum liquid one- and two-electron 
subspace defined in Ref. \cite{companion} is provided. 
There is a vacuum of general form given in that equation for 
each subspace with a constant finite value $2S_c$ for the number 
of rotated electrons that singly occupy sites. $S_c$ is 
also the eigenvalue of the generator of the hidden charge 
$U(1)$ global symmetry \cite{bipartite}. For both the model on the square and
one-dimensional (1D) lattice such a $S_c>0$ vacuum is invariant under 
the electron - rotated-electron unitary transformation.

For the square-lattice quantum liquid associated with the
Hubbard model in the one- and two-electron subspace only the charge $c$ fermions 
and spin-neutral two-spinon $s1$ bond particles play an active role \cite{companion,s1-bonds}.
In contrast to previous descriptions involving Jordan-Wigner transformations
\cite{Wang,Feng} or slave-boson representations \cite{2D-MIT,Fazekas,Xiao-Gang}
and referring to the model for large values of the on-site
repulsion $U$ or Heisenberg and $t-J$ models, whose spinless fermions
arise from spin-$1/2$ objects, the $s1$ fermions emerge from hard-core
spin-neutral two-spinon composite objects and are well defined 
for finite $U$ values of the on-site repulsion. As mentioned above, 
the spin-$1/2$ spinons are the spins of the rotated electrons that
singly occupied lattice sites and emerge from
a suitably electron - rotated-electron unitary transformation
introduced in Ref. \cite{companion}. Therefore, here the single-occupancy 
constraint is naturally fulfilled.

For the Hubbard model on the square lattice in the
one- and two-electron subspace as defined in that reference there are
no $\alpha\nu$ fermions other than none or one zero-momentum $s2$ fermion and 
$[S_c-S_s]$ or $[S_c-S_s-2]$ $s1$ fermions, respectively, where $S_s$ denotes the spin.
The interest of the description introduced in Ref. \cite{companion} lies in the fact that
for the subspaces of the one- and two-electron subspace
spanned by mutually neutral states where the $s1$ fermion operators act onto,
the two $s1$ translation generators $\vec{q}_{s1\,x_1}$ and $\vec{q}_{s1\,x_2}$  
in the presence of the fictitious magnetic field ${\vec{B}}_{s1}$  
associated with the $s1$ fermion Jordan-Wigner transformation
commute with each other and with both the Hamiltonian and momentum operator.
In contrast, the Hubbard model on the square lattice in the whole Hilbert
space does not commute with such translation generators \cite{companion}.
Therefore, for the square-lattice quantum liquid corresponding to the
Hubbard model on a square lattice in the one- and two-electron subspace
the $s1$ fermion discrete momentum values 
$\vec{q}_j=[{q_j}_{x_1},{q_j}_{x_2}]$ are good quantum numbers. 
Here ${q_j}_{x_1}$ and ${q_j}_{x_2}$ are eigenvalues of the two $s1$ translation generators
$\vec{q}_{s1\,x_1}$ and $\vec{q}_{s1\,x_2}$,
respectively. In addition, the description of Ref. \cite{companion} has been constructed
to inherently the $c$ fermion discrete momentum values being good quantum numbers for
the whole Hilbert space. In turn, our 
method involves approximations to derive the shape of the $s1$ momentum
band boundary and the form of the $c$ and $s1$ fermion energy dispersions. 
The fulfillment of such tasks is the main goal of this paper.
For the 1D model the discrete momentum values of the $c$ and $s1$ fermions in units of $2\pi/L$ 
are quantum numbers of the exact solution \cite{companion}.
For the model on the square lattice the nearest-neighboring components of 
the $c$ and $s1$ discrete momentum values $\vec{q}_j=[{q_j}_{x_1},{q_j}_{x_2}]$
are such that $[{q_j}_{xi}-{q_{j'}}_{xi}]=2\pi/L$ where $i=1$ or $i=2$ and the indices
$j$ and $j'$ refer to nearest neighboring discrete momentum values. The
shape of the $c$ band is that of the first Brillouin zone. Hence the main open issue is
the shape of the $s1$ boundary line for $x\geq 0$ and $m=0$ ground states, and that 
along with the $c$ and $s1$ fermion energy dispersions is 
one of the problems studied in this paper.

It follows from the results of Ref. \cite{companion} 
that for the Mott-Hubbard insulator at hole concentration $x=0$ and
spin density $m=0$ both the $c$ and $s1$ bands 
are full. For the Hubbard model on a square-lattice the $c$
band has for all energy eigenstates the same momentum area and 
shape as the electronic first Brillouin zone.
For $m=0$ ground states with a small finite hole concentration
$x$ there arises a circular $c$ Fermi line around $-\vec{\pi}=-[\pi,\pi]$
of ratio $\approx \sqrt{x\pi}\,2$, which encloses a $c$ fermion
unfilled momentum area $x\,4\pi^2$. Since the $c$ effective lattice equals
the original lattice and thus its spacing $a_c$ equals
the original lattice constant $a$, the $c$ fermion occupancy configurations
that generate the energy eigenstates conserve translational invariance.

In turn, consistently with the ground-state $s1$ effective-lattice
occupancies of Ref. \cite{companion}, for $x\geq 0$ and $m=0$ ground states the
$s1$ band is full. For the $x=0$ and $m=0$ absolute ground state
that band has a momentum area $2\pi^2$ and is found in this paper to 
coincide with an antiferromagnetic reduced Brillouin 
zone such that $\vert q_{x_1}\vert+\vert q_{x_2}\vert\leq\pi$,
which is enclosed by a boundary whose 
momenta ${\vec{q}}_{Bs1}$ belong to the lines
connecting $[\pm\pi,0]$ and $[0,\pm\pi]$. Such a $s1$ band
shape and momentum area are consistent with the result of Ref. \cite{companion}
that at $x=0$ and $m=0$ 
the square $s1$ effective lattice has spacing $a_{s1}=\sqrt{2}\,a$.
Indeed our studies confirm that its periodicity has increased relative to that of
the original lattice owing to the appearance of a long-range antiferromagnetic
order. Moreover, we find that the Fermi line
has for the $x=0$ and $m=0$ ground state the same
form as the $s1$ boundary line so that it refers to
the lines connecting $[\pm\pi,0]$ and $[0,\pm\pi]$.
For very small values of $x$, both the momentum areas
enclosed by the $s1$ boundary line and the Fermi line 
decrease to $(1-x)2\pi^2$. Then the latter line becomes
hole like and both such lines remain near 
the lines connecting $[\pm\pi,0]$ and $[0,\pm\pi]$. This is consistent with the 
recent experimental results of Ref. \cite{k-r-spaces}, which reveal that
for small hole concentrations the momentum-space
region near such lines plays a major role 
in the Fermi-line physics of the hole-doped cuprate
superconductors. 

Upon further increasing the hole concentration $x$ the $s1$
momentum band remains full for $x\geq 0$ and $m=0$ 
ground states and encloses a smaller momentum area $(1-x)2\pi^2$, alike for small $x$. For $0< x<x_*$ and 
$m=0$ there is a short-range spin order. Here $x_*\in (0.23,0.28)$ for approximately
$u_0\leq U/4t\leq u_1$ where $u_0\approx 1.3$ and $u_1\approx 1.6$
is a critical concentration $x_*$ introduced in Subsection IV-F, below which such an order prevails. 
For $0<x\ll 1$ it is a short-range incommensurate-spiral spin
order, consistently with the spacing of the square $s1$ effective lattice
reading $a_{s1}=\sqrt{2/(1-x)}\,a$. 
Such and related types of spin orders have been observed
in the cuprate superconductors \cite{nematic-order,spiral-order}.
The schemes introduced in Refs. \cite{cuprates0,cuprates} involve modified versions of 
the square-lattice quantum liquid investigated in this paper and contribute to the
further understanding of the unusual properties of the hole-doped cuprate superconductors  
\cite{Kam,PSI-ANI-07,LSCO-resistivity,Y-resistivity}.

Our study focus on the Hubbard model on the square lattice.
The reason why often our analysis refers to the same model on 
the 1D lattice as well is that in contrast to real-space 
dimensions $D>1$ there is an exact solution for 1D 
\cite{Lieb,Takahashi,Martins}. In spite of then the model
referring to a qualitatively different physics,
our quantum-object description also applies to 1D.
For instance, in the $U/4t\rightarrow\infty$ limit the $c$
fermions become both for the model on a 1D and square
lattices non-interacting spinless fermions. In turn, 
the $s1$ fermion occupancy configurations that generate the
spin degrees of freedom of spin-density
$m=0$ ground states become in that limit for 1D and the square lattice
those of the spins of the spin-charge factorized wave function introduced both
by Woynarovich \cite{Woy} and Ogata and Shiba \cite{Ogata} and within  
a suitable mean-field approximation for a fictitious magnetic field brought about
by the electronic correlations those of a full lowest Landau level with
$N_{s1}=N_{a_{s1}}^2=N/2$ one-$s1$-fermion degenerate states of the 2D quantum  
Hall effect (QHE), respectively. Here $N_{a_{s1}}^2$ is the number of 
both sites of the square $s1$ effective lattice and $s1$ band discrete
momentum values.

Consistently with the validity of the $c$ fermion,
$2\nu$-$\eta$-spinon $\eta\nu$ fermion, and $2\nu$-spinon $s\nu$ fermion
description of Ref. \cite{companion} where $\nu=1,2,...$ gives the
number of $\eta$-spinon or spinon pairs, which for the model in the one- and
two-electron subspace considered in this paper involves only the $c$ fermions
and composite two-spinon $s1$ fermions, it is shown in that reference 
that for 1D and the limit $N_a\gg1$, which such a description refers to, the
discrete momentum values of the $c$ fermions and $\alpha\nu$
fermions where $\alpha =\eta,s$ coincide with the quantum numbers of the exact solution. 

In contrast to the original strongly-correlated electron problem, 
the interactions of the $c$ and $s1$ fermions are residual 
owing to their momentum values being good quantum numbers. 
Consistently, the non-perturbative and involved problem concerning the
effects of the electronic correlations of the Hubbard model
on the square lattice simplifies when expressed in 
terms of the $c$ and $s1$ fermion
residual interactions. Our results and those of Ref. \cite{cuprates0} 
reveal that for the model on that lattice the residual interactions  
recombine the charge and spin degrees 
of freedom to such an extent that one cannot speak of 
a spin-charge separation as that occurring in 1D.
For instance, one cannot express the electronic spectral functions
as a simple convolution of $c$ fermion and $s1$ fermion
spectral functions. Therefore, the concept of spin-charge separation
does not apply to the model on the square lattice, at least with the meaning it has in
1D correlated systems.

The square-lattice quantum liquid studied in this paper is non-perturbative
in terms of electron operators so that, in contrast to a 3D isotropic
Fermi liquid \cite{Landau,Pines}, rewriting its theory in terms of
the standard formalism of many-electron physics is an extremely
complex problem. Fortunately, the description of the physics of such a quantum liquid simplifies 
when it is expressed in terms of the $c$ and $s1$ fermion operators.
In the $x>0$ studies of Refs. \cite{cuprates0,cuprates} the effects of 
the residual interactions of the $c$ and $s1$ fermions play a major
role. At $x=0$ the residual interactions of the $s1$ fermions have
no direct effects on the spin spectrum. Within the operator description
used in this paper the study of the $x=0$ spin spectrum refers to an effectively non-interacting
problem. In turn, in terms of electrons it is an involved many-body problem.
The studies of Subsection VI-B confirm the validity of our description 
for the model on the square lattice: At $x=0$ our analytical expressions
for the spin-wave dispersion, which corresponds to the  
coherent part of the $x=0$ spin spectrum, fully agree with the 
controlled numerical results of 
Ref. \cite{LCO-Hubbard-NuMi}, obtained by summing up an infinite set of electronic ladder 
diagrams. Moreover, an excellent 
quantitative agreement with the inelastic neutron scattering of
the La$_{2-x}$Sr$_x$CuO$_4$ (LSCO) Mott-Hubbard insulator parent 
compound La$_2$CuO$_4$ (LCO) \cite{LCO-neutr-scatt} is reached. 

The paper is organized as follows. The model, its $c$ and $s1$ fermion description
and related rotated-electron representation, the limitations and advantages 
of such a description, and the emergence of the 
$s1$ fermions from the hard-core spin-neutral two-spinon $s1$ 
bond particles are the subjects of Section II.  
In Section III further evidence is provided that for the
Hubbard model on the square lattice in the one- and two-electron
subspace the $c$ and $s1$ band discrete momentum values are good quantum numbers, 
consistently with the results of Ref. \cite{companion} 
on that issue. In addition, the Fermi line is expressed in terms momenta
belonging to the $c$ Fermi line and $s1$ boundary line and its anisotropy is investigated.
The derivation of the $c$ and $s1$ fermion
energy dispersions of the square-lattice quantum liquid first-order
energy functional associated with the ground-state normal-ordered Hamiltonian
of the Hubbard model in the one- and two-electron
subspace is the subject of Section IV. 
The $c$ and $s1$ fermion velocities and analysis of their
relation to the velocities associated with the one- and two-electron 
excitations is the subject of Section V. Section VI presents a brief discussion
concerning the combination of the $c$ and $s1$ fermion description 
with the exact Bethe-ansatz solution to study the
dynamical and correlation functions of the
1D Hubbard model. Moreover, 
in that section we study the spin excitations of the 
half-filling Hubbard model on the square lattice and
find excellent agreement between the
results obtained by use of the square-lattice quantum
liquid of $c$ and $s1$ fermions and the standard formalism of many-electron 
physics \cite{LCO-Hubbard-NuMi}. Finally, the concluding 
remarks are presented in Section VII.

\section{The model and the $c$ and $s1$ fermion description}

Here we address the problem of the description of the Hubbard model 
on a square lattice in the one- and two-electron subspace in terms of charge 
$c$ fermions and spin-singlet two-spinon $s1$ fermions introduced in Ref. \cite{companion}. 
In that subspace the model refers to the square-lattice quantum liquid, 
which is expected to refer to a wider class of 
many-electron problems with short-range interactions
on the square lattice belonging to the same universality class. 
We start by introducing the model, its global 
symmetry, and the rotated-electron representation and discuss the
limitations and advantages of the corresponding $c$ and $s1$ fermion description.

\subsection{The Hubbard model and its global symmetry, rotated electrons, and the
limitations of our description}

The Hubbard model on the two-dimensional (2D) square lattice with 
torus periodic boundary conditions and the same model on the 1D
lattice with periodic boundary conditions, spacing $a$, 
$N_a^D\equiv [N_a]^D$ sites where $D=1$ and $D=2$ 
for the 1D and square lattices, respectively, $N_a^D\gg1$ even 
and large, and lattice edge length $L=N_a\,a$
for 2D and chain length $L=N_a\,a$ for 1D is given by,
\begin{equation}
\hat{H} = -t\sum_{\langle\vec{r}_j\vec{r}_{j'}\rangle}\sum_{\sigma =\uparrow
,\downarrow}[c_{\vec{r}_j,\sigma}^{\dag}\,c_{\vec{r}_{j'},\sigma}+h.c.] + 
U\,[N_a^D-\hat{Q}]/2 \, .
\label{H}
\end{equation}
Here ${\langle\vec{r}_j\vec{r}_{j'}\rangle}$ refers to nearest neighboring sites, the operator
$c_{\vec{r}_j,\sigma}^{\dag}$ creates an electron of spin projection $\sigma$ at the
site of real-space coordinate $\vec{r}_j$, and the operator, 
\begin{equation}
{\hat{Q}} = \sum_{j=1}^{N_a^D}\sum_{\sigma =\uparrow
,\downarrow}\,n_{\vec{r}_j,\sigma}\,(1- n_{\vec{r}_j,-\sigma}) \, ,
\label{Q-op}
\end{equation}
where $n_{{\vec{r}}_j,\sigma} = c_{\vec{r}_j,\sigma}^{\dag} c_{\vec{r}_j,\sigma}$
and $-\sigma=\uparrow$ (and $-\sigma=\downarrow$)
for $\sigma =\downarrow$ (and $\sigma =\uparrow$)
counts the number of electron singly occupied sites.
Hence the operator ${\hat{D}}=[{\hat{N}}-{\hat{Q}}]/2$
counts the number of electron doubly
occupied sites where ${\hat{N}} = \sum_{\sigma}
{\hat{N}}_{\sigma}$ and ${\hat{N}}_{\sigma}=\sum_{j=1}^{N_a^D}
n_{{\vec{r}}_j,\sigma}$. We 
denote the $\eta$-spin (and spin) value of the energy 
eigenstates by $S_{\eta}$ (and $S_s$) and the corresponding 
projection by $S^z_{\eta}= -[N_a^D-N]/2$ (and $S^z_s= -[N_{\uparrow}-
N_{\downarrow}]/2$). We focus our attention onto initial ground states with 
a hole concentration $x= [N_a^D-N]/N_a^D\geq 0$ and spin
density $m= [N_{\uparrow}-N_{\downarrow}]/N_a^D=0$ and
their excited states belonging to the one- and two-electron
subspace defined in Ref. \cite{companion}.

The unitary operator $\hat{V}=\hat{V}(U/4t)$ associated with
the electron - rotated-electron unitary transformation plays
a key role in the construction of the general description
of the Hubbard model on a square lattice introduced in 
Ref. \cite{companion}. Out of the manyfold of unitary 
operators of Refs. \cite{Stein,bipartite},
the studies of that paper consider a unique choice
for ${\hat{V}}={\hat{V}}(U/4t)$
such that the states $\vert \Psi_{U/4t}\rangle =
{\hat{V}}^{\dag}\vert\Psi_{\infty}\rangle$ are
energy and momentum eigenstates for $U/4t>0$. It corresponds
to a suitable chosen set $\{\vert\Psi_{\infty}\rangle\}$
of $U/4t\rightarrow\infty$ energy eigenstates.
The states $\vert \Psi_{U/4t}\rangle ={\hat{V}}^{\dag}\vert\Psi_{\infty}\rangle$
(one for each value of $U/4t>0$) that are 
generated from the same initial state $\vert\Psi_{\infty}\rangle$ 
belong to the same {\it $V$ tower}. 
The generator ${\tilde{S}}_c$ of the hidden $U(1)$ symmetry 
of the global $SO(3)\times SO(3)\times U(1)=[SO(4)\times U(1)]/Z_2$ symmetry found
for the $U/4t>0$ Hubbard model on a square lattice in Ref. \cite{bipartite} 
reads ${\tilde{S}}_c= {\hat{V}}^{\dag}\,{\hat{S}}_c\,{\hat{V}}$
where ${\hat{S}}_c= {\hat{Q}}/2$ and the operator
${\hat{Q}}$ is given in Eq. (\ref{Q-op}). Its eigenvalue $S_c$
equals one-half the number of rotated-electron singly occupied
sites $2S_c$. The unitary transformation associated with the operator 
${\hat{V}}^{\dag}$ maps the electron operators
$c_{\vec{r}_j,\sigma}^{\dag}$ and $c_{\vec{r}_j,\sigma}$
onto rotated-electron creation and annihilation operators 
${\tilde{c}}_{\vec{r}_j,\sigma}^{\dag} =
{\hat{V}}^{\dag}\,c_{\vec{r}_j,\sigma}^{\dag}\,{\hat{V}}$ and 
${\tilde{c}}_{\vec{r}_j,\sigma} =
{\hat{V}}^{\dag}\,c_{\vec{r}_j,\sigma}\,{\hat{V}}$, respectively.
In terms of the latter operators the expression of the
generator whose application onto the $N=0$ and $S_c=0$ electron vacuum
generates the energy eigenstates $\vert \Psi_{U/4t}\rangle$
belonging to the same $V$ tower are the same 
for the whole range of $U/4t>0$ values:
It has the same expression as the
generator of the initial state $\vert\Psi_{\infty}\rangle$
in terms of electron creation and annihilation operators. 
(The $N=0$ and $S_c=0$ electron vacuum corresponds to that of Eq. (\ref{23}) of Appendix A
for $N_{a_{\eta}}^D=N_a^D$ and $N_{a_{s}}^D=2S_c=0$.)

However, for $U/4t>0$ (and $U/4t\rightarrow\infty$) the expression of the generators
of the energy eigenstates from the electron vacuum are very complex in terms of rotated-electron (and electron) 
operators. Indeed, those are not the ultimate objects whose occupancy configurations
that generate such states have a simple expression. The studies of Ref. \cite{companion}
considered a complete set $\{\vert \Phi_{U/4t}\rangle\}$ of general momentum
eigenstates $\vert \Phi_{U/4t}\rangle ={\hat{V}}^{\dag}\vert\Phi_{\infty}\rangle$
given in Eq. (\ref{non-LWS}) of Appendix A. On the right-hand side of that equation the 
lowest-weight-state (LWS) momentum eigenstates 
$\vert \Phi_{LWS;U/4t}\rangle$ are those given in Eq. (\ref{LWS-full-el}) of such an Appendix.
These states can be generated from corresponding $U/4t\rightarrow\infty$
momentum eigenstates as $\vert \Phi_{U/4t}\rangle={\hat{V}}^{\dag}\vert \Phi_{\infty}\rangle$.
Moreover, $\vert \Phi_{LWS;U/4t}\rangle={\hat{V}}^{\dag}\vert \Phi_{LWS;\infty}\rangle$
where ${\hat{V}}^{\dag}$ is the electron - rotated-electron unitary operator
that also generates the energy and momentum eigenstates
$\vert \Psi_{U/4t}\rangle={\hat{V}}^{\dag}\vert \Psi_{\infty}\rangle$ from
the corresponding $U/4t\rightarrow\infty$ energy and momentum eigenstates $\vert \Psi_{\infty}\rangle$.

The generators onto the vacuum of Eq. (\ref{23}) of Appendix A
of the LWS momentum eigenstates $\vert \Phi_{LWS;U/4t}\rangle$
are Slatter-determinant products of $c$ and $\alpha\nu$ fermion creation
operators. For the 1D Hubbard model one has that $\vert \Phi_{U/4t}\rangle =\vert \Psi_{U/4t}\rangle$ 
so that the states $\vert \Phi_{U/4t}\rangle$ of Eq. (\ref{non-LWS}) of Appendix A are both momentum and energy eigenstates whereas
for the model on the square lattice the energy and momentum
eigenstates $\vert \Psi_{U/4t}\rangle ={\hat{V}}^{\dag}\vert\Psi_{\infty}\rangle$ are
a superposition of a well-defined set of momentum eigenstates  
$\vert \Phi_{U/4t}\rangle ={\hat{V}}^{\dag}\vert\Phi_{\infty}\rangle$ with the
same momentum eigenvalue, values of
$S_{\eta}$, $S_{\eta}^z$, $S_s$, $S_s^z$, $S_c$, $C_{\eta}=\sum_{\nu}\nu\,N_{\eta,\nu}$,
$C_{s}=\sum_{\nu}\nu\,N_{s,\nu}$, and $c$ fermion momentum distribution function 
$N_c ({\vec{q}})$ \cite{companion}. In some cases that set of states refers
to a single state so that $\vert \Psi_{U/4t}\rangle =\vert \Phi_{U/4t}\rangle$ and
the state $\vert \Phi_{U/4t}\rangle$ under consideration is both an energy
and momentum eigenstate. This is so for the momentum eigenstates $\vert \Phi_{U/4t}\rangle$ that span
the one- and two-electron subspace as defined in Ref. \cite{companion}. 
That $\vert \Phi_{U/4t}\rangle =\vert \Psi_{U/4t}\rangle$ for the 1D Hubbard model in the
whole Hilbert space follows from its integrability being for $N_a\gg 1$ associated with an infinite number of
conservation laws \cite{Martins}. According to the results of Ref. \cite{Prosen} such laws are equivalent to the independent 
conservation of the set of numbers $\{N_{\alpha\nu}\}$ of $\alpha\nu$ fermions, which for that model are good quantum
numbers. 

The $\alpha\nu$ fermion operators
$f_{\vec{q}_j,\alpha\nu}^{\dag}$ appearing in the Slatter-determinant products 
of the LWS momentum eigenstates $\vert \Phi_{LWS;U/4t}\rangle$ given in Eq. (\ref{LWS-full-el}) of Appendix A  
correspond to an independent problem in each subspace with constant values of $S_c$, $S_s$, and
number $N_{a_{\alpha\nu}}^D = [N_{\alpha\nu} + N^h_{\alpha\nu}]$ of both sites of the $\alpha\nu$ 
effective lattice and discrete momentum values of the $\alpha\nu$ band.
The operators $f_{\vec{q}_j,\alpha\nu}^{\dag}$ and $f_{\vec{q}_j,\alpha\nu}$ act on
subspaces spanned by mutually neutral states, which are transformed into each other
by $\alpha\nu$ fermion particle-hole processes. As discussed in Ref. \cite{companion}, that assures that for the 
model on the square lattice the components $q_{x_1}$ and $q_{x_2}$ of the microscopic momenta 
$\vec{q}=[q_{x_1},q_{x_2}]$ refer to commuting $s1$ translation generators $\hat{q}_{s1\,x_1}$ and $\hat{q}_{s1\,x_2}$.
In turn, for the operators $f_{\vec{q}_j,c}^{\dag}$ appearing in the Slatter-determinant products 
of the LWS momentum eigenstates there are in 1D two types of quantum
problems depending on the even or odd character of the number $[B_{\eta}+B_s]=\sum_{\alpha\nu}N_{\alpha\nu}$.
Indeed, in spite of the $c$ effective lattice being identical to the original lattice, 
both for the model on the 1D and square lattices the $c$ fermions feel the Jordan-Wigner 
phases of the $\alpha\nu$ fermions created or annihilated under subspace transitions that
do not conserve the number $[B_{\eta}+B_s]=\sum_{\alpha\nu}N_{\alpha\nu}$.

The $\alpha\nu$ fermion operators $f^{\dag}_{{\vec{q}},\alpha\nu}$ 
and $f_{{\vec{q}},\alpha\nu}$ act onto subspaces spanned by neutral states.
Creation of one $\alpha\nu$ fermion is however a well-defined process whose generator
is the product of an operator that fulfills small changes in the $\alpha\nu$ effective lattice
and corresponding $\alpha\nu$ momentum band and the operator 
$f^{\dag}_{{\vec{q}},\alpha\nu}$ suitable to the excited-state subspace.
In the  Slatter-determinant products of Eq. (\ref{LWS-full-el}) of Appendix A it is implicitly
assumed that the $\alpha\nu$ momentum bands are those of the state under consideration
so that the corresponding generators on the vacua are simple products of
$f^{\dag}_{{\vec{q}},\alpha\nu}$ operators.
The $c$ effective lattice and number $N_a^D$ of $c$ band discrete momentum
values remain unaltered for the whole Hilbert space. In turn,
concerning transitions between subspaces with slightly different discrete values for such momenta
again creation or annihilation of one $c$ fermion is a well-defined process whose generator
is the product of an operator that fulfills the corresponding small changes in the $c$ momentum band and the operator 
$f^{\dag}_{{\vec{q}},c}$ or $f_{{\vec{q}},c}$, respectively, appropriate to the excited-state subspace.

Following the results of Ref. \cite{companion},
the bad news is that for the Hubbard model on the square lattice the microscopic momenta 
carried by the $\alpha\nu$ fermions are not in general good quantum numbers, in contrast to the
integrable 1D Hubbard model. That results from the lack of integrability of the Hubbard 
model on the square lattice, which is behind the set of numbers $\{N_{\alpha\nu}\}$ of $\alpha\nu$ fermions
not being in general conserved, yet the numbers $C_{\eta}=\sum_{\nu}\nu\,N_{\eta,\nu}$
and $C_{s}=\sum_{\nu}\nu\,N_{s,\nu}$ and the $c$ fermion momentum distribution function 
$N_c ({\vec{q}})$ are. It follows that for the model on the square
lattice the generators onto the vacuum 
of Eq. (\ref{23}) of Appendix A of the energy and momentum eigenstates
are not simple Slatter-determinant products of $c$ and $\alpha\nu$ fermion creation operators 
as those of Eq. (\ref{LWS-full-el}) of Appendix A 
and such states are in general different from the momentum eigenstates 
$\vert \Phi_{U/4t}\rangle ={\hat{V}}^{\dag}\vert\Phi_{\infty}\rangle$. 
The good news is that for the model on the square lattice
in the one- and two-electron subspace defined in Ref. \cite{companion} one has that
$N_{\alpha\nu}=0$ except for the $\alpha\nu =s1$ and $\alpha\nu =s2$ fermion
branches and owing to symmetries specific to that subspace $N_{s1}=[S_c-S_s-2N_{s2}]$
and $N_{s2}=0,1$ are conserved so that the microscopic momenta of the $s1$ 
fermions are good quantum numbers and the
states $\vert \Phi_{U/4t}\rangle = \vert \Psi_{U/4t}\rangle$ are energy eigenstates.

Addition of chemical-potential and magnetic-field operator 
terms to the Hamiltonian (\ref{H}) lowers its symmetry. 
A property of key importance follows from such 
operator terms commuting with that Hamiltonian.
Such a property reveals that the $c$ fermion and $\alpha\nu$ fermion occupancy 
configurations and independent $\eta$-spinon and spinon occupancies 
that generate the $4^{N_a^D}$ momentum eigenstates 
$\vert \Phi_{U/4t}\rangle ={\hat{V}}^{\dag}\vert\Phi_{\infty}\rangle$ 
of general form given in Eq. (\ref{non-LWS}) of Appendix A correspond
to state representations of the $SO(3)\times SO(3)\times U(1)$ group for 
all values of the densities $n=(1-x)$ and $m$, as confirmed explicitly in
Ref. \cite{companion}.

The unitary operator ${\hat{V}}$ commutes with the momentum 
operator $\hat{\vec{P}}$, three generators
of the spin $SU(2)$ symmetry, and three generators
of the $\eta$-spin $SU(2)$ symmetry \cite{bipartite,companion}. These two $SU(2)$
symmetries are contained in the model global
$SO(3)\times SO(3)\times U(1)=[SU(2)\times SU(2)\times U(1)]/Z_2^2$
symmetry. Hence, the momentum operator and such six generators 
have the same expression in terms of electron and 
rotated-electron creation and annihilation operators. 
In contrast, the generator ${\tilde{S}}_c$ of the $U(1)$ symmetry 
also contained in that global symmetry and the Hamiltonian (\ref{H}) do
not commute with such a unitary operator.

The general rotated-electron description introduced in
Ref. \cite{companion} associates with any operator ${\hat{O}}$ a rotated
operator ${\tilde{O}}={\hat{V}}^{\dag}\,{\hat{O}}\,{\hat{V}}$, which
has the same expression in terms of rotated-electron creation and 
annihilation operators as ${\hat{O}}$ in terms of electron creation and 
annihilation operators, respectively. Any operator ${\hat{O}}$ expressed
in terms of electron creation and annihilation operators can 
then be written in terms of rotated-electron creation and annihilation
operators as,
\begin{eqnarray}
{\hat{O}} & = & {\hat{V}}\,{\tilde{O}}\,{\hat{V}}^{\dag}
= {\tilde{O}}+ [{\tilde{O}},\,{\hat{S}}\,] + {1\over
2}\,[[{\tilde{O}},\,{\hat{S}}\,],\,{\hat{S}}\,] + ... 
=  {\tilde{O}}+ [{\tilde{O}},\,{\tilde{S}}\,] + {1\over
2}\,[[{\tilde{O}},\,{\tilde{S}}\,],\,{\tilde{S}}\,] + ... \, ,
\nonumber \\
{\hat{S}} & = & -{t\over U}\,\left[\hat{T}_{+1} -\hat{T}_{-1}\right] 
+ {\cal{O}} (t^2/U^2) \, ; \hspace{0.25cm}
{\tilde{S}} = -{t\over U}\,\left[\tilde{T}_{+1} -\tilde{T}_{-1}\right] 
+ {\cal{O}} (t^2/U^2) \, .
\label{OOr}
\end{eqnarray}
The operator $\hat{S}$ appearing in this equation 
is such that ${\hat{V}}^{\dag} = e^{{\hat{S}}}$, 
${\hat{V}} = e^{-{\hat{S}}}$, and then
$[\hat{S},{\hat{V}}]=0$ and $\hat{S}=\tilde{S}$. 
The expression of ${\hat{S}}$ involves only the 
kinetic operators $\hat{T}_0$, $\hat{T}_{+1}$, and $\hat{T}_{-1}$ 
so that that of ${\tilde{S}}$ involves only the rotated
kinetic operators $\tilde{T}_0$, $\tilde{T}_{+1}$, and $\tilde{T}_{-1}$.
The former kinetic operators are related to the kinetic-energy operator $t\,\hat{T}=
-t\sum_{\langle\vec{r}_j\vec{r}_{j'}\rangle}\sum_{\sigma =\uparrow
,\downarrow}[c_{\vec{r}_j,\sigma}^{\dag}\,c_{\vec{r}_{j'},\sigma}+h.c.]$
of the Hamiltonian (\ref{H}), which can be expressed as
$\hat{T}= \hat{T}_0 + \hat{T}_{+1} + \hat{T}_{-1}$. (The expressions of the three kinetic operators
are provided in Ref. \cite{companion}.) The operator $\hat{T}_0$ does not change electron double 
occupancy. In turn, the operators $\hat{T}_{+1}$ and $\hat{T}_{-1}$ do it by $+1$ 
and $-1$, respectively. For $U/4t>0$ the operator $\hat{S}$ can be expanded in 
a series of $t/U$ and the corresponding first-order
term has the universal form given in Eq. (\ref{OOr}). To arrive
to the expression in terms of the operator
${\tilde{S}}$ also given in that equation, the above property 
that $\hat{S}=\tilde{S}$ so that both the operators
${\hat{V}}$ and ${\hat{S}}$ have the same expression in terms of
electron and rotated-electron creation and annihilation operators is used. 
This is behind the expansion 
${\tilde{S}} = -(t/U)\,[\tilde{T}_{+1} -\tilde{T}_{-1}] + {\cal{O}} (t^2/U^2)$
given in that equation for the operator ${\tilde{S}}$
whose higher-order terms involve only products of the rotated
kinetic operators $\tilde{T}_0$, $\tilde{T}_{+1}$, and $\tilde{T}_{-1}$. 

Since the Hamiltonian $\hat{H}$ of Eq. (\ref{H}) does not commute with 
the unitary operator ${\hat{V}} = e^{-{\hat{S}}}$, when expressed in terms 
of rotated-electron creation and annihilation operators it has an infinite 
number of terms. According to Eq. (\ref{OOr}) it reads, 
\begin{equation}
{\hat{H}} = {\hat{V}}\,{\tilde{H}}\,{\hat{V}}^{\dag}
= {\tilde{H}} + [{\tilde{H}},\,{\tilde{S}}\,] + {1\over
2}\,[[{\tilde{H}},\,{\tilde{S}}\,],\,{\tilde{S}}\,] + ... \, .
\label{HHr}
\end{equation}
The commutator $[{\tilde{H}},\,{\tilde{S}}\,]$ does not vanish
except for $U/4t\rightarrow\infty$ so that for finite values
of $U/4t$ one has that ${\hat{H}} \neq {\tilde{H}}$. 
For $U/4t\gg 1$ the Hamiltonian of Eq. (\ref{HHr}) 
expressed in terms of rotated-electron creation and annihilation
operators corresponds to a simple rotated-electron $t-J$ model. The higher-order 
$t/U$ terms, which become increasingly important upon decreasing $U/4t$,
generate effective rotated-electron hopping between second, 
third, and more distant neighboring sites. 
In spite of the operators 
$\tilde{T}_0$, $\tilde{T}_{+1}$, and $\tilde{T}_{-1}$ generating 
only rotated-electron hoping between nearest-neighboring sites,
their products contained in the higher-order terms of 
${\tilde{S}} = -(t/U)\,[\tilde{T}_{+1} -\tilde{T}_{-1}] + {\cal{O}} (t^2/U^2)$
generate effective hoping between for instance second and
third neighboring sites whose real-space distance in units of
the lattice spacing $a$ is for the model on the square lattice $\sqrt{2}\,a$ and $2\,a$, 
usually associated with transfer integrals $t'$ and $t''$, respectively \cite{Tiago}. 

In spite that when expressed in terms of rotated-electron operators the
Hamiltonian has an infinite number of terms, as given in Eq. (\ref{HHr}),
for intermediate and large $U/4t$ values obeying for the model on the square lattice approximately 
the inequality $U/4t\geq u_0 \approx 1.3$, besides the original nearest-neighboring hoping processes  
only those involving second and third neighboring sites are
relevant for the square-lattice quantum liquid 
described by the Hamiltonian of Eqs. (\ref{H}) and (\ref{HHr}) in the
one- and two-electron subspace \cite{companion}.
Hence for such a range of $U/4t$ values, only the first few Hamiltonian terms 
on the right-hand side of Eq. (\ref{HHr}) play
an active role in the physics of the Hubbard model on the square lattice
in that subspace. The results of Ref. \cite{companion} reveal that for
such a model the usual large-$U$ physics corresponding to energy 
contributions of the order of $t^2/U$ is only dominant for $U/4t\gg 19$, so that 
the higher-order $t/U$ terms become important for quite large $U/4t$ values.

It follows from the above analysis that for intermediate and large values of $U/4t$ 
such a square-lattice quantum liquid can be mapped onto an effective $t-J$ model on
a square lattice with $t$, $t'=t'(U/4t)$, and $t''=t''(U/4t)$ transfer integrals
where the role of the processes associated with 
$t'=t'(U/4t)$ and $t''=t''(U/4t)$ becomes increasingly important
upon decreasing the $U/4t$ value. For approximately $U/4t\geq u_0$
the latter model is equivalent to the Hubbard model on the square lattice
(\ref{HHr}) in the subspace under consideration and  
expressed in terms of rotated-electron creation and annihilation operators. 
Indeed, the $t-J$ model constraint against double occupancy is in that subspace
equivalent to expressing the Hubbard model in terms of rotated-electron 
creation and annihilation operators.

As discussed in Ref. \cite{companion}, the general operator description introduced in 
that reference for the Hubbard model on the square lattice and used in the studies 
of this paper has two main limitations: 
\vspace{0.25cm}

i) For small and intermediate values of $U/4t$ the explicit form of the unitary operator $\hat{V}$ 
associated with the rotated-electron operators as defined in Ref. \cite{companion}
remains an open problem. It is known that such a unitary operator has for $U/4t>0$ the 
above-mentioned general
form ${\hat{V}} = e^{-{\hat{S}}}$ where the expression of $\hat{S}$ involves only the three kinetic 
operators $\hat{T}_0$, $\hat{T}_{+1}$, and $\hat{T}_{-1}$. The finding 
of the explicit form of the $U/4t$-dependent functional $\hat{S}$ in terms of the latter three operators,
valid for the whole range of finite $U/4t$ values, is though a very 
involved quantum problem beyond the reach of our approach current status. 
Indeed, the operator description used in the studies of this paper has been constructed to inherently the solution 
of that problem being equivalent to the solution of the Hubbard model on the square lattice. 
\vspace{0.25cm}

ii) It turns out that the quantum problem under consideration is non-perturbative
in terms of electron operators so that, in contrast to a 3D isotropic
Fermi liquid \cite{Landau,Pines}, rewriting the square-lattice quantum liquid theory emerging
from the general description used in the studies of this paper for the
model in the one- and two-electron subspace in terms of
the standard formalism of many-electron physics is an extremely
complex problem. Fortunately, such a quantum liquid simplifies 
when expressed in terms of the $c$ and $s1$ fermion operators. 
The point is that their momentum values are good quantum 
numbers so that the interactions of these objects are residual.
\vspace{0.25cm}

A third limitation beyond those discussed in Ref. \cite{companion} is related to the results 
obtained in this paper:
\vspace{0.25cm}

iii) The phase factor arising from the extended Jordan-Wigner
transformation considered below corresponds to that created by a 
gauge field whose effective vector potential generates long-range 
interactions between the $s1$ fermions emerging from the $s1$ bond particles.
Alike previous theories by many authors involving a gauge theory formulation 
of the Hubbard or t-J model, our method cannot give a fully controllable approximation 
concerning the effects of the interactions brought about by the effective vector 
potential associated with the gauge field. 
\vspace{0.25cm}

We start by emphasizing that upon expressing the problem in terms of $c$ and
$s1$ fermions the higher-order contributions associated with the effective
$t'=t'(U/4t)$ and $t''=t''(U/4t)$ transfer integrals are incorporated
in the $U/4t$ dependence of the $c$ and $s1$ fermion parameters studied
in this paper and in Ref. \cite{cuprates0}.
Furthermore, concerning limitation (ii), the microscopic processes corresponding to the effective
$t'=t'(U/4t)$ and $t''=t''(U/4t)$ transfer integrals are important to characterize
the type of order associated with the phases of the square-lattice quantum liquid. Fortunately,
it is found in Ref. \cite{companion} that for general one- and two-electron 
operators ${\hat{O}}$ other than the Hamiltonian
the leading operator term ${\tilde{O}}$ on the right-hand side of Eq. (\ref{OOr}) 
generates nearly the whole spectral weight. Hence in spite of the limitations (i) and (ii),
such a description provides useful information 
about the physics contained in the model on the square lattice.
Indeed, there are several reasons why, in spite of both the explicit form
of the unitary operator $\hat{V}$ being known only for
large values of $U/4t$ and the difficulties in rewriting 
the theory emerging from the description introduced here
in terms of the standard formalism of many-electron physics, that 
description is rather useful to extracting valuable information on
the quantum problem for values of $U/4t$ approximately in the
range $U/4t\geq u_0$. 

Indeed, some of the effects of 
the unitary operator $\hat{V}$ are associated with the operator terms 
of the general expression (\ref{OOr}) containing
commutators involving the related operator $\hat{S}={\tilde{S}}$,
which for one- and two-electron operators ${\hat{O}}$ 
are found in Ref. \cite{companion} to generate very little spectral weight.
Hence one can reach a quite faithful representation for
one- and two-electron operators ${\hat{O}}$ by expressing
the corresponding operator ${\tilde{O}}$ in terms of the operators
of the objects used in the studies of this paper whose occupancy configurations
generate the energy eigenstates that span the one- and two-electron
subspace. 

Also concerning the limitation (ii), in Subsection VI-B we consider one of the
few physical limits where there are reliable and controlled studies
of the Hubbard model on the square lattice by means of
the standard formalism of many-electron physics \cite{LCO-Hubbard-NuMi}: 
The spin excitations at half filling. The results of that section 
reveal an excellent quantitative agreement between the results obtained within
the square-lattice quantum liquid and those of Ref.
\cite{LCO-Hubbard-NuMi}, which profit from the usual
many-electron machinery and are reached by summing up an infinite set of ladder 
diagrams. 

Whether our description
provides the correct $x=0$ and $m=0$ spin spectrum is
a valuable checking of its validity and of the residual character
of the $c$ - $s1$ fermion interactions.
Indeed, within such a description the derivation of the spin spectrum is in that
physical limit equivalent to a non-interacting problem. That results from
both the $c$ and $s1$ bands being full for the initial
$x=0$ and $m=0$ ground state. As found in Ref. \cite{companion}, 
the spin excitations generate the emergence of two holes in
the $s1$ momentum band but phase-space, exclusion-principle, and energy 
and momentum conservation restrictions prevent inelastic collisions
between the two emerging $s1$ fermion holes. In contrast, in
terms of the original electrons it is a complex many-electron
problem whose solution requires summing up an infinite number
of diagrams \cite{LCO-Hubbard-NuMi}. The excellent agreement
reached in Section VI between the spin spectrum as expressed
in terms of the $s1$ fermion dispersions and as obtained
by the usual many-electron machinery then confirms
the validity of the $c$ and
$s1$ fermion description associated with the square-lattice quantum liquid.

Finally, concerning the limitation (iii), in our case
the momentum band of the $s1$ fermions is full for the
$x\geq 0$ and $m=0$ ground states and has none, one, or two holes
for the excited states that span the one- and two-electron subspace.
The extended Jordan-Wigner transformation $f^{\dag}_{{\vec{r}}_{j},s1} = e^{i\phi_{j,s1}}\,
g^{\dag}_{{\vec{r}}_{j},s1}$ that maps the $s1$ bound-particle operator
onto the $s1$ fermion operator involves the 
operator phase factor $\phi_{j,s1} = \sum_{j'\neq j}f^{\dag}_{{\vec{r}}_{j'},s1}
f_{{\vec{r}}_{j'},s1}\,\phi_{j',j,s1}$, which corresponds to that created by a 
gauge field. Fortunately, in our case the operator phase factor $\phi_{j,s1}$ can 
for the study of many properties be replaced by the
trivial phase factor $\phi^0_{j,s1} = \sum_{j'\neq j} \phi_{j',j,s1}$ \cite{cuprates0}.
That results from replacing within the present $N_a^2\rightarrow\infty$ limit the
operators $f^{\dag}_{{\vec{r}}_{j'},s1}f_{{\vec{r}}_{j'},s1}$
by their average $\langle f^{\dag}_{{\vec{r}}_{j'},s1}f_{{\vec{r}}_{j'},s1}\rangle\approx 1$.
Indeed, for the $s1$ fermion occupancies of the energy eigenstates that span the 
one- and two-electron subspace there are none, one, or two
unoccupied sites and the total number of sites of the $s1$ effective lattice is
$N_{a_{s1}}^2\approx N/2=(1-x)N_a^2/2$. 
Moreover, the spin degrees of freedom of such states 
are generated by occupancy configurations in the $s1$ momentum band
with none, one, or two holes whereas the total number of discrete momentum
values in that band is also $N_{a_{s1}}^2\approx N/2=(1-x)N_a^2/2$.
Since the average value $\langle f^{\dag}_{{\vec{r}}_{j'},s1}f_{{\vec{r}}_{j'},s1}\rangle$
can be expressed as a superposition of expectation values
$\langle f^{\dag}_{{\vec{q}}_{j'},s1}f_{{\vec{q}}_{j'},s1}\rangle$,
which involves summations over all momenta ${\vec{q}}_{j'}$,
one then finds $\langle f^{\dag}_{{\vec{r}}_{j'},s1}f_{{\vec{r}}_{j'},s1}\rangle\approx 1$ 
within the present thermodynamic limit. As discussed below in Subsection II-C,
replacing $f^{\dag}_{{\vec{r}}_{j'},s1}f_{{\vec{r}}_{j'},s1}$
by $\langle f^{\dag}_{{\vec{r}}_{j'},s1}f_{{\vec{r}}_{j'},s1}\rangle\approx 1$ renders
the form of the effective vector potential associated with the extended Jordan-Wigner 
transformation $f^{\dag}_{{\vec{r}}_{j},s1} = e^{i\phi_{j,s1}}\,g^{\dag}_{{\vec{r}}_{j},s1}$
that of a Chern-Simons vector potential \cite{Giu-Vigna}.

Moreover, phase-space, exclusion-principle, and energy 
and momentum conservation restrictions dramatically limit
the effects of the long-range $s1$ - $s1$ fermion interactions generated
by the effective vector potential associated with 
the above gauge field. Indeed, these restrictions follow from within the description
used here the $s1$ band being full for the initial ground state and displaying 
one and two holes for the one-electron and spin excitations, respectively. This
prevents such long-range interactions
leading to inelastic $s1$ fermion - $s1$ fermion scattering.
Hence the direct effects of such interactions are somehow frozen. 
In turn, an indirect side effect of the latter interactions is
the occurrence of residual interactions between the $s1$ fermions and $c$ 
fermions. Fortunately, the studies of Refs. \cite{cuprates0,cuprates} reveal that
the description of such interactions is within the range of the
theory and that they play an important role in the scattering properties
of the square-lattice quantum liquid perturbed by small 3D 
anisotropy effects. In turn, the direct $c$ - $c$ fermion interactions either
vanish or are very weak.

For $U/4t>0$, the investigations of Ref. \cite{companion} accessed 
the transformation laws and/or 
invariance of several operators and quantum objects under the
electron - rotated-electron unitary transformation associated with the operator
$\hat{V}$, what provides valuable information about the physics of the Hubbard model
on the square lattice. The suitable use of such properties and the
combination of the general description introduced in that reference
with different methods, to extract additional information on the
quantum problem, allows the further study in this paper and 
Ref. \cite{cuprates0} of the square-lattice quantum liquid and its relation
to and usefulness for the physics of real materials. 

That quantum liquid refers to the Hubbard model
on the square lattice in the one- and two-electron subspace 
as defined in Ref. \cite{companion}. For the study of some properties one
may consider the smaller subspace spanned by the LWSs of both the $\eta$-spin and
spin algebras, whose values
of $S_{\alpha}$ and $S^z_{\alpha}$ are such that 
$S_{\alpha}=-S^z_{\alpha}$ for $\alpha =\eta,s$. (The
LWS one- and two-electron subspace is a subspace
of the {\it LWS-subspace}.) In reference \cite{companion} it is shown that the whole 
physics of the model (\ref{H}) can be extracted from it in 
the large LWS-subspace of the full Hilbert space. In Appendix A some basic information
on the one- and two-electron subspace needed for the 
studies of this paper is provided.
The quantum liquid of $c$ and $s1$ fermions 
studied in the following is expected to play the same role for
many-electron problems with short-range interactions on a square
lattice as a Fermi liquid for 3D isotropic metals \cite{Landau,Pines,Ander07}. 

\subsection{The algebra of the $c$ fermion operators and 
hard-core spin-neutral two-spinon $s1$ bond-particle operators}

Within the LWS representation, the 
$c$ fermion creation operator can be 
expressed in terms of the rotated-electron operators as 
follows \cite{companion},
\begin{equation}
f_{\vec{r}_j,c}^{\dag} =
{\tilde{c}}_{\vec{r}_j,\uparrow}^{\dag}\,
(1-{\tilde{n}}_{\vec{r}_j,\downarrow})
+ e^{i\vec{\pi}\cdot\vec{r}_j}\,{\tilde{c}}_{\vec{r}_j,\uparrow}\,
{\tilde{n}}_{\vec{r}_j,\downarrow} \, ;
\hspace{0.5cm}
{\tilde{n}}_{\vec{r}_j,\sigma}
= {\tilde{c}}_{\vec{r}_j,\sigma}^{\dag}{\tilde{c}}_{\vec{r}_j,\sigma} \, ,
\label{fc+}
\end{equation}
where here and throughout the remaining of this paper
${\vec{\pi}}$ denotes the momentum of components
$[\pi,\pi]$ and $\pi$ for the model on the square and 1D lattices, 
respectively, and $e^{i\vec{\pi}\cdot\vec{r}_j}$ is $\pm 1$ depending on which
sublattice site $\vec{r}_j$ is on. As mentioned in Subsection II-A, the 
rotated-electron operators are related to the original electron operators as,
\begin{equation}
{\tilde{c}}_{\vec{r}_j,\sigma}^{\dag} =
{\hat{V}}^{\dag}\,c_{\vec{r}_j,\sigma}^{\dag}\,{\hat{V}} \, ; \hspace{0.5cm}
{\tilde{c}}_{\vec{r}_j,\sigma} =
{\hat{V}}^{\dag}\,c_{\vec{r}_j,\sigma}\,{\hat{V}} \, ,
\label{tilde-c-c}
\end{equation}
where ${\hat{V}}$ is the operator associated with the electron -
rotated-electron unitary transformation. The unitary character of that
transformation implies that the operators 
${\tilde{c}}_{\vec{r}_j,\sigma}^{\dag}$ and ${\tilde{c}}_{\vec{r}_j,\sigma}$
have the same anticommutation relations as 
$c_{\vec{r}_j,\sigma}^{\dag}$ and $c_{\vec{r}_j,\sigma}$.
Straightforward manipulations based on Eq. (\ref{fc+}) then lead
to the following algebra for the $c$ fermion operators,
\begin{equation}
\{f^{\dag}_{\vec{r}_j,c}\, ,f_{\vec{r}_{j'},c}\} = \delta_{j,j'} \, ;
\hspace{0.35cm}
\{f_{\vec{r}_j,c}^{\dag}\, ,f_{\vec{r}_{j'},c}^{\dag}\} =
\{f_{\vec{r}_j,c}\, ,f_{\vec{r}_{j'},c}\} = 0 \, .
\label{albegra-cf}
\end{equation}

Let us introduce the corresponding momentum-dependent $c$ fermion operators, 
\begin{equation}
f_{\vec{q}_j,c}^{\dag} =
{1\over{\sqrt{N_a^D}}}\sum_{j'=1}^{N_a^D}\,e^{+i\vec{q}_j\cdot\vec{r}_{j'}}\,
f_{\vec{r}_{j'},c}^{\dag} 
\, ; \hspace{0.50cm}
f_{\vec{q}_j,c} = {1\over{\sqrt{N_a^D}}}\sum_{j'=1}^{N_a^D}\,e^{-i\vec{q}_j\cdot\vec{r}_{j'}}\,
f_{\vec{r}_{j'},c} 
\, ; \hspace{0.35cm}
j = 1,...,N_a^D \, ; \hspace{0.50cm} L = a\,N_{a} \, ,
\label{fc-q-x}
\end{equation}
which refer to the conjugate variable $\vec{q}_j$
of the $c$ effective lattice real-space coordinate $\vec{r}_{j}$ introduced
in Refs. \cite{companion,s1-bonds}. 

The $s1$ bond particles are more complex than the $c$ fermions.
According to the studies of these references 
they have an internal structure associated with a spin-neutral 
superposition of two-spinon bond occupancy configurations. 
Therefore, the $s1$ bond particle operators $g^{\dag}_{{\vec{r}}_{j},s1}$
and $g_{{\vec{r}}_{j},s1}$ involve a sum of $N_{s1}$ two-site one-bond 
operators, including $N_{s1}/2D$ of such operators per link family.
The concepts of two-site link family and type, which characterize the set of 
two-site one-bond operators whose superposition defines a $s1$ bond-particle
operator, are introduced in Ref. \cite{s1-bonds}. Alike the remaining $\alpha\nu$
bond particles and as confirmed in that reference, the $s1$ bond particles 
have been constructed to inherently being hard-core objects
so that upon acting onto the $s1$ effective lattice the $s1$ bond-particle 
operators anticommute on the same site of that lattice,
\begin{equation}
\{g^{\dag}_{{\vec{r}}_{j},s1},g_{{\vec{r}}_{j},s1}\} = 1 \, ;
\hspace{0.35cm}
\{g^{\dag}_{{\vec{r}}_{j},s1},g^{\dag}_{{\vec{r}}_{j},s1}\} =
\{g_{{\vec{r}}_{j},\alpha\nu},g_{{\vec{r}}_{j},s1}\}=0 \, ,
\label{g-local}
\end{equation}
and commute on different sites,
\begin{equation}
[g^{\dag}_{{\vec{r}}_{j},s1},g_{{\vec{r}}_{j'},s1}] =
[g^{\dag}_{{\vec{r}}_{j},s1},g^{\dag}_{{\vec{r}}_{j'},s1}]
= [g_{{\vec{r}}_{j},s1},g_{{\vec{r}}_{j'},1}] = 0 \, ; \hspace{0.5cm}
j\neq j' \, .
\label{g-non-local}
\end{equation}

The anti-commutation and commutation relations
of Eqs. (\ref{g-local}) and (\ref{g-non-local}),
respectively, follow in part from the algebra of the spinon operators
$s^{\pm}_{\vec{r}_j}$ and $s^{x_3}_{\vec{r}_j}$ introduced
in Ref. \cite{s1-bonds}, 
which are the building blocks of the two-site
one-bond operators of that reference.
Such spinon operators obey the usual spin-$1/2$
operator algebra. Indeed and except for the 
$U/4t\gg 1$ limit, for finite values $U/4t$
rather than to electronic spins they refer to the 
spins of rotated electrons and according to the studies of
Ref. \cite{companion} are given by, 
\begin{equation}
s^l_{\vec{r}_j} = n_{\vec{r}_j,c}\,q^l_{\vec{r}_j} \, ; \hspace{0.50cm}
p^{l}_{\vec{r}_j} = (1-n_{\vec{r}_j,c})\,q^{l}_{\vec{r}_j} 
\, ; \hspace{0.50cm}
n_{\vec{r}_j,c} = f_{\vec{r}_j,c}^{\dag}\,f_{\vec{r}_j,c} 
\, , \hspace{0.35cm} l =\pm,x_3 \, .
\label{sir-pir}
\end{equation}
Here we provide also the expression of
the $\eta$-spinon operator $p^{l}_{\vec{r}_j}$
associated with the $\eta$-spin $SU(2)$ algebra.
Within the LWS representation the 
rotated quasi-spin operators $q^l_{\vec{r}_j}$ 
can be expressed in terms of the rotated-electron 
operators as follows,
\begin{equation}
q^+_{\vec{r}_j} = ({\tilde{c}}_{\vec{r}_j,\uparrow}^{\dag}
- e^{i\vec{\pi}\cdot\vec{r}_j}\,{\tilde{c}}_{\vec{r}_j,\uparrow})\,
{\tilde{c}}_{\vec{r}_j,\downarrow} 
\, ; \hspace{0.50cm}
q^-_{\vec{r}_j} = (q^+_{\vec{r}_j})^{\dag} \, ,
\hspace{0.50cm}
q^{x_3}_{\vec{r}_j} = {1\over 2} - {\tilde{n}}_{\vec{r}_j,\downarrow} \, ,
\label{rotated-quasi-spin}
\end{equation}
and the $c$ fermion operators are given in Eq. (\ref{fc+}). 
Straightforward manipulations based on Eq. (\ref{rotated-quasi-spin}) 
confirm the validity of the following usual algebra for the spinon operators 
$s^{\pm}_{\vec{r}_j}$ and $s^{x_3}_{\vec{r}_j}$,
\begin{equation}
\{s^{+}_{\vec{r}_j},s^{-}_{\vec{r}_j}\} = 1 \, ,
\hspace{0.5cm}
\{s^{\pm}_{\vec{r}_j},s^{\pm}_{\vec{r}_j}\} = 0 \, ,
\label{albegra-s-p-m}
\end{equation}
\begin{equation}
[s^{+}_{\vec{r}_j},s^{-}_{\vec{r}_{j'}}] =
[s^{\pm}_{\vec{r}_j},s^{\pm}_{\vec{r}_{j'}}]=0 
\, ; \hspace{0.5cm}
j\neq j' \, ,
\label{albegra-s-com}
\end{equation}
and,
\begin{equation}
[s^{x_3}_{\vec{r}_j},s^{\pm}_{\vec{r}_{j'}}] =
\pm\delta_{j,j'}s^{\pm}_{\vec{r}_{j}} \, .
\label{albegra-s-sz-com}
\end{equation}
Hence the spinon operators $s^{\pm}_{\vec{r}_j}$ anticommute 
on the same site and commute on different sites. Consistently 
with the rotated-electron singly-occupied site projector
expressed in terms of $c$ fermion operators 
$n_{\vec{r}_j,c}$ appearing in the expression of the spinon 
operators $s^{\pm}_{\vec{r}_j}$ and $s^{x_3}_{\vec{r}_j}$
provided in Eq. (\ref{sir-pir}), within the $N_a^D\rightarrow\infty$ limit their real-space coordinates 
$\vec{r}_j$ can be identified with those of the spin effective lattice.

The hard-core character of the two-spinon $s1$ bond-particle operators
associated with the algebra of Eqs. (\ref{g-local}) and (\ref{g-non-local}) is 
used in the following in the introduction of the $s1$ fermions. 

\subsection{The $s1$ fermions emerging from the extended
Jordan-Wigner transformation and relation to the two-dimensional quantum Hall effect}

\subsubsection{The extended Jordan-Wigner transformation}

The present study focus mainly on the 
model on the square lattice, yet the corresponding 1D problem
is also often considered. For instance, the original 
Jordan-Wigner transformation
\cite{J-W} transforms 1D spin-$1/2$ spin operators into spinless
fermion operators. The extension to the square lattice of that
transformation has been considered previously again for  
spin-$1/2$ spin operators \cite{Fradkin,Wang}.   
In turn, here we apply it to the hard-core spin-neutral
two-spinon $s1$ bond-particle operators of Eqs. (\ref{g-local}) and (\ref{g-non-local}).

We follow the method of Ref. \cite{Wang} for the spin-$1/2$ spin operators
of both the 1D and isotropic Heisenberg model on the square-lattice,
which was used in Ref. \cite{Feng} in studies of the $t-J$ model.
We recall that in terms of the creation and annihilation rotated-electron operators the
spinon operators given in Eq. (\ref{sir-pir}) refer for $U/4t>0$ 
to singly occupied sites. The corresponding 
two-spinon $s1$ bond-particle operators act onto the 
$s1$ effective lattice defined in Refs. \cite{companion,s1-bonds}.
Each of such objects corresponds to rotated-electron occupancies involving 
two sites of both the original lattice and spin effective lattice defined
in Ref. \cite{companion}.
 
As a result of the $s1$ bond-particle operator 
algebra given in Eqs. (\ref{g-local}) and 
(\ref{g-non-local}) for the $s1$ effective lattice of both the 
square-lattice and 1D problems, one can perform a 
Jordan-Wigner transformation that maps the 
$s1$ bond particles onto $s1$ fermions with 
creation operators $f^{\dag}_{{\vec{r}}_{j},s1}$ and 
annihilation operators $f_{{\vec{r}}_{j},s1}$. 
Such operators are related to the corresponding 
$s1$ bond-particle operators studied in Ref. \cite{s1-bonds} as follows,
\begin{equation}
f^{\dag}_{{\vec{r}}_{j},s1} = e^{i\phi_{j,s1}}\,
g^{\dag}_{{\vec{r}}_{j},s1} 
\, ; \hspace{0.50cm} 
f_{{\vec{r}}_{j},s1} = e^{-i\phi_{j,s1}}\,
g_{{\vec{r}}_{j},s1} \, ,
\label{JW-f+}
\end{equation}
where 
\begin{equation}
\phi_{j,s1} = \sum_{j'\neq j}f^{\dag}_{{\vec{r}}_{j'},s1}
f_{{\vec{r}}_{j'},s1}\,\phi_{j',j,s1} 
\, ; \hspace{0.50cm}
\phi_{j',j,s1} = \arctan \left({{x_{j'}}_2-{x_{j}}_2\over {x_{j'}}_1-{x_{j}}_1}\right) 
\, ; \hspace{0.35cm} 
0 \leq \phi_{j',j,s1}\leq 2\pi \, .
\label{JW-phi}
\end{equation}
Let $z_{j}={x_{j}}_1+i\,{x_{j}}_2$ (with ${x_j}_2=0$ for 1D) be the complex coordinate
of the $s1$ bond particle at the $s1$ effective-lattice site of real-space
coordinate ${\vec{r}}_{j}$ and $j=1,...,N_{s1}^D$. For the model on the square 
lattice the real-space vector ${\vec{r}}_{j}=[{x_{j}}_1,{x_{j}}_2]$ of
Cartesian coordinates ${x_{j}}_1$ and ${x_{j}}_2$ 
and the complex number $z_{j}={x_{j}}_1+i\,{x_{j}}_2$
are alternative representations of the same quantity. The vector difference
$[{\vec{r}}_{j'}-{\vec{r}}_{j}]=[{x_{j'}}_1-{x_{j}}_1,{x_{j'}}_2-{x_{j}}_2]$
can be written as,
\begin{equation}
[{\vec{r}}_{j'}-{\vec{r}}_{j}] = \vert{\vec{r}}_{j'}-{\vec{r}}_{j}\vert\,{\vec{e}}_{\phi_{j',j,s1}}
\label{r-r}
\end{equation}
where $\phi_{j',j,s1}$ is the phase of Eq. (\ref{JW-phi}) and here and in the remaining 
of this paper we denote by,
\begin{equation}
{\vec{e}}_{\phi}  =
\left[\begin{array}{c}
\cos \phi \\
\sin \phi
\end{array} \right] \, ,
\label{unit-vector}
\end{equation}
a unit vector whose direction is defined by the angle $\phi$.

It follows from the Jordan-Wigner transformation (\ref{JW-f+}) that the operators 
$f^{\dag}_{\vec{r}_{j},s1}$ and $f_{{\vec{r}}_{j},s1}$ have anticommuting relations
similar to those given in Eq. (\ref{albegra-cf}) for the
$c$ fermion operators,
\begin{equation}
\{f^{\dag}_{\vec{r}_{j},s1}\, ,f_{\vec{r}_{j'},s1}\} =
\delta_{j,j'} 
\, ; \hspace{0.50cm}
\{f^{\dag}_{\vec{r}_{j},s1}\, ,f^{\dag}_{\vec{r}_{j'},s1}\} =
\{f_{\vec{r}_{j},s1}\, ,f_{\vec{r}_{j'},s1}\} = 0 \, ,
\label{1D-anti-com}
\end{equation}
and the $c$ fermion operators commute with the $s1$ fermion operators. 

The $s1$ fermion operators $f^{\dag}_{\vec{r}_{j},s1}$ and $f_{{\vec{r}}_{j},s1}$
act onto subspaces spanned by mutually neutral states \cite{companion} corresponding to
constant values of $S_c$, $S_s$, and number $N_{a_{s1}}^D$ of sites
of the $s1$ effective lattice whose expression in terms of the number 
of sites of the spin effective lattice $N_{a_{s}}^D = (1-x)\,N_a^D$
is for the one- and two-electron subspace given in Eq. (\ref{68}) of Appendix A.
Neutral states are defined below in terms of the occupancy configurations
of $s1$ fermions carrying microscopic momenta. 

\subsubsection{The extended Jordan-Wigner transformation for the
model on the square lattice and relation to the 2D quantum Hall effect}

For the model on the square lattice the problem is much more
complex than for the 1D model. The phase factor in the expressions
given in Eq. (\ref{JW-f+}) corresponds to that created by a 
gauge field whose effective vector potential ${\vec{A}}_{s1} ({\vec{r}}_j)$
and corresponding fictitious magnetic field ${\vec{B}}_{s1} ({\vec{r}}_j)$ read,
\begin{equation}
{\vec{A}}_{s1} ({\vec{r}}_j) = \Phi_0\sum_{j'\neq j}
n_{\vec{r}_{j'},s1}\,{{\vec{e}}_{x_3}\times {\vec{e}}_{\phi_{j',j,s1}}
\over \vert{\vec{r}}_{j'}-{\vec{r}}_{j}\vert} \, ; \hspace{0.35cm} 
{\vec{B}}_{s1} ({\vec{r}}_j) = {\vec{\nabla}}_{\vec{r}_j}\times {\vec{A}}_{s1} ({\vec{r}}_j)
= \Phi_0\sum_{j'\neq j} n_{\vec{r}_{j'},s1}\,\delta ({\vec{r}}_{j'}-{\vec{r}}_{j})\,{\vec{e}}_{x_3} \, ,
\label{A-j-s1-3D}
\end{equation}   
where we use units such that the fictitious magnetic flux quantum is given by $\Phi_0=1$ and
\begin{equation}
{\vec{e}}_{x_3}  =
\left[\begin{array}{c}
0 \\
0 \\
1
\end{array} \right] 
\, ; \hspace{0.35cm} 
{\vec{e}}_{\phi_{j',j,s1}}  =
\left[\begin{array}{c}
\cos \phi_{j',j,s1} \\
\sin \phi_{j',j,s1} \\
0
\end{array} \right] \, ,
\label{e-phi-e-3}
\end{equation}   
are unit vectors perpendicular to and contained in the square-lattice plane,
respectively. Moreover, 
\begin{equation}
n_{\vec{r}_j,s1} = f_{\vec{r}_j,s1}^{\dag}\,f_{\vec{r}_j,s1} \, ,
\label{n-j-s1}
\end{equation}
is the $s1$ fermion local density operator. 
The effective vector potential (\ref{A-j-s1-3D}) can be written in terms of 2D 
coordinates alone as,
\begin{equation}
{\vec{A}}_{s1} ({\vec{r}}_j) = \Phi_0\sum_{j'\neq j}
{n_{\vec{r}_{j'},s1}
\over \vert{\vec{r}}_{j}-{\vec{r}}_{j'}\vert}\,
{\vec{e}}_{\phi_{j',j,A_{s1}}} 
\, ; \hspace{0.35cm}
{\phi_{j',j,A_{s1}}}\equiv {\phi_{j',j,s1}+\pi/2} 
=  -\arctan \left({{x_{j}}_1-{x_{j'}}_1\over {x_{j}}_2-{x_{j'}}_2}\right) \, ,
\label{A-j-s1}
\end{equation}   
where the unit vector ${\vec{e}}_{\phi_{j',j,s1}+\pi/2}$
has the general form (\ref{unit-vector}) and is perpendicular to
the unit vector ${\vec{e}}_{\phi_{j',j,s1}}$ of Eq. (\ref{r-r}). This 
effective potential generates long-range interactions between
the $s1$ fermions. 

The two components of the microscopic momenta of the $s1$ fermions are eigenvalues
of the two $s1$ translation generators $\hat{q}_{s1\,x_1}$ and $\hat{q}_{s1\,x_2}$ 
in the presence of the fictitious magnetic field ${\vec{B}}_{s1} ({\vec{r}}_j)$ of Eq. (\ref{A-j-s1-3D})
given below. Due to the non-commutativity of such $s1$ translation generators whose
eigenvalues are the components
$q_{x_1}$ and $q_{x_2}$, respectively, of $s1$ fermion microscopic momenta $\vec{q}=[q_{x_1},q_{x_2}]$
of the model on the square lattice one expects that it is impossible to classify the states
in terms of such momenta. However, in a subspace spanned by mutually neutral states
such translation generators commute \cite{companion,Giu-Vigna}. Neutral states are transformed
into each other by $s1$ fermion particle-hole processes.
The contributions of the $s1$ fermion and $s1$ fermion hole to the commutator of
the two $s1$ translation generators $\hat{q}_{s1\,x_1}$ and $\hat{q}_{s1\,x_2}$ cancel
in the case of such particle-hole excitations so that the two components
$q_{x_1}$ and $q_{x_2}$ of $s1$ fermion microscopic momenta $\vec{q}=[q_{x_1},q_{x_2}]$
can be simultaneously specified. The $s1$ fermion operators are defined in subspaces
spanned by mutually neutral states which correspond to constant $S_c$, $S_s$, and $N_{a_{s1}}^D$
values. Moreover, the values of the set of $N_{a_{s1}}^D$ discrete 
momenta $\vec{q}_j=[q_{j\,x_1},q_{j\,x_2}]$ where $j=1,...,N_{a_{s1}}^D$ are well defined for 
each such a subspace where the $s1$ fermion operators of Eq. (\ref{JW-f+}) act onto.
The corresponding momentum-dependent $s1$ fermion operators are given in
terms of the operators labelled by real-space coordinates provided in that equation as, 
\begin{equation}
f_{\vec{q}_j,s1}^{\dag} = 
{1\over{\sqrt{N_{a_{s1}}^D}}}\sum_{j'=1}^{N_{a_{s1}}^D}\,e^{+i\vec{q}_j\cdot\vec{r}_{j'}}\,
f_{\vec{r}_{j'},s1}^{\dag} 
\, ; \hspace{0.50cm}
f_{\vec{q}_j,s1} = {1\over{\sqrt{N_{a_{s1}}^D}}}\sum_{j'=1}^{N_{a_{s1}}^D}\,e^{-i\vec{q}_j\cdot\vec{r}_{j'}}\,
f_{\vec{r}_{j'},s1} 
\, ; \hspace{0.35cm}
j = 1,...,N_{a_{s1}}^D \, ; \hspace{0.50cm} L = a_{s1}\,N_{a_{s1}} \, .
\label{fs1-q-x}
\end{equation}
The $s1$ band microscopic discrete momentum values $\vec{q}_j$ where
$j = 1,...,N_{a_{s1}}^D$ play the role
of conjugate variables of the $s1$ effective-lattice real-space coordinates $\vec{r}_{j}$
which label the corresponding operators of Eq. (\ref{JW-f+}). Often we omit the index
$j = 1,...,N_{a}^D$ or $j = 1,...,N_{a_{s1}}^D$ of the discrete momentum values $\vec{q}_j$
and denote them by $\vec{q}$. According to the results
of Ref. \cite{companion}, the $c$ and $s1$ translation generators read
${\hat{{\vec{q}}}}_c  = \sum_{{\vec{q}}}{\vec{q}}\, \hat{N}_c ({\vec{q}})$ and
${\hat{{\vec{q}}}}_{s1} = \sum_{{\vec{q}}}{\vec{q}}\, \hat{N}_{s1} ({\vec{q}})$, respectively, where,
\begin{equation}
\hat{N}_{c}({\vec{q}}) = f^{\dag}_{{\vec{q}},c}\,f_{{\vec{q}},c} \, ;
\hspace{0.35cm}
\hat{N}_{s1}({\vec{q}}) = f^{\dag}_{{\vec{q}},s1}\,f_{{\vec{q}},s1} \, ,
\label{Nc-s1op}
\end{equation}
are the momentum distribution-function operators. It follows that the
$s1$ translation generators $\hat{q}_{s1\,x_1}$ and $\hat{q}_{s1\,x_2}$
considered above are given by,
\begin{equation}
\hat{q}_{s1\,x_i} = \sum_{{\vec{q}}}q_{x_i}\, \hat{N}_{s1} ({\vec{q}}) 
\hspace{0.35cm} i = 1, 2 \, .
\label{q-s1-xi}
\end{equation}

The one- and two-electron subspace contains several subspaces with 
constant $S_c$, $S_s$, and $N_{a_{s1}}^D$ values. It turns out that besides
neutral states, which transform into each other by particle-hole processes
generated by operators of the form $f^{\dag}_{{\vec{q}},s1}\,f_{{\vec{q}}\,',s1}$ or
$f_{{\vec{q}},s1}\,f^{\dag}_{{\vec{q}}\,',s1}$, also spin-singlet excited states generated 
by application onto the $m=0$ and $x\geq 0$ initial ground state of the operator 
$f^{\dag}_{0,s2}\,f_{{\vec{q}},s1}\,f_{{\vec{q}}\,',s1}$ where ${\vec{q}}$ and
${\vec{q}}\,'$ are the momenta of the two emerging $s1$ fermion holes are
neutral states, which conserve $S_c$, $S_s$, and $N_{a_{s1}}^D$. For
the model on the square lattice the role of the $s2$ fermion creation operator
$f^{\dag}_{0,s2}$ is exactly canceling the contributions of the
annihilation of the two $s1$ fermions of momenta ${\vec{q}}$ and ${\vec{q}}\,'$   
to the commutator of the $s1$ band two $s1$ translation generators
$\hat{q}_{s1\,x_1}$ and $\hat{q}_{s1\,x_2}$ of Eq. (\ref{q-s1-xi}) 
in the presence of the fictitious magnetic field ${\vec{B}}_{s1}$ of Eq. (\ref{A-j-s1-3D}),
so that the overall excitation is neutral. Since the $s2$ fermion has vanishing
energy and momentum and the $s1$ momentum band and its number $N_{a_{s1}}^2$ of
discrete momentum values remain unaltered \cite{companion}, one can effectively consider 
that the generator of such an excitation is $f_{{\vec{q}},s1}\,f_{{\vec{q}}\,',s1}$ and omit the $s2$
fermion creation, whose only role is assuring that the overall excitation is neutral
and the $s1$ fermion microscopic momenta can be specified. It follows
that for the one- and two-electron subspace the operators $f_{{\vec{q}},s1}\,f_{{\vec{q}}\,',s1}$,
$f^{\dag}_{{\vec{q}}\,',s1}\,f^{\dag}_{{\vec{q}},s1}$, $f^{\dag}_{{\vec{q}},s1}\,f_{{\vec{q}}\,',s1}$,
and $f_{{\vec{q}},s1}\,f^{\dag}_{{\vec{q}}\,',s1}$ generate neutral excitations.

Transitions between states with different values for $S_c$, $S_s$, and/or $N_{a_{s1}}^D$
involve creation or annihilation of single $s1$ fermions. The $s1$ fermion operators $f^{\dag}_{{\vec{q}},s1}$ 
and $f_{{\vec{q}},s1}$ refer to subspaces spanned by neutral states. However, as mentioned above
creation or annihilation of one $s1$ fermion is a well-defined process whose generator
is the product of an operator that fulfills small changes in the $s1$ effective lattice
and corresponding $s1$ momentum band and the operator 
$f^{\dag}_{{\vec{q}},s1}$ or $f_{{\vec{q}},s1}$, respectively, appropriate to the excited-state subspace.

Upon replacing $n_{\vec{r}_j,s1} = f_{\vec{r}_j,s1}^{\dag}\,f_{\vec{r}_j,s1}$
by its average $\langle n_{\vec{r}_j,s1}\rangle \approx 1$ the effective vector 
potential of Eq. (\ref{A-j-s1-3D}) becomes,
\begin{equation}
{\vec{A}}_{s1} ({\vec{r}}_j) \approx \Phi_0\sum_{j'\neq j}
{{\vec{e}}_{x_3}\times {\vec{e}}_{\phi_{j',j,s1}}
\over \vert{\vec{r}}_{j'}-{\vec{r}}_{j}\vert} \, .
\label{A-j-s1-3D-average}
\end{equation}   
This is a Chern-Simons like vector potential \cite{companion,Giu-Vigna} so that
each spin-neutral two-spinon $s1$ fermion has on average under the Jordan-Wigner 
transformation a flux tube of one flux quantum attached to it. That the number
of flux quanta attached to each $s1$ fermion is odd is consistent with the $s1$ bond-particle operators of the 
Jordan-Wigner transformation $f^{\dag}_{{\vec{r}}_{j},s1} = e^{i\phi_{j,s1}}\,g^{\dag}_{{\vec{r}}_{j},s1}$ 
of Eq. (\ref{JW-f+}) being associated with Bose statistics and the corresponding
$s1$ fermion operators obeying Fermi statistics, respectively. The field strength corresponds
on average to one flux quantum per elementary plaquette of the square $s1$ effective lattice. 
Thus for the Hubbard model on a square lattice in the one- and two-electron
subspace a composite $s1$ fermion consists of a spin-singlet spinon pair
plus an infinitely thin flux tube attached to it. It follows that within our representation,
each $s1$ fermion appears to carry a magnetic solenoid with it as it moves 
around the $s1$ effective lattice. The $s1$ fermions interact with
each other via the effective vector potential that they create.

For the model on the square lattice in the one- and two-electron subspace the
$s1$ effective lattice spacing magnitude $a_{s1}=L/N_{a_{s1}}\approx \sqrt{2/(1-x)}\,a$ is 
controlled by that of the magnetic length $l_{s1}$ associated with the fictitious magnetic 
field of Eq. (\ref{A-j-s1-3D}). Indeed, for such a subspace one has that
$\langle n_{\vec{r}_j,s1}\rangle\approx 1$ and the fictitious magnetic field reads
${\vec{B}}_{s1} ({\vec{r}}_j) \approx \Phi_0\delta ({\vec{r}}_{j'}-{\vec{r}}_{j})\,{\vec{e}}_{x_3}$.
It acting on one $s1$ fermion differs from zero only at the positions
of other $s1$ fermions. In the mean field approximation one replaces it
by the average field created by all $s1$ fermions at position $\vec{r}_j$. This gives,
\begin{equation}
{\vec{B}}_{s1} ({\vec{r}}_j) \approx \Phi_0\,n_{s1} (\vec{r}_j)\,{\vec{e}}_{x_3}
\approx \Phi_0{N_{a_{s1}}^2\over L^2}\,{\vec{e}}_{x_3} = {\Phi_0\over 2\pi\,l_{s1}^2}\,{\vec{e}}_{x_3}
\approx {\Phi_0(1-x)\over 2a^2}\,{\vec{e}}_{x_3} 
\, ; \hspace{0.35cm} a_{s1}^2=2\pi\,l_{s1}^2 \, ,
\label{ficti-B}
\end{equation}
where $l_{s1}$ is the above fictitious magnetic-field length. Hence the number $N_{a_{s1}}^2$
of $s1$ band discrete momentum values equals $[B_{s1}\,L^2]/\Phi_0$ and the magnitude of the
$s1$ effective lattice spacing $a_{s1}$ is determined by the fictitious 
magnetic-field length $l_{s1}$ as $a_{s1}^2=2\pi\,l_{s1}^2$.
This is consistent with for such states each $s1$ fermion having a flux
tube of one flux quantum on average attached to it. 

We find below that the $s1$ fermions have a momentum dependent energy dispersion
and only in the $U/4t\rightarrow\infty$ limit their energy bandwidth vanishes and
the $N_{a_{s1}}^2$ one-$s1$-fermion states corresponding to the $N_{a_{s1}}^2=[B_{s1}\,L^2]/\Phi_0$
discrete momentum values are degenerate. Hence, in that limit $N_{a_{s1}}^2=[B_{s1}\,L^2]/\Phi_0$
plays the role of the number of degenerate states in each Landau level of the 2D QHE. For the
one- and two-electron subspace the $s1$ filling factor $\nu_{s1}=N_{s1}/N_{a_{s1}}^2$
reads $\nu_{s1}=1$ for the $m=0$ ground state and $\nu_{s1}=1$, $\nu_{s1}=1-1/N_{a_{s1}}^2$,
or $\nu_{s1}=1-2/N_{a_{s1}}^2$ for the excited states where $N_{a_{s1}}^2\approx [(1-x)/2]\,N_a^2$.
Therefore, within the suitable mean-field approximation (\ref{ficti-B}) for the fictitious magnetic field 
${\vec{B}}_{s1}$ of Eq. (\ref{A-j-s1-3D}) such a ground state corresponds to a full lowest Landau level and in
the $N_a^2\rightarrow\infty$ limit that our description refers to that applies as
well to the excited states. Only for the $U/4t\rightarrow\infty$ limit there is
fully equivalence between the $s1$ fermion occupancy configurations of the
one- and two-electron subspace of the Hubbard model on a square lattice and 
the 2D QHE with a full lowest Landau level. In spite of
the lack of state degeneracy emerging upon decreasing the value of $U/4t$, for finite $U/4t$ values
there remains though some relation to the 2D QHE. The occurrence of QHE-type behavior 
in the square-lattice quantum liquid shows that a magnetic field is not 
essential to the 2D QHE physics. Indeed, here the fictitious magnetic field
arises from expressing the effects of the electronic correlations in terms of the
$s1$ fermion interactions. The $s1$ fermion description leads to the intriguing situation
where the $s1$ fermions interact via long-range forces 
of Eqs. (\ref{A-j-s1-3D}), (\ref{A-j-s1}), and (\ref{A-j-s1-3D-average}) while all interactions
in the original Hamiltonian (\ref{H}) are onsite. That combination of our description with the
small effects of 3D anisotropy and intrinsic disorder is shown in Refs.
\cite{cuprates,cuprates0} to successfully describe the unusual properties
of several families of hole-doped cuprate superconductors is an indication
that QHE-type behavior with or without magnetic field may be ubiquitous in nature. 

A site of the $s1$ effective lattice occupied by a $s1$ fermion 
corresponds to two-sites of the spin effective lattice. Moreover,
the $N_{a_s}^D=2S_c$ sites of the spin effective lattice describe the spin
degrees of freedom of $2S_c$ rotated electrons that
singly occupy $2S_c$ sites of the original lattice. 
In turn, the charge degrees of freedom of such
rotated electrons are described by the $2S_c$ sites
of the $c$ effective lattice that are occupied by $c$
fermions \cite{companion}.

Since the $c$ fermions and $s1$ fermions describe the charge
and spin degrees of freedom, respectively, of the same
sites singly occupied by rotated electrons, the long-range 
interactions between the $s1$ fermions generated by
the effective vector potential of Eqs. (\ref{A-j-s1-3D}), (\ref{A-j-s1}),
and (\ref{A-j-s1-3D-average}) are felt by the $c$
fermions and lead to residual interactions between
those and the $s1$ fermions, whose effective
interaction energy is derived in Ref. \cite{cuprates0}. For instance, according to
Table \ref{tableIV} of Appendix A upon creation and 
annihilation of one electron, one $c$ fermion is
created and annihilated, respectively, and one
$s1$ fermion hole is created. In contrast to the 1D case,
the residual interactions brought about by the extended
Jordan-Wigner transformation are for $x>0$ behind inelastic
collisions between for instance a $s1$ fermion going over to the
momentum value of the $s1$ band hole created upon
creation or annihilation of one electron 
and the $c$ fermions with momenta near the $c$ Fermi  
line introduced in this paper. 

\subsubsection{The effects of the $s1$ fermion Jordan-Wigner transformation in 1D}
   
The above discussion is different from 1D where the quantum problem is 
integrable and due to the occurrence
for $N_a\gg1$ of an infinite number of conservation laws,
the $c$ and $s1$ fermions have zero-momentum forward-scattering
only. That the physical consequences of the extended Jordan-Wigner  
transformation through which the $s1$ fermions 
emerge from the $s1$ bond particles are different for 1D and 2D
is consistent with for 1D the number $z_{j}={x_{j}}_1+i\,{x_{j}}_2$ involving the coordinates
${x_{j}}_1$ and ${x_{j}}_2$ appearing in Eq. (\ref{JW-phi}) being such that ${x_{j}}_2=0$ and hence 
then reducing to the real-space coordinate ${x_{j}}_1$ of the $s1$ bond particle 
in its effective lattice. Therefore, 
for 1D the phase $\phi_{j',j,s1}=
\arctan ([{x_{j'}}_2-{x_{j}}_2]/[{x_{j'}}_1-{x_{j}}_1])$ in
that equation can have the values
$\phi_{j',j,s1}=0$ and
$\phi_{j',j,s1}=\pi$ only. Indeed, the
relative angle between two sites of the 1D $s1$
effective lattice can only be one of the 
two values. It follows that in 1D 
$e^{i(\phi_{j+1,s1}-\phi_{j,s1})} 
= e^{i\pi\,f^{\dag}_{x_{j},s1}\,f_{x_{j},s1}}$.
The $s1$ fermion discrete momentum values
$q_j$ are eigenvalues of the $s1$ translation generator
$\hat{q}_{s1}  = \sum_{q} q\,\hat{N}_{s1} (q)$
in the presence of the fictitious magnetic field ${\vec{B}}_{s1}$ 
of Eq. (\ref{A-j-s1-3D}) associated with the Jordan-Wigner transformation.
Their values are given by the exact solution and can be 
expressed as a sum of a bare momentum of the usual
form $[2\pi / L]\,{\cal{N}}^{s1}_j$ where ${\cal{N}}^{s1}_j=0,\pm 1, \pm 2,...$
are integers and a small deviation $q_{s1}^0/N_{s1}$ that
for a subspace spanned by mutually neutral states
has a constant value. This gives, 
\begin{equation}
q_j = {2\pi\over L}\,{\cal{N}}^{s1}_j + q_{s1}^0/N_{s1} 
\, ; \hspace{0.50cm}
{\cal{N}}^{s1}_j= j - {N_{a_{s1}}\over 2} = 0,\pm 1, \pm 2, ... 
\, ; \hspace{0.35cm} j=1,...,N_{a_{s1}} \, .
\label{q-j-f-Qs1-0}
\end{equation}
Concerning the value of $q_{s1}^0$ it turns out that since
for 1D the above phase $\phi_{j',j,s1}$ can have the values
$\phi_{j',j,s1}=0$ and $\phi_{j',j,s1}=\pi$ only, there are as well two types of subspaces only
where it is given either by $q_{s1}^0=0$ or $q_{s1}^0= 
\pi [N_{s1}/L]$, respectively, for all $j=1,...,N_{a_{s1}}$ discrete momentum values.
The $s1$ effective lattice length $L=N_{a_{s1}}\,a_{s1}$ equals that of the original lattice. Here
$a_{s1}=L/N_{a_{s1}}=[N_a/N_{a_{s1}}]\,a$ is the $s1$ effective lattice 
constant. It follows that in 1D the transitions between subspaces with
different $s1$ fermion discrete momentum values $q_j$ are associated with deviations given by 
$\delta q_{s1}^0=\pm\pi [N_{s1}/L]$: Under such subspace transitions all the discrete momentum values 
$q_j$ of Eq. (\ref{q-j-f-Qs1-0}) are shifted by the same overall momentum 
$\delta q_{s1}^0/N_{s1}=\pm\pi/L$. 

Each site of the $s1$ (and $s2$) effective lattice 
occupied by a $s1$ (and $s2$) fermion refers to the spin degrees of freedom of two (and four) 
sites of the original lattice singly occupied by rotated electrons whose charge degrees of freedom
are described by two (and four) $c$ fermions. In 1D the sharing of the same rotated-electron
sites by the $s1$ (and $s2$) fermions and $c$ fermions, respectively, leads to residual interactions
between such objects whose main effect is on the boundary conditions of the momentum values of
the charge $c$ fermions: Due to such a residual interaction the latter objects feel the Jordan-Wigner
phase of the $s1$ and $s2$ fermions created or annihilated under a subspace transition 
so that  their discrete momentum values can be written in the form,  
\begin{equation}
q_j = {2\pi\over L}\,{\cal{N}}^{c}_j + q_{c}^0/N_{c} 
\, ; \hspace{0.50cm}
{\cal{N}}^{c}_j= j - {N_{a}\over 2} = 0,\pm 1, \pm 2, ... 
\, ; \hspace{0.35cm} j=1,...,N_{a} \, .
\label{q-j-f-Q-c-0}
\end{equation}
Again, there are only two types of subspaces where $q_{c}^0$ is given either by $q_{c}^0=0$ or 
$q_{c}^0=\pi [N_c/L]$, respectively, for all $j=1,...,N_a$ discrete momentum values of the $c$ band.
For the subspaces of the one- and two-electron subspace there are finite occupancies of $c$ and $s1$ fermions plus one or
none $s2$ fermion of vanishing energy and momentum so that the
general equation for the phase factor $e^{iq_j\,L}=e^{i\pi [B_{\eta}+B_s]}=e^{i\pi\sum_{\alpha\nu}N_{\alpha\nu}}$ 
involving the $q_j$ $c$ band momenta considered in Ref. \cite{companion} simplifies to 
$e^{iq_j\,L} = e^{i\pi [N_{s1}+N_{s2}]}$. It follows that the exact-solution quantum numbers 
$I^{c}_j\equiv [{\cal{N}}^c_j + {q_{c}^0\over 2\pi}{L\over N_c}]$ considered in that reference are 
integers and half-odd integers for subspaces spanned by states with $[N_{s1}+N_{s2}]$ 
even and odd, respectively, where consistently with the values of 
the number $[2S_s + 2N_{s2}]$ given for the one- and
two-electron subspace in Eq. (\ref{46}) of Appendix A one has $N_{s2}=0$ for
subspaces of spin $S_s=1/2$ or $S_s=1$ and $N_{s2}=0,1$ for subspaces 
of spin $S_s=0$. In Ref. \cite{companion} it is confirmed
that such an effect results indeed from the phase $\phi_{j,s1}$ 
of Eq. (\ref{JW-phi}), which controls the $s1$ fermion
Jordan-Wigner transformation. In turn, the energy
is not affected, in contrast to the model on the square lattice.

Analysis of the quantum numbers associated with the exact solution, which are
related to the $c$ and $\alpha\nu$ description in Ref. \cite{companion},
reveals that in 1D the good quantum numbers are not the integer
numbers ${\cal{N}}^{s1}_j= j - N_{a_{s1}}/2  =0,\pm 1, \pm 2,...$ and
${\cal{N}}^c_j= j - N_{a}/2 = 0,\pm 1, \pm 2,...$ 
of Eqs. (\ref{q-j-f-Qs1-0}) and (\ref{q-j-f-Q-c-0}),
respectively, but rather the corresponding shifted
numbers  $I^{s1}_j\equiv [{\cal{N}}^{s1}_j + {q_{s1}^0\over 2\pi}{L\over N_{s1}}]$
and $I^{c}_j\equiv [{\cal{N}}^c_j + {q_{c}^0\over 2\pi}{L\over N_c}]$
where  $j=1,...,N_{a_{s1}}$ and $j=1,...,N_a$, respectively.
For a given subspace the numbers $I^{c}_j$ (and
$I^{s1}_j$) are either consecutive integers $I^{c}_j= {\cal{N}}^c_j$
(and $I^{s1}_j={\cal{N}}^{s1}_j$) or half-odd integers
$I^{c}_j= [{\cal{N}}^c_j-1/2]$
(and $I^{s1}_j=[{\cal{N}}^{s1}_j-1/2]$). Such an
effect is associated with boundary conditions controlled by the
Jordan-Wigner phases.

Hence in 1D the eigenvalues $q_j = ([2\pi/L]\,{\cal{N}}^{s1}_j + q_{s1}^0/N_{s1})$ 
of the $s1$ translation generator in the presence of the fictitious magnetic field are the good quantum
numbers. Such conserving numbers are the $s1$ band discrete momentum values
of Eq. (\ref{q-j-f-Qs1-0}). However, that as a side effect the $s1$ fermion Jordan-Wigner phase
of a $s1$ (and $s2$) fermion created or annihilated (and created) under subspace transitions is
felt by the $c$ fermions so that all $c$ band discrete momentum values are shifted
by $\delta q_{c}^0/N_c=\pm\pi/L$ is not a trivial effect. (Note that creation
or annihilation of two $s1$ fermions or creation or annihilation of one
$s1$ fermion plus creation of one $s2$ fermion does not lead to changes
in the $c$ band discrete momentum values.) This confirms that for 1D the
Jordan-Wigner transformation that generates the
$s1$ fermions from the $s1$ bond particles gives
rise to zero-momentum forward-scattering 
interactions between the emerging $s1$ fermions
and the pre-existing $c$ fermions.

That the Jordan-Wigner phases of the $s1$ fermions
are felt by the $c$ fermions under $s1$ fermion creation
or annihilation through the residual interactions
associated with their sharing of the same sites of the original
lattice occurs for the model on the square
lattice as well. In 1D such a residual interactions 
involve only zero-momentum forward-scattering. In contrast, for
the square lattice they lead to inelastic collisions involving 
exchange of energy and momentum, as confirmed in 
Refs. \cite{cuprates0,cuprates}.

\section{Quantum numbers of the Hubbard model in the one- and 
two-electron subspace: the $c$ and $s1$ band discrete momentum values}

In this section we address the problem of the $c$ and $s1$ band discrete momentum values,
including that of the relation of the Fermi line to the $c$ Fermi line and $s1$
boundary line introduced in the following.

The $s1$ bond particles studied in Refs. \cite{companion,s1-bonds} 
correspond to well-defined occupancy configurations in
a spin effective lattice that for the model
on the square lattice is a square lattice. The $s1$ bond-particle
description of Ref. \cite{s1-bonds} involves a change of gauge structure 
\cite{s1-bonds,Xiao-Gang}
so that the real-space coordinates of the sites of the 
$s1$ effective lattice correspond to one of the two 
sub-lattices of the square spin effective lattice. 
In Appendix B it is confirmed that for the 
$N_a^D\gg 1$ limit that our study refers to and for hole concentrations $x$ such that
the density $n=(1-x)$ is finite the two choices of $s1$ effective lattice
lead to the same description. 

\subsection{The states generated by $c$ and
$s1$ fermion occupancy configurations}

For the model in the one- and two-electron subspace the form of the
general states $\vert \Phi_{U/4t}\rangle={\hat{V}}^{\dag}\vert \Phi_{\infty}\rangle$
of Eq. (\ref{non-LWS}) of Appendix A whose LWSs  
$\vert \Phi_{LWS;U/4t}\rangle$ also appearing in that equation 
are those given in Eq. (\ref{LWS-full-el}) of that Appendix
simplifies so that such a subspace is spanned by a $x\geq 0$ and $m=0$ ground state
and its one- and two-electron excited states of the form,
\begin{equation}
\vert \Phi_{U/4t}\rangle =\frac{({\hat{S}}^{\dag}_{s})^{L_{s,\,-1/2}}}{
\sqrt{{\cal{C}}_{s}}}\vert \Phi_{LWS;U/4t}\rangle
\, ; \hspace{0.35cm}
\vert \Phi_{LWS;U/4t}\rangle = \vert 0_{\eta};N_{a_{\eta}}^D\rangle
[\prod_{j'=1}^{N_{s1}}f^{\dag}_{{\vec{q}}_{j'},s1}\vert 0_{s};N_{a_{s}}^D\rangle]
[\prod_{j=1}^{2S_c}f^{\dag}_{{\vec{q}}_{j},c}\vert GS_c;0\rangle] \, .
\label{LWS-1-2-el}
\end{equation}
Here the coefficient ${\cal{C}}_{s}$ is provided in Eq. (\ref{non-LWS}) of Appendix A,
according to Table \ref{tableIV} of that Appendix the number of $-1/2$ independent spinons is 
given by $L_{s,\,-1/2}=0$ for the initial ground state and some of its excited states,
$L_{s,\,-1/2}=1$ for one-electron excited states, and
$L_{s,\,-1/2}=2$ both for spin-triplet excited states and excited states
involving addition or removal of two electrons with parallel
spin projections, and
$\prod_{j=1}^{2S_c}f^{\dag}_{{\vec{q}}_{j},c}\vert GS_c;0\rangle$
is a generalization of the state given in Eq. (\ref{GS-c}) of Appendix A.
Indeed, only for the spin density $m=(1-x)$ of that equation are
the $c$ fermions invariant under the electron - rotated-electron 
unitary transformation for all values of $U/4t$. In turn, for $m=0$ such an invariance
occurs only for $U/4t\rightarrow\infty$.

As shown in Ref. \cite{companion}, for the Hubbard model on the square lattice in
the one- and two-electron subspace as defined in that reference the numbers $N_{s1}$ of $s1$ 
fermions, $N_{a_{s1}}^D$ of sites of the $s1$ effective lattice and thus of discrete momentum
values of the $s1$ fermion momentum band, and $N_{s2}$ of zero-momentum and vanishing
energy spin-neutral four-spinon $s2$ fermions are good quantum 
numbers. They read $N_{s1} = [S_c -S_s -2N_{s2}]$, $N_{a_{s1}}^D = [N_{a_s}^D/2 +S_s]=[S_c+S_s]$,
and the $s2$ fermion number is given by $N_{s2}=0$ for all excited states of spin
$S_s=0$, $S_s=1/2$, and $S_s=1$ except for the $S_s=0$ spin states for which 
$N_{s2}=1$. Hence the number of $s1$ fermion holes $N_{s1}^h=[N_{a_{s1}}^D-N_{s1}]=[2S_s +2N_{s2}]$
is also conserved for the square-lattice model in such a subspace. For the 1D model all such
numbers are good quantum numbers in the whole Hilbert space.

Moreover, according to the results of Ref. \cite{companion} the general states 
$\vert \Phi_{U/4t}\rangle$ and $\vert \Phi_{LWS;U/4t}\rangle$
of Eqs. (\ref{non-LWS}) and (\ref{LWS-full-el}), respectively, of Appendix A 
are for the model on the square lattice momentum eigenstates yet in
contrast to 1D they are not in general energy eigenstates. Indeed,
for the model on the square lattice neither the set of $\alpha\nu$ 
fermion numbers $\{N_{\alpha\nu}\}$ nor the corresponding numbers $\{N_{a_{\alpha\nu}}^2\}$
of $\alpha\nu$ band discrete momentum values are in general conserved.
Also in contrast to 1D, in general the corresponding set of $\alpha\nu$ translation 
generators ${\hat{\vec{q}}}_{\alpha\nu}$ in the presence of the fictitious magnetic fields ${\vec{B}}_{\alpha\nu}$ 
considered in Ref. \cite{companion} do not commute 
with the Hamiltonian, yet commute with the momentum operator.
Since the components $q_{j_{x1}}$ and $q_{j_{x2}}$ of the $\alpha\nu$ band discrete momentum 
values ${\vec{q}}_j$ are eigenvalues of the corresponding $\alpha\nu$ translation 
generators $\hat{q}_{\alpha\nu\,x_1}$ and $\hat{q}_{\alpha\nu\,x_2}$, such momentum values are
not good quantum numbers and thus are not in general conserved,
again in contrast to 1D. Consistently, the states of Eqs. (\ref{non-LWS}) and (\ref{LWS-full-el})
of Appendix A generated by momentum occupancy configurations of such 
$\alpha\nu$ fermions are not in general energy eigenstates.

In turn, since for the model on the square lattice in the one- and two-electron
subspace the set of $\alpha\nu$ fermion numbers $\{N_{\alpha\nu}\}$
are conserved and read $N_{s1} = [S_c -S_s -2N_{s2}]$, $N_{s2}=0,1$,
$N_{s\nu}=0$ for $\nu\geq 3$, and $N_{\eta\nu}=0$ for $\nu\geq 1$ and
the two $s1$ translation generators $\hat{q}_{s1\,x_1}$ and $\hat{q}_{s1\,x_2}$ of Eq. (\ref{q-s1-xi}) 
in the presence of the fictitious magnetic 
field ${\vec{B}}_{s1}$ of Eq. (\ref{A-j-s1-3D}) commute with the Hamiltonian,
the states (\ref{LWS-1-2-el}) are both momentum and energy eigenstates \cite{companion}.
Consistently, it follows from the algebra of the $c$ and $s1$ fermion
operators of Eqs. (\ref{albegra-cf}) and (\ref{1D-anti-com}) and the 
relations of Eqs. (\ref{fc-q-x}) and (\ref{fs1-q-x})
that for subspaces with constant values of $S_c$, $S_s$, and $N_{s2}$
the set of states of form (\ref{LWS-1-2-el}) are
orthogonal and normalized. In addition, both for the model on the square 
and 1D lattices they span the subspaces of the one- and two-electron subspace 
for which the numbers $(S_c,S_s)$ are given by $(N/2,0)$ and $N_{s2}=0$ or $N_{s2}=1$, 
as well as its subspaces with $N_{s2}=0$ and either $(N/2\pm 1/2,1/2)$, $(N/2,1)$, $(N/2\pm 1,0)$, or $(N/2\pm 1,1)$ where
$N$ is the electron number of the initial $m=0$ ground state.
In such subspaces the states of form (\ref{LWS-1-2-el}) have been constructed to inherently
referring to a complete basis. 

Creation or annihilation of one $s1$ fermion is a well-defined process whose generator
is the product of an operator which fulfills small changes in the $s1$ effective lattice
and corresponding $s1$ momentum band and the operator 
$f^{\dag}_{{\vec{q}},s1}$ or $f_{{\vec{q}},s1}$, respectively, suitable to the excited-state subspace.
In the Slatter-determinant products of Eq. (\ref{LWS-1-2-el}) it is implicitly
assumed that the $s1$ momentum band is that of the state under consideration
so that the corresponding generators on the vacua are simple products of
$f^{\dag}_{{\vec{q}},s1}$ operators.
The $c$ effective lattice and number $N_a^D$ of $c$ band discrete momentum
values remain unaltered for the whole Hilbert space.
Concerning transitions between subspaces with different values for such momenta,
again creation or annihilation of one $c$ fermion is a well-defined process whose generator
is the product of an operator that fulfills small changes in the $c$ band momentum values and the operator 
$f^{\dag}_{{\vec{q}},c}$ or $f_{{\vec{q}},c}$, respectively, suitable to the excited-state subspace.
The operators of the vanishing-energy and zero-momentum $+1/2$ independent $\eta$-spinons, 
$+1/2$ independent spinons, and $s2$ fermion do not appear explicitly in the expression of the
generators onto the vacua of Eq. (\ref{LWS-1-2-el}) because the effects of such operators
are take into account by the above-mentioned two operators that fulfill small changes both in the $s1$ effective lattice
and corresponding $s1$ momentum band and in the $c$ momentum band, respectively.
Since in the case of the general states (\ref{LWS-1-2-el}) the operators are created onto the $c$ and
$s1$ momentum bands of the state under consideration such effects are accounted for implicitly.

In the particular case of the one- and two-electron
subspace as defined in Ref. \cite{companion}, the general
momentum expression introduced in that reference 
simplifies to, 
\begin{equation}
\vec{P} =\sum_{{\vec{q}}}{\vec{q}}\, N_c ({\vec{q}})
+ \sum_{{\vec{q}}}
{\vec{q}}\, N_{s1} ({\vec{q}}) \, .
\label{P-1-2-el-ss}
\end{equation}
Here $N_{c}({\vec{q}})$ and $N_{s1}({\vec{q}})$ 
are the expectation values of the momentum distribution-function operators of Eq. (\ref{Nc-s1op}).
That for the Hubbard model on the square lattice in
the one- and two-electron subspace the discrete momentum values
${\vec{q}}_j$ of the $c$ and $s1$ fermions are for $U/4t>0$ good quantum numbers implies that
the interactions of such objects have a residual character. That
allows one to express the excited-state quantities
in terms of the corresponding deviations $\delta N_{c} ({\vec{q}}_j)$
and $\delta N_{s1} ({\vec{q}}_j)$ of the $c$ and $s1$ fermion momentum distribution functions,
respectively, relative to the initial ground-state values. Alike in a Fermi liquid \cite{Landau,Pines}, one can then 
construct an energy functional whose first-order terms in such deviations
refer to the dominant processes and whose second-order
terms in the same deviations are associated with the
$c$ and $s1$ fermion residual interactions. Such a functional 
refers to the square-lattice quantum liquid whose $c$ and $s1$ fermion
energy dispersions are studied in this paper. 

The momentum area $(2\pi/L)^2\,N_{a_{s1}}^2$ and 
discrete momentum values number $N_{a_{s1}}^2=[S_c+S_s]$ of the $s1$ band 
are known. In turn, its shape remains an open issue. 
The shape of the $c$ band is the same as for the first Brillouin zone, yet 
the discrete momentum values may have small overall shifts under variations 
of $[N_{s1}+N_{s2}]$ generated by subspace transitions. Indeed, alike in 1D
each $s1$ (and $s2$) effective lattice site occupied by one $s1$ (and $s2$)
fermion corresponds to the spin degrees of freedom of the rotated-electron
occupancy of two (and four) sites of the original lattice whose charge
degrees of freedom are described by two (and four) $c$ fermions.
Such a sharing of the original-lattice sites by $c$ fermions and
$s1$ or $s2$ fermions, respectively, is behind the $c$ fermions feeling the Jordan-Wigner
phases (\ref{JW-phi}) of the $s1$ or $s2$ fermions created or annihilated under
subspace transitions. In the studies of this paper we use several approximations to obtain information on 
the $c$ and $s1$ bands momentum values and the corresponding $c$ and $s1$ 
fermion energy dispersions of the model on the square lattice in the one- and 
two-electron subspace, for which the $c$ and $s1$ fermion discrete momentum
values are good quantum numbers. In turn, the $s2$ band does not exist
for $N_{s2}=0$ subspaces such that $N_{a_{s2}}^D=0$ and for the $N_{s2}=1$ subspace the $s2$ momentum band
reduces to a single vanishing momentum value. The corresponding
$s2$ fermion has both vanishing momentum and energy.

The main difference relative to the usual
Landau's Fermi liquid theory \cite{Landau,Pines} is that 
upon adiabatically turning off the interaction $U$ the $c$ and $s1$
fermions do not become electrons. Indeed, the non-interacting limit of the
present objects corresponds instead to $t^2/U\rightarrow 0$. For the $c$ and $s1$
fermions the second-order terms of the above energy
functional are associated with residual interactions. However, the 
effects of the original electronic correlations are accounted
for both by its first-order and second-order terms. 
In Section IV the ground-state 
normal-ordered Hamiltonian is described in terms of 
the square-lattice quantum liquid of $c$ and $s1$ fermions
with residual interactions and the corresponding energy functional
is studied up to first order in the $c$ and $s1$ fermion
hole momentum distribution-function deviations. The
second-order energy terms of that functional are discussed in Ref. \cite{cuprates0}
for an extended square-lattice quantum liquid perturbed by small 3D anisotropy
effects. The studies of this paper
do not include the effects of the phase fluctuations
associated with the superconducting order of that
extended quantum problem, which are
found in that reference to lead to a new first-order
contribution. 

\subsection{Subspace-dependent deviations of the $c$ and $s1$
band discrete momentum values in 1D and 2D}

Since according to the analysis of Ref. \cite{companion} the $c$ and $s1$ band 
discrete momentum values are for the Hubbard model on a square lattice
in the one- and two-electron subspace good quantum numbers they deserve further investigations. 
In contrast to the momentum values of the quasiparticles of a
Fermi liquid, the number $N_{a_{s1}}^D =[S_c+S_s]$ of $s1$ fermion
discrete momentum values may change upon state transitions
involving variations of $S_c$ and/or $S_s$. In addition to changes
in the number of discrete momentum values, subspaces spanned
by states with different $N_{a_{s1}}^D =[S_c+S_s]$ values may
have slightly different sets of discrete momentum values. The
number $N_a^D$ of discrete momentum values of the
$c$ band remains unaltered for the whole Hilbert space.
However, alike in 1D such momentum values may also slightly change upon variations of
the number $[B_s+B_{\eta}]=\sum_{\alpha\nu}N_{\alpha\nu}$, which
for subspaces contained in the one- and two-electron subspace
simplifies to $[B_s+B_{\eta}]=[N_{s1}+N_{s2}]=[S_c -S_s -N_{s2}]$. Alike in 1D, for the Hubbard 
model on the square lattice the $c$ fermions feel the Jordan-Wigner 
phases of the $s1$ (and $s2$) fermions created or annihilated (and created) under subspace transitions that 
do not conserve the number $[B_{\eta}+B_s]=\sum_{\alpha\nu}N_{\alpha\nu}$.
The $c$ and $s1$ band momentum shifts originated by transitions between such
subspaces considered in the following refer both to the model on the 1D and square lattices. In turn, the 
$s1$ band boundary line deformations, which in the present $N_a^2\rightarrow\infty$ limit 
can be described by $s1$ band rotations, as well as the concept of $s1$ fermion doublicity 
studied in the ensuing subsection are specific
to the one-electron excited states of the model on the square lattice and thus to the
corresponding square-lattice quantum liquid.

The $c$ fermion operators $f^{\dag}_{{\vec{q}},c}$ and
$f_{{\vec{q}},c}$ refer to a $c$ momentum band 
whose momenta are related to the real-space coordinates of
the $c$ effective lattice through Eq. (\ref{fc-q-x}).
The latter lattice is identical to the original lattice and in 
the local $c$ fermion occupancies whose superposition gives $c$ fermion
momentum occupancy configurations that generate the states
(\ref{LWS-1-2-el}) the $c$ fermions occupy the sites
singly occupied by the rotated electrons in the latter lattice. Consistently, for all 
energy eigenstates the momentum area 
$(2\pi/L)^2\,N_a^2$ and shape 
(and momentum width $(2\pi/L)\,N_a$) of the $c$ 
band are (and is) for the model on the square (and 1D) lattice the same as for the 
first Brillouin zone. Out of the $N_a^D$ discrete
momentum values ${\vec{q}}_{j}$ where
$j=1,...,N_a^D$, $N_c=2S_c$ are occupied and 
$N_c^h=[N_a^D-2S_c]$ are unoccupied and
correspond to $c$ fermions and $c$ fermion holes,
respectively. 

In turn, the $s1$ fermion operators $f^{\dag}_{{\vec{q}},s1}$ and
$f_{{\vec{q}},s1}$ are associated with a $s1$ momentum band
related to the $s1$ effective lattice through Eq. (\ref{fs1-q-x}).
As confirmed below, such a $s1$ momentum
band is exotic. It is full for the $x\geq 0$ and $m=0$ 
ground states considered here so that there is no $s1$
Fermi line and as given in Table \ref{tableIV} of Appendix A for 
one- and two-electron excited states it has one hole and none or 
two holes, respectively. Moreover, for the model on the square lattice the shape of its boundary line is subspace dependent. 
That line is strongly anisotropic for that model, its energy dependence being $d$-wave
like: As confirmed in Section IV, it vanishes for excitation momenta pointing in 
the nodal directions and reaches its maximum magnitude for 
those pointing in the anti-nodal directions.

Let ${\vec{q}}^{\,d}_{Bs1}$ (and $\pm q_{Bs1}$) be momentum values belonging to
the boundary line (and being the two boundary points). For the model on the
square (and 1D) lattice such a line encloses (and such points limit) the $s1$ band $N^D_{a_{s1}}$ 
discrete momentum values. The quantity $d=\pm 1$ in the boundary-line momentum 
${\vec{q}}^{\,d}_{Bs1}$ of the model on the square lattice is the doublicity.
It refers to two alternative subspace-dependent shapes of that line, which for 
the subspace spanned by one-electron excited
states result from well-defined $s1$ boundary line deformations
relative to the initial ground-state $s1$ band, as discussed below 
in Subsection III-C. The $s1$ fermions whose occupancy
configurations generate a given one-electron excited state have
the same doublicity $d$. In spite of the $c$ band not being deformed
upon one-electron excitations, for $x>0$ the doublicity of $c$ fermions at or near the 
$c$ Fermi line is considered to be that
of the $s1$ fermion holes created along with the creation or annihilation
of the $c$ fermion under consideration. Hence, the doublicity $d$ is
the same for the $s1$ fermion holes and $c$ fermions created or annihilated upon an one-electron
excitation. The doublicity $d=\pm 1$ and the corresponding two types 
of $s1$-band boundary line shapes occur only for one-electron excited states
whereas both for $x\geq 0$ and $m=0$ ground states and their
two-electron excited states one finds in that subsection
that ${\vec{q}}^{\,+1}_{Bs1}={\vec{q}}^{\,-1}_{Bs1}\equiv {\vec{q}}_{Bs1}$.

For the model on the square lattice the $s1$ boundary-line momentum values
${\vec{q}}^{\,d}_{Bs1}$ are well defined for each subspace spanned by states with constant
$U/4t$, $S_c$, $S_s$, and thus $N_{a_{s1}}^2=[S_c +S_s]$ values. The precise form of 
such a boundary line is in general an open problem, yet we know
that it encloses a momentum area given by
$(2\pi/L)^2\,N^2_{a_{s1}}$. Fortunately, the problem
can be solved for the $x=0$ and $m=0$ absolute
ground state, as confirmed in Section IV-E. Below we also address
it for some of the subspaces spanned by excited states generated from application
of one- and two-electron operators onto $x>0$ and $m=0$ 
ground states.

Alike in 1D, one can in general decouple the components ${q_j}_{x_1}$ and 
${q_j}_{x_2}$ of the discrete momentum values ${\vec{q}}_j$ of the $s1$ fermions 
into a bare part of the form $[2\pi/L]\,{\cal{N}}_i$ where $i=1,2$
and  ${\cal{N}}_i=0,\pm 1,\pm 2,...$ and a small momentum deviation whose value is the
same for all momentum values. As in 1D, the separation of ${q_j}_{xi}$ in a bare 
component $[2\pi/L]\,{\cal{N}}_i$ and a small deviation 
$({q_j}_{xi}-[2\pi/L]\,{\cal{N}}_i)$ is just a matter of convenience. The same separation can be
fulfilled for the components of the $c$ band microscopic momenta so that for the model on the square 
lattice the $c$ and $s1$ band discrete momentum values may be written in following general form,
\begin{eqnarray}
{\vec{q}}_j & = & {\vec{k}}_j + {\vec{q}}_{\gamma}^{\,0}/N_{\gamma}
= \left[\begin{array}{c}
{q_j}_{x_1} \\
{q_j}_{x_2}
\end{array} \right] 
\, ; \hspace{0.5cm} 
{\vec{k}}_j = {2\pi\over L}
\left[\begin{array}{c}
{\cal{N}}_1^{\gamma} \\
{\cal{N}}_2^{\gamma}
\end{array} \right]
\, ; \hspace{0.5cm} 
{\vec{q}}_{\gamma}^{\,0} = 
\left[\begin{array}{c}
q_{\gamma\,x_1}^{0} \\ 
q_{\gamma\,x_2}^{0}
\end{array} \right] \, ,
\nonumber \\
{q_j}_{xi} & =  & {2\pi\over L}{\cal{N}}^{\gamma}_i +
q_{\gamma\,xi}^{0}/N_{\gamma}
\, ; \hspace{0.5cm}
{\cal{N}}_i^{\gamma} 
= 0,\pm 1, \pm 2,... \, ,
\, ; \hspace{0.5cm} 
\gamma = c, s1
\, ; \hspace{0.50cm}
j = 1,...,N_{a_{\gamma}}^2 
\, ; \hspace{0.50cm} i=1,2 \, ,
\label{q-j-f-Q-c-0-s1-2D}
\end{eqnarray}
where $N_{a_{c}}^2=N_{a}^2$.

Since for all energy and momentum eigenstates (\ref{LWS-1-2-el})
the momentum area and shape of the $c$ momentum band 
are the same as for the original Brillouin zone, the integer numbers 
${\cal{N}}^{c}_i$ associated with the Cartesian components of the bare 
momenta ${\vec{k}}_j$ appearing in this equation read,
\begin{equation}
{{\cal{N}}}^{c}_i = j_i - {N_{a}\over 2}
=0,\pm 1, \pm 2,... \, ,
\, ; \hspace{0.50cm}
j = (j_2 -1) N_{a} + j_1 = 1,...,N_{a}^2 
\, ; \hspace{0.50cm}
j_i = 1,...,N_{a}
\, ; \hspace{0.50cm} i=1,2 \, ,
\label{numbers-Nci}
\end{equation}
and are such that $\vert{\cal{N}}^{c}_i\vert\leq N_a/2$. 
In turn, the shape of the $s1$ band is subspace dependent so that
the limiting values of the integer numbers 
${\cal{N}}^{s1}_i$ associated with the Cartesian components of the bare 
momenta ${\vec{k}}_j$ of Eq. (\ref{q-j-f-Q-c-0-s1-2D}) depend
on the shape of its boundary line for each subspace and value of $U/4t$. In turn,
the number of discrete momentum values ${\vec{q}}_j$
of that equation is always given by $N_{a_{s1}}^2=[S_c +S_s]$ so that
$j = 1,...,N_{a_{s1}}^2$ for all subspaces with constant $[S_c +S_s]$ value.  

Subspaces with different values of $S_c$, $S_s$, and $N_{a_{s1}}^2=[S_c +S_s]$
have in general different $s1$ band discrete momentum values ${\vec{q}}_j$ as well so that in addition
to a change in the number $N_{a_{s1}}^2=[S_c +S_s]$ of such values the corresponding
subspace transitions may involve effects such as $s1$ band overall shifts by the same
small momentum $\delta{\vec{q}}_{s1}^{\,0}/N_{a_{s1}}^2$ 
and $s1$ band boundary line deformations, which in the present $N_a^2\rightarrow\infty$ 
limit may be described by $s1$ band rotations. For momenta near the $s1$ boundary line 
such rotations are defined by the $F$ angle introduced in Subsection III-C. 
Subspace transitions involving changes in the $[S_c -S_s-N_{s2}]$ value
may involve $c$ band overall shifts by the same small momentum $\delta{\vec{q}}_{c}^{\,0}/N_{c}^2$.
Both for the model on the 1D and square lattices symmetry implies that the momentum shift 
$\delta{\vec{q}}_{s1}^{\,0}/N_{s1}$ originated by subspace transitions is such that,
\begin{equation}
\sum_{j=1}^{N_{a_{s1}}^D}{\vec{q}}_j=0 \, ,
\label{sum-q-0}
\end{equation}
both for the initial and final subspaces. Consistently, for the model on the square (and 1D) lattice the $s1$ band refers to a compact 
momentum area $(2\pi/L)^2\,[N_{a_{s1}}]^2$ (and momentum width
$(2\pi/L)\,[N_{a_{s1}}]$) centered at zero momentum. The physical meaning of Eq. (\ref{sum-q-0})
is that when a subspace transition leads to a deviation
$\delta N_{a_{s1}}^D = \delta [S_c + S_s]=\pm 1$ associated
with addition $(+1)$ or removal $(-1)$ of a discrete momentum
value ${\vec{q}}_{j_0}$ to or from the $s1$ momentum band,
respectively, the corresponding momentum shift 
$\delta {\vec{q}}_{s1}^{\,0}/N_{s1}$ is such that
$\delta {\vec{q}}_{s1}^{\,0}=\mp{\vec{q}}_{j_0}$.
That exact cancelation of the momentum 
$\pm {\vec{q}}_{j_0}$ by the momentum shift
generated by the subspace transition
is a symmetry equivalent to the sum-rule (\ref{sum-q-0}) always 
being fulfilled.

The momentum ${\vec{q}}_{c}^{\,0}$ of Eq. 
(\ref{q-j-f-Q-c-0-s1-2D}) is well defined for the subspaces of the model on the square 
lattice spanned by states with the same $[S_c -S_s-N_{s2}]$ value. However,
one cannot in general access it by the use of symmetry alone, 
except in some cases. For instance, for hole concentrations
$x\geq 0$ its shift $\delta{\vec{q}}_{c}^{\,0}$
can be derived for transitions from $x>0$ and $m=0$ ground states
to excited states generated by one-electron processes. For an excited state generated
by addition of one electron onto such a ground state one has
that $\delta {\vec{q}}_{c}^{\,0}=0$. In turn, for an excited state generated
by removal of one electron from the same initial ground state
it is found in the ensuing subsection that the value of the 
corresponding shift $\delta {\vec{q}}_{c}^{\,0}$ is such 
that the Fermi line has the same form as for 
one-electron addition. 

\subsection{Fermi line of the square-lattice quantum liquid}

\subsubsection{Three hole concentration ranges according to the Fermi-line anisotropy}

According to Luttinger's theorem, the momentum area (and width) enclosed (and
limited) by the Fermi line (and points), when centered at 
zero momentum, is for the Hubbard model on the square
(and 1D) lattice $(1-x)2\pi^2$ (and $(1-x)\pi$). Given the non-perturbative 
character in terms of electrons of the Hubbard model on both the
square and 1D lattices, such Fermi line and points, respectively,
could correspond to the generalized {\it Luttinger line} 
and {\it points} defined in Ref. \cite{Lutt-Theorem}. That line (and points) 
is (and are) determined not only by poles in the one-electron Green's functions, 
but also by their zeros. It is argued below that for $x>x_{c1}$ where
$x_{c1}\approx 0.13$ for intermediate $U/4t$ values the Fermi line
of the model on the square lattice obeys Luttinger's theorem.
 
The non-perturbative effects are strongest in the case of the model on 
the 1D lattice. Fortunately, for the 1D Hubbard model one can profit 
from its exact solution to study the problem. 
Then we define the Fermi points as the
momenta where the real part of the one-electron Green's function
has strongest singularities at very small excitation energy
$\omega$. For $0\leq x\leq 1$, $0\leq m\leq (1-x)$, and $U/4t>0$ such
singularities occur at the $U=0$ Fermi momenta $\pm k_{F\sigma}$,
as given in Eqs. (25) and (26) of Ref. \cite{CPP}. For the ranges 
$0\leq x\leq 1$, $m=0$, and $U/4t>0$ considered here 
the two exponents provided in Eq. (26) of that reference are
identical and given in Table II of the same paper for magnetic field
$H\rightarrow 0$. Then the singularities occur for $U/4t>0$ at the 
$U=0$ Fermi momenta $\pm k_{F}=\pm (1-x)\pi/2$ so that the 
Fermi points limit a momentum width $(1-x)\pi$ for all values of $U/4t$, 
as required by the usual Luttinger's theorem. 

For the 1D Hubbard model the one-electron renormalization factor
vanishes at the Fermi points as defined above. For general correlated
2D problems, if such a factor is finite for the whole Fermi line 
one defines that line in the usual way. In turn, if it is finite for some 
Fermi-line parts and vanishes for other parts, it could be defined as for 1D
in the latter parts. Finally, if the one-electron renormalization factor
vanishes for the whole Fermi line it could be defined as the set of
momenta belonging to a line in momentum space where the real 
part of the one-electron Green's function has strongest singularities 
at very small excitation energy $\omega$, alike in 1D. 

Let us consider a first Brillouin zone of the model on the square lattice centered at
the momentum $-\vec{\pi}=[-\pi,-\pi]$ and define
the Fermi line in terms of a hole Fermi momentum 
$\vec{k}_F^{\,h}$ related to the Fermi momentum
$\vec{k}_F$ as, 
\begin{eqnarray}
\vec{k}_F^{\,h} & = & \vec{k}_F +\vec{\pi} = k_F^h (\phi)\,{\vec{e}}_{\phi} \, ;
\hspace{0.50cm}
\vec{\pi} = \left[\begin{array}{c}
\pi \\
\pi
\end{array} \right] 
\, ; \hspace{0.50cm}
\phi = \arctan \left({k^h_{Fx_2}\over 
k^h_{Fx_1}}\right) 
\, ; \hspace{0.50cm}
\phi \in (0,2\pi) \, , \hspace{0.25cm} 0 < x < x_h \, ,
\nonumber \\
\phi & \in & \left(\phi_{AN}, {\pi\over 2} -\phi_{AN}\right) \, ;
\hspace{0.10cm} 
\left( {\pi\over 2} +\phi_{AN}, \pi -\phi_{AN}\right) \, ;
\nonumber \\
& \in & \left(\pi + \phi_{AN}, {3\pi\over 2} -\phi_{AN}\right)  \, ;
\hspace{0.10cm} 
\left({3\pi\over 2} +\phi_{AN}, 2\pi -\phi_{AN}\right) 
\, , \hspace{0.25cm} x_h < x < x_* \, ,
\label{kF}
\end{eqnarray}
where $\phi$ is the Fermi-line angle that defines the
direction of $\vec{k}_F^{\,h}$ and we assume that the
hole concentration $x_h$ above which the Fermi line is
particle like belongs to the range $x_{c2}\leq x_h\leq x_*$.
Here the hole concentration $x_{c2}$ obeys the inequality $2x_*/3\leq x_{c2}\leq x_*$
and is defined in Ref. \cite{cuprates0} upon application 
of a uniform magnetic field aligned perpendicular to the square-lattice plane yet it has 
physical meaning in the absence of such a field as well,
the critical concentration $x_*$ in Subsection IV-F,
and the angle $\phi_{AN}=\phi_{AN} (x)\geq 0$ vanishes for
$x\leq x_h$ and is small (or vanishes if $x_h\geq x_*$) for $x\in (x_h,x_*)$. 

Consistently with the numbers and number deviations given in
Table \ref{tableIV} of Appendix A for one-electron excited states and results of Ref. \cite{companion},
creation of one electron at the Fermi line leads to creation of one $c$ fermion
at the $c$ Fermi line and one hole at the $s1$ boundary line
so that the hole Fermi momentum $\vec{k}_F^{\,h}$ of Eq. (\ref{kF}) can be written as,
\begin{equation}
\vec{k}_F^{\,h} = [{\vec{q}}_{Fc}^{\,h\,d} - {\vec{q}}^{\,d}_{Bs1}]
= k^h_{F} (\phi)\,{\vec{e}}_{\phi} \, .
\label{kF-qFc-qBs1}
\end{equation}
Here the $c$ fermion Fermi momentum ${\vec{q}}_{Fc}^{\,h\,d}$ and
$s1$ boundary-line momentum ${\vec{q}}^{\,d}_{Bs1}$ are given below
and as confirmed in Section V the doublicity $d=\pm 1$ refers to two types of one-electron excited
states with the same hole Fermi momentum $\vec{k}_F^{\,h}$ and
Fermi energy $E_F$ but different Fermi velocity $V_F^d$.

For $0<x<x_*$ and intermediate $U/4t$ values obeying approximately 
the inequality $u_0<U/4t<u_1$, where $u_0\approx 1.302$ is the $U/4t$ value 
at which the spinon pairing energy $2\Delta_0$ introduced in Ref. \cite{companion} corresponding to $0<x\ll 1$ 
has its absolute maximum magnitude and $u_1\approx 1.6$, the Fermi line level of anisotropy is controlled by 
the interplay of the $s1$ boundary line anisotropy and $c$ Fermi line isotropy, 
as discussed in the following. The degree of Fermi-line anisotropy
has impact on the form of the spectrum of the energy eigenstates. For instance, for the reduced 
one-electron subspace defined below the doublicity $d=\pm 1$ 
labels one-electron excited states with the same energy and 
momentum but different $d$-dependent velocities. However, we find in Subsection
V-B that for hole concentrations where the Fermi-line anisotropy is small these
two velocities are $d$ independent and equal. Since the doublicity effects on the
spectrum are for the electronic-operator description non-perturbative,
that non-perturbative character is strongest for
hole concentrations where the Fermi-line anisotropy is largest. We find below
that the Fermi-velocity anisotropy is small for $x>x_{c1}$ where $x_{c1}\approx 0.13$
for the intermediate $U/4t$ values $U/4t\in (u_0,u_1)$.
According to our scheme, that the Fermi-velocity anisotropy is small implies that 
for $x>x_{c1}$ the Fermi line encloses the momentum area $(1-x)2\pi^2$ predicted by
Luttinger's theorem. This is consistent with the experimental studies of Ref. \cite{k-r-spaces}, which for some 
hole-doped cuprates find that the Fermi line obeys Luttinger's theorem for approximately $x>0.10$.
In turn, we find below in Subsection IV-E that for the $x=0$ and $m=0$ Mott-Hubbard insulating phase 
the Fermi line has a shape independent of $U/4t$ and encloses the expected momentum area 
$2\pi^2$, in spite of the Mott-Hubbard gap. However, our results
are inconclusive concerning the validity of Luttinger's theorem for the
range $x\in (0,x_{c1})$ where the Fermi-line anisotropy is largest. Indeed, our 
investigations about that Fermi line do not focus on that range. 

The anisotropy of the Fermi line results from that of the $s1$ boundary line.
For $s1$ band momenta belonging to that line the absolute value of the $s1$ fermion velocity 
$V^{\Delta}_{Bs1} \equiv V^{\Delta}_{s1} ({\vec{q}}^{\,d}_{Bs1})$ derived in Section V
is for $0<x<x_*$ dependent on $\phi$. The corresponding Fermi energy has
the form $E_F=\mu +\delta E_F$ where $\mu$ is the zero-temperature
chemical potential and $\delta E_F=\vert\Delta\vert\vert\cos 2\phi\vert$ the anisotropic part
arising from the $s1$ boundary-line energy $\delta E_F$.
It reaches its maximum magnitude ${\rm max}\,\{\delta E_F\}=\vert\Delta\vert$ for hole Fermi momenta 
pointing in the anti-nodal directions and vanishes when such momenta
point in the nodal directions. Here $\vert\Delta\vert$ is the energy scale
introduced in Ref. \cite{companion} whose $0<x\ll 1$ expression $\vert\Delta\vert\approx \Delta_0 (1-x/x_*)$
given in that reference is argued in Subsection IV-F to be valid for
the hole concentrations $0<x<x_*$ in the approximate range $u_0\leq U/4t\leq u_{\pi}$ where
$x_*$ increases smoothly from $x_*\approx 0.23$ at $U/4t=u_0$ 
to $x_*\approx 0.32$ at $U/4t =u_{\pi}>u_1$. 
Whether for such a $U/4t$ range the hole concentration $x_h$ at which that line becomes particle like
is given by $x_h\approx x_{c2}$, is such that $x_{c2}<x_h<x_*$, is given by $x_h\approx x_*$,
or is larger than $x_*$ remains an open question. 
In the studies of this paper and Refs. \cite{cuprates0,cuprates}
we assume that $x_{c2}\leq x_h\leq x_*$. For $x>x_h$ the
Fermi line is particle like. Hence one has that ${\rm max}\,\{\delta E_F\}=\vert\Delta\vert\vert\cos 2\phi_{AN}\vert$
where $\phi_{AN}=\phi_{AN}(x)$ is the angle of Eq. (\ref{kF}), which 
vanishes for $x\leq x_h$ and is small for $x\in (x_h,x_*)$. As discussed
in Section V, we consider only corrections of first order in $\phi_{AN}$ so 
that ${\rm max}\,\{\delta E_F\}=\vert\Delta\vert$ for $0<x<x_*$ 
and $u_0\leq U/4t\leq u_{\pi}$, consistently with
${\rm max}\,\{\delta E_F\}\approx \vert\Delta\vert [1-2\phi_{AN}^2]\approx \vert\Delta\vert$
for $x_h<x<x_*$. In turn, $V^{\Delta}_{Bs1} \equiv V^{\Delta}_{s1} ({\vec{q}}^{\,d}_{Bs1})$
reaches its maximum magnitude $\vert\Delta\vert/\sqrt{2}$ for momenta 
pointing in the nodal directions and for $0<x<x_*$ vanishes for momenta
pointing in the anti-nodal directions. If $x_h<x_*$ one finds instead that 
for $x_h< x\leq x_*$ the minimum value of $V^{\Delta}_{Bs1}$ is small but finite and given 
by $\vert\Delta\vert\vert\sin 2\phi_{AN}\vert/\sqrt{2}
\approx [\phi_{AN}\,\sqrt{2}]\vert\Delta\vert$. 

The anisotropic Fermi energy part $\delta E_F$ is found in Section V to equal the $\phi$-dependent energy of the 
$s1$ fermion dispersion at the $s1$ boundary line. The absolute value $V_F^d$ of the Fermi velocity   
is found to be a function of both the $s1$ fermion velocity 
$V^{\Delta}_{Bs1} \equiv V^{\Delta}_{s1} ({\vec{q}}^{\,d}_{Bs1})$ and
$c$ fermion velocity $V_{Fc} \equiv V_{c} ({\vec{q}}^{\,h\,d}_{Fc})$ such that
at the hole concentration range for which ${\rm max}\,V^{\Delta}_{Bs1}/V_{Fc}\ll 1$
one has that $V_F^d\approx V_F=V_{c}$. Such an equality is not met 
the Fermi velocity $V_F^d$ is dependent on both $\phi$ and $d=\pm 1$ and the Fermi line
is that of an anisotropic system. Such a dependence on $\phi$ follows from that of the $s1$ boundary-line velocity
$V^{\Delta}_{Bs1}$ on that angle. In turn, for approximately $x>x_{c1}$ one finds that 
${\rm max}\,V^{\Delta}_{Bs1}/V_{Fc}\ll 1$ and $V_F^d\approx V_F=V_{Fc}$,
where the $c$ fermion Fermi velocity absolute value is $\phi$ independent and given by $V_{Fc}
\approx [\sqrt{x\pi}\,2/m_c^*]$ so that the corresponding Fermi line
is that of an isotropic system. Here $m^{*}_{c}$ is the $c$ fermion energy-dispersion mass  
introduced below in Subsection IV-D. The Fermi-velocity anisotropy
coefficient $\eta_{\Delta}\equiv {\rm max}\,V^{\Delta}_{Bs1}/V_{Fc}$ provides
a measure of such a Fermi-line anisotropy, that line fully becoming that of
an isotropic system as $\eta_{\Delta}\rightarrow 0$ for $x\rightarrow x_*$.

A complementary coefficient, which also provides a measure of the Fermi-line anisotropy 
is the Fermi-energy anisotropy coefficient $\eta_0\equiv \vert\Delta\vert/W_c^h$.
The zero-temperature chemical potential $\mu =[E_F-\delta E_F]$ is for $0<x<x_*$ 
and $u_0\leq U/4t\leq u_1$ approximately given by $\mu\approx \mu^0 +W_c^h$.
This expression is exact for small hole concentrations $x\ll 1$ and a good approximation for
the range $0<x<x_*$ provided that $U/4t\in (u_0,u_1)$. In addition to $\mu^0=\lim_{x\rightarrow 0}\mu$, it 
involves the energy scale $W_c^h=[8t-W_c^p]$ where $W_c^p$ is the energy 
bandwidth of the $c$ fermion filled sea of Ref. \cite{companion}. 
The energy bandwidth $W_c^h$ vanishes in the limit $x\rightarrow 0$ and
corresponds to a contribution to the zero-temperature chemical potential of
the rotated-hole kinetic energy. Indeed, we recall that the $c$ fermion holes describe the
charge degrees of freedom of the rotated-electron holes whose concentration $x$
equals that of the electronic holes. Hence $W_c^h$ is the isotropic Fermi energy
measured from the upper level of the unfilled band. Indeed, the chemical potential
range $\mu\in (-\mu_0,\mu_0)$ refers to half filling so that for $x>0$ one has that
$\mu>\mu_0$. Therefore, for $x>0$ rather than $\vert\Delta\vert/\mu$
with $\mu\approx \mu^0 +W_c^h$, the coefficient of interest to provide a measure the Fermi-energy 
anisotropy is $\eta_0= \vert\Delta\vert/W_c^h$.

It is found in Section IV that in the units used here the mass $m_c^*$ appearing
in the $c$ Fermion energy dispersion reads $m_c^*=1/2r_c t$
where the ratio $r_c=m_*^{\infty}/m_c^*$ involves the $U/4t\rightarrow\infty$
bare $c$ fermion mass $m_c^{\infty}=1/2t$. For approximately the range 
$u_0\leq U/4t\leq u_1$ the magnitude of such a ratio is given by $r_c\approx 2r_s$ where the spin ratio 
$r_s=\Delta_0/4W_{s1}^0$ is introduced in Ref. \cite{companion} and studied below in Subsection IV-F.
For such a range of $U/4t$ values they read $r_c\approx 2r_s \approx \pi x_*$
where the critical hole concentration $x_*$ is studied in that subsection.
The expressions of the anisotropy coefficients given in the following can be
extended to the range $U/4t\in (u_0, u_{\pi})$ where $u_{\pi}>u_1$
is the $U/4t$ value considered below in Subsection IV-F at which 
$x_*=1/\pi\approx 0.32$. However, here we provide such expressions for the approximate range
$U/4t\in (u_0, u_1)$ where their form simplifies. For $0<x<x_*$ and the latter $U/4t$ range the
coefficients are given by, 
\begin{eqnarray}
\eta_{\Delta} & = & {\rm max}\,r_{\Delta} \approx \sqrt{x\pi\over 2}\eta_0
\, ; \hspace{0.35cm} 
r_{\Delta} = {V^{\Delta}_{Bs1}\over V_{Fc}}
\, ; \hspace{0.35cm} \eta_0 =
{\vert\Delta\vert\over W_c^h} \approx \sqrt{2x_{\Delta}\over \pi}\left({1\over x}-{1\over x_*}\right) 
\, ; \hspace{0.35cm}
x_{\Delta} = {1\over 2\pi}\left({\Delta_0\over 4r_c t}\right)^2 \, ,
\nonumber \\
x_0 & = & {\Delta_0\over 4r_c^2t+\Delta_0}\,x_*
\, ; \hspace{0.35cm}
x_{c1} = x_*\,[F_c -\sqrt{ F_c^2-1}] 
\, ; \hspace{0.35cm} 
F_c = 1 + {4r_c\over \pi^2(1+\Delta_0/4r_c^2t)^2}
\, ; \hspace{0.35cm} 
2x_*/3\leq x_{c2}\leq x_* \, .
\label{x-h}
\end{eqnarray}
The hole concentrations $x_{\Delta}$ and $x_0$ also given here are those at which $\eta_{\Delta} =1$
and $\eta_0=1$, respectively, and the hole concentrations $x_{c1}$ that at which $\eta_{\Delta}=2x_0$. 
For hole concentrations above $x_{c1}$ the Fermi-velocity coefficient $\eta_{\Delta}$ is considered small. 
In the present context the hole concentration $x_{c2}$ defined in Ref. \cite{cuprates0} is that
above which both the anisotropy coefficients $\eta_{\Delta}$ and $\eta_0$ are considered yo be small.
The expressions $V_{Fc}\approx [\sqrt{x\pi}\,2/m_c^*]$ and $W_c^h\approx [x\pi 2/m_c^*]$ used
to derive the expression of such coefficients given in Eq. (\ref{x-h}) are 
a good approximation for the range $0<x<x_*$ provided that $u_0\leq U/4t\leq u_1$. Indeed, for small $U/4t$ the 
$c$ fermion mass $m_c^*$ introduced in Subsection IV-D becomes large and behaves as 
$1/m_c^*\rightarrow 0$ for $U/4t\rightarrow 0$ so that for $0<U/4t\ll 1$ the expressions 
$V_{Fc}\approx [\sqrt{x\pi}\,2/m_c^*]$ and $W_c^h\approx [x\pi 2/m_c^*]$ are valid for
$x\ll 1$ only whereas for $U/4t>u_1$ the hole concentration $x_*$ is larger than $0.28$.

The magnitudes of $x_{\Delta}$, $x_0$, and $x_{c1}$ and that of $x_*\approx 2r_s/\pi\approx r_c/\pi$ 
are provided in Table I for $U/4t=u_0$ and $U/4t=u_*$. The value $U/4t=u_*= 1.525$ is found in Refs. \cite{cuprates0,cuprates} to be suitable for the 
description of several families of hole-doped cuprates with superconducting zero-temperature 
critical hole concentrations $x=x_c\approx 0.05$ and $x=x_*\approx 0.27$. The magnitudes of the 
coefficients $\eta_{\Delta}$ and $\eta_0$ are for $U/4t\approx u_*$ provided in Table II for several values 
of the hole concentration. To reach those and the $x_{\Delta}$ and $x_0$ values given above we used the 
magnitude $\Delta_0\approx 0.285\,t$ found in Subsection VI-B for $U/4t\approx u_*$.
 \begin{table}
\begin{tabular}{|c|c|c|c|c|c|} 
\hline
$U/4t$ & $x_{\Delta}$ & $x_0$ & $x_{c1}$ & $x_*$ \\
\hline
$u_0$ & $1.9\times 10^{-3}$ & $3.0\times 10^{-2}$ & 0.12 & $0.23$ \\
\hline
$u_*$ & $1.1\times 10^{-3}$ & $2.4\times 10^{-2}$ & 0.13 & $0.27$ \\
\hline
\end{tabular}
\caption{Magnitudes of the hole concentrations $x_{\Delta}$, $x_0$, and $x_{c1}$ given in Eq.
(\ref{x-h}) and critical hole concentration $x_*$ introduced in Subsection IV-F for 
$U/4t=u_0\approx 1.302$ and $U/4t= u_* =1.525$. 
The interplay of the Fermi-line anisotropy and the physics of the different hole-concentration 
ranges discussed in the text refers to intermediate $U/4t$ values $u_0\leq U/4t\leq u_1$.}
\label{tableI}
\end{table} 

We emphasize that the hole concentrations $x_{\Delta}$, $x_0$, and $x_{c1}$ do not mark sharp phase
transitions. They refer instead to crossovers between different levels of Fermi-line anisotropy occurring
for approximately the range $u_0\leq u\leq u_1$. Whether $x_{c2}$ marks a crossover or refers to a critical point remains
an open question \cite{cuprates}. The Fermi line
anisotropy has important effects on the physics of the square-lattice quantum liquid and, therefore, these
hole concentrations mark crossovers between regimes where some of the physical properties are
different. Often we focus mainly our analysis on
the value $U/4t\approx u_* =1.525\in (u_0,u_1)$ of interest for the physics of several families of hole-doped 
cuprate superconductors \cite{cuprates0,cuprates}. For $u_0\leq U/4t\leq u_1$ it is considered that the Fermi-velocity 
anisotropy is small when $\eta_{\Delta}<2x_0$ and thus $x>x_{c1}$ where $\eta_{\Delta}\approx 0.048$
and $x_{c1}\approx 0.13$ for $U/4t\approx u_*$ and that above the hole concentration 
$x_{c2}$ both the Fermi-velocity and Fermi-energy anisotropies are very small. The hole concentration
$x_{c2}$ marks the beginning of the crossover to a particle-like Fermi line and
the accurate value $x_h\geq x_{c2}$ at which the Fermi line becomes actually particle like
remains an unsolved problem.
If as assumed above $x_h<x_*$ then for $0<x<x_h$ 
the Fermi line angle $\phi$ belongs to the range given in Eq. (\ref{kF}),
where $\phi_{AN}=\phi_{AN} (x)$ is a small angle that vanishes 
for $x\leq x_h$.

The above analysis leads to three main ranges of hole concentrations where the system has different levels of 
Fermi-line anisotropy. The first range refers to hole concentrations smaller than 
$x_{c1}\approx 0.13$. Then according to Table II both coefficients are larger than $2x_0\approx 0.05$ and the Fermi-line 
anisotropy is not small and plays an important role in the physics. This is the anisotropic Fermi-line hole-concentration
range. It may be further divided 
into three subranges corresponding to $0<x<x_{\Delta}$, $x_{\Delta}<x<x_0$, and $x_0<x<x_{c1}$, respectively.
For $0<x<x_{\Delta}$ both ${\rm max}\,V^{\Delta}_{Bs1}> V_{Fc}$ and $\delta E_F>W_c^h$ so that
the Fermi-line anisotropy is huge. In the range $x_{\Delta}<x<x_0$ it remains large because 
$\delta E_F>W_c^h$ yet one has that ${\rm max}\,V^{\Delta}_{Bs1}< V_{Fc}$. Finally, the range
$x_0<x<x_{c1}$ marks the transition of a system with large Fermi-velocity anisotropy at $x\approx x_0$
to one with small Fermi-velocity anisotropy at $x\approx x_{c1}$. The strong Fermi-line anisotropy
of the range $0<x<x_{c1}$ renders the study of that line a complex problem, which we do not address here.

Fortunately, the study of the Fermi line simplifies for the isotropic Fermi-velocity hole-concentration range
$x_{c1}<x<x_{c2}$, where the Fermi-velocity anisotropy is small and that line hole like. For that hole-concentartion 
range the coefficient $\eta_{\Delta}$ is small whereas $\eta_0$ is not as small. 
This range can be further divided into the subranges corresponding to $x_{c1}<x<x_{op}$ and $x_{op}\leq x<x_{c2}$
where the Fermi-energy anisotropy is not small and starts being small, respectively. Here $x_{op}=0.16$ for $U/4t\approx 1.525$. 
The hole concentration $x_{op}$ that separates such subranges is found in Ref. \cite{cuprates0} to play the role of optimal hole
concentration, where the critical temperature of the superconducting phase of the
related square-lattice quantum liquid perturbed by small 3D anisotropic effects is
largest. (Such effects refer to weak plane coupling whereas the anisotropy considered here 
is that of the 2D Fermi line.)  The Fermi velocity is for the isotropic Fermi-velocity hole-concentration range nearly 
independent of the Fermi-line angle $\phi$ so that concerning quantities involving 
mainly that velocity the system behaves as isotropic. There remains though some level of anisotropy, which
has effects on some properties. The Fermi-energy coefficient $\eta_{0}$ provides a measure
of such a remaining isotropy. Indeed, that coefficient decreases slower upon
increasing $x$ than $\eta_{\Delta}$, yet both coefficients vanish in the 
limit $x\rightarrow x_*$, where the Fermi line fully becomes that of an isotropic system.
That nearly occurs in the third range considered here, $x_{c2}\leq x<x_*$, where 
both coefficients are small and the Fermi line is approximately that of an isotropic system. 
This is the isotropic Fermi-line hole-concentration range. (For $U/4t\approx 1.525$ one
has that $x_*\approx 0.27$.) 

For intermediate $U/4t$ values $u_0\leq U/4t\leq u_1$ the expressions introduced below for the Fermi line are valid for 
the isotropic Fermi-velocity hole-concentration range $x_{c1} <x<x_{c2}$ where the angles 
that define the directions of the momenta ${\vec{q}}_{Fc}^{\,h\,d}$ and 
${\vec{q}}^{\,d}_{Bs1}$ in expression (\ref{kF-qFc-qBs1}) have a simple relation to the Fermi-line 
angle $\phi$. As mentioned above, the experimental studies of Ref. \cite{k-r-spaces} 
find that for some hole-doped cuprates 
the Fermi line does not obey Luttinger's theorem for approximately $x<0.10$. For
the hole-doped cuprate value $U/4t\approx 1.525$ our study of the
Fermi line is restricted to the range $x_{c1} <x<x_{c2}$ where $x_{c1}\approx 0.13$ 
so that we cannot judge on that
interesting issue. In turn, our results agree with those of Ref. \cite{k-r-spaces} 
in that for approximately $x>10^{-1}$ the Fermi line obeys Luttinger's theorem. 
Anyway, for smaller hole concentrations the above issue concerns mostly how one defines the
Fermi line. The zeros of the one-electron Green's function on the real axis 
signify poles in the self-mass operator and the derivation by Luttinger 
involves integration of it over frequencies. The trouble is that the
result depends on how these singular integrals are regularized.
The Fermi line is mostly a zero-temperature concept. Nevertheless
the regularization used in Ref. \cite{Lutt-Theorem} starts, before
turning the temperature to zero, by considering
finite temperatures and performs the Luttinger calculations
in the Euclidean sector of the energy-momentum space where
there are no poles in self mass. However, there are at
zero temperature other possible regularizations as for instance
that considered in Ref. \cite{Altshuler}.

Within our approach, for $x>x_{c1}$ the Fermi line encloses
the momentum area $(1-x)2\pi^2$ predicted by Luttinger's theorem. In turn, the simple relation
of the angles that define the directions of ${\vec{q}}_{Fc}^{\,h\,d}$ and 
${\vec{q}}^{\,d}_{Bs1}$ in expression (\ref{kF-qFc-qBs1}) to the Fermi-line 
angle $\phi$ is valid only for the isotropic Fermi-velocity range $x_{c1}<x<x_{c2}$.
For such a $x$ range the Fermi line is hole like and each $s1$ band momentum value ${\vec{q}}$ at or near
the $s1$ boundary line referring to states belonging 
to the reduced one-electron subspace defined below has a simple
relation to exactly one hole momentum value $\vec{k}^{\,h} = \vec{k} +\vec{\pi}$ 
at or near the Fermi line. Indeed,
$\vec{q}$ is the momentum of the hole emerging in the
$s1$ band upon removal from a $m=0$ initial ground state of one electron of hole momentum $\vec{k}^{\,h}$ 
at or near the Fermi line. As discussed below, for that hole-concentration range
each momentum of the Fermi line is generated from a 
$s1$ band momentum ${\vec{q}}$ at the $s1$ boundary line
and a $c$ band hole momentum ${\vec{q}}^{\,h}$ at the 
$c$ Fermi line. 
\begin{table}
\begin{tabular}{|c|c|c|c|c|c|c|c|c|c|c|} 
\hline
x & $x_0$ & $0.11$ & $0.13$ & $0.15$ & $0.16$ & $0.21$ & $0.23$ & $0.24$ & $0.27$ \\
\hline
$\eta_{\Delta}$ & $0.20$ & $0.06$ & $0.05$ & $0.04$ & $0.03$ & $0.02$ & $0.01$ & $0.01$ & $0$ \\
\hline
$\eta_0$ & $1.00$ & $0.14$ & $0.11$ & $0.08$ & $0.07$ & $0.03$ & $0.02$ & $0.01$ & $0$ \\
\hline
\end{tabular}
\caption{Magnitudes of the Fermi-velocity anisotropy coefficient $\eta_{\Delta}$ and
Fermi-energy anisotropy coefficient $\eta_0$ of Eq. (\ref{x-h}) for different hole concentrations $x$ 
and $U/4t=u_*= 1.525$, where $x_0\approx 0.024$. It is considered that the 
Fermi-velocity anisotropy is small when $\eta_{\Delta}<2x_0\approx 0.048$ and thus $x>x_{c1}\approx 0.13$.}
\label{tableII}
\end{table} 

\subsubsection{The reduced one-electron subspace and the doublicity}

An important exact property is that one-electron (and two-electron) excitations do not couple
to excited states with an even (and odd) number of $s1$ fermion holes \cite{companion}. 
The transformation laws of the eigenvalue 
spectrum of the $s1$ translation generators ${\hat{\vec{q}}}_{s1}$ in the presence of the fictitious magnetic field 
${\vec{B}}_{s1}$ of Eq. (\ref{A-j-s1-3D}) under subspace transitions from a $m=0$ ground state to
excited states involving creation of an odd number of $s1$ fermion holes
are such that the $s1$ boundary line is deformed. We emphasize that the 
momentum shifts $\delta{\vec{q}}_{s1}^{\,0}/N_{s1}$ of the momenta ${\vec{q}}_{\gamma}^{\,0}/N_{\gamma}$ 
of Eq. (\ref{q-j-f-Q-c-0-s1-2D}) for the $\gamma =s1$ band momenta under subspace transitions do not deform the
$s1$ boundary line in the present $N_a^2\rightarrow\infty$ limit. Indeed,
the shift of each such momenta is of the order of $1/N_a^2$ and vanishes, only
the resulting macroscopic momentum ${\vec{q}}_{s1}^{\,0}$ due to the shift of all
$N_{s1}$ occupied $s1$ band discrete momentum values being of physical
importance, assuring that the sum-rule (\ref{sum-q-0}) is fulfilled.

In turn, the changes in the eigenvalue spectrum of the $s1$ translation generators ${\hat{\vec{q}}}_{s1}$
considered here are of another type, leading to deformations of the $s1$ boundary
line. Subspace transitions from a generating $m=0$ ground state lead
to such $s1$ boundary line deformations only when the final subspace
is spanned by excited states with an odd number of $s1$ fermion holes. 
Indeed, the transformation laws of the eigenvalue 
spectrum of the $s1$ translation generators ${\hat{\vec{q}}}_{s1}$ are such that the contributions from
each of the emerging $s1$ band holes to the $s1$ boundary line deformation cancel when the number
of such holes is even. In turn, when it is odd one of such contributions remains uncompensated
so that the $s1$ boundary line is deformed. The latter subspace transitions involve
creation of an odd number of $c$ fermions or $c$ fermion
holes as well and when only one $s1$ fermion hole is created, the resulting excited
states couple (and do not couple) to one-electron (and two-electron) excitations.

Here we address the occurrence of such deformations of the $s1$ boundary
line for the particular case of subspace transitions involving creation or annihilation of one
electron. In the present $N_a^2\rightarrow\infty$ limit, $u_0\leq U/4t\leq u_1$, and 
for momenta near the $s1$ boundary line such deformations
can be described by $s1$ band rotations, which for hole
concentrations in the range $x_{c1}<x<x_{c2}$ preserve the absolute values 
of such momenta but change their angles. Similar deformations occur for
$x<x_{c1}$ but owing to the larger Fermi-line anisotropy their study is a more 
involved problem, which we do not address here. Similar mechanisms occur
for $x>x_{c2}$ but again are nor studied here due to complications associated
with the transition from a hole like to a particle-like Fermi line. For the
present hole-concentration range $x\in (x_{c1},x_{c2})$
such rotations occur in the square-lattice quantum liquid 
under subspace transitions from the initial $m=0$ and $(N_{\uparrow},N_{\downarrow})=(N/2,N/2)$-electron 
ground state to the subspace spanned by a $(N/2+1,N/2)$-electron or 
$(N/2,N/2-1)$-electron ground state
and its excited states of small momentum and low energy. All such
states have a single hole in the $s1$ momentum band, as confirmed
by the numbers provided in Table \ref{tableIV} of Appendix A.
They are one-electron excited states of the 
$m=0$ ground state, being generated from it
by removal (or addition) of a spin-down (or spin-up)
electron. Relative to the latter ground state such states
may have finite momentum and finite energy.
In turn, for two-electron excited sates of the initial $m=0$
ground state such rotation-like $s1$ band deformations do not occur.

Most of the results of this subsection refer to the square-lattice
quantum liquid in a well-defined subspace of its one- and two-electron
subspace: The {\it reduced one-electron subspace}. Some of our
results are valid only for the isotropic Fermi-velocity range $x\in (x_{c1},x_{c2})$.
For each ground state with $(N/2,N/2-1)$ electrons there is a corresponding
initial $m=0$ and $(N/2,N/2)$-electron ground state with one more  
spin-down electron. We say that such a ground state is the {\it $m=0$ 
generating ground state} of both the $(N/2,N/2-1)$-electron ground state 
and its excited states of small momentum and low energy. For $x>0$ the latter states
are generated from the $(N/2,N/2-1)$-electron ground state by particle-hole
processes in the $c$ band. The reduced one-electron subspace is spanned
by such a $(N/2,N/2-1)$-electron ground state plus its
excited states of small momentum and low energy. (Similar results are
obtained for a $(N/2+1,N/2)$-electron ground state.) The reduced one-electron subspace
is a subspace of the one-electron subspace where the 
square-lattice quantum liquid is defined.

All sites of the $s1$ effective lattice of a $m=0$ 
generating ground state are occupied so that the $s1$ momentum
band is full. In turn, the $s1$ effective lattice of both a 
$(N/2+1,N/2)$-electron or $(N/2,N/2-1)$-electron ground state
and its excited states of small momentum and low
energy have a single unoccupied site and thus the 
$s1$ momentum band has a single hole. In the present
$N_a^2\rightarrow\infty$ limit, the deformation of the $s1$ boundary line
generated under the transition from the $m=0$ generating ground state to 
the reduced one-electron subspace can be
described by a well-defined rotation of the $s1$ band about
an axis perpendicular to the plane, crossing the zero momentum. 
Symmetries imply that the corresponding rotation angle is proportional to the
hole concentration $x$ so that the $s1$ band remains unchanged for the one-electron
excited states generated from the $x=0$ and $m=0$
ground state.
  
For finite hole concentrations one finds within a continuous representation
of the $s1$ fermion discrete momentum values corresponding to
$(2\pi/L)^2\rightarrow 0$ within the $N_a^2\rightarrow\infty$ limit that  
the $s1$ band momenta $\vec{q}$ of the reduced 
one-electron subspace are related to the momenta 
${\vec{q}}_{0}$ of the corresponding $m=0$ 
generating ground state as follows,
\begin{equation}
\vec{q} = A^d_{s1}\,{\vec{q}}_{0}
+ [2/(1-x)][\delta {\vec{q}}_{s1}^{\,0}/N^2_{a}] \, ; \hspace{0.35cm}
\lim_{x\rightarrow 0}A^d_{s1}={\bf I} \, ; \hspace{0.25cm} d=\pm 1 \, .
\label{q-A-general}
\end{equation}
Here $A^d_{s1}$ is a $2\times 2$ matrix, ${\bf I}$ the unit $2\times 2$
matrix, $d=\pm 1$ the doublicity, and $[2/(1-x)][\delta {\vec{q}}_{s1}^{\,0}/N^2_{a}]$ the small
momentum shift $\delta {\vec{q}}_{s1}^{\,0}/N_{s1}$ of Eq. (\ref{q-j-f-Q-c-0-s1-2D}) for $\gamma =s1$.
The transformation (\ref{q-A-general}) describes a deformation of the
$s1$ boundary line under a subspace transition from the $m=0$ 
generating ground state to the reduced one-electron subspace. 
It refers to momentum values $\vec{q}$ at or near the $s1$ boundary line 
and for hole concentrations in the isotropic Fermi-velocity range $x_{c1}< x<x_{c2}$
and $u_0\leq U/4t\leq u_1$
it is described by a $s1$ band rotation by a well-defined angle $\phi_{F}^d$, 
as discussed below.

Here we consider the particular case of states with none, one, or
two unoccupied sites in the $s1$ effective lattice. 
The $2\times 2$ matrix $A^d_{s1}$ is well defined though for states with an arbitrary
finite number of unoccupied sites. Due to symmetry it reduces to the
unit matrix, $A^d_{s1}= {\bf I}$, for states with an even number of
unoccupied sites in the $s1$ effective lattice. Therefore, such a matrix
is the unit matrix for subspace transitions from a $m=0$ ground state 
to subspaces spanned by energy eigenstates with a finite even number of
holes in the $s1$ band. In turn, the derivation of its form for
states with an odd number of holes in the $s1$ band is in general a
complex problem. For the one- and two-electron subspace of the
square-lattice quantum liquid the states with an odd number of 
holes in such a band are those having a single hole and therefore
here we limit our analysis to the reduced one-electron subspace, which 
is spanned by one-$s1$-band-hole states.

The momentum shift $[2/(1-x)][\delta {\vec{q}}_{s1}^{\,0}/N^2_{a}]$
vanishes in the thermodynamic limit. The small $c$ band momentum shift
$\delta {\vec{q}}_{c}^{\,0}/N_{c}=[1/(1-x)][\delta {\vec{q}}_{c}^{\,0}/N^2_{a}]$
of the momenta of Eq. (\ref{q-j-f-Q-c-0-s1-2D}) has a similar limiting behavior. 
When one considers a single $s1$ band discrete momentum 
value ${\vec{q}}$ or a single $c$ band hole momentum value
${\vec{q}}^{\,h}$ one may ignore the shifts 
$[2/(1-x)][\delta {\vec{q}}_{s1}^{\,0}/N^2_{a}]$ or
$[1/(1-x)][\delta {\vec{q}}_{c}^{\,0}/N^2_{a}]$, respectively, since
they refer to contributions of order $1/N^2_{a}$, which vanish
in the thermodynamic limit. However, the corresponding 
excitation momenta $\delta {\vec{q}}_{s1}^{\,0}$
and $\delta {\vec{q}}_{c}^{\,0}$ resulting from the contributions
of all $N_{s1}$ $s1$ fermions and
$N_c$ $c$ fermions, respectively, are macroscopic momenta, which
must be taken into account. For instance, for the $s1$ band 
the main role of the momentum $\delta {\vec{q}}_{s1}^{\,0}$
is to assure that the $s1$-band momentum sum rule 
(\ref{sum-q-0}) is fulfilled for the excited state under consideration.
Therefore, the discrete momentum values are shifted by
$[2/(1-x)][\delta {\vec{q}}_{s1}^{\,0}/N^2_{a}]$ after the $s1$ band transformation generated by
the matrix $A^d_{s1}$ is performed. 

The matrix $A^d_{s1}$ of Eq. (\ref{q-A-general}) is for $x\rightarrow 0$ 
given by $\lim_{x\rightarrow 0}A^d_{s1}={\bf I}$. For hole concentrations 
$x\in (x_{c1},x_{c2})$ it can be evaluated for 
momentum values at and near the $s1$ boundary line. The
generation of the excited states of small momentum and low
energy from the above $(N/2+1,N/2)$ or $(N/2,N/2-1)$ ground state 
involves only changes in the occupancy configurations of
the $s1$ fermions at and near the $s1$ boundary 
line. 

For a $m=0$ generating ground state the $s1$ boundary-line momenta
have the following general form,
\begin{equation} 
{\vec{q}}_{Bs1} = q_{Bs1} (\phi)\,{\vec{e}}_{\phi_{s1}} 
\, ; \hspace{0.25cm} \phi_{s1} = \phi + \pi \, .
\label{q-Bs1-m=0}
\end{equation}
This result is valid for the whole range of hole concentrations $0<x<x_*$ 
considered in the investigations of this paper.
The term $\pi$ of $\phi_{s1} = \phi + \pi$ follows from the $s1$ band momenta belonging to the quadrant such that 
$k_{x_1}<0$, $k_{x_2}<0$, and $0<\phi<\pi/2$ of the Brillouin zone centered 
at $-\vec{\pi}=[-\pi,-\pi]$ pointing in a direction defined
by an angle $\phi_{s1}$ in the range $\pi< \phi_{s1}<3\pi/2$. 

Since the $s1$ boundary line encloses a momentum area
$(1-x)\,2\pi^2$, the momentum absolute value $q_{Bs1} =q_{Bs1} (\phi)$ of 
Eq. (\ref{q-Bs1-m=0}) obeys the following normalization condition and has the
following limiting magnitudes, 
\begin{equation} 
\int_{0}^{2\pi}{d\phi\over 2\pi}\,
\pi\left[q_{Bs1} (\phi)\right]^2 = (1-x)\,2\pi^2 
\, ; \hspace{0.35cm}
q_{Bs1} (\phi) = {\pi\over\vert\cos\phi +\sin\phi\vert} \, ,
\hspace{0.25cm} x\ll 1 \, ; \hspace{0.5cm}
q_{Bs1} (\phi) = \sqrt{(1-x)2\pi} \, ,
\hspace{0.25cm} (1-x)\ll 1 \, .
\label{qBs1}
\end{equation}
It is confirmed below that at $x=0$ the $s1$ band 
coincides with an antiferromagnetic reduced Brillouin 
zone such that $\vert q_{x_1}\vert+\vert q_{x_2}\vert\leq\pi$,
consistently with the limiting behavior given in Eq. (\ref{qBs1})
for the corresponding boundary line when $x\ll 1$.

That upon increasing the value of $x$ the $s1$ band remains full for $x\geq 0$ and $m=0$ 
ground states yet encloses a smaller momentum area
$(1-x)2\pi^2$ is associated with the short-range incommensurate-spiral spin
order addressed in Ref. \cite{companion}. As discussed in
Section I, this is consistent with the spacing $a_{s1}=\sqrt{2/(1-x)}\,a$ found
for the square $s1$ effective lattice in that reference
such that the finite-energy real-space excitations 
involving $s1$ fermions with momenta pointing near
the anti-nodal directions break the translational symmetry.

As mentioned above, the momentum area enclosed 
by the $s1$ boundary line when centered at zero momentum
is $(1-x)2\pi^2$ and by $c$ Fermi line when centered at zero momentum
is $(1-x)4\pi^2$  (and on the momentum $-\vec{\pi}=-[\pi,\pi]$
is $x4\pi^2$). Such properties are valid for the hole-concentration range
$0<x<x_*$ of interest for the square-lattice quantum liquid. 
However, that the momentum area enclosed by the
Fermi line when centered at zero momentum
is $(1-x)2\pi^2$ (and on the momentum $-\vec{\pi}=-[\pi,\pi]$
is $(1+x)2\pi^2$) is a property that holds at least for $x>x_{c1}$.
These properties of such three lines together with (i) the Fermi-velocity 
anisotropy being small and (ii) the Fermi line hole like imply that for 
$s1$ band momenta ${\vec{q}}$ at or near the $s1$ boundary line the 
matrix $A^d_{s1}$ of Eq. (\ref{q-A-general}) is for $x_{c1}< x<x_{c2}$ 
orthogonal and generates a rotation by a well-defined $x$ dependent 
angle $\phi_{F}^d$. For each electronic hole momentum $\vec{k}^{\,h}=\vec{k}+\vec{\pi}$ 
at or near the Fermi line of Eqs. (\ref{kF}) and (\ref{kF-qFc-qBs1})
there are two values of $\vec{q}$ at or near
the $s1$ boundary line corresponding to two different
one-electron states whose $s1$ fermions
have different doublicity $d=\pm 1$, as discussed below.

For a $s1$ band momentum $\vec{q}_0 = q (\phi)\,{\vec{e}}_{\phi_{s1}}$ of
the initial $m=0$ ground-state subspace at or near the $s1$ boundary line
the corresponding momentum $\vec{q}$ of Eq. (\ref{q-A-general}) of the 
reduced one-electron subspace $s1$ band is at or near that line as 
well. The derivation of the $F$ angle relating these two momenta 
takes into account the emergence of a single hole in the
$s1$ band of the reduced one-electron subspace. From the interplay
of the hole momentum $\vec{k}^{\,h}$ at or near the Fermi line of
the removed or added electron, momentum of the $c$ fermion
removed or added, respectively, and momentum of the hole
emerging in the $s1$ band one finds by imposing that the Fermi line,  
$c$ Fermi line, and $s1$ boundary line enclose the momentum areas given above
that for hole concentrations in approximately the isotropic Fermi-velocity range $x_{c1}< x<x_{c2}$ 
the $s1$ band momenta $\vec{q}$ at or near the $s1$ boundary line
and corresponding momenta $\vec{q}_0$ of the initial generating ground state
$s1$ band have the general form,
\begin{equation}
\vec{q} = q (\phi)\,{\vec{e}}_{\phi_{s1}+\phi_{F}^d} 
+ [2/(1-x)][\delta{\vec{q}}_{s1}^{\,0}/N^2_{a}] \, ; \hspace{0.5cm}
\vec{q}_0 = q (\phi)\,{\vec{e}}_{\phi_{s1}} \, ; \hspace{0.35cm}
\phi^{d}_{F} (\phi) = d \arctan 
\left({\sqrt{k^h (\phi)^2-q (\phi)^2}\over q (\phi)}\right) \, ; \hspace{0.25cm} d=\pm 1 \, ,
\label{q-A}
\end{equation}
where $\phi^{d}_{F} (\phi)$ is the $F$ angle.

Alike the general expression given in Eq. (\ref{q-A-general}),
those provided in Eq. (\ref{q-A}) are valid for the reduced one-electron subspace.
For $\vec{q}$ at or near the $s1$ boundary line the expression provided in the latter equation reveals that
the matrix $A^d_{s1}$ of Eq. (\ref{q-A-general}) is orthogonal for the range $x_{c1}< x<x_{c2}$ 
where the Fermi-velocity anisotropy is small and the Fermi
line hole like. It reads,
\begin{equation}
A^{d}_{F} = \left[
\begin{array}{cc}
\cos \phi^{d}_{F} & -\sin \phi^{d}_{F}  \\
\sin \phi^{d}_{F} & \cos \phi^{d}_{F} 
\end{array}\right] \, .
\label{A-c-s1}
\end{equation}
Here the $F$ angle given in Eq. (\ref{q-A}) can have two values, $\phi_{F}^{-1}$ and $\phi_{F}^{+1}$. 
The two corresponding rotations refer to the doublicity $d=-1$ and $d=+1$, respectively. 
The expression provided in Eq. (\ref{q-A}) 
refers to $s1$ band momentum values at or near the $s1$ boundary line.
The doublicity also labels the $s1$ fermions for all remaining momenta of
the $s1$ band, which are also of the form (\ref{q-A-general})
yet the precise form of the matrix $A^d_{s1}$ is in general unknown. 
Therefore, all $N_{s1}$ $s1$ fermions whose occupancy
configurations generate the same state have the same doublicity.  

The $s1$ boundary line momenta have the following general form,
\begin{equation} 
{\vec{q}}^{\,d}_{Bs1} = q_{Bs1} (\phi)\,{\vec{e}}_{\phi^{d}_{s1}} \, .
\label{q-Bs1}
\end{equation}
Such a boundary line is rotated relative to
that of the corresponding $m=0$ generating ground
state with one more electron by the angle $\phi^{d}_{F}=\phi^{d}_{F}(\phi)$
so that the $s1$ angle $\phi_{s1}$ of Eq. (\ref{q-Bs1-m=0})
is for $x_{c1}<x<x_{c2}$ replaced by the angle $\phi^{d}_{s1}$ on the 
right-hand side of Eq. (\ref{q-Bs1}) given by,
\begin{equation}
\phi^{d}_{s1} = [\phi_{s1} + \phi^{d}_{F}] = [\phi + \pi + \phi^{d}_{F}] \, .
\label{phiF-s1}
\end{equation}

In turn, for $x>0$ and $m=0$ generating ground states and approximately
$u_0\leq U/4t\leq u_{\pi}$ the $c$ Fermi line is hole 
like for hole concentrations below $x\approx r_s >x_*$ where $x_*=2r_s/\pi$, as found in Section IV. 
Since the square-lattice quantum liquid
scheme refers to the range $x\in (0,x_*)$, for it the $c$ Fermi line is hole like. 
Such a $c$ Fermi line exists for $x>0$ and we define it in terms of hole $c$ Fermi momenta
${\vec{q}}_{Fc}^{\,h}$ centered at $-\vec{\pi}=[-\pi,-\pi]$ 
and given by,
\begin{equation} 
{\vec{q}}_{Fc}^{\,h} = {\vec{q}}_{Fc} + \vec{\pi} 
= q^h_{Fc} (\phi)\,{\vec{e}}_{\phi_c} 
\, ; \hspace{0.25cm} \phi_{c} = \phi \, .
\label{q-Fc-h-m0}
\end{equation}
As mentioned above, when centered at $-\vec{\pi}=[-\pi,-\pi]$ the $c$ Fermi line
encloses a momentum area $x\,4\pi^2$ so that for $0\leq x\leq x_*$ 
the momentum absolute value $q^h_{Fc}=q^h_{Fc} (\phi)$ obeys 
the sum rule,
\begin{equation} 
\int_{0}^{2\pi}{d\phi\over 2\pi}\,
\pi\left[q^h_{Fc} (\phi)\right]^2 = x\,4\pi^2 \, ;
\hspace{0.50cm} x\leq x_* \, .
\label{qFc-sum-rule}
\end{equation}

As also mentioned above, for the hole-concentration range $x_{c1}< x<x_{c2}$ 
and $u_0\leq U/4t\leq u_1$ one can 
label the $c$ fermions created or annihilated at or near the $c$ Fermi
line upon one-electron excitations by the doublicity value $d=-1$ or $d=+1$
of the $s1$ fermion hole created upon the same excitation. 
According to the numbers and number deviations given in Table \ref{tableIV} of Appendix A, 
removal of one electron from a $m=0$ generating ground state leads to creation
of a $c$ fermion hole and a $s1$ fermion hole. In turn, addition of one electron to 
such a ground state leads to creation of a $c$ fermion and a $s1$ fermion hole. The
latter process does not change the $c$ band discrete momentum values. Moreover, 
one can suitably choose the $S_c=N/2$ value of the initial $m=0$ ground-state subspace 
so that ${\vec{q}}_{c}^{\,0}=0$ in Eq. (\ref{q-j-f-Q-c-0-s1-2D}). 

The reduced one-electron subspace $s1$ band momenta ${\vec{q}}$ of Eqs. (\ref{q-A-general})
and (\ref{q-A}) are in these equations related to the
corresponding momentum values ${\vec{q}}_0$ of the
$m=0$ generating ground state. The changes in the $s1$ band originated by subspace transitions
involving removal of one electron are simpler to describe than those 
involving one electron addition. Indeed, as confirmed by analysis of the
numbers and number deviations given in Table \ref{tableIV} of Appendix A,
for one electron removal the $s1$ band of the $m=0$ and $(N/2,N/2)$ generating ground state has the same number of discrete momentum
values as that of the $(N/2,N/2-1)$ ground-state subspace, so that
Eq. (\ref{q-A-general}) refers exactly to an one-to-one relation. In turn,
the $s1$ band of the generating $(N/2,N/2)$ ground state has one 
discrete momentum value less than that of the $(N/2+1,N/2)$ ground
state. Nevertheless, in the present thermodynamic limit the number of sites 
of the $s1$ effective lattice and thus that of discrete
momentum values of the $s1$ band is $N_{a_{s1}}^2\approx N/2$
so that for $x<x_*$ such a difference is a $1/N_a^D$ effect and 
the momentum values ${\vec{q}}$ of Eqs. (\ref{q-A-general})
and (\ref{q-A}) could as well be expressed in terms of those
of the $m=0$ generating ground state with less one electron than
the corresponding reduced one-electron subspace.

We confirm below that within the $c$ and $s1$ fermion
description of the one-electron problem an identical hole Fermi line is generated
by removal and addition of one electron from and to a $m=0$ generating ground
state, respectively. Since the Fermi line generated by either
process is identical, we start by considering addition
of one electron. We emphasize that independently of the actual 
exact form of the momentum absolute value 
functions $q^h_{Fc} (\phi)$ and $q_{Bs1} (\phi)$, for $u_0\leq U/4t\leq u_1$ and the
hole-concentration range $x\in (x_{c1},x_{c2})$ where the Fermi velocity anisotropy
is small, Luttinger's Theorem holds, and the Fermi line is hole like, the hole Fermi line 
generated by creation of one electron and
thus creation of one $c$ fermion at the
$c$ Fermi line and one $s1$ fermion hole at the $s1$
boundary line encloses the correct area provided
that the momenta of the two created objects
are perpendicular to each other. It follows that for the model 
on the square lattice the Fermi line results in either case from $s1$ fermion and $c$ fermion excitations
that involve a well-defined relation between the angles
defining the directions of the momenta belonging to the $s1$ boundary line
and the $c$ Fermi-line hole momenta, respectively.

That for instance for electron addition the $c$ fermion created under the subspace transition
has doublicity follows from for the hole-concentration range $x\in (x_{c1},x_{c2})$
the hole momentum of such a $c$ fermion and the momentum of the $s1$ fermion hole
created under the one-electron excitation at the $c$ Fermi line and $s1$
boundary line, respectively, being perpendicular to each other. Indeed,
depending on the doublicity of the $s1$ fermion hole, the $c$ fermion
hole momentum points in different directions. Those correspond
to two alternative $c$ Fermi momenta ${\vec{q}}_{Fc}^{\,h\,d}$
where $d=\pm 1$ has the same value as the momentum
${\vec{q}}^{\,d}_{Bs1}$ of the $s1$ fermion hole under consideration 
given in Eq. (\ref{q-Bs1}). Hence in spite of the $c$ Fermi line remaining unaltered 
under the subspace transition associated with the one-electron excitation,
for hole concentrations $x_{c1}< x<x_{c2}$
there are for $d=\pm 1$ two alternative $c$ Fermi momenta given by,
\begin{equation} 
{\vec{q}}_{Fc}^{\,h\,d} = q^h_{Fc} (\phi^{d}_c)\,{\vec{e}}_{\phi^{d}_c} \, ,
\label{q-Fc-h}
\end{equation}
where the momentum absolute value is defined by the same
function $q^h_{Fc}=q^h_{Fc} (\phi')$
as in Eqs. (\ref{q-Fc-h-m0}) and (\ref{qFc-sum-rule}) but 
now with $\phi'=\phi^{d}_c$ rather than
$\phi'=\phi$ where the angle $\phi^{d}_c=\phi^{d}_c  (\phi)$ reads,
\begin{equation}
\phi^{d}_c = [\phi -d\pi/2 + \phi^{d}_{F}] \, ; \hspace{0.25cm}
d = \pm 1 \, .
\label{phiF-c}
\end{equation}

As confirmed in Section IV, for small
hole concentrations $0<x\ll 1$ the hole-momentum
absolute value $q^h_{Fc}$ is given by $q^h_{Fc} = \sqrt{x\pi}\,2$.
That expression remains a good approximation for
$x_{c1}< x<x_{c2}$ and $u_0\leq U/4t\leq u_1$,
\begin{equation} 
{\vec{q}}_{Fc}^{\,h\,d} \approx \sqrt{x\pi}\,2\,{\vec{e}}_{\phi^{d}_c} 
\, ; \hspace{0.5cm} x_{c1}< x<x_{c2} \, .
\label{qFc}
\end{equation}

\subsubsection{The Fermi line for the hole-concentration range $x\in (x_{c1},x_{c2})$}

For hole concentrations in the range $x_{c1}< x<x_{c2}$ and approximately $u_0\leq U/4t\leq u_1$
the general $s1$ band momentum values ${\vec{q}}$ of states belonging to the reduced one-electron subspace are related to the 
corresponding momentum value of the $m=0$ generating ground state as given in
Eq. (\ref{q-A-general}). The form of the matrix $A^{d}_{s1}$ appearing in
that equation is not in general known. The exception is for ${\vec{q}} $ at or near the $s1$ boundary line.
Then it is given by $A^{d}_{s1} = A^{d}_{F}$ where $A^{d}_{F}$ is the $F$ rotation matrix (\ref{A-c-s1}).
It follows from the value of the relative angle of the hole $c$ Fermi momentum ${\vec{q}}_{Fc}^{\,h\,d}$
and $s1$-boundary-line momentum ${\vec{q}}^{\,d}_{Bs1}$ of the hole Fermi
momentum ${\vec{k}}_{F}^{\,h}$ of Eq. (\ref{kF-qFc-qBs1}) that the absolute value $k^h_{F} (\phi)$ of
the latter momentum reads,
\begin{equation}
k^h_{F} (\phi) = \sqrt{[q^h_{Fc}(\phi)]^2+[q_{Bs1}(\phi)]^2} \, ,
\label{KF-h-absolute}
\end{equation}
so that on using the integrals of Eqs. (\ref{qBs1}) and (\ref{qFc-sum-rule}) one finds,
\begin{equation}
\int_{0}^{2\pi}{d\phi\over 2\pi}\,
\pi\left[k^h_{F} (\phi)\right]^2 = \int_{0}^{2\pi}{d\phi\over 2\pi}\,
\pi\left(\left[q^h_{Fc}(\phi)\right]^2 + 
\left[q_{Bs1}(\phi)\right]^2\right) = x\,4\pi^2 + (1-x)\,2\pi^2 = (1+x)\,2\pi^2 \, .
\label{KF-sum-rule}
\end{equation} 
Hence the Fermi line centered at $-\vec{\pi}=-[\pi ,\pi]$, which the hole momentum 
(\ref{kF-qFc-qBs1}) belongs to, encloses the correct momentum area $(1+x)\,2\pi^2$.
Expression (\ref{KF-h-absolute}) is valid for the isotropic Fermi-line range
$x_{c1}<x<x_{c2}$ and approximately $u_0\leq U/4t\leq u_1$. 
For these ranges of $x$ and $U/4t$ values the Fermi line has the simple relation to
the $c$ Fermi line and $s1$ boundary line described above such that the hole
momentum ${\vec{q}}_{Fc}^{\,h\,d}$ and momentum ${\vec{q}}^{\,d}_{Bs1}$ of 
Eq. (\ref{kF-qFc-qBs1}) are perpendicular. However, the general expression
given in Eq. (\ref{kF-qFc-qBs1}) is valid for the whole range $0<x<x_*$ of the square-lattice
quantum liquid provided that $u_0\leq U/4t\leq u_{\pi}$ yet for small hole concentration values we could not derive the angle 
between the directions where the $s1$ band momentum and $c$ band hole momentum
of the $s1$ fermion hole and $c$ fermion, respectively, created under the one-electron excitation
point. 

Moreover, one finds from the use of
Eq. (\ref{KF-h-absolute}) that in the $\phi^{d}_{F} (\phi)$ expression of Eq. (\ref{q-A}) 
the quantity $[k^h (\phi)^2-q (\phi)^2]$ is given by $[k^h (\phi)^2-q (\phi)^2]=q^h (\phi)^2$ 
where $q^h (\phi)$ is the absolute
value of the hole momentum ${\vec{q}}^{\,h}\approx {\vec{q}}_{Fc}^{\,h\,d}$
at or near the $c$ Fermi line of a $c$ fermion
created along with a $s1$ fermion of hole of momentum ${\vec{q}}\approx {\vec{q}}^{\,d}_{Bs1}$
upon creation of one electron of hole momentum $\vec{k}^{\,h} \approx \vec{k}_F^{\,h}$. 
As a result, for $x_{c1}<x<x_{c2}$ the $F$ angle $\phi^{d}_{F} (\phi)$ defined 
in Eq. (\ref{q-A}) can be expressed as,
\begin{equation}
\phi^{d}_{F} (\phi) = d \arctan 
\left({q^h_{Fc} (\phi)\over q_{Bs1} (\phi)}\right) =
d \arcsin\left({q^h_{Fc} (\phi)\over \sqrt{[q^h_{Fc}(\phi)]^2+[q_{Bs1}(\phi)]^2}}\right) 
= \arccos\left({q_{Bs1}(\phi)\over \sqrt{[q^h_{Fc}(\phi)]^2+[q_{Bs1}(\phi)]^2}}\right) \, .
\label{phi-c-s1}
\end{equation}

Consistently with expression (\ref{phiF-s1}) 
for the angle $\phi^{d}_{s1}$, the matrix of 
Eq. (\ref{A-c-s1}) is such that for $x_{c1} <x<x_{c2}$ the momenta ${\vec{q}}_{Bs1}$ of Eq. (\ref{q-Bs1-m=0})
and ${\vec{q}}^{\,d}_{Bs1}$ of Eq. (\ref{q-Bs1}) are related as follows,
\begin{equation}
{\vec{q}}_{Bs1} = A^{-d}_{F}\,{\vec{q}}^{\,d}_{Bs1}
\, ; \hspace{0.35cm}
{\vec{q}}^{\,d}_{Bs1} = A^{d}_{F}\,{\vec{q}}_{Bs1} \, .
\label{qFc-qBs1}
\end{equation}
Here we have ignored contributions of order $1/N_a^2$.
For $x\geq 0$ and $m=0$ generating ground states and their 
two-electron excited states the $s1$ boundary-line momentum
${\vec{q}}_{Bs1}$ is that given here and in Eq. (\ref{q-Bs1-m=0}).
In turn, for the states that span the reduced one-electron subspace
it is instead the momentum ${\vec{q}}^{\,d}_{Bs1}$ of Eq. (\ref{q-Bs1}) and
that of Eq. (\ref{q-Bs1-m=0}) is called {\it auxiliary momentum}.

The doublicity $d=\pm 1$ corresponds to two types of 
one-electron elementary processes that generate the same Fermi line.
In such processes the hole Fermi momentum $\vec{k}_F^{\,h}$ has an angle relative to the 
corresponding momentum $-{\vec{q}}^{\,-1}_{Bs1}$ 
or $-{\vec{q}}^{\,+1}_{Bs1}$  (and 
${\vec{q}}^{\, h\,-1}_{Fc}$ or ${\vec{q}}^{\, h\,+1}_{Fc}$) given by 
$\phi^{+1}_{F}$
or $\phi^{-1}_{F}=-\phi^{+1}_{F}$ (and 
$[\phi^{+1}_{F}-\pi/2]$ or $[\phi^{-1}_{F}+\pi/2]$), 
respectively. It is confirmed below in Subsection V-B that such two
processes also reach the same Fermi 
energy but involve electrons with different 
velocities. Therefore, they refer to two 
degenerate ground states for the same electron numbers,
which are obtained from the same
$m=0$ generating ground state. Consistently, each value
of the doublicity $d=\pm 1$ corresponds to a different reduced
one-electron subspace.

It follows that whether the $F$ angle is given by
$\phi^{+1}_{F}$ or $\phi^{-1}_{F}=-\phi^{+1}_{F}$ 
depends only on the excited state and the reduced subspace it
belongs to. Indeed and alike for the
momentum shifts considered in the previous
subsection, the value of the $F$ angle depends for $x_{c1} <x<x_{c2}$
on the subspace transition doublicity and the shape of the $s1$ boundary line is different
for one-electron excited states associated with the $F$ angles
given by $\phi^{+1}_{F}$ and $\phi^{-1}_{F}=-\phi^{+1}_{F}$,
respectively. Note that since,
\begin{equation}
[A^{-d}_{F}]^2\,{\vec{q}}^{\,d}_{Bs1} =
{\vec{q}}^{-d}_{Bs1} \, ,
\label{A-2}
\end{equation}
such shapes are transformed into each other under a rotation 
by the angle $2\phi^{-d}_{F}$.

We now clarify the mechanism through which the $c$ and $s1$ fermion
description leads to the same Fermi line for one-electron addition and removal. 
In contrast to electron addition, electron removal does not conserve the values of the $c$ band momenta of Eq. (\ref{q-j-f-Q-c-0-s1-2D}),
leading to a small overall finite momentum shift $\delta{\vec{q}}_{c}^{\,0}/N_c$.
One can choose the hole concentration value 
of the initial $m=0$ generating ground state so that  
${\vec{q}}_{c}^{\,0}=0$ for the $c$ band of that state subspace. In that case one finds
that the hole Fermi momentum given in Eq. (\ref{kF})
can be expressed as,
\begin{equation}
\vec{k}_F^{\,h} = -[{\vec{q}}_{c}^{\,0} - {\vec{q}}_{Fc}^{\,h\,d} - {\vec{q}}^{\,d}_{Bs1}]
= [-{\vec{q}}_{c}^{\,0} + {\vec{q}}_{Fc}^{\,h\,d} + {\vec{q}}^{\,d}_{Bs1}] \, ,
\label{kF-qFc-qBs1-removal}
\end{equation}
where ${\vec{q}}_{c}^{\,0}\neq 0$ refers to $c$ band of the $(N/2,N/2-1)$ ground state
and we used that the excitation momentum of the one-electron removal 
process is $-\vec{k}_F^{\,h}$. For the reduced one-electron subspace 
the macroscopic momentum shift $\delta {\vec{q}}_{c}^{\,0}={\vec{q}}_{c}^{\,0}$ and corresponding
$c$ band momentum shift $\delta {\vec{q}}_{c}^{\,0}/N_c={\vec{q}}_{c}^{\,0}/N_c$
of Eq. (\ref{q-j-f-Q-c-0-s1-2D}) are such that the hole Fermi momenta given in
Eqs. (\ref{kF-qFc-qBs1}) and (\ref{kF-qFc-qBs1-removal})
associated with the one-electron addition and removal
processes, respectively, are the same,
\begin{equation}
\delta {\vec{q}}_{c}^{\,0}={\vec{q}}_{c}^{\,0} = 2{\vec{q}}^{\,d}_{s1} =
2q_{Bs1}(\phi)\,{\vec{e}}_{\phi^{d}_{s1}} 
\, ; \hspace{0.50cm}
{{\vec{q}}_{c}^{\,0}\over N_c} = {2\over N_a^2}{q_{Bs1}(\phi)\over (1-x)}\,
{\vec{e}}_{\phi^{d}_{s1}} \, .
\label{qc-shift}
\end{equation}
The expressions given in Eqs. (\ref{kF-qFc-qBs1-removal}) and (\ref{qc-shift})
are valid for $0<x<x_*$. In turn, the particular form of the angle $\phi^{d}_{s1}$ of
Eq. (\ref{phiF-s1}) refers only to the isotropic Fermi-velocity range $x_{c1}<x<x_{c2}$.

The inverse of the relation given in Eq. (\ref{q-A-general}) plays an important
role in the expressions of the $s1$ energy dispersion
studied in Section IV. Neglecting the terms of order $1/N^2_{a}$ one finds,
\begin{equation}
{\vec{q}}_0 = [A^d_{s1}]^{-1}\,{\vec{q}} \, ; \hspace{0.50cm}
\lim_{x\rightarrow 0}[A^d_{s1}]^{-1}={\bf I} 
\, ; \hspace{0.25cm} d=\pm 1 \, ,
\label{q-pm}
\end{equation}
where $[A^d_{s1}]^{-1}$ is the inverse matrix of $A^d_{s1}$.
For ${\vec{q}}_0$ at or near the $s1$ boundary line
and $x_{c1}< x<x_{c2}$ such a matrix reads $[A^d_{s1}]^{-1} = A^{-d}_{F}$
where $A^{d}_{F}$ is the orthogonal matrix given in Eq. (\ref{A-c-s1}).

As discussed above, the contribution to the $s1$ boundary line deformation
of the two $s1$ band holes emerging in the case of subspace transitions from
a $m=0$ ground state to two-electron excitations vanish so that the matrix 
$A^d_{s1}$ of Eq. (\ref{q-A-general}) reduces to the $2\times 2$ unit matrix, $A^d_{s1}={\bf I}$. 
In turn, for subspace transitions from such a ground state to excited 
states involving creation of one $s1$ fermion hole at the $s1$ boundary line that matrix
is such that the $s1$ band hole momentum leads to the 
correct hole Fermi momentum ${\vec{k}}_F^{\,h}$ through Eq. (\ref{kF}). 
As also discussed above, the $s1$ boundary line is deformed under general subspace
transitions from a generating $m=0$ ground state to excited 
states involving creation of an odd number of $s1$ fermion holes. In contrast,
such a matrix reads $A^d_{s1}={\bf I}$ and the $F$ angle vanishes for subspace
transitions to excited states under which an even number of $c$ fermions or $c$
fermion holes and an even number of $s1$ fermion holes, respectively, are created. 
It also vanishes for subspace transitions to excited states
involving creation of both an even number of $c$ fermions or $c$
fermion holes and of $s1$ fermion holes.

According to the number and number deviations of Table \ref{tableIV} of Appendix A,
the charge and spin excitations are an example of subspace transitions
under which none or two $s1$ fermion holes emerge so that the
$s1$ boundary line remains unaltered and $A^d_{s1}={\bf I}$. 
Furthermore, also $A^d_{s1}={\bf I}$ for excitations generated by creation or annihilation of two 
electrons, as confirmed by the number and number deviations given in that table. 

Since one-electron (and two-electron) excitations do not couple
to excited states with an even (and odd)
number of $s1$ fermion holes, one has for 
$m=0$ generating ground states and their above 
two-electron excited states with $N_{s1}^h=0,2$ holes in  
the $s1$ band that,
\begin{equation}
{\vec{q}}_{Bs1} = {\vec{q}}^{\,d}_{Bs1}
= q_{Bs1}(\phi)\,{\vec{e}}_{\phi_{s1}} 
\, ; \hspace{0.25cm}
\phi_{s1}  = \phi+\pi
\, ; \hspace{0.25cm} 
({\rm for}\hspace{0.1cm}N_{s1}^h = 0,2
\hspace{0.1cm}{\rm states}) \, .
\label{qh-q-qFc-qBs1-0,2}
\end{equation}
Here the angle $\phi_{s1}  = \phi+\pi$ is that of the auxiliary
momenta belonging to the $s1$ boundary line. For $x_{c1}<x<x_{c2}$ it corresponds to the limiting value of 
the angle $\phi_{s1}^d$ of Eq. (\ref{phiF-s1}) reached for $\phi^{d}_{F}=0$. 

Finally, we emphasize that the doublicity $d=\pm 1$ introduced
in this subsection is only well defined for the square-lattice quantum liquid in the reduced one-electron subspace. 
Furthermore, our above results and expressions refer to the $N_a^2\rightarrow\infty$ limit
whereas for a finite system the $s1$ boundary line deformations studied here
cannot be described by simple $s1$ band rotations.

\section{The square-lattice quantum liquid of $c$ and $s1$ fermions}

\subsection{The general problem of expressing the model in terms
of $c$ and $s1$ fermion operators}

Expressing the Hamiltonian ${\hat{H}}$ of Eq. (\ref{H}) 
in the one- and two-electron subspace in terms of
$c$ fermion and $s1$ fermion operators is a very complex
problem. First one should write it in terms of rotated-electron 
creation and annihilation operators. Since for finite values of
$U/4t$ that Hamiltonian does not commute with the operator ${\hat{V}}$
associated with the electron - rotated-electron unitary
transformation, its formal expression in terms of such operators 
is given in Eq. (\ref{HHr}) and has an infinite number of terms.
Fortunately, however, following the discussions of Subsection
II-A only a small finite number of such terms is for $U/4t\geq u_0\approx 1.302$
relevant to the physics of the square-lattice quantum liquid.

Next one should invert the expressions given in Eqs. (\ref{fc+}) 
and (\ref{rotated-quasi-spin}) to express the
rotated-electron operators in terms of $c$ fermion operators
and the rotated quasi-spin operators, respectively. The 
obtained expressions refer to the LWS-subspace and read, 
\begin{equation}
{\tilde{c}}_{\vec{r}_j,\uparrow}^{\dag} =
f_{\vec{r}_j,c}^{\dag}\,\left({1\over 2} +
q^z_{\vec{r}_j}\right) + e^{i\vec{\pi}\cdot\vec{r}_j}\,
f_{\vec{r}_j,c}\,\left({1\over 2} - q^z_{\vec{r}_j}\right) 
\, ; \hspace{0.50cm}
{\tilde{c}}_{\vec{r}_j,\downarrow}^{\dag} =
q^-_{\vec{r}_j}\,(f_{\vec{r}_j,c}^{\dag} -
e^{i\vec{\pi}\cdot\vec{r}_j}\,f_{\vec{r}_j,c}) \, .
\label{c-up-c-down}
\end{equation}
The use of these expressions on the right-hand side of
Eq. (\ref{HHr}) leads to an expression of the Hamiltonian
in terms of $c$ fermion operators and rotated quasi-spin operators.

An expression of that Hamiltonian for the subspace without
rotated-electron doubly occupancy where the one- and
two-electron subspace is contained is simply obtained 
by replacement in the Hamiltonian expression 
in terms of $c$ fermion operators and rotated quasi-spin operators
of the latter operators $q^l_{\vec{r}_j}$
by the spin operators $s^l_{\vec{r}_j}$ given in Eq. (\ref{sir-pir})
where $l=\pm,z$. 
Indeed, since the occupied sites of the $c$ effective
lattice correspond to the rotated-electron singly-occupied sites,
the operator $n_{\vec{r}_j,c} = f_{\vec{r}_j,c}^{\dag}\,f_{\vec{r}_j,c}$
appearing in that equation
counts the number of the latter sites. Therefore,
the spin operator $s^l_{\vec{r}_j} = n_{\vec{r}_j,c}\,q^l_{\vec{r}_j}$
and the $\eta$-spin operator
$p^l_{\vec{r}_j} = (1-n_{\vec{r}_j,c})\,q^l_{\vec{r}_j}$ of
Eq. (\ref{sir-pir}) are such that
$q^l_{\vec{r}_j}=s^l_{\vec{r}_j}+p^l_{\vec{r}_j}$
and refer to rotated-electron singly occupied sites
and rotated-electron doubly and unoccupied sites, 
respectively. Since for the subspace under consideration all Hamiltonian 
terms involving the $\eta$-spin operator
$p^l_{\vec{r}_j} = (1-n_{\vec{r}_j,c})\,q^l_{\vec{r}_j}$
must be eliminated, replacement in the above Hamiltonian 
expression of the rotated quasi-spin operators 
$q^l_{\vec{r}_j}=s^l_{\vec{r}_j}+p^l_{\vec{r}_j}$
by the spinon operators $s^l_{\vec{r}_j}$ leads
indeed to the desired expression.

Hence expression of the Hamiltonian in terms of $c$ fermion
operators $f_{\vec{r}_j,c}^{\dag}$ and $f_{\vec{r}_j,c}$
and spinon operators $s^l_{\vec{r}_j}$ where $l=\pm,z$
is a well-defined yet quite involved procedure. However,
its ultimate expression in terms of $c$ fermion and
$s1$ fermion operators requires further expression of 
the contributions associated with its spin degrees of freedom in terms of two-spinon $s1$
bond-particle operators followed by the above extended
Jordan-Wigner transformation of the latter operators. 
That problem is addressed in Ref. \cite{cuprates0} for a
particular class of important terms of the Hamiltonian rotated-electron
operator expression (\ref{HHr}), which control the quantum fluctuations 
of the square-lattice quantum liquid and hence play a major role in
its physics. Such Hamiltonian terms are consistent with the result of 
Ref. \cite{companion} that the $x\geq 0$ and $m=0$ ground 
states are spin-singlet states containing $N_{s1}=N/2$ $s1$ bond
particles. The two spinons of each of such spin-neutral composite objects refer to
rotated electrons that singly occupied sites. Therefore, in the subspace without 
rotated-electron doubly occupancy where such ground states are
contained and that the square-lattice quantum liquid refers to there is an energetic 
preference for the formation of spin-singlet rotated-electron bonds. 
Consistently with the number of independent bonds being $N/2$,
the corresponding simplest Hamiltonian terms are of the form,
\begin{equation}
{\hat H}^{bonds} = \sum_{j=1}^{N/2}{\hat H}^{bonds}_j + {(\rm h. c.)} \, ;
\hspace{0.5cm}
{\hat H}^{bonds}_j =
\sum_{j',j''[j-const]}\Delta_{j'j''}[{\tilde{c}}^{\dag}_{\vec{r}_{j'},\uparrow}\,
{\tilde{c}}^{\dag}_{\vec{r}_{j''},\downarrow} -
{\tilde{c}}^{\dag}_{\vec{r}_{j'},\downarrow}\,
{\tilde{c}}^{\dag}_{\vec{r}_{j''},\uparrow}] \, ,
\label{H-r-el}
\end{equation}
where the complex gap function $\Delta_{j'j''}$ replaces the corresponding 
pairing operator, the summation $\sum_{j',j''[j-const]}$ is over
all $N/2$ spin-singlet rotated-electron bonds centered 
at a point of real-space coordinate $[\vec{r}_{j'}+\vec{r}_{j''}]/2$
near $\vec{r}_{j}=[\vec{r}_{j'}+\vec{r}_{j''}]/2-l\,[a_s/2]\,{\vec{e}}_{x_d}$,
and $j=1,...,N/2=N_{s1}$. Here the two-site bond indices $l\pm 1$ and $d=1,2$
are those used within the notation of Ref. \cite{s1-bonds}
and $a_s=a/\sqrt{1-x}$ is the spacing of the spin effective lattice.

After expression of (\ref{H-r-el}) in terms of $c$ and $s1$
fermion operators one finds in Ref. \cite{cuprates0} that the
action describing the quantum fluctuations of the square-lattice quantum
liquid perturbed by small 3D anisotropy effects  
is controlled by contributions from the summation $\sum_{j',j''}$
over nearest neighboring sites. However, in order to
arrive to that result one must take into account the contributions 
from all possible bonds involving rotated electrons at arbitrarily distant
sites associated with the summation $\sum_{j',j''[j-const]}$ of
Eq. (\ref{H-r-el}). 

The expression of (\ref{H-r-el}) in terms of $c$ fermions
and $s1$ fermions is performed in Ref.
\cite{cuprates0}. The expression in terms of such operators
of other Hamiltonian contributions is a very complex task,
which we do not address in this paper.
Fortunately, there is another path to handle the
problem for some of these contributions, as described in 
the following. That allows to take into account the effects
of the quantum fluctuations of the square-lattice quantum liquid
associated with the short-range spin correlations, whereas
those associated with the long-range superconducting
order of the related extended quantum problem
perturbed by small 3D anisotropy effects
are addressed in Ref. \cite{cuprates0}.

For the study of the Hubbard model in the one- and two-electron subspace 
the quantum problem can be described by the Hamiltonian 
in normal order relative to the initial $x\geq 0$ and $m=0$ ground state.
Such a problem simplifies when expressed in terms of
$c$ and $s1$ fermion operators owing to the discrete
momentum values $\vec{q}_j$ of such objects being 
good quantum numbers \cite{companion}. Indeed, that
implies that the interactions between them have a residual character. 

\subsection{The general first-order energy functional}

The general energy spectrum of the ground-state normal-ordered Hamiltionian
of the Hubbard model on the square lattice in the
one- and two-electron subspace describes the square-lattice
quantum liquid of $c$ and $s1$ fermions
studied in this paper and in Ref. \cite{cuprates0}. Here we study such an energy spectrum up
to first order in the $c$ and $s1$ fermion hole momentum distribution-function
deviations for the $U/4t$ range $u_0\leq U/4t\leq u_{\pi}$ for which $x_*\leq 1/\pi$. 
That excludes the first order contributions found in Ref. \cite{cuprates0}
to emerge for the hole-concentration range $x_c<x<x_*$ where $x_c\approx 10^{-2}$
from the superconducting fluctuations of the extended square-lattice quantum liquid 
perturbed by small 3D anisotropy effects.
As justified below, it is convenient to consider the $c$ and $s1$ fermion
hole momentum distribution-function number operators rather than the 
related operators of Eq. (\ref{Nc-s1op}). The former operators read,
\begin{equation}
\hat{N}^h_{c}({\vec{q}}^{\,h}) = f_{{\vec{q}}^{\,h},c}\,f^{\dag}_{{\vec{q}}^{\,h},c} \, ;
\hspace{0.50cm}
\hat{N}^h_{s1}({\vec{q}}) = f_{{\vec{q}},s1}\,f^{\dag}_{{\vec{q}},s1} \, ,
\label{Nc-s1-h-op}
\end{equation}
where for a $c$ band centered at the momentum $-\vec{\pi}$ 
the hole momentum values ${\vec{q}}^{\,h}$ are
given by ${\vec{q}}^{\,h}=[{\vec{q}}+\vec{\pi}]$.
Let $N^h_{c}({\vec{q}}^{\,h}$ and $N^h_{s1}({\vec{q}})$ denote the
expectation values of such operators. We limit our study to the model
in the one- and two-electron subspace defined in Ref. \cite{companion} 
for which the hole momentum values ${\vec{q}}^{\,h}$ and momentum values 
${\vec{q}}$ are good quantum numbers so that the corresponding hole 
momentum distributions $N^h_{c}({\vec{q}}^{\,h})$ and $N^h_{s1}({\vec{q}})$ are eigenvalues
of the operators given in Eq. (\ref{Nc-s1-h-op}), which read
$1$ and $0$ for unfilled and filled, respectively, 
hole momentum values and momentum
values. For the model in such a subspace the energy dispersion
$\epsilon_{c} (\vec{q}^{\,h})$ [and $-\epsilon_{s1} ({\vec{q}})$] 
associated with creation of one $c$ fermion (and 
$s1$ fermion hole) of hole momentum ${\vec{q}}^{\,h}$
(and momentum ${\vec{q}}$) onto the initial ground state is
well defined. One of the goals of this paper is deriving such
energy dispersions.

For the subspace with a constant number $N_c =2S_c$ of $c$ fermions,
which the vacuum given in Eq. (\ref{23}) of Appendix A
refers to, the value of the number $N_{a_{s1}}^D=[S_c+S_s]$ of discrete 
momentum values of the $s1$ band provided in Eq. (\ref{68})
of that Appendix depends on the
spin $S_s$ and thus corresponding number $L_s =2S_s$ of independent spinons.
Therefore, for subspace transitions leading to finite deviations
$\delta S_s$ so that $\delta N_{a_{s1}}^D=\delta S_s\neq 0$ the 
usual relation $\delta N^h_{s1}({\vec{q}})=-\delta N_{s1}({\vec{q}})$ 
is not fulfilled. In turn, the number of discrete momentum
values of the $c$ fermion band is given by $N_a^D$ so that the usual relation
$\delta N^h_{c}({\vec{q}}^{\,h})=-\delta N_{c}({\vec{q}}^{\,h})$ holds
for the whole Hilbert space.

Some of the reasons why it is convenient to express the energy functional associated with the
ground-state normal-ordered Hamiltionian in terms of the
$c$ and $s1$ fermion hole momentum distribution-function deviations 
$\delta N^h_{c}({\vec{q}})$ and $\delta N^h_{s1}({\vec{q}})$
rather than the usual momentum distribution-function deviations 
$\delta N_{c}({\vec{q}})$ and $\delta N_{s1}({\vec{q}})$, respectively,
are that for $x>0$ and $m=0$ ground states the $s1$ band is full,
for their one- and two-electron excitations that band has one and
none or two holes, respectively, and in addition the concentration 
of $c$ fermion holes $x$ in the $c$ band equals that of holes.

The ground-state normal-ordered $c$ and $s1$ fermion hole 
momentum distribution-function deviations read,
\begin{equation}
\delta N^h_{c}({\vec{q}}^{\,h}) = [N^h_{c}({\vec{q}}^{\,h}) -
N^{h,0}_{c}({\vec{q}}^{\,h})] 
\, ; \hspace{0.50cm}
\delta N^h_{s1}({\vec{q}}) = [N^h_{s1}({\vec{q}}) -
N^{h,0}_{s1}({\vec{q}})] \, ,
\label{DNq}
\end{equation}
where $N^{h,0}_{c}({\vec{q}}^{\,h})$ and $N^{h,0}_{s1}({\vec{q}})$
are the initial-ground-state values.

In the following we consider a range 
of hole concentrations $x\in (0,x_*)$ where $x_*$ is the critical hole concentration
found below above which within the square-lattice
quantum liquid there is no short-range spin order at zero temperature. 
To first order in the $c$ and $s1$ fermion hole momentum distribution-function 
deviations of Eq. (\ref{DNq}) the energy functional has the
following general form,
\begin{equation}
\delta E = -\sum_{\vec{q}^{\,h}}\epsilon_{c} (\vec{q}^{\,h})\delta N^h_{c}({\vec{q}}^{\,h}) -
\sum_{{\vec{q}}}\epsilon_{s1} ({\vec{q}})\delta N^h_{s1}({\vec{q}}) \, .
\label{DE}
\end{equation}
Such an energy functional refers to the model in the one- and two-electron
subspace for which the discrete hole momentum values
${\vec{q}}^{\,h}$ and discrete momentum values ${\vec{q}}$
are good quantum numbers so that
the first-order terms (\ref{DE}) and those found in Ref. \cite{cuprates0}
for an extended quantum problem perturbed by small 3D anisotropy effects
refer to the dominant contributions.

The corresponding excitation momentum spectrum 
is linear in the $c$ and $s1$ fermion hole momentum distribution-function 
deviations and reads,
\begin{equation}
\delta {\vec{P}} = \delta {\vec{q}}_{c}^{\,0}
+\sum_{\vec{q}^{\,h}}[\vec{\pi} -\vec{q}^{\,h}] \delta N^h_{c}({\vec{q}}^{\,h}) -
\sum_{{\vec{q}}}{\vec{q}}\,\delta N^h_{s1}({\vec{q}}) \, ,
\label{DP}
\end{equation}
where $\delta {\vec{q}}_{c}^{\,0}$ is the deviation in the $c$ band
momentum of Eq. (\ref{q-j-f-Q-c-0-s1-2D}). Fortunately, the general 
excitation momentum given in Eq. (\ref{DP}) does
not involve the $s1$ band momentum deviation ${\vec{q}}_{s1}^{\,0}$
of the $s1$ band momenta of Eq. (\ref{q-j-f-Q-c-0-s1-2D}). The reason is
that its role is to assure that the excited-state $s1$ momentum
band obeys the sum-rule $\sum_{{\vec{q}}}{\vec{q}}=0$
of Eq. (\ref{sum-q-0}). Hence the net excitation momentum 
arising from the $s1$ momentum band is given by   
$-\sum_{{\vec{q}}}{\vec{q}}\,\delta N^h_{s1}({\vec{q}})$ only.

Another advantage of expressing the momentum functional
(\ref{DP}) in terms of the $s1$ fermion hole momentum
distribution-function deviation $\delta N^h_{s1}({\vec{q}})$
rather than in the corresponding deviation
$\delta N_{s1}({\vec{q}})$ is that the absence
of the momentum deviation ${\vec{q}}_{s1}^{\,0}$ from that
momentum functional occurs only when one expresses it in
terms of $\delta N^h_{s1}({\vec{q}})$. That follows
from the above exotic property according to which
$\delta N^h_{s1}({\vec{q}})\neq -\delta N_{s1}({\vec{q}})$
for state transitions leading to finite spin deviations
$\delta S_s$.

\subsection{The $c$ band Fermi line and
$s1$ band boundary line}

Based on several symmetries and approximations one can reach useful information about
the $c$ and $s1$ energy dispersions $\epsilon_{c} (\vec{q}^{\,h})$
and $\epsilon_{s1} ({\vec{q}})$, respectively, on the right-hand
side of Eq. (\ref{DE}). As a result of a symmetry such that the unitary 
operator $\hat{V}$ preserves nearest hopping only for rotated electrons, 
as occurs for electrons within the model (\ref{H}), the hole momentum
(and momentum) dependence of the energy dispersion $\epsilon_{c} (\vec{q}^{\,h})$
(and $\epsilon_{s1} ({\vec{q}})$) must for the model on a square lattice be fully
determined by that of related elementary functions $e_{c} (q_{x_1}^{h})$ and 
$e_{c} (q_{x_2}^{h})$ (and $e_{s1} (q_{0x_1})$ and 
$e_{s1} (q_{0x_2})$) whose arguments are the Cartesian 
components of the corresponding hole momentum  
${\vec{q}}^{\,h}=[q_{x_1}^{h},q_{x_2}^{h}]$
(and auxiliary momentum ${\vec{q}}_0=[A_{s1}^d]^{-1}\,{\vec{q}}=[q_{0x_1},q_{0x_2}]$).
The variables of the elementary functions $e_{c} (q_{x_1}^{h})$ and 
$e_{c} (q_{x_2}^{h})$ are the components of the
hole momentum ${\vec{q}}^{\,h}$. Those of the elementary functions $e_{s1} (q_{0x_1})$ and 
$e_{s1} (q_{0x_2})$ are the components of the auxiliary momentum ${\vec{q}}_0$ 
of Eq. (\ref{q-pm}) of the $m=0$ generating ground-state
associated with the momentum ${\vec{q}}$ under consideration. 
(For $x\geq 0$ and $m=0$ ground states and their 
two-electron excited states one has that ${\vec{q}}={\vec{q}}_0$.)

We often consider hole momenta ${\vec{q}}^{\,h}$ at or near the $c$ Fermi line 
and momenta ${\vec{q}}$ at or near the $s1$ boundary line
such that  $A_{s1}^d=A_{F}^d$ and $[A_{s1}^d]^{-1}=A_{F}^{-d}$ where the 
$2\times 2$ matrix $A_{s1}^d$ and the $2\times 2$ $F$ rotation matrix $A_{F}^{d}$ 
are those of Eqs. (\ref{q-A-general}) and (\ref{A-c-s1}), respectively.
We recall that for one-electron excited states the $F$ angle $\phi_{F}^{d} (\phi)$ 
vanishes at $x=0$ so that the corresponding $s1$ band rotation 
occurs for subspace transitions from a $m=0$ ground state
to a reduced one-electron subspace provided that the hole concentration 
$x$ of the initial ground state is finite. Furthermore, for subspace transitions to $N_{s1}^h=0,2$ states 
the momentum values whose components appear in the argument of 
the $s1$ elementary functions $e_{s1} (q_{0x_1})$ and 
$e_{s1} (q_{0x_2})$ are not rotated since the $F$ angle vanishes for such states
so that $q_{0x_1}=q_{x_1}$ and $q_{0x_2}=q_{x_2}$. 
However, for the sake of generality we use the
components $q_{0x_1}$ and $q_{0x_2}$ in the arguments 
of the elementary functions $e_{s1} (q_{0x_1})$ and 
$e_{s1} (q_{0x_2})$, respectively, including when 
$q_{0x_1}=q_{x_1}$ and $q_{0x_2}=q_{x_2}$. 

For the model on the square lattice 
the expression of the energy dispersions 
$\epsilon_{c} (\vec{q}^{\,h})$
and $\epsilon_{s1} ({\vec{q}})$ involves auxiliary 
energy dispersions 
$\epsilon^0_{c} (\vec{q}^{\,h})$
and $\epsilon^0_{s1} ({\vec{q}})$ given by,
\begin{eqnarray}
\epsilon^0_{c} (\vec{q}^{\,h}) & = & 
e_{c} (q_{x_1}^{h}) + e_{c} (q_{x_2}^{h}) 
-e_{c} (q_{Fc x_1}^{h\,d}) -e_{c} (q_{Fc x_2}^{h\,d}) \, ; \hspace{0.25cm}
\epsilon^0_{s1} ({\vec{q}}) = 
\epsilon^{0,\parallel}_{s1} ([A_{s1}^d]^{-1}\vec{q})
= \epsilon^{0,\parallel}_{s1} (\vec{q}_0) \, ,
\nonumber \\ 
\epsilon^{0,\parallel}_{s1} (\vec{q}_0) & = & 
e_{s1} (q_{0x_1}) + e_{s1} (q_{0x_2}) -
e_{s1} (q_{Bs1 x_1}) -e_{s1} (q_{Bs1 x_2}) \, .
\label{general-epsilon}
\end{eqnarray}
For one-electron excited states the Cartesian components
$[q_{Bs1 x_1},q_{Bs1 x_2}]$ appearing here correspond to the auxiliary 
momentum ${\vec{q}}_{Bs1}$ generated from the momentum ${\vec{q}}_{Bs1}^{\,d}$
belonging to the $s1$ band boundary line by the transformation
(\ref{qFc-qBs1}). In turn, for a $x\geq 0$ and $m=0$ ground state and 
its two-electron excited states such components refer to momenta 
${\vec{q}}_{Bs1}={\vec{q}}_{Bs1}^{\,d}$ belonging to the $s1$ boundary line. 
The shapes of the $c$ Fermi line and
$s1$ boundary line are fully determined by the 
form of the auxiliary energy dispersions 
defined in Eqs. (\ref{general-epsilon}) as follows,
\begin{eqnarray}
{\vec{q}}_{Fc}^{\,h\,d} & \in & {\rm hole} 
\hspace{0.10cm} c \hspace{0.10cm} {\rm Fermi} 
\hspace{0.10cm} {\rm line}
 \hspace{0.35cm} \Longleftrightarrow  \hspace{0.35cm}
\epsilon^0_{c} ({\vec{q}}_{Fc}^{\,h\,d}) = 0 \, ,
\nonumber \\
{\vec{q}}^{\,d}_{Bs1} & \in & s1 \hspace{0.10cm} {\rm boundary} 
\hspace{0.10cm} {\rm line}
\hspace{0.35cm} \Longleftrightarrow  \hspace{0.35cm}
\epsilon^0_{s1} ({\vec{q}}^{\,d}_{Bs1})
= \epsilon^{0,\parallel}_{s1} ({\vec{q}}_{Bs1}) = 0 \, .
\label{g-FS}
\end{eqnarray}

The $c$ and $s1$ band discrete momentum
values whose occupancy configurations generate the states belonging to the same
$V$ tower remain the same for the whole $U/4t>0$ range. Since, as discussed
below, for that range of $U/4t$ values the $x=0$, $\mu=0$, and $m=0$ absolute ground state corresponds to 
a single $V$ tower, its $c$ and $s1$ band discrete momentum
values remain the same for $U/4t>0$. Such a state refers to full $c$ and $s1$ 
momentum bands with numbers $N_c=2S_c=N_a^2$ 
and $N_{s1}=N_{a_{s1}}^2=N_a^2/2$ and thus $N^h_c=N^h_{s1}=0$. 
These properties are consistent with an exact result according to which the
momentum eigenvalues given in Eq. (\ref{P-1-2-el-ss}) are the same for all energy 
eigenstates belonging to a given $V$ tower.

On the contrary, the energy eigenvalues of states belonging to the same
$V$ tower are in general $U/4t$ dependent both for the model on the 1D and square lattices. 
For the model on the square lattice and different $U/4t$ values 
the ground state of a canonical ensemble associated
with a given hole concentration $x\geq 0$ and spin density $m=0$ 
does not belong in general to the same $V$ tower. This is consistent with the shape of the $c$ 
fermion auxiliary energy dispersion $\epsilon^0_{c} ({\vec{q}}^{\,h}_j)$ and thus that of
the hole $c$ Fermi line of Eq. (\ref{g-FS}) depending on $U/4t$,
as confirmed below.

The expression of the $\gamma =c,s1$ fermion energy dispersions in terms of the elementary functions $e_{\gamma} (q)$ 
on the right-hand side of Eq. (\ref{general-epsilon}) plays an important role in the square-lattice quantum liquid.
In general we denote the 1D variable of the arguments of the elementary functions $e_{\gamma} (q)$ and elementary velocities 
$v_{\gamma} (q)=d e_{\gamma} (q)/dq$ by $q$. For 
$N_a^2\gg1$ such elementary functions and the corresponding 
elementary velocities $v_{\gamma} (q)$ have universal properties. 
The momentum $\vec{P}$ given in Eq.
(\ref{P-1-2-el-ss}) being additive in the $\gamma =c,s1$ band momentum values 
$\vec{q}=[q_{x_1},q_{x_2}]$ together with the general properties of the
matrix $A_{\gamma}^{d}$ of Eq. (\ref{q-A-general}) reveals that due to symmetry 
the following relations hold,
\begin{equation}
e_{\gamma}(q) = e_{\gamma} (-q) \, ; \hspace{0.35cm}
v_{\gamma} (q) = -v_{\gamma} (-q) 
\, ; \hspace{0.50cm}
v_{\gamma} (0) = 0 \, ; \hspace{0.35cm}
{\rm sgn}\{v_{\gamma} (q)\} = {\rm sgn}\{q\} \, .
\label{pairf}
\end{equation}
These symmetry relations imply corresponding symmetries for the $c$ Fermi line 
and $s1$ boundary line of Eq. (\ref{g-FS}). For instance, the shape of 
the $c$ Fermi line (and $s1$ boundary line)
for the quadrant of the $c$ band (and $s1$ band) whose hole momentum values 
${\vec{q}}^{\,h}=[q_{x_1}^{h},q_{x_2}^{h}]$
(and auxiliary momentum values $[A_{s1}^d]^{-1}\,\vec{q}=\vec{q}_0=[q_{0x_1},q_{0x_2}]$) 
are such that $q_{x_1}^{h}\geq 0$ and $q_{x_2}^{h}\geq 0$
(and $q_{0x_1}\leq 0$ and $q_{0x_2}\leq 0$) provides full 
information about its shape in the remaining three quadrants. Therefore,
we are free to restrict the expressions of momentum dependent
quantities to a single quadrant.

Among the $s1$ boundary-line momenta ${\vec{q}}^{\,d}_{Bs1}$ of the
model on the square lattice it is useful to consider those whose corresponding 
auxiliary $s1$ boundary-line momenta ${\vec{q}}_{Bs1}$ related to them
by the transformation Eq. (\ref{qFc-qBs1}) point in the nodal and anti-nodal directions.
Indeed, by definition the nodal and anti-nodal $s1$ boundary-line momenta
${\vec{q}}^{\,d\,N}_{Bs1}$ and ${\vec{q}}^{\,d\,AN}_{Bs1}$, respectively, are those
whose corresponding auxiliary momenta  ${\vec{q}}^{N}_{Bs1}$ and ${\vec{q}}^{AN}_{Bs1}$
have for instance for the quadrant such that $q_{0x_1}\leq 0$ and $q_{0x_2}\leq 0$ 
the following Cartesian components,
\begin{equation} 
{\vec{q}}^{N}_{Bs1} =  
-\left[\begin{array}{c}
q^N_{Bs1}/\sqrt{2} \\ 
q^N_{Bs1}/\sqrt{2}
\end{array} \right]
\, ; \hspace{0.5cm}
{\vec{q}}^{AN}_{Bs1} =
- \left[\begin{array}{c}
q^{AN}_{Bs1} \\ 
0
\end{array} \right] \, ; 
- \left[\begin{array}{c}
0 \\
q^{AN}_{Bs1} 
\end{array} \right] \, .
\label{q-A-q-AN-c-s1}
\end{equation}
Here $q^N_{Bs1}$ and $q^{AN}_{Bs1}$ are the
absolute values of both the auxiliary momenta ${\vec{q}}^{N}_{Bs1}$ and ${\vec{q}}^{AN}_{Bs1}$
and corresponding momenta ${\vec{q}}^{\,d\,N}_{Bs1}$ and ${\vec{q}}^{\,d\,AN}_{Bs1}$,
respectively. As confirmed below, for $U/4t>0$, $m=0$, and $x\rightarrow 0$ one has that, 
\begin{equation}
q^{AN}_{Bs1} = \pi
\, ; \hspace{0.50cm} q^N_{Bs1}=\pi/\sqrt{2}
\, ; \hspace{0.50cm}
x\rightarrow 0 \, .
\label{limits-FN-FAN}
\end{equation}
The minus signs in the components of the expressions of the
momenta ${\vec{q}}^{N}_{Bs1}$ and ${\vec{q}}^{AN}_{Bs1}$ given in
Eq. (\ref{q-A-q-AN-c-s1}) refer to $\phi=\pi/4$ and $\phi=0,\pi/2$,
respectively. Indeed, the angle of such momenta
is that provided in Eq. (\ref{qh-q-qFc-qBs1-0,2}), which
reads $\phi_{s1}=\phi +\pi$. It corresponds to a limiting value of 
the angle $\phi_{s1}^d$ of Eq. (\ref{phiF-s1}) reached at $\phi^{d}_{F}=0$. 

As a result of the limiting values $q_{Fc}^h\rightarrow 0$ plus those reported in Eq. 
(\ref{limits-FN-FAN}), all reached for $x\rightarrow 0$, together with
the $s1$ momentum band being full for $x\geq 0$ and $m=0$ ground states, one finds that
the maximum ranges of the momentum components in
the argument of the elementary functions $e_c (q)$ and $e_{s1} (q')$ appearing in 
Eq. (\ref{general-epsilon}) are for $m=0$ given by 
$q\in (-\pi ,\pi)$ and $q'\in (-q^{AN}_{Bs1} ,q^{AN}_{Bs1})$, 
respectively. Moreover, for $m=0$ and $x>0$ the zero-temperature chemical potential
$\mu$ is fully determined by the charge $c$ fermion energy dispersion and
is given by the last two terms of the $\epsilon^{0}_{c} (\vec{q}^{\,h})$ expression provided in
Eq. (\ref{general-epsilon}),
\begin{equation}
\mu= [-e_c(q_{Fcx_1}^{h\,d})-e_c(q_{Fcx_2}^{h\,d})] \, ,
\label{mux}
\end{equation}
where within our LWS representation we use the
convention that for $x>0$ the chemical-potential sign is that of the hole concentration $x$.
This general expression refers to $x>0$. According 
to the results reported in Appendix A for $x=0$ and $m=0$ the
range of the chemical potential is $\mu \in (-\mu^0,\mu^0)$ 
where $\mu^0\equiv \lim_{x\rightarrow 0}\mu$ equals one half 
the Mott-Hubbard gap, whose magnitude is finite for $U/4t>0$. 
For $0<x\leq 1$ and $m=0$ the chemical potential (\ref{mux}) is 
an increasing function of the hole concentration $x$ such that $\mu^0<\mu (x)\leq\mu^1$ 
where $\mu^1$ is given in Eq. (\ref{37}) of Appendix A and $\mu^0$ 
has the limiting behaviors provided in Eq. (\ref{38}) of that Appendix.
Hence $\mu^0\rightarrow 0$ as
$U/4t\rightarrow 0$ and $\mu^0\rightarrow\infty$ for
$U/4t\gg 1$. (This also applies to the 1D model.) The energy scale $\mu^0$ 
appears here associated with the charge degrees of freedom yet in the studies of
Ref. \cite{companion} and below it is found to play an important role in the
square-lattice quantum liquid spin degrees of freedom as well.

The ground-state filled $c$ fermion energy dispersion bandwidth
$W^p_{c}=[\epsilon^0_{c} ({\vec{q}}_{Fc}^{\,h\,d})-
\epsilon^0_{c} (\vec{\pi})]$, corresponding ground-state unfilled 
$c$ fermion energy bandwidth 
$W^h_{c}=[\epsilon^0_{c} (0)-\epsilon^0_{c} ({\vec{q}}_{Fc}^{\,h\,d})]$,
and $s1$ fermion auxiliary energy dispersion bandwidth
$W_{s1}=[\epsilon^0_{s1} ({\vec{q}}^{\,d}_{Bs1})-
\epsilon^0_{Bs1} (0)]$ can for the model
on the square lattice be written in terms of the
elementary functions as,
\begin{equation}
W^p_{c} /2 = [e_{c} (q^{h}_{Fc})/2 - e_{c} (\pi)] 
\, ; \hspace{0.35cm}
W^h_c/2 =  [e_c (0) - e_{c} (q^{h}_{Fc})/2] 
\, ; \hspace{0.35cm}
W_{s1} /2 = [e_{s1} (\pm q^N_{Bs1}/\sqrt{2}) - e_{s1} (0)] \, .
\label{Wg} 
\end{equation}
Moreover, the relations,
\begin{equation}
e_c (q^{h}_{Fc}) = [e_{c} (q^{h}_{Fc}/2) + 4t - W^p_{c} /2] = [e_{c} (0) - 8t + W^p_{c}] \, ,
\label{P-FS-h} 
\end{equation}
for the hole-like $c$ band Fermi line 
where we have used that $[W^p_{c}+W^h_{c}]=8t$ and,
\begin{equation}
e_{s1} (\pm q^{AN}_{Bs1}) = [e_{s1} (\pm q^{N}_{Bs1}/\sqrt{2}) + W_{s1} /2] =
[e_{s1} (0) + W_{s1}] \, , 
\label{P-FS} 
\end{equation}
for the $s1$ boundary line result from the relations of Eq. (\ref{g-FS}), which
define the $c$ Fermi line and $s1$ boundary line, respectively.

According to Eq. (\ref{42}) of Appendix A, the ranges of
the energy $\epsilon_{c}$ for addition onto a 
$x\geq 0$ and $m=0$ ground state of 
one $c$ fermion and the energy $-\epsilon_{c}$ 
for removal from that state of one $c$ fermion 
are controlled by the energy bandwidths $W_c^h$ and
$W^p_c$, respectively, of Eq. (\ref{Wg}). Here $W^h_c=[4Dt - W^p_c]$ where
$D=1,2$. According to Eq. (\ref{42}) of Appendix A such ranges are defined by the 
inequalities $0\leq\epsilon_{c}\leq W^h_c$ and 
$0\leq -\epsilon_{c}\leq W^p_c$, respectively, for both the model
on the square and 1D lattice. Moreover, 
one has that $W^p_c =4Dt$ for $x = 0$ and $W^p_c =0$ for $x = 1$.
Indeed, for $U/4t>0$ the energy bandwidth 
$W^p_c\in (0,4Dt)$ (and $W^h_c\in (0,4Dt)$) 
decreases (and increases) monotonously for increasing 
values of hole concentration $x\in (0,1)$.

According to Eq. (\ref{43}) of Appendix A, the 
range of the energy $-\epsilon_{s1}$ for adding to a
$m=0$ ground state one $s1$ fermion hole is for the square-lattice model controlled by the $s1$ fermion
auxiliary energy dispersion bandwidth $W_{s1}$ or maximum $s1$ fermion pairing 
energy per spinon $\vert\Delta\vert$ according to the inequality $0 \leq -\epsilon_{s1} \leq {\rm max}\,\{W_{s1},\vert\Delta\vert\}$.
The limiting value $W_{s1}^0 = \lim_{x\rightarrow 0}W_{s1}$
of that energy bandwidth plays an important role in the physics of the square-lattice quantum liquid. 
According to Eq. (\ref{51}) of Appendix A, upon increasing the value 
of $U/4t$ it decreases from $W_{s1}^0 = 2Dt$ at $U/4t = 0$
to $W_{s1}^0 \approx 2D\pi\,t^2/U$ for $U/4t\rightarrow\infty$.
In turn, as discussed in Ref. \cite{companion}, the $s1$ fermion pairing energy per
spinon $\vert\Delta\vert$ has a singular behavior
at $x=0$ and for $0<x\ll 1$, being given by $\mu^0/2$ and
$\Delta_0 =\lim_{x\rightarrow 0}\vert\Delta\vert$, respectively,
where $\Delta_0<\mu_0/2$ for $U/4t>0$. 
Such a behavior is due to the sharp quantum phase transition occurring
at $x=0$, such that there is long-range antiferromagnetic order
and short-range spiral-incommensurate spin order at $x=0$
and for $0<x\ll 1$, respectively. 

\subsection{The $c$ momentum band and $c$ fermion energy dispersion}

As discussed in the previous section, for the model on the square
lattice and $U/4t>0$ the $c$ band hole momentum values ${\vec{q}}^{\,h}$ 
are good quantum numbers. The energy $\epsilon_{c} ({\vec{q}}^{\,h})$ 
[and $-\epsilon_{c} ({\vec{q}}^{\,h})$] associated with addition to (and removal from)
the ground state of one $c$ fermion of hole momentum ${\vec{q}}^{\,h}$ 
has for $U/4t>0$, $x\geq 0$, $m=0$, and both the model on the
square and 1D lattices the general form,
\begin{equation}
\epsilon_c ({\vec{q}}^{\,h}) = {\rm sgn}\,(\epsilon^0_c ({\vec{q}}^{\,h}))
\mu^0\,\delta_{x,0} +\epsilon^0_c ({\vec{q}}^{\,h}) \, .
\label{c-band}
\end{equation}
Here $\epsilon^0_c ({\vec{q}}^{\,h})$ is the $c$ fermion auxiliary
energy dispersion given in Eq. 
(\ref{general-epsilon}) and the $x=0$ chemical-potential 
zero level corresponds to the middle of the Mott-Hubbard gap
whose magnitude is $2\mu^0$. (We recall that its limiting behaviors 
are given in Eq. (\ref{38}) of Appendix A.) For $U/4t\ll 1$ and $U/4t\gg1$ they
are such that $2\mu^0\rightarrow 0$ and 
$2\mu^0\rightarrow \infty$, respectively, both for
the model on the square and 1D lattices.

It follows from Eq. (\ref{c-band}) that $\epsilon_c ({\vec{q}}^{\,h}) = \epsilon^0_c ({\vec{q}}^{\,h})$
for $x>0$ and $m=0$ where consistently with the inequalities provided
in Eq. (\ref{42}) of Appendix A,
\begin{equation}
\epsilon^0_c (0) = [4Dt - W^p_c] 
\, ; \hspace{0.50cm} D=1,2 
\, ; \hspace{0.50cm}
\epsilon^0_c ({\vec{q}}_{Fc}^{\,h\,d}) = 0 \, ; \hspace{0.50cm}
\epsilon^0_c (\vec{\pi}) = - W^p_c \, .
\label{WWW}
\end{equation}

For $U/4t\gg 1$ the electrons that singly occupy sites are insensitive to the on-site 
repulsion and behave like the spinless $c$ fermions. It follows that the corresponding $c$ fermion
occupancy configurations that generate the charge degrees of freedom of the energy eigenstates
have a non-interacting character. Therefore, in such a limit 
for which the rotated electrons become electrons the $c$
band hole momenta ${\vec{q}}^{\,h}$ are associated with electron 
hopping and the corresponding $c$ fermion hole-momentum occupancy configurations
are behind the whole kinetic energy. One then finds
that in such a limit the elementary function $e_{c}(q)$ has an non-interacting form 
and reads $e_{c}(q)=-U/2+2t\cos q$. From its use in the
general expression provided in Eq. (\ref{general-epsilon}) one finds
that the $c$ fermion auxiliary energy dispersion $\epsilon^0_{c} (\vec{q}^{\,h})$ defined
in that equation is for $U/4t\gg 1$ given by,
\begin{equation}
\epsilon^0_{c} (\vec{q}^{\,h})= 2t \sum_{i=1}^D\,
[\cos (q^{h}_{x_i})-\cos (q^{h\,d}_{Fcx_i})] \, ; \hspace{0.50cm} 
U/4t\gg 1 \, .
\label{EcGen-U-inf}
\end{equation}
(For 1D one has that $q^h_{x_1}\equiv q^h$ 
and $q^h_{Fcx_1}\equiv q^h_{Fc}$.) 
For small values of $\vert{\vec{q}}^{\,h}\vert^2-\vert{\vec{q}}^{\,h\,d}_{Fc}\vert^2$ 
(and $(q^h)^2 -(q_{Fc}^h)^2$) and $0<x\ll 1$ this gives,
\begin{equation}
\epsilon_{c} (\vec{q}^{\,h}) \approx  -{\vert{\vec{q}}^{\,h}\vert^2-\vert{\vec{q}}^{\,h\,d}_{Fc}\vert^2\over 
2m^{\infty}_{c}} \, , 
\label{Ec-U-inf}
\end{equation}
for the model on the square lattice and $U/4t\gg 1$ (and
$\epsilon_{c} (q^h)= -[(q^h)^2 -(q_{Fc}^h)^2]/2m^{\infty}_c$
for 1D and $U/4t\gg 1$) where,
\begin{equation}
m^{\infty}_{c} = 1/2t \, .
\label{m-inf}
\end{equation}

For the limit $U/4t\rightarrow\infty$, which expressions (\ref{Ec-U-inf}) and
(\ref{m-inf}) refer to, the $c$ fermions are invariant under the
electron - rotated-electron unitary transformation. Then they are the
non-interacting spinless fermions that describe the charge degrees
of freedom of the electrons. Also for maximum spin density $m=(1-x)$ 
and arbitrary $U/4t$ values the $c$ fermions are invariant under that
transformation and describe the charge degrees
of freedom of the electrons of the fully-polarized state (\ref{23}) of
Appendix A. In that limit their energy dispersion is for $x>0$ of the
form given in Eq. (\ref{EcGen-H-c}) of that Appendix. Transformation
of the momenta of that dispersion into hole momenta gives expression 
(\ref{EcGen-U-inf}). For $m=(1-x)$ and arbitrary $U/4t$ values the $c$ Fermi 
line equals the Fermi line. 

For spin density $m=0$ the expression (\ref{Ec-U-inf}) can be generalized
to finite values of $U/4t$  by the use of the relation between the $U/4t\gg 1$ 
and $U/4t>0$ physics. That relation is controlled by the electron -
rotated-electron unitary transformation 
associated with the operator $\hat{V}$. For $0<x\ll 1$ and $U/4t>0$ the 
unitary character of such a transformation preserves for the model
on the square (and 1D) lattice the value
of the hole momentum ${\vec{q}}_{Fc}^{\,h\,d}$ of Eq. (\ref{qFc}) 
(and hole momentum values $\pm q^h_{Fc}=\pm x\,\pi$) and that of
$\epsilon_{c} (\vec{q}^{\,h})$ 
(and $\epsilon_{c} (q^h)$)
with the bare mass $m^{\infty}_{c} = 1/2t$ replaced by a 
renormalized mass $m^*_c$,
\begin{equation}
\epsilon_{c} (\vec{q}^{\,h}) \approx  -{\vert{\vec{q}}^{\,h}\vert^2-\vert{\vec{q}}^{\,h\,d}_{Fc}\vert^2\over 
2m^{*}_{c}} \, , 
\label{c-band-x-small}
\end{equation}
(and $\epsilon_{c} (q^h)= -[(q^h)^2 -(q_{Fc}^h)^2]/2m^{*}_c$).
From the form of the $c$ fermion energy dispersion (\ref{c-band-x-small}) one confirms 
that for the model on the square lattice, $0\leq x\ll 1$, and $U/4t>0$ the absolute value
$q^h_{Fc} (\phi)$ of the $c$ hole Fermi momentum 
${\vec{q}}_{Fc}^{\,h\,d}$ of Eq. (\ref{q-Fc-h}) is indeed 
independent of the angle $\phi$ and given by $q^h_{Fc} (\phi)=\sqrt{x\pi}\,2$, 
as provided in Eq. (\ref{qFc}). 
It follows that $\vert{\vec{q}}_{Fc}^{\,h\,d}\vert^2=x\,4\pi$
and for $U/4t>0$ and $0\leq x\ll 1$ the hole momentum values
that belong to the $c$ Fermi line of the model on
the square lattice have indeed the form given
in Eq. (\ref{qFc}). The expression of the $c$ fermion energy dispersion 
provided in Eq. (\ref{c-band-x-small}) is valid for some finite energy
bandwidth $\vert\epsilon_{c} (\vec{q}^{\,h})\vert$ measured from the $c$ Fermi
level. For the model on the square lattice at intermediate $U/4t$ values 
$U/4t\in (u_0,u_1)$ and $0<x<x_*$ that
expression is a good  approximation for $\vert\epsilon_{c} (\vec{q}^{\,h}) 
\vert<W_c^h\vert_{x=x_*}$ where $W_c^h =[8t -W_c^p]$ (and $W_c^p$) is the energy 
bandwidth of the ground-state unfilled (and filled) $c$ fermion sea.

Moreover, from the use of Eqs. (\ref{mux}), (\ref{c-band}), and (\ref{c-band-x-small}) one 
finds that for $U/4t>0$ and hole concentrations $0\leq x\ll 1$ the zero-temperature 
chemical potential dependence on $x$ is given by,
\begin{equation}
\mu (x)\approx \mu^0 + W_c^h \, ; \hspace{0.35cm}
W_c^h \approx {\vert{\vec{q}}^{\,h\,d}_{Fc}\vert^2\over 2m^*_c} = {x\,2\pi\over m^*_c} \, ,
\label{mu-Hm}
\end{equation}
where $\mu^0$ is the chemical potential $\mu^0=\lim_{x\rightarrow 0}\mu$ of
Eq. (\ref{38}) of Appendix A. For intermediate $U/4t$ values $U/4t\in (u_0,u_1)$ the 
expressions given in Eq.  (\ref{mu-Hm}) remain good approximations for $0<x<x_*$.

Alike the quasiparticle mass ratios in Fermi liquid theory \cite{Landau,Pines}, 
the $U/4t$-dependent charge mass ratio $r_c =m^{\infty}_c/m^{*}_{c}$
involving the $c$ fermion mass appearing in the expression provided in Eq. (\ref{c-band-x-small})  
and the spin ratio $r_s=\Delta_0/4W_{s1}^0$ considered below
play an important role in the quantum liquid physics. However, in contrast 
to the former liquid here the mass $m^{\infty}_c=1/2t$ of Eq. (\ref{m-inf}), which 
refers to the limit of infinite on-site interaction plays the role 
of that of the non-interacting Fermi liquid. Moreover, 
in contrast to the Fermi-liquid quasi-particles the
$c$ fermions do not evolve into electrons upon turning
off adiabatically the interaction $U$. Instead, upon adiabatically turning
off the parameter $4t^2/U$ they evolve into
the spinless fermions that describe the charge degrees
of freedom of the electrons that singly occupy sites within the energy eigenstate configurations. In
that limit all finite-energy electronic configurations that
generate the energy eigenstates involve such electrons only. In turn, 
the $m=0$ ground state $s1$ fermion occupancy configurations that generate the
spin degrees of freedom of such electrons are degenerate and for the 1D model 
become in that limit those of the spins of the spin-charge factorized wave function introduced 
independently by Woynarovich \cite{Woy} and by Ogata and Shiba \cite{Ogata},
respectively. For the model on a square lattice such $s1$ fermion occupancy configurations
are within the suitable mean-field approximation (\ref{ficti-B}) for the fictitious magnetic field 
${\vec{B}}_{s1}$ of Eq. (\ref{A-j-s1-3D}) those of a full lowest Landau level with 
$N_{s1}=N_{a_{s1}}^2=N/2$ one-$s1$-fermion degenerate states of the 
2D QHE.

For the latter model the ratio $m^{\infty}_c/m^{*}_{c}$ increases smoothly for 
increasing values of $U/4t$ and has the following limiting 
behaviors,
\begin{eqnarray}
r_c & \equiv & {m^{\infty}_c \over m^{*}_c} = 0
\, , \hspace{0.25cm} U/4t\rightarrow 0
\nonumber \\
& \approx & 2\,e^{-1} \approx 0.736 
\, , \hspace{0.25cm} U/4t \approx u_0\approx 1.302 \, ,
\nonumber \\
& = &  0.27\pi \approx 0.848 \, , \hspace{0.25cm} U/4t \approx u_* = 1.525 \, ,
\nonumber \\ 
& = & 1
\, , \hspace{0.25cm} U/4t\rightarrow
\infty \, .
\label{m*c/mc-UL}
\end{eqnarray}
This charge mass ratio and the $s1$ fermion energy dispersion 
given below for $x=0$ and $m=0$ are well-defined for $U/4t>0$. In turn, the $x$ dependence of the energy
parameter $\vert\Delta\vert$ appearing in $s1$ fermion
energy dispersion for $x>0$ and $m=0$ derived below is a good approximation for
$u_0\leq U/4t\leq u_1$ and thus the studies of the remaining of this paper are
often restricted either to such a $U/4t$ range or to $U/4t\geq u_0$. 

In contrast to the quasiparticles of a Fermi liquid, for the $c$ and $s1$ fermions of the square-lattice
quantum liquid the non-interacting  
$U/4t\rightarrow 0$ limit is not trivial. The values obtained in that limit for some physical 
quantities are well defined. However, in some
cases one must be careful, since the physics associated with the
$U/4t\rightarrow 0$ limit and that of the $U/4t=0$ non-interacting system may not
coincide. For instance, for the model on the square lattice the expression 
(\ref{c-band-x-small}) of the $c$ fermion energy dispersion for
small $\vert\vec{q}^{\, h}\vert$ and $0<x\ll 1$ and the
corresponding chemical-potential formula (\ref{mu-Hm}) are
valid for finite values of $U/4t>0$ but do not apply at
$U/4t=0$. Indeed, they apply provided 
that $1/m^{*}_{c}$ is finite at $x=0$ so that expressions (\ref{c-band-x-small}) and
(\ref{mu-Hm}) are the dominant first-order terms in the hole concentration
$x$ of expansions in $x$ valid for $0<x\ll 1$.
However, for $U/4t\rightarrow 0$ one has that $1/m^{*}_{c}=0$.

Study of the spectrum of the model (\ref{H}) on the square lattice 
at $U/4t=0$ reveals that for hole concentrations $x\ll 1$
the chemical potential and its derivative with
respect to $x$ are given by $\mu (x) \approx x\,2\pi/w (x)$ 
and $\partial\mu (x)/\partial x \approx 2\pi/[w (x)-2/(\pi t)]$,
respectively, where the inverse function of $w (x)$ reads 
$x (w) =  [8wt/\pi]\,e^{-{\pi wt\over 2}+{3\over 2}}$
so that $\lim_{x\rightarrow 0} w(x)=\infty$. Since
for $U/4t\rightarrow 0$ one has $\mu^0\rightarrow 0$,
the $U/4t=0$ chemical-potential expression $\mu (x) \approx x\,2\pi/w (x)$ 
is to be compared with the general expression
$\mu (x) \approx x\,2\pi/m_c^*$ of Eq. (\ref{mu-Hm})
for $U/4t\rightarrow 0$. From such a comparison one finds that the $U/4t=0$ quantity
$w(x)$ plays the role of the effective mass $m_c^*$. For 
$0<x\ll 1$ and $U/4t= 0$ the $x$ dependence of $w(x)$ 
is defined by the equation $w (x) = x [\pi/8t]\,e^{w(x)t{\pi\over 2}-{3\over 2}}$
so that it is large but finite for small values of $x$. It becomes the
mass $m_c^*$ in the limit $x\rightarrow 0$ for which
one finds that $\lim_{x\rightarrow 0} w(x)=\infty$, consistently
with the limiting behavior $r_c=m^{\infty}_c/m^{*}_c\rightarrow 0$ 
as $U/4t\rightarrow 0$ of Eq. (\ref{m*c/mc-UL}) where
according to Eq. (\ref{m-inf}) $m^{\infty}_c=1/2t$. 
The result $\lim_{U/4t\rightarrow 0} 
m^*_c=\lim_{x\rightarrow 0}w(x)=\infty$ is
consistent with the $U/4t\rightarrow 0$ limiting behavior 
of the ratio $r_c$ given in 
Eq. (\ref{m*c/mc-UL}). For $0<x\ll 1$ and $U/4t=0$
the $U/4t>0$ chemical-potential formula (\ref{mu-Hm})
must be replaced though by $\mu (x) \approx x\,2\pi/w (x)$,
where rather than $w (0)=m_c^*=\infty$ one must
use the large but finite value 
of $w (x)$ corresponding to the small hole
concentration $x$ under consideration. This confirms how careful one must
be concerning the interplay of the physical quantities magnitudes at $U/4t=0$
and for $U/4t\rightarrow 0$, respectively. 

For the model on the 1D lattice the mass $m^*_c$ 
is that called $m^*_h$
in Ref. \cite{renor}. Its exact expression is given in Eq.
(48) of that reference and reads $m_c^*=c_0^2/2c_1 t$ 
for all values of $U/4t$ so that,
\begin{equation}
r_c \equiv {m^{\infty}_c \over m^{*}_c} = 
{c_1\over c_0^2} \, ; \hspace{0.5cm}
c_j = 1-2\int_0^{\infty} d\omega 
{J_j (\omega)\over 1+e^{\omega U\over 2t}}
\, ; \hspace{0.35cm} j=0,1 
\, ; \hspace{0.35cm} D = 1 \, ,
\label{rU-1D}
\end{equation}
where $J_j (\omega)$ are Bessel
functions. In 1D the charge mass ratio $1/r_c=m^{*}_{c}/m^{\infty}_c$ 
has a behavior opposite to that of the model on the square lattice:
For the square (and 1D) lattice it 
decreases (and increases) smoothly upon increasing $U/4t$
from $m^{*}_{c}/m^{\infty}_c=\infty$ (and $m^{*}_{c}/m^{\infty}_c=0$)
for $U/4t\rightarrow 0$ to $m^{*}_{c}/m^{\infty}_c=1$ as 
$U/4t\gg 1$. Such a qualitatively different behavior 
is related to the different value of the chemical-potential derivative
$\partial\mu (x)/\partial x\vert_{x=0}$ at $U/4t=0$. For $U/4t=0$
one has that $\mu (0)=0$ both for the model on the square and 1D
lattices. In contrast to 1D, where the
derivative $\partial\mu (x)/\partial x\vert_{x=0}=\pi\,t$
is finite, for the model on the square lattice it vanishes,
$\partial\mu (x)/\partial x\vert_{x=0}=2\pi/[w (0)-2/(\pi t)]=0$.

\subsection{The $s1$ momentum band and $s1$ fermion energy dispersion
at $x=0$ and $m=0$}

\subsubsection{Invariance under the electron - rotated-electron
unitary transformation of the $\mu=0$ and $m=0$ absolute ground state}

The expression $\vec{k}_F^{\,h} = [{\vec{q}}_{Fc}^{\,h\,d} - {\vec{q}}^{\,d}_{Bs1}]$ 
given in Eq. (\ref{kF-qFc-qBs1}) for the hole Fermi momentum $\vec{k}_F^{\,h}$ of Eq. (\ref{kF})
refers to electron addition and is valid for $U/4t>0$. (That the directions
of ${\vec{q}}_{Fc}^{\,h\,d}$ and ${\vec{q}}^{\,d}_{Bs1}$ are perpendicular is a result
that applies to $x\in (x_{c1},x_{c2})$ and approximately $U/4t\in (u_0,u_1)$
and otherwise the relative directions of such
momenta remains an unsolved problem.) Furthermore, according to Eq. 
(\ref{kF-qFc-qBs1-removal}), for one-electron removal one has 
$\vec{k}_F^{\,h} = [-{\vec{q}}_{c}^{\,0} + {\vec{q}}_{Fc}^{\,h\,d} + {\vec{q}}^{\,d}_{Bs1}]$
so that ${\vec{q}}_{c}^{\,0} = 2{\vec{q}}^{\,d}_{s1}$, as given in Eq. (\ref{qc-shift}).
  
At $x=0$ the $F$ angle provided in Eq. (\ref{phi-c-s1})
vanishes for subspace transitions from the $m=0$ ground
state to one-electron excited states
so that ${\vec{q}}^{\,d}_{Bs1}={\vec{q}}_{Bs1}$ in Eq. (\ref{qFc-qBs1}). 
Moreover, expression (\ref{kF-qFc-qBs1}) for the Fermi
hole momentum $\vec{k}_F^{\,h}$ simplifies
at $x=0$ and $m=0$ since then ${\vec{q}}^{\,h}_{Fc}=0$ so that,
\begin{equation} 
\vec{k}_F^{\,h} = - {\vec{q}}_{Bs1} \, ; \hspace{0.35cm}
\vec{k}_F = [-\vec{\pi} - {\vec{q}}_{Bs1}] \, , \hspace{0.25cm} 
x = 0  \, , \hspace{0.15cm} m = 0 \, ,
\label{kF-x0-m0}
\end{equation} 
where the $\vec{k}_F$ expression follows from Eq. (\ref{kF}). 

Importantly, the $x=0$, $\mu=0$, and $m=0$ absolute ground state is the only ground state that for $U/4t>0$
belongs to a single $V$ tower. Hence for its subspace the $s1$ boundary-line momenta 
${\vec{q}}_{Bs1}$ are independent of $U/4t$. This is consistent with the above $x=0$ and $m=0$ expressions 
given in Eq. (\ref{kF-x0-m0}) and the requirement of the particle-hole symmetry specific to
$x=0$, $\mu=0$, and $m=0$ that $\vec{k}_F^{\,h} = \vec{k}_F$. The latter requirement is fulfilled provided that
for the absolute ground state the $s1$ band coincides with an antiferromagnetic reduced Brillouin 
zone such that $\vert q_{x_1}\vert+\vert q_{x_2}\vert\leq\pi$, which is enclosed by a boundary line whose 
momenta ${\vec{q}}_{Bs1}$ have Cartesian components $q_{Bs1x_1}$ and $q_{Bs1x_2}$ obeying the equations,
\begin{equation}
{\vec{q}}_{Bs1} \in s1 \hspace{0.10cm} {\rm boundary} 
\hspace{0.10cm} {\rm line}
\hspace{0.10cm} \Longleftrightarrow\hspace{0.10cm}
q_{Bs1x_1}\pm q_{Bs1x_2}=\pi \hspace{0.10cm} {\rm or}
\hspace{0.10cm} q_{Bs1x_1}\pm q_{Bs1x_2}=-\pi \, .
\label{g-FS-x-0}
\end{equation}
This implies that the absolute value $q_{Bs1} (\phi)$ has the form given in 
Eq. (\ref{qBs1}) for $x\ll 1$. Hence the $s1$ boundary line refers
to the lines connecting $[\pm\pi,0]$ and $[0,\pm\pi]$. Such a $s1$ band
shape and area are consistent with the result of Ref. \cite{companion}
according to which at $x=0$ and $m=0$  the square $s1$ effective lattice has 
spacing $a_{s1}=\sqrt{2}\,a$. Its periodicity increases relative to that of
the original lattice owing to the appearance of the long-range antiferromagnetic order,
consistently with the results of that reference.

This result reveals that not only the boundary-line momenta ${\vec{q}}_{Bs1}$ 
are the same for the whole range of $U/4t>0$ values but also in the 
$U/4t\rightarrow 0$ limit are associated with the correct $U/4t=0$ Fermi line. 
It follows that the shape of that line is for
the Hubbard model on a square lattice independent
of $U/4t$ and equals that of its $U/4t=0$ Fermi line, consistently
with the absolute ground state corresponding to a
single $V$ tower of energy eigenstates. 

Only for $x=0$ and $m=0$ is the Fermi line invariant under the 
electron - rotated-electron transformation: its Fermi momentum values 
have then the form $\vec{k}_F=\vec{k}_F^{\,h} = - {\vec{q}}_{Bs1}$
where ${\vec{q}}_{Bs1}$ is $U/4t$ independent and has
Cartesian components obeying Eq. (\ref{g-FS-x-0}).
Our results then provide strong evidence that the $x=0$,
$\mu=0$, and $m=0$ absolute ground state with numbers
$N_{a_{\eta}}^2=0$, $N_{a_{s}}^2=N_a^2=2N_{s1}$,
and $2S_c=N_a^2$ is invariant under
the electron - rotated-electron transformation. 
The theory vacuum of Eq. (\ref{23}) corresponds to
a fully polarized state and is indeed invariant under that transformation.

The form of the $x=0$ and $m=0$ Fermi line found here 
together with the property that for $U/4t>0$ the $c$ and $s1$ 
bands of the absolute ground state are full is consistent 
with the discussions of Ref. \cite{companion} 
according to which the Hubbard model on a square lattice 
is for $x=0$, $\mu =0$, and $m =0$ a Mott-Hubbard insulator 
with long-range antiferromagnetic order for all 
finite values of $U/4t$. This is for instance consistent with the
behavior reported for the Mott-Hubbard gap
$2\mu^0$ in Eq. (\ref{38}) of Appendix A,
such that it is finite for finite values of $U/4t$ and
behaves as $2\mu^0\rightarrow 0$ for $U/4t\rightarrow 0$ and 
$2\mu^0\rightarrow\infty$ for $U/4t\gg 1$. The studies
of Ref. \cite{cuprates0} on the quantum fluctuations of the present quantum problem
confirm the occurrence of zero-temperature long-range antiferromagnetic order for 
$x=0$, $m =0$, and all finite values of $U/4t$.

Interestingly, our results reveal that the singular
physics associated with the $x=0$, $\mu=0$, and $m=0$ absolute ground state
and the corresponding Mott-Hubbard insulator phase
is related to that state being invariant under the electron -
rotated-electron unitary transformation. However, since for finite
values of $U/4t$ the Hamiltonian and the generator
of the global $U(1)$ symmetry are not invariant under
that transformation the physics of the Mott-Hubbard
insulator phase is not trivial.

\subsubsection{The $s1$ fermion energy dispersion
for $x=0$ and $m=0$}

The expression of the $s1$ fermion energy dispersion $\epsilon_{s1} ({\vec{q}})$  
involves the auxiliary energy dispersion $\epsilon^0_{s1} ({\vec{q}})$ of Eq. (\ref{general-epsilon})
whose momentum dependence defines the shape of the $s1$ band
boundary line through Eq. (\ref{g-FS}). Since for $m=0$ ground states
the $s1$ band is full it follows from Eq. (\ref{Wg}) that 
$[\epsilon^0_{s1} ({\vec{q}}_{Bs1})-\epsilon^0_{s1} (0)]=W_{s1}^0$
for $x=0$ and $m=0$. However, the dispersion $\epsilon^0_{s1} ({\vec{q}})$
does not include the $s1$ fermion spinon-pairing energy.
According to Eq. (\ref{g-FS}), $\epsilon^0_{s1} ({\vec{q}}_{Bs1})=0$,
so that the energy that is left for momentum values belonging to the $s1$
boundary line is the $s1$ fermion pairing energy per spinon.
Hence for $x=0$ and $m=0$ the $s1$ fermion energy dispersion $\epsilon_{s1} ({\vec{q}})$ reads
for momenta at the boundary line,
\begin{equation}
\epsilon_{s1} ({\vec{q}}_{Bs1}) = -\vert\Delta_{s1} ({\vec{q}}_{Bs1})\vert \, ; 
\hspace{0.50cm} 0 \leq \vert\Delta_{s1} ({\vec{q}}_{Bs1})\vert\leq \mu^0/2 \, .
\label{pairing-en-x0-m0}
\end{equation}
Here $\vert\Delta_{s1} ({\vec{q}}_{Bs1})\vert$ denotes the pairing energy
per spinon of a $s1$ fermion of momentum ${\vec{q}}_{Bs1}$,
which may be different for different $s1$ boundary-line momentum 
directions and may vanish for some of these directions so that it
obeys the above inequality. 

The maximum $s1$ fermion spinon-pairing energy $\mu^0$ 
equals the excitation energy below which the long-range antiferromagnetic 
order survives for $x=0$, $m=0$, and zero  
temperature $T=0$. In reference \cite{companion} evidence 
is provided that the Mermin and Wagner Theorem \cite{MWT}
applies to the half-filling Hubbard model on a square lattice
for $U/4t>0$. This is consistent with the existence of long-range antiferromagnetic
order only at $x=0$, $m=0$, $T=0$, and $U/4t>0$, in agreement with the numerical results of Refs. 
\cite{2D-A2,2D-NM}, and its replacement by a short-range spin
order both for (i) $x=0$, $T>0$, and $U/4t>0$ and
(ii) $0<x\ll 1$, $T\geq 0$, and $U/4t>0$. 

The electron occupancy configurations associated with
the long-range antiferromagnetic order refer to a subspace
that for $U/4t$ finite exists below the energy 
scale $\mu^0$, which equals one half the magnitude of the Mott-Hubbard
gap. Indeed, the Mott-Hubbard gap $2\mu^0$, which refers 
to the charge degrees of freedom, also affects the spin 
degrees of freedom. In reference \cite{companion} it
is argued that for the model on the square lattice
the occurrence of long-range antiferromagnetic order requires
the absence of a $\eta$-spin effective lattice
and the spin effective lattice being identical to the original lattice. That happens only
for a subspace with numbers $N_{a_{\eta}}^2=0$ and $N_{a_{s}}^2=N_a^2$ so 
that indeed there is no $\eta$-spin effective lattice and
the spin lattice has as many sites as the original lattice and is identical
to it. In such a subspace there are neither rotated-electron 
doubly occupied nor unoccupied sites, so that
the Hubbard-model spin effective lattice is
identical to that of the spin-$1/2$ isotropic 
Heinsenberg model on a square lattice.

For $U/4t\gg 1$ the spin degrees of freedom
of the half-filling Hubbard model are described by the Heisenberg model and
the above subspace for which $N_{a_{s}}^2=N_a^2$ describes 
the whole finite-energy physics, since $\mu^0 = U/2\rightarrow\infty$.  
However, for finite $U/4t$ the spin occupancy 
configurations behind the long-range antiferromagnetic order 
that refer to a spin effective lattice with $N_{a_{s}}^2=N_a^2$
sites exist only for excitation energy below $\mu^0$.
This justifies why for the Hubbard model 
on a square lattice $\mu^0$ is the energy below which the long-range 
antiferromagnetic order survives for $x= 0$, $\mu=0$, $m=0$, 
and temperature $T=0$.  

In the absence of any spin order the $s1$ fermion spinon-pairing 
energy would vanish and the energy dispersion $\epsilon_{s1} ({\vec{q}})$ of 
the $s1$ fermions would be given by the auxiliary energy dispersion
$\epsilon^0_{s1} ({\vec{q}})$ of Eq. (\ref{general-epsilon}).
An important point for the derivation of the $s1$ fermion energy dispersion 
$\epsilon_{s1} ({\vec{q}})$ in the presence of such an order
is that for excitation energy below the energy scale 
$\mu^0$ one can ignore the amplitude fluctuations of the order 
parameter and the problem can be handled by a suitable mean-field theory 
where the occurrence of that order is described for the $x=0$ 
and $m=0$ problem by a $s1$ energy dispersion of 
the general form $-\sqrt{\vert\epsilon^0_{s1} ({\vec{q}})\vert^2+\vert\Delta_{s1} ({\vec{q}})\vert^2}$
where $\vert\Delta_{s1} ({\vec{q}})\vert$ is the $s1$ fermion pairing energy per
spinon appearing in Eq. (\ref{pairing-en-x0-m0}).
Indeed, the $s1$ boundary-line condition of that equation
imposes that the gap $\vert\Delta_{s1} ({\vec{q}})\vert$ is 
the pairing energy per spinon of a $s1$ fermion of momentum 
${\vec{q}}$. 

There are both bad and good news. The bad news
is that according to Eq. (\ref{38}) of Appendix A 
for small $U/4t$ values the energy scale $\mu^0/2$ becomes 
small and given by $\mu^0/2 \approx 16\,t\,e^{-\pi\sqrt{4t/U}}$,
so that the amplitude fluctuations of the order parameter 
cannot be ignored and thus the energy dispersion 
$-\sqrt{\vert\epsilon^0_{s1} ({\vec{q}})\vert^2+\vert\Delta_{s1} ({\vec{q}})\vert^2}$ may not
be a good approximation. Indeed and as discussed 
above, within the $c$ and $s1$ fermion description
the small-$U/4t$ physics corresponds to a non-trivial problem.
This applies to the derivation of the $s1$ fermion
dispersion $\epsilon_{s1} ({\vec{q}})$ for small values
of $U/4t$, which remains an unsolved issue. Fortunately, 
the above energy dispersion is 
expected to be a good approximation 
for $U/4t\geq u_0$. The good news is then that according to the studies of 
Subsection VI-B the relation of the present quantum problem
to the unusual physics of the Mott-Hubbard insulator parent 
compounds such as LCO refers to a value
$U/4t\approx u_* =1.525>u_0$ so that the description introduced
here is of interest for the study of such materials.
The $U/4t$ value $u_0$ corresponds
to the maximum magnitude of the energy scale $2\Delta_0$ whose $U/4t$
dependence is investigated in Ref. \cite{companion}. 

We recall that for transitions from the $x=0$ and $m=0$ ground state
to all subspaces of the corresponding one- and two-electron subspace
the matrix $A_{s1}^d$ equals the $2\times 2$ unit matrix ${\bf I}$ 
so that the elementary functions can be expressed as 
$e_{s1} (q_{0x_1})=e_{s1} (q_{x_1})$ 
and $e_{s1} (q_{0x_2})=e_{s1} (q_{x_2})$.
The same symmetry arguments that imply that the $s1$ fermion
energy dispersion $\epsilon^0_{s1} ({\vec{q}})$ has for the model on the
square lattice the general
form provided in Eq. (\ref{general-epsilon}) impose that
$\vert\Delta_{s1} ({\vec{q}})\vert$ is a superposition of the elementary functions $e_{s1} (q_{x_1})$ 
and $e_{s1} (q_{x_2})$. Since a function of the form 
$\vert\Delta_{s1} ({\vec{q}})\vert\propto [\mu^0/2]\,\vert e_{s1} (q_{x_1}) + e_{s1} (q_{x_2})\vert$
would lead to a mere change in the $s1$ energy bandwidth, the only remaining possibility 
for the model on the square lattice is,
\begin{equation}
\vert\Delta_{s1} ({\vec{q}})\vert\propto {\mu^0\over 2}\,\vert e_{s1} (q_{x_1}) - e_{s1} (q_{x_2})\vert \, ,
\label{Ds1-q-x0-m0}
\end{equation}
where $\mu^0/2$ is the maximum $s1$ fermion pairing energy per spinon. 

That for the $x=0$, $\mu =0$, and $m=0$ absolute ground state the 
components of the boundary-line momenta ${\vec{q}}_{Bs1}$
obey the relation given in Eq. (\ref{g-FS-x-0}) and for all values of $U/4t$ that line
coincides with the limits of an antiferromagnetic reduced Brillouin zone 
is consistent with a elementary function $e_{s1} (q)\approx -[W^0_{s1}/2]\,\cos q$
whose $U/4t$ dependence occurs through that of the energy bandwidth $W^0_{s1}$.
Hence combining all above results one finds, as confirmed in Subsection VI-B, 
for approximately $U/4t\geq u_0$, excitation energy $\omega$ below $\mu^0$, and temperatures
$T$ below $\omega/k_B$, that the following $s1$ fermion energy dispersion
is a good approximation for $x=0$ and $m=0$, 
\begin{eqnarray}
\epsilon_{s1} ({\vec{q}}) & = & -\sqrt{\vert\epsilon^0_{s1} ({\vec{q}})\vert^2
+\vert\Delta_{s1} ({\vec{q}})\vert^2} 
\, ; \hspace{0.50cm}
\epsilon^0_{s1} (\vec{q}) = - {W^0_{s1}\over 2}[\cos q_{x_1} +\cos q_{x_2}] \, , 
\nonumber \\
\vert\Delta_{s1} (\vec{q})\vert & = & {\mu^0\over 2}\,F_{s1} ({\vec{q}}) 
\, ; \hspace{0.50cm}
F_{s1} ({\vec{q}}) = {\vert \cos q_{x_1} -\cos q_{x_2}\vert
\over 2} \, .
\label{Esx0}
\end{eqnarray}
Here the maximum gap magnitude $\mu^0/2$ of the $s1$ 
fermion energy dispersion equals the maximum $s1$ 
fermion pairing energy per spinon and corresponds to the $s1$ band 
momentum values belonging to the $s1$ boundary line and
pointing in the anti-nodal directions,
\begin{equation}
\mu^0/2 =\vert\Delta_{s1} ({\vec{q}}^{\,AN}_{Bs1})\vert = \vert\Delta_{s1} (\pi,0)\vert
= \vert\Delta_{s1} (0,\pi)\vert
\, ; \hspace{0.35cm} x=0 \, , \hspace{0.25cm} m=0 \, .
\label{D-x0-m0}
\end{equation}
According to Eq. (\ref{38}) of Appendix A, $\mu^0/2 \approx 16\,t\,e^{-\pi\sqrt{4t/U}}$ for
$U/4t\ll 1$ whereas $\mu^0/2\approx [U/4 - 2t]$ for $U/4t\gg1$. However, only for approximately  
$U/4t\geq u_0$ is is the expression for the energy dispersion $\epsilon_{s1} ({\vec{q}})$ provided in
Eq. (\ref{Esx0}) valid and thus can one identify the $s1$ fermion anti-nodal
energy gap $\vert\Delta_{s1} ({\vec{q}}^{\,AN}_{Bs1})\vert$
with $\mu^0/2$. Indeed, owing to the effects of the 
amplitude fluctuations of the order parameter, for 
small $U/4t$ values such a $s1$ fermion energy dispersion 
is expected to be different from that provided in Eq. (\ref{Esx0}).

We recall that for excited states of the $x=0$, $\mu=0$, and $m=0$ 
absolute ground state the one- and two-electron subspace considered here
corresponds to excitation energy $\omega <\mu^0$.
For excitation energy below such an energy scale there is both 
no rotated-electron and rotated-hole double occupancy. Indeed, for
the $x=0$, $\mu=0$, and $m=0$ absolute ground state with the
chemical-potential zero level at the middle of the Mott-Hubbard gap
the energy scale $\mu^0$ is the smallest energy required for creation
of both one rotated-electron doubly occupied site and
one rotated-hole doubly occupied site. Hence 
that the $s1$ energy dispersion (\ref{Esx0}) is valid 
for excitation energy below $\mu^0$ reveals that 
provided that $U/4t\geq u_0$ it is fully consistent with our general 
$c$ and $s1$ fermion description, which refers to the 
one- and two-electron subspace.   

Moreover, that according to Eq. (\ref{Esx0})
the $s1$ fermion spinon-pairing energy is at $x=0$, $m=0$,
and $U/4t\geq u_0$ given by $2\vert\Delta_{s1} (\vec{q})\vert = [\mu^0/2]\,\vert \cos q_{x_1} -\cos q_{x_2}\vert$
reveals that the $s1$ fermion spinon pairing has $d$-wave
symmetry.  An important point is that at half filling one may have 
for the model on the square lattice a ground state with both 
long-range antiferromagnetic order and $s1$
fermion spinon pairing with $d$-wave symmetry. 
According to Ref. \cite{s1-bonds}, for values of the 
two-site bond length $\xi_{g}$ given in that
reference not too small the 
absolute value $\vert h_{g}\vert$
of the coefficients $h_{g}$ of the $s1$ bond-particle
operators of Eq. (\ref{46-b}) of Appendix A falls off as 
$\vert h_{g}\vert \approx  C\,(\xi_{g})^{-\alpha_{s1}}$.
Provided that for the $x=0$ and $m=0$ absolute 
ground state the value of the exponent $\alpha_{s1}$ 
is approximately in the range $\alpha_{s1}\leq 5$, the corresponding
spin-singlet spinon $s1$ bond pairing of the 
$N_a^2/2$ two-spinon $s1$ bond particles associated with such
operators can for $U/4t\geq u_0$ have $d$-wave symmetry 
yet the corresponding spin occupancy configurations have 
long-range antiferromagnetic order   
\cite{Auerbach}. The upper value $5$ is that obtained
from numerical results on spin-singlet two-spin bonds  
\cite{Fazekas,Fradkin,Auerbach}.

The construction of the $s1$ bond-particle spin configurations
fulfilled in Ref. \cite{s1-bonds}
takes into account the result found here that for
$x=0$ and $m=0$ the $s1$ fermion energy
dispersion has for $U/4t\geq u_0$ the form
$2\vert\Delta_{s1} (\vec{q})\vert = [\mu^0/2]\,\vert \cos q_{x_1} -\cos q_{x_2}\vert$
given in Eq. (\ref{Esx0})
so that the $s1$ fermion spinon pairing has $d$-wave symmetry. 
Let us denote by $g^0_{\vec{r}_{j},s1}$ the primary bond operator 
obtained by restricting the summation in the expression
for the $s1$ bond particle operator $g_{\vec{r}_{j},s1}$ of
Eq. (\ref{46-b}) of Appendix A to $g=0$ only. The corresponding
$s1$ fermion operator is according to Eq. (\ref{JW-f+}) 
given by $f_{{\vec{r}}_{j},s1} = e^{-i\phi_{j,s1}}\,
g_{{\vec{r}}_{j},s1}$ where the phase
$\phi_{j,s1}$ is provided in Eq. (\ref{JW-phi}). By 
combining Eq. (\ref{46-b}) of Appendix A 
with Eq. (\ref{JW-f+})  
one can then introduce a primary $s1$ fermion operator 
$f^0_{{\vec{r}}_{j},s1} = e^{-i\phi_{j,s1}}\,
g^0_{{\vec{r}}_{j},s1}$. One then finds from the results of Ref. \cite{s1-bonds} that 
for the model on the square lattice such a primary $s1$ fermion operator reads,
\begin{eqnarray}
f^0_{\vec{r}_{j},s1} & = & e^{-i\phi_{j,s1}}\,h_{0}\,\sum_{d=1}^{2}\sum_{l=\pm1}
\,b_{\vec{r}_{j}+{\vec{r}_{d,l}}^{\,0},s1,d,l,0} 
\nonumber \\
& = & e^{-i\phi_{j,s1}}\,{h_{0}\over\sqrt{2}}\,\sum_{d=1}^{2}\sum_{l=\pm1}
(-1)^{d-1}\left(s^+_{\vec{r}_{j}+2{\vec{r}_{d,l}}^{\,0}}
\left[{1\over 2}+s^z_{\vec{r}_{j}}\right]
- s^+_{\vec{r}_{j}}
\left[{1\over 2}+s^z_{\vec{r}_{j}+2{\vec{r}_{d,l}}^{\,0}}\right]\right) 
\, .
\label{a-b-+-primary}
\end{eqnarray}
(Rather than doublicity $d=\pm 1$, here $d=1$ and $d=2$ refer according to the
notation of Ref. \cite{s1-bonds} to two types of bonds.)
The spin operators of the second expression given here are those
of Eqs. (\ref{sir-pir}) and (\ref{albegra-s-p-m})-(\ref{albegra-s-sz-com})
and the index values $d=1$ and $d=2$ refer to the horizontal
and vertical primary two-site one-link bonds, respectively, with the $d=1$ (and $d=2$) 
two-site bond whose index $l$ reads $l=-1$ and $l=+1$ being that 
whose link connects
the site of real-space coordinate $\vec{r}_{j}$ with the nearest-neighboring
site on its left-hand and right-hand side (and above and below it), 
respectively. The summation and phase 
factor $\sum_{d=1}^{2}\sum_{l=\pm1}
(-1)^{d-1}$ of Eq. (\ref{a-b-+-primary}) then imply that for the
square lattice the primary $s1$ fermion operator $f^0_{\vec{r}_{j},s1}$ 
involves a spinon pairing with $d$-wave symmetry, consistently
with the corresponding $s1$ fermion energy dispersion having the form
$\epsilon_{s1} ({\vec{q}}) = -\sqrt{\vert\epsilon^0_{s1} ({\vec{q}})\vert^2
+\vert\Delta_{s1} ({\vec{q}})\vert^2}$ given in Eq. (\ref{Esx0}) where
$2\vert\Delta_{s1} (\vec{q})\vert = [\mu^0/2]\,\vert \cos q_{x_1} -\cos q_{x_2}\vert$.

\subsection{The $s1$ momentum band and $s1$ fermion energy dispersion
for $x>0$ and $m=0$}

\subsubsection{The general $s1$ fermion energy dispersion for $x>0$ and $m=0$}

Alike for $x=0$, for $0<x<x_*$, $m=0$, and approximately $U/4t\geq u_0$ the auxiliary energy 
dispersion $\epsilon^0_{s1} ({\vec{q}})$ does not contain the $s1$ fermion pairing energy 
per spinon associated with the short-range spin order. Again since
according to Eq. (\ref{g-FS}) the corresponding auxiliary energy dispersion
vanishes at the $s1$ boundary line $\epsilon^0_{s1} ({\vec{q}}^{\,d}_{Bs1})
=\epsilon^{0,\parallel}_{s1} ({\vec{q}}_{Bs1})=0$, 
the energy this is left for momentum values 
belonging to the $s1$ boundary line is the $s1$ fermion pairing energy per
spinon so that the following result holds for $x>0$ and $m=0$,
\begin{equation}
\epsilon_{s1} ({\vec{q}}^{\,d}_{Bs1}) = -\vert\Delta_{s1} ({\vec{q}}^{\,d}_{Bs1})\vert 
= -\vert\Delta^{\parallel}_{s1} ({\vec{q}}_{Bs1})\vert \, ;
\hspace{0.25cm} 
\vert\Delta_{s1} ({\vec{q}})\vert 
= \vert\Delta^{\parallel}_{s1} ([A_{s1}^d]^{-1}{\vec{q}})\vert
= \vert\Delta^{\parallel}_{s1} ({\vec{q}}_0)\vert \, ;
\hspace{0.25cm} 
0 \leq \vert\Delta^{\parallel}_{s1} ({\vec{q}}_{Bs1})\vert\leq \vert\Delta\vert \, .
\label{pairing-en-x>0-m0}
\end{equation}
Alike in Eq. (\ref{pairing-en-x0-m0}), here 
$\vert\Delta_{s1} ({\vec{q}})\vert 
= \vert\Delta^{\parallel}_{s1} ({\vec{q}}_0)\vert$
is the pairing energy per spinon of a $s1$ fermion of momentum 
${\vec{q}}$.

As discussed for the $x=0$ and $m=0$ problem and
except that for $x>0$ the $F$ angle is finite for transitions from
a $x>0$ and $m=0$ ground state to the reduced one-electron subspace 
yet vanishes for transitions to two-electron excited states,
in the absence of spin order our analysis in
terms of the elementary function $e_{s1} (q_0)$ would lead to a $s1$
fermion energy dispersion of the general form $\epsilon_{s1} ({\vec{q}}) 
= \epsilon^0_{s1} ({\vec{q}}) = \epsilon^{0,\parallel}_{s1} ({\vec{q}}_0)
= e_{s1} (q_{0x_1}) + e_{s1} (q_{0x_2})+{\rm constant}$.
The amplitude of the order parameter of the short-range spin 
correlations is frozen below the energy $2\vert\Delta\vert$. Hence the
occurrence of the corresponding short-range spin order leads for 
$m=0$ initial ground states corresponding to the hole concentration
range $0<x<x_*$ and below an excitation energy and
a temperature for which the amplitude fluctuations of the 
corresponding order parameter can be ignored to an effective
mean field theory associated with a $s1$ fermion energy
dispersion of the general form,
\begin{equation}
\epsilon_{s1} ({\vec{q}}) = -\sqrt{\vert\epsilon^0_{s1} ({\vec{q}})\vert^2 +\vert\Delta_{s1} ({\vec{q}})\vert^2} \, ,
\label{bands}
\end{equation}
where $\epsilon^0_{s1} ({\vec{q}})$ is the $s1$ fermion auxiliary energy dispersion 
given in Eq. (\ref{general-epsilon}) and the gap function $\vert\Delta_{s1} ({\vec{q}})\vert$
is derived below.

Similarly to $\mu^0/2$, also the energy scale $\Delta_0$
becomes very small and reads $\Delta_0\approx \mu_0/2\approx 16t\,e^{-\pi\sqrt{4t/U}}$
for small values of $U/4t$ so that then the amplitude fluctuations of the order parameter
cannot be ignored and the $s1$ fermion energy dispersion
is not expected to have the form given in Eq. (\ref{bands}).
Since according to the investigations of Ref. \cite{companion}
the energy scale $\Delta_0$ also vanishes
for $U/4t\gg 1$, for hole concentrations
$0<x<x_*$ the range of validity of the $s1$ fermion energy
dispersion expression (\ref{bands})
corresponds approximately to $U/4t\geq u_0$. 
Indeed, all spin energy scales vanish for 
$U/4t\gg 1$ so that $\Delta_0$ vanishing
in that limit has less severe consequences
than it being small for small values of $U/4t$. 
Consistently, the $s1$ fermion energy dispersion
provided in Eq. (\ref{bands}) remains valid
provided that its energy bandwidth is finite.
A qualitatively different 2D QHE physics is reached in
the limit $U/4t\gg 1$ when the
spectrum of the two-spinon $s1$ fermions
becomes dispersionless and all one $s1$ fermion states are degenerate. 
Provided that the energy scale $4t^2/U$ is finite that
energy bandwidth is also finite and the
above description associated with the
$s1$ fermion energy dispersion of Eq. (\ref{bands}) applies.   

Hence provided that the value of $U/4t$ is larger
than $u_0\approx 1.302$ and finite the $s1$ fermion energy 
dispersion is of the form given in Eq. (\ref{bands})
and the physics is that described by the square-lattice 
quantum liquid of $c$ and $s1$ fermions considered
in this paper. However, that for $x\in (0,x_*)$ the magnitude of the energy parameter
$2\vert\Delta\vert$ decreases linearly in $x$ is a result found
below for the smaller range $U/4t\in (u_0,u_{\pi})$
corresponding to $x_*$ values obeying the inequality
$x_*\leq 1/\pi$. Therefore, our expressions of physical quantities involving
such a linear-$x$ behavior of $2\vert\Delta\vert$ are
valid only for that $U/4t$ range. Moreover, often we consider the smaller range
$U/4t\in (u_0,u_1)$ for which $r_c\approx 2r_s\approx \pi\,x_*$.
Fortunately, the studies of Ref.
\cite{cuprates} reveal that alike for the parent
compounds discussed in Subsection VI-B,
the values of the effective parameters $U$ and $t$ that
lead to good agreement between the square-lattice quantum liquid
perturbed by small 3D anisotropy effects considered in Ref. \cite{cuprates0} and
experiments on hole-doped cuprates with 
superconducting zero-temperature hole concentrations
$x_c\approx 0.05$ and $x_*\approx 0.27$
is reached at the value $U/4t\approx u_*=1.525$ belonging to the range
$U/4t\in (u_0,u_1)$. This is consistent with
the critical hole concentration $x_*$ being found below 
to read $x_*\approx 0.27$ for $U/4t\approx 1.525$.

Furthermore, the symmetry arguments implying that the 
half-filling $s1$ fermion energy gap $\vert\Delta_{s1} ({\vec{q}})\vert$ 
has the general form provided in Eq. (\ref{Ds1-q-x0-m0}), indicate here that
for hole concentrations $0<x<x_*$ the gap
$\vert\Delta_{s1} ({\vec{q}})\vert$ has a similar form but with
$[\mu^0/2]$ replaced by $\vert\Delta\vert$. 
The use of arguments similar to those considered 
for the $x=0$ and $m=0$ problem suggests that for $0<x<x_*$, 
$m=0$, and finite values of $U/4t\geq u_0$
the gap function $\vert\Delta_{s1} ({\vec{q}})\vert$ appearing in the $s1$ fermion energy dispersion
$\epsilon_{s1} ({\vec{q}})$ of Eq. (\ref{bands}) and related quantities are given by,
\begin{eqnarray}
\vert\Delta_{s1} ({\vec{q}})\vert & = &
\vert\Delta^{\parallel}_{s1} ({\vec{q}}_0)\vert = \vert\Delta\vert\,
F^{\parallel}_{s1} ({\vec{q}}_0) 
\, ; \hspace{0.35cm}
F^{\parallel}_{s1} ({\vec{q}}_0) = {\vert e_{s1} (q_{0x_1}) - e_{s1} (q_{0x_2})\vert\over W_{s1}} 
\, ; \hspace{0.25cm} 0<x<x_* \, , \hspace{0.25cm} m=0 \, .
\label{bands-bipt}
\end{eqnarray}
Note that $\vert\Delta_{s1} ({\vec{q}}^{\,d\,N}_{Bs1})\vert=\vert\Delta^{\parallel}_{s1} ({\vec{q}}^{\,N}_{Bs1})\vert=0$ and
$\vert e_{s1} (q^{AN}_{Bs1}) - e_{s1} (0)\vert =W_{s1}$ and thus,
\begin{equation}
F^{\parallel}_{s1} ({\vec{q}}_0) \in (0,1) \, ; \hspace{0.35cm}
F^{\parallel}_{s1} ({\vec{q}}_{Bs1}) =
\vert\cos 2\phi\vert \, ; \hspace{0.35cm}
F^{\parallel}_{s1} ({\vec{q}}^{\,N}_{Bs1}) = 0
\, ; \hspace{0.35cm}
F^{\parallel}_{s1} ({\vec{q}}^{\,AN}_{Bs1}) = 1\, ,
\label{F-s-limits}
\end{equation}
where the expression $F^{\parallel}_{s1} ({\vec{q}}_{Bs1}) =
\vert\cos 2\phi\vert$ is found below for $0<x\ll 1$ and is expected to be a
good approximation for $0<x<x_*$ provided that $U/4t\in (u_0,u_{\pi})$. 
Hence the maximum energy gap magnitude of Eq. (\ref{bands-bipt}) of 
the $s1$ fermion energy dispersion equals the maximum $s1$ fermion 
pairing energy per spinon $\vert\Delta\vert$ and corresponds to 
momentum values belonging to the $s1$ boundary line 
whose auxiliary momenta of Eq. (\ref{qFc-qBs1})
point in the anti-nodal directions.
Such behaviors are consistent with the $d$-wave symmetry of the $x=0$ and $m=0$
problem surviving at least in part for finite values of the hole concentration 
$x$ in spite of the $x=0$ long-range antiferromagnetic order being replaced by
a short-range spin order. 

\subsubsection{Shape of the $s1$ band for $x>0$ and $x$ dependence of the energy parameter $2\vert\Delta\vert$}

For $x>0$ and $m=0$ ground states the $s1$ band 
is full and enclosed by the $s1$ boundary line whose  
momenta are defined by the relation given in Eq. (\ref{g-FS}). Only
for the $x=0$ and $m=0$ Mott-Hubbard insulator 
does such a line contain the nodal momentum $-[\pi/2,\pi/2]$ 
and anti-nodal momenta $-[\pi,0]$ and $-[0,\pi]$. For $x>0$ and $m=0$
these momenta are replaced by the nodal momentum and
anti-nodal momentum related to the auxiliary momentum values ${\vec{q}}^{N}_{Bs1}$ 
and ${\vec{q}}^{AN}_{Bs1}$
of Eq. (\ref{q-A-q-AN-c-s1}) with components
${\vec{q}}^{N}_{Bs1} =-[q^N_{Bs1}/\sqrt{2}, q^N_{Bs1}/\sqrt{2}]$ and
${\vec{q}}^{AN}_{Bs1}= -[q^{AN}_{Bs1},0]$,
respectively, as given in Eq. (\ref{qFc-qBs1}). For $x>0$ and $m=0$
one has according to Eq. (\ref{limits-FN-FAN}),
$q^N_{Bs1}<\pi/\sqrt{2}$ and $q^{AN}_{Bs1}< \pi$,
consistently with the momentum area enclosed by the
$s1$ boundary line being given by $(2\pi/L)^2\,N_{a_{s1}}^2
=(2\pi/L)^2\,[(1-x)/2]N_a^2=2\pi^2(1-x)$ and thus decreasing upon
increasing $x$. This deformation and contraction of the $s1$ band
boundary line upon increasing $x$ is found in 
Ref. \cite{cuprates0} to play an important role in the
square-lattice quantum liquid physics. For instance,
strong evidence is given in that reference that it is behind the positions in momentum space of
the incommensurate low-energy peaks observed
by neutron scattering in LSCO \cite{2D-MIT,pseudogap-review}.

Fortunately, owing to symmetry arguments the $x$ dependence of the 
absolute values $q^N_{Bs1}$ and $q^{AN}_{Bs1}$ of the nodal and anti-nodal $s1$ 
boundary momenta, respectively, can be estimated for approximately $x<x_{c1}$ and 
$U/4t\geq u_0$. It is useful to consider the $s1$ band auxiliary nodal arc momentum 
${\vec{q}}^{\,N}_{s1,\,arc}$ such that,
\begin{equation}
{\vec{q}}^{\,N}_{s1,\,arc} \equiv 
-\left[\begin{array}{c}
\pi/2 \\ 
\pi/2
\end{array} \right] - {\vec{q}}^{\,N}_{Bs1} 
\label{q-N-s-arc}
\end{equation}
and ${\vec{q}}^{\,N}_{Bs1}= -[\pi/2, \pi/2]-{\vec{q}}^{\,N}_{s1,\,arc}$. 

The starting points of the study of the $x$ dependence of $q^N_{Bs1}$ and $q^{AN}_{Bs1}$
are that the area enclosed by the $s1$ boundary line is exactly given by $2\pi^2(1-x)$,
the $s1$ band momentum values of states belonging to the same $V$ tower are
independent of $U/4t$, the $0<x\ll 1$ and $m=0$ ground state belongs to the
same $V$ tower except for very small $U/4t$ values alike the $x=0$ and $m=0$
ground state does for the whole range $U/4t>0$, and the momenta belonging to the 
ground-state $s1$ boundary line are determined by the $s1$ fermion auxilary
energy dispersion $\epsilon^0_{s1} ({\vec{q}}) = 
\epsilon^{0,\parallel}_{s1} ([A_{s1}^d]^{-1}\vec{q})
= \epsilon^{0,\parallel}_{s1} (\vec{q}_0)$ through the relation given in
Eq. (\ref{g-FS}). According to Eq. (\ref{general-epsilon}),
such an energy dispersion can be expressed in terms of
the elementary function $e_{s1} (q)=e_{s1} (-q)$ whose
1D momentum $q$ belongs to the domain $q\in (-q^{AN}_{Bs1},q^{AN}_{Bs1})$. 
At $x=0$ and $m=0$ such an elementary function is found above
in Subsection IV- to be given by $e_{s1} (q)\approx -[W_{s1}^0/2]\,\cos q$. Concerning 
physical quantities involving the elementary-function sum 
$e_{s1} (q_{0x_1})+e_{s1} (q_{0x_2})$ such as the dispersion $\epsilon^0_{s1} ({\vec{q}})$ of 
Eq. (\ref{general-epsilon}) and the related expressions found in the following for $q^N_{Bs1}$ and $q^{AN}_{Bs1}$,
it is a good approximation to consider that 
$e_{s1} (q)\approx -[W_{s1}^0/2]\,\cos q$ for $0<x\ll 1$, $m=0$,
and the whole $q$ range $q\in (-q^{AN}_{Bs1},q^{AN}_{Bs1})$. 

For approximately $U/4t\geq u_0$ and $x<x_{c1}$ the shape of the $m=0$ ground-state $s1$ boundary line and thus
the magnitudes of both $q^N_{Bs1}$ and $q^{AN}_{Bs1}$ are nearly
independent of $U/4t$. Indeed, that such a shape and the eigenvalues of the momentum 
operator are the same for states belonging to the
same $V$ tower implies that there is an interplay between the $U/4t$ dependence of the 
ground-state $c$ Fermi line and $s1$ boundary line. Specifically, if for some region of parameter space 
one of such lines does not depend on $U/4t$ implies in general a similar behavior for the other.
In turn, that the shapes of such $m=0$ ground-state lines depends on $U/4t$ means that upon varying $U/4t$
the ground state belongs to different $V$ towers. 

Consistently, if for some range of hole 
concentrations $x$ and $U/4t$ values the $c$ Fermi line is $U/4t$ independent 
provides strong evidence that the ground state refers to the same $V$ tower for such a parameter-space region.
If so the $s1$ boundary line is $U/4t$ independent as well. It turns out that for
$U/4t\in (u_i (x),\infty)$ and small $x$ where $u_i (x)\approx 0$ for $0<x\ll 1$ 
and $u_i (x)\approx u_0$ for hole concentrations $x\approx x_{c1}$ and
smaller than $x_{c1}$ the $c$ Fermi hole momentum reads
$q^h_{Fc}\approx \sqrt{x\pi}\,2$ so that the ground state belongs to the same
$V$ tower or neighboring $V$ towers and the $c$ Fermi line shape does not depend on $U/4t$. 
Then that strongly suggests that the same occurs for the $s1$ boundary line. 

We recall that the $x=0$ and $m=0$ ground state refers exactly to a single $V$ tower for 
the whole range $U/4t>0$. Consistently, the physics of such a behavior is then for very small hole 
concentrations $0<x\ll 1$ that remains true for $U/4t\in (u_i (x),\infty)$ where $u_i (x)\approx 0$ and upon further increasing
$x$ that remains true as well for some range $U/4t\in (u_i (x),\infty)$ where we estimate
that $u_i (x)\in (0,u_0)$ for approximately $x\in (0, x_{c1})$. For the range $U/4t\in (u_i (x),\infty)$
for which the $x>0$ and $m=0$ ground state refers approximately to a single $V$ tower
the ground-state $c$ Fermi line and $s1$ boundary line shapes are nearly independent of $U/4t$.
It is expected that for $x<x_{c1}$ and $U/4t\geq u_0$ the ground state belongs to neighboring $V$ 
towers in Hilbert space so that both the $c$ Fermi and $s1$ boundary line are independent of $U/4t$
or nearly independent of $U/4t$. Concerning some properties, it is a good approximation to consider that 
$q^h_{Fc}\approx \sqrt{x\pi}\,2$ remains nearly independent of $U/4t$ for $u_0\leq U/4t\leq u_1$ and $x\in (0,x_*)$.
Note however that for $x\in (x_{c1},x_*)$ the ground-state hole-like $c$ Fermi line may not be a perfect circle anymore,
its shape and that of the ground-state $s1$ boundary line having some $U/4t$ dependence. 

The energy parameter $\vert\Delta\vert$ of Eq. (\ref{bands-bipt}) is the maximum energy bandwidth 
of the $s1$ boundary line of the $s1$ fermion dispersion $\epsilon_{s1} ({\vec{q}})$ of Eq. (\ref{bands}) and
thus refers directly to that line. We recall that for the $s1$ fermion auxiliary dispersion
$\epsilon^0_{s1} ({\vec{q}})$ of Eq. (\ref{general-epsilon}) such a line has vanishing energy bandwidth, yet
this is not so for the $s1$ fermion energy dispersion $\epsilon_{s1} ({\vec{q}})$, owing to its $d$-wave-like
structure. For $U/4t\rightarrow 0$ the energy parameter $\Delta_0=\lim_{x\rightarrow 0}\vert\Delta\vert$ equals 
one half the chemical potential $\mu_0/2=\lim_{x\rightarrow 0}\mu/2$. 
For approximately $U/4t\geq u_0$ and $0<x\ll 1$ the $s1$ fermion-pairing-energy per spinon shift $\vert \vert\Delta\vert -\Delta_0\vert$
plays concerning the $s1$ boundary line a role similar to that of the chemical-potential shift $\vert \mu -\mu_0\vert=
[-e_c(q_{Fcx_1}^{h\,d})-e_c(q_{Fcx_2}^{h\,d})-\mu_0]$ for the Fermi line. Indeed, consistently with
$\vert\vert\Delta\vert -\Delta_0\vert$ being the shift of the maximum energy bandwidth of the $s1$
boundary line upon slightly increasing $x$, for a small hole concentration 
$0<x\ll 1$ such a shift controls the $s1$ boundary line energy 
of the $s1$ fermion auxiliary dispersion $\epsilon^0_{s1} ({\vec{q}})$ and $s1$ boundary line 
nodal energy of the corresponding dispersion $\epsilon_{s1} ({\vec{q}})$, respectively, as follows,
\begin{equation}
[e_{s1} (q_{Bs1\,x_1}) + e_{s1} (q_{Bs1\,x_2})] = -\delta\mu_{s1} 
\, ; \hspace{0.35cm} \delta\mu_{s1} = {1\over 4}\,\vert\vert\Delta\vert-\Delta_0\vert \, ,
\hspace{0.25cm} 0<x\ll 1 \, , \hspace{0.15cm} U/4t\geq u_0 \, ,
\label{relation-ep-Delta}
\end{equation}
where the factor $1/4$ is justified below and is related to the $U/4t\rightarrow\infty$ ratio $W_{s1}^0/\Delta_0=1/4$.
The $s1$ boundary energy $\delta\mu_{s1}=-[e_{s1} (q_{Bs1\,x_1}) + e_{s1} (q_{Bs1\,x_2})]$
of Eq. (\ref{relation-ep-Delta}) vanishes for 
$x=0$ and is small for $ 0<x\ll 1$. For approximately $U/4t\geq u_0$ and $0<x\ll 1$ one
can estimate its magnitude. The chemical potential shift $\vert \mu -\mu_0\vert$ obeys the inequality 
$\vert \mu -\mu_0\vert=[-e_c(q_{Fcx_1}^{h\,d})-e_c(q_{Fcx_2}^{h\,d})-\mu_0]\leq x\,\pi [W_c/2]$
where $W_c=[W_c^p+W_c^h]=8t$ is the $c$ fermion energy dispersion bandwidth, the chemical 
potential $\mu$ is given in Eq. (\ref{mux}), and the equality $[-e_c(q_{Fcx_1}^{h\,d})-e_c(q_{Fcx_2}^{h\,d})
-\mu_0]= x\,\pi [W_c/2]$ is reached for $U/4t\rightarrow\infty$. Similarly, symmetry implies that the $s1$ 
boundary-line energy $\delta\mu_{s1}\equiv [-e_{s1} (q_{Bs1\,x_1}) - e_{s1} (q_{Bs1\,x_2})]$ 
obeys a corresponding inequality $\delta\mu_{s1}\leq x\,\pi [W_{s1}^0/2]$ and is given by 
$\delta\mu_{s1}= x\,\pi [W_{s1}^0/2]$ for $U/4t\rightarrow\infty$. Furthermore, that the $s1$ 
boundary line is nearly $U/4t$ independent for approximately $U/4t\geq u_0$ and $x<x_{c1}$ 
requires that $\delta\mu_{s1}\approx x\,\pi [W_{s1}^0/2]$ for such a parameter-space region so
that the only $U/4t$ dependence of $\delta\mu_{s1}$ occurs through that of the 
energy bandwidth $W_{s1}^0$. Hence,
\begin{equation}
[e_{s1} (q_{Bs1\,x_1}) + e_{s1} (q_{Bs1\,x_2})] = -\delta\mu_{s1} 
\, ; \hspace{0.35cm} \delta\mu_{s1} \approx x\,\pi {W_{s1}^0\over 2}  
\, , \hspace{0.25cm} x<x_{c1} \, , \hspace{0.15cm} U/4t\geq u_0 \, ,
\label{EBL-s1}
\end{equation}
so that according to the relation of Eq. (\ref{relation-ep-Delta}), $\vert \vert\Delta\vert-\Delta_0\vert/4 \approx  x\,\pi W_{s1}^0/2$ 
for $0<x\ll 1$ and approximately $U/4t\geq u_0$.

The fulfillment of Eq. (\ref{EBL-s1}) implies both that $2e_{s1} (-q^N_{Bs1})=-\delta\mu_{s1}$ 
for $q_{0x_1}=q_{0x_2}=-q^N_{Bs1}$ and $[e_{s1} (0)+e_{s1} (-q^{AN}_{Bs1})]=-\delta\mu_{s1}$ for the
auxiliary $s1$ boundary-line momentum pointing in the nodal and
anti-nodal directions, respectively. Such conditions are met provided that, 
\begin{equation}
q^N_{Bs1} \approx \sqrt{2}\,{\pi\over 2}(1-x)  
\, ; \hspace{0.35cm}
q^N_{s1,\,arc} \approx \sqrt{2}\,{\pi\over 2}x 
\, ; \hspace{0.35cm} 
q^{AN}_{Bs1}\approx [\pi -\sqrt{x\,2\pi}]
\, , \hspace{0.25cm} x<x_{c1} \, , \hspace{0.15cm} U/4t\geq u_0 \, ,
\label{q-N-Fs}
\end{equation}
where $q^N_{s1,\,arc}$ is the absolute value of the $s1$ band auxiliary nodal arc momentum ${\vec{q}}^{\,N}_{s1,\,arc}$
of Eq. (\ref{q-N-s-arc}). The Cartesian components of auxiliary momenta belonging to the $s1$  boundary line and pointing 
in or near the nodal directions then obey the following equations,
\begin{equation}
q_{Bs1x_1} \pm q_{Bs1x_2} = [\pi - \delta q^N_{s1}] \hspace{0.15cm}
{\rm or}  \hspace{0.15cm} q_{Bs1x_1} \pm q_{Bs1x_2} = -[\pi - \delta q^N_{s1}] 
\, ; \hspace{0.25cm} \delta q^N_{s1} \equiv \sqrt{2}\,q^N_{s1,\,arc} 
\, , \hspace{0.25cm} x<x_{c1} \, , \hspace{0.15cm} U/4t\geq u_0 \, .
\label{qx_1-qx_2}
\end{equation}
Importantly, the $U/4t$ independent expressions given in Eqs. (\ref{q-N-Fs}) and (\ref{qx_1-qx_2})
for momenta belonging to the $s1$ boundary line confirm that the above expression 
$\delta\mu_{s1}= x\,\pi [W_{s1}^0/2]$ is consistent with the requirement of the $s1$ 
boundary line being $U/4t$ independent for approximately $U/4t\geq u_0$ and $x<x_{c1}$.
Since $\delta\mu_{s1}=\vert\vert\Delta\vert-\Delta_0\vert/4$ for $0<x\ll 1$ where the factor
$1/4$ is confirmed below, such a requirement
is fulfilled provided that $\vert\vert\Delta\vert-\Delta_0\vert\approx  x\,2\pi W_{s1}^0$ for small
hole concentrations and approximately $U/4t\geq u_0$.

If $\vert\Delta\vert =0$ there would be no short-range spin order and the $s1$ boundary line would be a vanishing 
energy line for all its momenta. However, that only holds for auxiliary momenta pointing in the nodal directions, the 
energy scale $\vert\Delta\vert$ being the maximum energy bandwidth of the $s1$ boundary line reached for auxiliary 
momenta pointing in the anti-nodal directions. The $s1$ fermion auxiliary
energy dispersion $\epsilon^{0}_{s1} ({\vec{q}})$ of Eq. (\ref{general-epsilon}) 
defines the momentum shape of the $s1$ boundary line through the
second relation provided in Eq. (\ref{g-FS}). In turn, the $s1$ fermion pairing energy per spinon
$\vert\Delta_{s1} ({\vec{q}})\vert$ given in Eq. (\ref{bands-bipt})
defines the energy bandwidth of that line within the dispersion $\epsilon_{s1} ({\vec{q}})$ of
Eq. (\ref{bands}). Furthermore, that for small $x$ the shift $\vert\vert\Delta\vert-\Delta_0\vert$
of its maximum magnitude $\vert\Delta\vert$ reads 
$\vert\vert\Delta\vert-\Delta_0\vert = 4\delta\mu_{s1}\approx  x\,2\pi W_{s1}^0$ 
assures that the $s1$ boundary line is $U/4t$ independent for approximately $U/4t\geq u_0$,
as Eqs. (\ref{q-N-Fs}) and (\ref{qx_1-qx_2}) confirm.

For $0<x\ll 1$ such a maximum magnitude $\vert\Delta\vert$ of the $s1$ fermion pairing energy per spinon 
given in Eq. (\ref{bands-bipt}) can be expressed as,
\begin{equation}
\vert\Delta\vert = \vert\Delta^{\parallel}_{s1} ({\vec{q}}_{Bs1}^{\,AN})\vert 
\approx \Delta_0\,{W_{s1}\over W^0_{s1}} = 
{\Delta_0\over W^0_{s1}}\vert e_{s1} (0) -e_{s1} (-q_{Bs1}^{AN})\vert 
\, ; \hspace{0.35cm} 0<x\ll 1 \, .
\label{Delta-x-small}
\end{equation}
Here the energy parameter $W_{s1} = 2\,[e_{s1} (-q^N_{Bs1}/\sqrt{2}) 
- e_{s1} (0)]$ of Eq. (\ref{Wg}), which according to Eq. (\ref{P-FS}) can be expressed both as
$W_{s1} = 2\,[e_{s1} (-q^{AN}_{Bs1})-e_{s1} (-q^N_{Bs1}/\sqrt{2})]$ and $W_{s1} = [e_{s1} (-q^{AN}_{Bs1})-e_{s1} (0)]$,
refers to the energy bandwidths $W_{s1}=[\epsilon^0_{s1} ({\vec{q}}^{\,d}_{Bs1})-\epsilon^0_{s1} (0)]$ and
$W_{s1}=[\epsilon_{s1} ({\vec{q}}^{\,N\,d}_{Bs1})-\epsilon_{s1} (0)]$ of the $s1$ fermion auxiliary dispersion
$\epsilon^0_{s1} ({\vec{q}})$ of Eq. (\ref{general-epsilon}) and between the nodal energy level and
zero-momentum energy level of the $s1$ fermion dispersion $\epsilon_{s1} ({\vec{q}})$ of Eq. (\ref{bands}), respectively.
     
Within the square-lattice quantum liquid of $c$ and $s1$ fermions the $U/4t$ dependence of the
charge mass ratio $r_c=m_c^{\infty}/m_c^*$ of Eq. (\ref{m*c/mc-UL}) controls the effects of electronic
correlations on physical quantities related to the charge degrees of freedom, including 
the $U/4t$ dependence of the $c$ fermion energy dispersion of Eq. (\ref{c-band-x-small}). 
In turn, the spin ratio $r_s=\Delta_0/4W_{s1}^0$ controls the effects of such correlations on 
quantities associated with the spin degrees of freedom. According to the results of Ref. \cite{companion},
upon increasing $U/4t$ the energy parameter $\Delta_0$
interpolates between $\Delta_0\approx \mu^0/2\approx 16te^{-\pi\sqrt{4t/U}}$ for $U/4t\rightarrow 0$ and 
$\Delta_0\approx  4W_{s1}^0\approx \pi (4t)^2/U$ for $U/4t\gg 1$. Hence both $\mu^0/2$ and $\Delta_0$ 
vanish for $U/4t\rightarrow 0$, whereas for $U/4t\gg 1$ the energy
parameter $\Delta_0$ becomes the energy scale $4W_{s1}^0\approx \pi (4t)^2/U$
associated with antiferromagnetic correlations. Within the square-lattice quantum liquid
scheme, the spin ratio $r_s=\Delta_0/4W_{s1}^0$ is often parametrize as $r_s=e^{-\lambda_s}$ where 
$\lambda_s = \vert\ln (\Delta_0/4W_{s1}^0)\vert$. For $U/4t>0$ it controls the interpolation 
behavior of the energy scale $\Delta_0$, which for the whole range of $U/4t$ values reads,
\begin{equation}
\Delta_0 = r_s\,4W_{s1}^0 = 4W_{s1}^0\,e^{-\lambda_s} 
\, ; \hspace{0.35cm} 
\lambda_s = \vert\ln (\Delta_0/4W_{s1}^0)\vert \, ,
\label{Delta-0-gen}
\end{equation}
where $\lambda_s$ has the limiting behaviors \cite{companion},
\begin{eqnarray}
\lambda_s & = & \pi\sqrt{4t/U} \, , \hspace{0.25cm} U/4t \ll 1 
\, ; \hspace{0.50cm}
\lambda_s \approx 4t\,u_0/U \, , \hspace{0.25cm} 
u_{00} \leq U/4t \leq u_1 \, ; \hspace{0.50cm}
\lambda_s = 0 \, , \hspace{0.25cm} U/4t \rightarrow\infty \, ,
\nonumber \\
u_{00} & \approx & (u_0/\pi )^2 \approx 0.171  \, ; \hspace{0.35cm} 
u_0 \approx 1.302 \, ; \hspace{0.35cm} u_1 \approx 1.600 \, .
\label{lambda-s}
\end{eqnarray}
The $u_0$ magnitude provided here is obtained taking into account that according to the investigations of Section VI,
$r_s=0.135\,\pi\approx 0.424$ for $U/4t\approx 1.525$. Indeed within the approximate 
expression $r_s\approx e^{-4t\,u_0/U}$ it is then uniquely determined by the equation 
$0.135\,\pi = e^{-u_0/1.525}$ and reads $u_0 \approx 1.302$,
as given in Eq. (\ref{Delta-0-gen}). That for approximately $u_{00} \leq U/4t \leq u_1$
the $U/4t$ exponential dependence of $\lambda_s$ involves $t/U$ rather than $\sqrt{t/U}$, as it does for $U/4t\ll 1$ when
$r_s= e^{-\pi\sqrt{4t/U}}$, is consistent with for large $U/4t$ values $r_s$ being 
of the form $r_s\approx [1-{\cal{C}}\,t/U]$. For large $U/4t$ only the zeroth order term of
the expansion $e^{-4t\,u_0/U}\approx [1-4t\,u_0/U]$ is exact, $4u_0\approx 5.2$
being an approximate value for the unknown coefficient ${\cal{C}}>0$ of 
the first order term $-{\cal{C}}\,t/U$. Nevertheless, the expression $r_s\approx e^{-4t\,u_0/U}$ is expected to
be a good quantitative approximation for the range $u_{00}\leq U/4t\leq u_1$ of Eq. (\ref{lambda-s}).
 
It follows from the above analysis that for $x<x_{c1}$ and the approximate range $U/4t\geq u_0$ 
the momentum shape of the ground-state $s1$ band boundary line is independent of $U/4t$. 
The factor $1/4$ of the relation $\delta\mu_{s1}=\vert\vert\Delta\vert -\Delta_0\vert/4$ provided in 
in Eq. (\ref{relation-ep-Delta}) is confirmed
from properties specific to the $U/4t\rightarrow\infty$ limit. Indeed, symmetries associated with that limit
imply that the $x=0$ elementary-function expression $e_{s1} (q)\approx -[W_{s1}^0/2]\,\cos q$ is for $0<x\ll 1$
and $U/4t\rightarrow\infty$ a good approximation for both
$[e_{s1} (q_{0x_1})+e_{s1} (q_{0x_2})]$ and $[e_{s1} (q_{0x_1})-e_{s1} (q_{0x_2})]$. This
is in contrast to finite $U/4t$ values and small finite hole concentrations
$0<x\ll 1$, for which it is a good approximation  
concerning the quantity $[e_{s1} (q_{0x_1})+e_{s1} (q_{0x_2})]$ but not 
$[e_{s1} (q_{0x_1})-e_{s1} (q_{0x_2})]$. It then follows that for $0<x\ll 1$ and $U/4t\rightarrow\infty$ 
the energy bandwidth ratio $W_{s1}/W_{s1}^0=\vert e_{s1} (0) -e_{s1} (-q_{Bs1}^{AN})\vert/W_{s1}^0$ 
appearing in the $\vert\Delta\vert$ expression of Eq. (\ref{Delta-x-small}) reads 
$\lim_{U/4t\rightarrow\infty} W_{s1}/W_{s1}^0\approx [1-\cos (\pi -q_{Bs1}^{AN})]/2= (1-x\,\pi/2)$.
This implies that the $s1$ fermion auxiliary dispersion-energy bandwidth 
deviation $[W_{s1}^0-W_{s1}]$ has for $0<x\ll 1$ the limiting behavior $\lim_{U/4t\rightarrow\infty}[W_{s1}^0-W_{s1}]=\delta\mu_{s1}$.
Furthermore, since $r_s=1$ for $U/4t\rightarrow\infty$ one finds from the use of Eqs. (\ref{Delta-x-small}) and (\ref{Delta-0-gen})
that $\lim_{U/4t\rightarrow\infty}[\Delta_0-\vert\Delta\vert]=\lim_{U/4t\rightarrow\infty}\vert\vert\Delta\vert -\Delta_0\vert=4\delta\mu_{s1}$ so that the 
relation $\delta\mu_{s1}=\vert\vert\Delta\vert -\Delta_0\vert/4$ of Eq. (\ref{relation-ep-Delta}) holds
in that limit. As confirmed above, that the $s1$ boundary line is nearly $U/4t$ independent for approximately $U/4t\geq u_0$ 
and $x<x_{c1}$ implies that for $0<x\ll 1$ and approximately $U/4t\geq u_0$ the ratio 
$\delta\mu_{s1}/\vert\vert\Delta\vert -\Delta_0\vert=1/4$ is independent of $U/4t$ and $x$ so that
the relation $\delta\mu_{s1}/\vert\vert\Delta\vert -\Delta_0\vert/4$ remains valid for such a range,
as given in Eq. (\ref{relation-ep-Delta}). That relation refers to $m=0$ and very small but finite values of $x$
and assures that the $s1$ boundary-line momenta expressions provided in Eqs.
Eqs. (\ref{q-N-Fs}) and (\ref{qx_1-qx_2}) are independent of $U/4t$. Indeed, 
since $\delta\mu_{s1}\propto \vert\vert\Delta\vert -\Delta_0\vert$, if the ratio
$\delta\mu_{s1}/\vert\vert\Delta\vert -\Delta_0\vert$ was not independent of $x$ and
$U/4t$ such a symmetry requirement would not be fulfilled. Hence its value $1/4$ found
here for $U/4t\rightarrow\infty$ is valid as well for approximately $U/4t\geq u_0$.

We recall that for $m=0$ the $s1$ fermion pairing energy per spinon $\vert\Delta\vert$ of
the model on the square lattice has a singular behavior at $x=0$, with the above energy
parameter $\Delta_0=\lim_{x\rightarrow 0}\vert\Delta\vert$ being for $U/4t>0$ different
from $\mu_0/2= \vert\Delta\vert\vert_{x=0}>\Delta_0$. This is due to a sharp quantum phase
transition between the $x=0$ and $m=0$ ground state with long-range antiferromagnetic order 
and the $m=0$ and $0<x\ll 1$ ground state with 
a incommensurate-spiral short-range spin order with strong antiferromagnetic
correlations \cite{companion}. At $m=0$ and both for small temperature $T>0$ and $x=0$ and 
for $T\geq 0$ and $0<x\ll 1$ the system is driven into
a renormalized classical regime where the 
$T=0$ and $x=0$ long-range antiferromagnetic 
order is replaced by a quasi-long-range spin order
as that studied in Ref. \cite{Principles} for simpler spin
systems.

Combination of expressions (\ref{relation-ep-Delta}), (\ref{Delta-x-small}), and (\ref{Delta-0-gen})
reveals that for small hole concentrations $0<x\ll 1$ and the approximate range $U/4t\geq u_0$ the spin ratio 
$r_s=e^{-\lambda_s}$ where $\lambda_s$ is given in Eq. (\ref{lambda-s}) 
controls the dependence on $x$ of the energy scale $\vert\Delta\vert$ and
energy bandwidth $W_{s1}$, which read,
\begin{eqnarray}
\vert\Delta\vert & \approx & 4r_s\vert e_{s1} (0) -e_{s1} (-q_{Bs1}^{AN})\vert \approx
\left(1-{x\over x_*^0}\right) 2\Delta_0 \, ; \hspace{0.35cm}
x_*^0 \approx {2r_s\over \pi}  = {2e^{-\lambda_s}\over \pi} \, ,
\nonumber \\
W_{s1} & = & \vert e_{s1} (0) -e_{s1} (-q_{Bs1}^{AN})\vert \approx \left(1-{x\over x_*^0}\right) W_{s1}^0 \, ,
\hspace{0.25cm} 0<x\ll 1 \, , \hspace{0.15cm} U/4t\geq u_0 \, ,
\label{D-D-HM-x<<1}
\end{eqnarray}
respectively. Since the spin ratio $r_s$ is a monotonous increasing function
of $U/4t$, for the range $U/4t\geq u_0$ the 
parameter $x_*^0$ is an increasing function of $U/4t$ as well. We emphasize that the 
$U/4t$ dependence of the energy scale $2\vert\Delta\vert$ and parameter $x_*^0$ 
does not contradict the $U/4t$ independence
of the $s1$ boundary line shape in momentum space. Such an independence
is related to the invariance under the electron - rotated-electron unitary transformation
of the momentum operator. Indeed, except for small values of $U/4t$ the $m=0$ ground
states refer for small $x$ to a single $V$ tower. In turn, the Hamiltonian is not invariant under
that transformation. Therefore, the $s1$ boundary line energy bandwidth $\vert\Delta\vert$
depends on $U/4t$. In turn, that as given in Eq. (\ref{relation-ep-Delta} ) 
$[\Delta_0-\vert\Delta\vert]/\delta\mu_{s1}=4$ is $U/4t$ independent for $0<x\ll 1$ and
the range $U/4t\geq u_0$ following indirectly from
the $U/4t$ independence of the $s1$ boundary line shape. 

In Appendix C it is shown that except for $U/4t\rightarrow\infty$ the elementary-function expression 
$e_{s1} (q)\approx -[W_{s1}^0/2]\,\cos q$, which is a good approximation for
$0<x\ll 1$ concerning physical quantities involving $[e_{s1} (q_{0x_1})+e_{s1} (q_{0x_2})]$,
does not lead to the correct expression of those involving the quantity $[e_{s1} (q_{Bs1\,x_1})-e_{s1} (q_{Bs1\,x_2})]$ 
such as the energy parameter $2\vert\Delta\vert$ and energy bandwidth $W_{s1}$ of Eq. (\ref{D-D-HM-x<<1}). The point
is that except for $U/4t\rightarrow\infty$ there are extra terms in $e_{s1} (q)$ that
do not contribute to $[e_{s1} (q_{Bs1\,x_1})+e_{s1} (q_{Bs1\,x_2})]$.
In that Appendix a corresponding $e_{s1} (q)$ expression valid up to first order in $x$ 
and given in Eq. (\ref{e-s1-q}) of that Appendix is evaluated. 

As discussed in the following, for the range $U/4t\geq u_0$ considered in our studies
the short-range spin order prevails at zero temperature for finite hole concentrations $0<x<x_*$ where 
the critical hole concentration $x_*$ is for approximately $u_0\leq U/4t\leq u_{\pi}$ given by
the parameter $x_*^0$ of Eq. (\ref{D-D-HM-x<<1}), consistently with the central role plaid
in the quantum liquid physics by the spin ratio $r_s =\Delta_0/4W_{s1}^0=e^{-\lambda_s}$.

\subsubsection{The critical hole concentration $x_*$}

For the square-lattice quantum liquid the linear dependence on $(1-x/x_*^0)$
of the maximum magnitude of the $s1$ fermion spinon-pairing energy $2\vert\Delta\vert$ given in 
Eq. (\ref{D-D-HM-x<<1}) is obtained here for small hole concentrations $0<x\ll 1$. 
The studies of Ref. \cite{cuprates0} reveal that the fluctuations of a phase denoted in that
reference by $\theta_1$ and the corresponding amplitude $g_1=\vert\langle e^{i\theta_{1}}\rangle\vert$
play an important role in the physics of the square-lattice quantum liquid. For zero temperature
that amplitude is denoted by ${\breve{g}}_1$ and for $U/4t$ intermediate values in approximately the
range $U/4t\in (u_0,u_{\pi})$ a consistent picture is
reached provided that the linear-$x$ decreasing of the energy scale
$2\vert\Delta\vert$ found above for $0<x\ll 1$ refers to the whole range $0<x<x_*$ . This
implies that for $U/4t\in (u_0,u_{\pi})$ the critical hole
concentration $x_*$ equals the parameter $x_*^0$ of Eq. (\ref{D-D-HM-x<<1}). Indeed, the investigations
of that reference lead to ${\breve{g}}_1\approx (1-x/x_*)^{z\nu}$ for $0<(x_*-x)\ll 1$ where
$x_*$ is a critical hole concentration above which there is no short-range spin order at zero temperature, 
$z=1$ the dynamical exponent, and owing to symmetry arguments the exponent $\nu$ is expected to
read $\nu =1$ both for $0<x\ll 1$ and $0<(x_*-x)\ll 1$, respectively. This is fulfilled
provided that $x_*\approx x_*^0$ and $\nu=1$. The fulfillment of symmetry requirements confirms that 
${\breve{g}}_1\approx (1-x/x_*)$ is a good approximation for $0<x<x_*$.  
In turn, for $U/4t>u_{\pi}$ and thus $1/\pi <x_*^0\leq 2/\pi\approx 0.64$ either $2\vert\Delta\vert$ 
decreases linearly or faster upon increasing $x$ so that $x_*\leq x_*^0$. Based on
the available information one then finds,
\begin{equation}
x_* \approx x_*^0 = {2r_s\over \pi}  = {2e^{-\lambda_s}\over \pi} \, , \hspace{0.25cm}{\rm for}\hspace{0.10cm}
u_0\leq U/4t\leq u_{\pi} \, ; \hspace{0.35cm}  x_*  \leq x_*^0 \, , \hspace{0.25cm}{\rm for}\hspace{0.10cm} U/4t\geq u_{\pi} \, ,
\label{x-c}
\end{equation}
where whether $x_*\approx x_*^0$ or $x_*< x_*^0$ for $U/4t\geq u_{\pi}$ depends on
$2\vert\Delta\vert$ decreasing linearly in $x$ or faster, upon increasing $x$.
It follows that the critical hole concentration $x_*$ has for $U/4t\geq u_0$ the limiting values,
\begin{eqnarray}
x_* & \approx & 0.23 \, , \hspace{0.50cm} U/4t = u_0 \approx 1.302 \, ,
\nonumber \\
& \approx &  0.27 \, , \hspace{0.50cm} U/4t = u_* = 1.525 \, ,
\nonumber \\
& \approx & 0.28 \, , \hspace{0.50cm} U/4t = u_1 \approx 1.600
\nonumber \\
& = & {1\over \pi} \approx 0.32 \, , \hspace{0.50cm} U/4t = u_{\pi}>u_1
\nonumber \\
& \leq & {2\over \pi} \approx 0.64 \, , \hspace{0.50cm} U/4t\rightarrow\infty  \, .
\label{xc-range}
\end{eqnarray}

The fluctuations of the $x$ dependent phase $\theta_1$ refer to 
short-range spin fluctuations. According to the investigations of Ref. \cite{cuprates0},
such fluctuations become large for $x\rightarrow x_*$.
For a temperature range $0\leq T<T^*$, $0<x<x_*$, and approximately $u_0\leq U/4t\leq u_{\pi}$ where
$T^*\approx {\breve{g}}_1 \Delta_0/k_B$ is the pseudogap temperature the fluctuations of the phase $\theta_1$ remain small
so that the amplitude $g_1=\vert\langle e^{i\theta_{1}}\rangle\vert$
remains finite. This confirms that the phase
$\theta_1$ and corresponding amplitude $g_1$
are directly related to the short-range spin correlations,
which prevail provided that $g_1>0$. For approximately $U/4t\geq u_0$ that occurs for
$0<x<x_*$ and $T<T^*$.

Since for $U/4t\in (u_0, u_{\pi})$ the prefactor ${\breve{g}}_1\approx (1-x/x_*)$ 
of $2\vert\Delta\vert =(1-x/x_*)\,2\Delta_0$
obeys for the hole concentration range $0<x<x_*$ the
amplitude properties $g_1=\vert\langle e^{i\theta_{1}}\rangle\vert\leq 1$
and $g_1=\vert\langle e^{i\theta_{1}}\rangle\vert\rightarrow 0$ for $x\rightarrow x_*$,
the studies of Ref. \cite{cuprates0} identify the zero-temperature magnitude 
${\breve{g}}_1\,2\Delta_0$ of $2\vert\Delta\vert =g_1\,2\Delta_0$
with the maximum magnitude of the $s1$ fermion spinon pairing energy
for the whole range of hole concentrations $0<x<x_*$ of the short-range spin ordered phase.
According to such an identification, for $u_0\leq U/4t\leq u_{\pi}$ the corresponding linear behavior in $(1-x/x_*)$ of
that energy scale remains dominant for $m=0$ and a hole-concentration range $0<x<x_*$
where $2\vert\Delta\vert =(1-x/x_*) 2\Delta_0>0$. 

The expression of Eq. (\ref{Delta-x-small}) is only valid for $0<x\ll 1$ 
and for larger values of $x$ a linear behavior of 
the energy scale $2\vert\Delta\vert$ does not imply a similar behavior for the
$x$ dependence of the energy bandwidth $W_{s1}$
given in Eq. (\ref{D-D-HM-x<<1}) for $0<x\ll 1$. That the dominant processes behind
the linear decreasing of the magnitude of the $s1$ 
fermion spinon-pairing energy $2\vert\Delta\vert$
given in Eq. (\ref{D-D-HM-x<<1}) remain dominant
for $u_0\leq U/4t\leq u_{\pi}$ and larger values of the hole concentration $x$
in the range $x\in (0,x_*)$ plays a central role in
the physics of the square-lattice quantum liquid:
It implies that for approximately $u_0\leq U/4t\leq u_{\pi}$ such a pairing energy 
decreases linearly with $x$ until it vanishes, as the limit $x\rightarrow x_*$ is reached. 
Indeed, extension to a larger hole-concentration range of the linear dependence 
on $(1-x/x_*)$ given in Eq. (\ref{D-D-HM-x<<1}) for that energy scale 
would lead to negative values of $2\vert\Delta\vert$ for $x>x_*$.
The results of Ref. \cite{cuprates0} confirm that $g_1=0$ for $u_0\leq U/4t\leq u_{\pi}$ and 
$x>x_*$ so that the zero-temperature expression $2\vert\Delta\vert ={\breve{g}}_1 2\Delta_0$ is 
only valid for $0<x<x_*$. Consistently with the energy scale $2\vert\Delta\vert$ being associated with the 
short-range spin order, it follows that the hole concentration $x_*$ of Eq. (\ref{x-c}) 
corresponds to the critical hole concentration above which 
the short-range spin order disappears at zero temperature
for the intermediate $U/4t$ range $U/4t\in (u_0,u_{\pi})$. Consistently with
both $g_1=0$ and $2\vert\Delta\vert =0$ for $x>x_*$. 
Often our analysis focuses on the smaller range $U/4t\in (u_0, u_1)$
for which the charge and spin ratios obey approximately the relation
$r_c\approx 2r_s\approx \pi\,x_*$ and $r_s\approx e^{-4t\,u_0/U}$. Indeed 
for finite $U/4t>u_{\pi}$ values one has 
that $r_s<r_c<2r_s$ and such a relation is not fulfilled. In addition, 
for such $U/4t$ values the approximate expression $r_s\approx e^{-4t\,u_0/U}$ is not
valid anymore and thus we
could not access the accurate value of $u_{\pi}>u_1$ at which the
$x_*$ magnitude reads $x_*=1/\pi$.  

For $m=0$ the $s1$ fermion spinon-pairing energy 
$2\Delta_0=\lim_{x\rightarrow 0} 2\vert\Delta\vert$ is the energy below which the short-range spin 
order with strong antiferromagnetic correlations survives for hole 
concentrations $0<x\ll 1$. The short-range spin order that emerges
for small hole concentrations and temperatures
$T\geq 0$ is similar to that occurring for $x=0$ and $T>0$ and was studied previously
as for instance in Ref. \cite{Hubbard-T*-x=0}. The results
of Ref. \cite{cuprates0} indicate that if one adds to the square-lattice
quantum liquid considered in this paper and in Ref. \cite{companion} a small perturbation associated with 3D anisotropy
effects brought about by weak plane coupling, for hole concentrations $x_c<x<x_*$ 
and temperatures $T<T_c\approx {\breve{g}} \Delta_0/2k_B$
the short-range spin order coexists in such a quasi-2D system with a long-range
superconducting order. Here ${\breve{g}}=[(x-x_c)/(x_*-x_c)]\,{\breve{g}}_1$
for $u_0\leq U/4t\leq u_1$ and the critical hole concentration $x_c$ reads $x_c\approx 10^{-2}$ for
the small 3D anisotropy effects considered in that reference.
The maximum energy gap magnitude of Eq. (\ref{bands-bipt}) of 
the $s1$ fermion energy dispersion equals the maximum $s1$ fermion 
pairing energy per spinon $\vert\Delta\vert$ and corresponds to 
momentum values belonging to the $s1$ boundary line 
whose auxiliary momenta of Eq. (\ref{qFc-qBs1})
point in the anti-nodal directions. For approximately $u_0\leq U/4t\leq u_{\pi}$ it reads,
\begin{equation}
\vert\Delta\vert =
\vert\Delta_{s1} ({\vec{q}}^{\,d\,AN}_{Bs1})\vert
= \vert\Delta^{\parallel}_{s1} ({\vec{q}}^{\,AN}_{Bs1})\vert
\approx  \Delta_0 \left(1-{x\over x_*}\right)
\, ; \hspace{0.35cm} 0<x<x_* \, , \hspace{0.15cm} m=0 
 \, , \hspace{0.15cm} u_0\leq U/4t\leq u_{\pi} \, .
\label{D-x}
\end{equation}
  
\section{$c$ and $s1$ fermion velocities and the 
one- and two-electron excitations}

In this section we derive the $c$ and $s1$ fermion velocities from the corresponding energy 
dispersions obtained above and
discuss their role in the one- and two-electron spectrum.

\subsection{The $c$ and $s1$ fermion velocities}

The $c$ and $s1$ fermion group velocities derived from the 
energy dispersions are the elementary
building blocks of the velocities associated with
one- and two-electron excitations. They read, 
\begin{equation}
{\vec{V}}_{s1} (\vec{q}) = {\vec{\nabla}}_{\vec{q}}\,\epsilon_{s1} (\vec{q})
\, ; \hspace{0.35cm}
{\vec{V}}^{\,0}_{s1} (\vec{q}) =
{\vec{\nabla}}_{\vec{q}}\,\epsilon^0_{s1} (\vec{q}) 
\, ; \hspace{0.35cm}
{\vec{V}}^{\,\Delta}_{s1} (\vec{q}) =
-{\vec{\nabla}}_{\vec{q}}\vert\Delta_{s1} 
({\vec{q}})\vert \, ; \hspace{0.35cm}
{\vec{V}}_{c} (\vec{q}^{\,h}) = {\vec{\nabla}}_{\vec{q}^{\,h}}\,\epsilon_{c} (\vec{q}^{\,h}) \, .
\label{g-velocities-def}
\end{equation}
Such velocities can be expressed as,
\begin{equation}
{\vec{V}}_{s1} (\vec{q}) = V_{s1} (\vec{q})\,{\vec{e}}_{\phi_{s1}(\vec{q})}
\, ; \hspace{0.35cm}
{\vec{V}}^{\,0}_{s1} (\vec{q}) = V^0_{s1} (\vec{q})\,{\vec{e}}_{\phi_{s1}^{0}(\vec{q})} 
\, ; \hspace{0.35cm}
{\vec{V}}^{\,\Delta}_{s1} (\vec{q}) =
V^{\Delta}_{s1} (\vec{q})\,{\vec{e}}_{\phi_{s1}^{\Delta}(\vec{q})}
\, ; \hspace{0.35cm}
{\vec{V}}_{c} (\vec{q}^{\,h}) = V_{c} (\vec{q}^{\,h})\,{\vec{e}}_{\phi_{c}(\vec{q}^{\,h})} \, ,
\label{velocities-mod-def}
\end{equation}
where $V_{s1} (\vec{q})$, $V^0_{s1} (\vec{q})$, $V^{\Delta}_{s1} (\vec{q})$,
$V^{\Delta}_{s1} (\vec{q})$, and $V_{c} (\vec{q}^{\,h})$ are the velocities absolute 
values and the unit vectors are obviously given by,
\begin{equation}
{\vec{e}}_{\phi_{s1}(\vec{q})} = {{\vec{V}}_{s1} (\vec{q})\over V_{s1} (\vec{q})}
\, ; \hspace{0.35cm}
{\vec{e}}_{\phi_{s1}^{0}(\vec{q})} = {{\vec{V}}^{\,0}_{s1} (\vec{q})\over V^0_{s1} (\vec{q})} 
\, ; \hspace{0.35cm}
{\vec{e}}_{\phi_{s1}^{\Delta}(\vec{q})} = {{\vec{V}}^{\,\Delta}_{s1} (\vec{q})\over 
V^{\Delta}_{s1} (\vec{q})}
\, ; \hspace{0.35cm}
{\vec{e}}_{\phi_{c}(\vec{q}^{\,h})} = {{\vec{V}}_{c} (\vec{q}^{\,h})\over V_{c} (\vec{q}^{\,h})} \, .
\label{unit-vec-velo-def}
\end{equation}
Within the notation used here a unit vector ${\vec{e}}_{\phi}$ is such that its
components are given in terms of the angle $\phi$ that defines
its direction as in Eq. (\ref{unit-vector}). 

The velocities ${\vec{V}}^{\,0}_{s1} (\vec{q})$ and  ${\vec{V}}^{\,\Delta}_{s1} (\vec{q})$
of Eq. (\ref{g-velocities-def}) can be expressed as,
\begin{equation}
{\vec{V}}^{\,0}_{s1} (\vec{q}) = ([A^d_{s1}]^{-1})^T\,{\vec{V}}^{\,0,\parallel}_{s1} (\vec{q}_0)
\, ; \hspace{0.50cm}
{\vec{V}}^{\,\Delta}_{s1} (\vec{q})
= ([A^d_{s1}]^{-1})^T\,{\vec{V}}^{\Delta,\parallel}_{s1} (\vec{q}_0) \, ,
\label{para-perp-v-gen}
\end{equation}
where $A^d_{s1}$ is the $2\times 2$ matrix appearing in Eq. (\ref{q-A-general}), $[A^d_{s1}]^{-1}$ its inverse matrix,
$([A^d_{s1}]^{-1})^T$ the transposition of the latter matrix, and
the velocities ${\vec{V}}^{\,0,\parallel}_{s1} (\vec{q}_0)$ and
${\vec{V}}^{\Delta,\parallel}_{s1} (\vec{q}_0)$ are given by,
\begin{equation}
{\vec{V}}^{\,0,\parallel}_{s1} (\vec{q}_0) =
{\vec{\nabla}}_{\vec{q}_0}\,\epsilon^{0,\parallel}_{s1} (\vec{q}_0)
\, ; \hspace{0.50cm}
{\vec{V}}^{\Delta,\parallel}_{s1} (\vec{q}_0) =
-{\vec{\nabla}}_{\vec{q}_0}\vert\Delta^{\parallel}_{s1} ({\vec{q}}_0)\vert \, .
\label{g-parallel-velocities-def}
\end{equation}
Here $\epsilon^{0,\parallel}_{s1} (\vec{q}_0)=\epsilon^{0,\parallel}_{s1} ([A_{s1}^d]^{-1}\vec{q})$ 
is the auxiliary energy dispersion of Eq. (\ref{general-epsilon})
and $\vert\Delta^{\parallel}_{s1} ({\vec{q}}_0)\vert= \vert\Delta^{\parallel}_{s1} ([A_{s1}^d]^{-1}{\vec{q}})\vert$ 
the auxiliary gap function of Eq. (\ref{pairing-en-x>0-m0}).

For $s1$ band momenta $\vec{q}$ at or near the $s1$ boundary line and hole
concentrations in the range $x_{c1}<x<x_{c2}$ one found above in Section III that
$A_{s1}^d=A^d_F$, $[A_{s1}^d]^{-1}=A^{-d}_F$, and $([A^d_{s1}]^{-1})^T=A^d_F$
where $A^d_F$ is the $2\times 2$ matrix given in Eq. (\ref{A-c-s1})
so that expressions (\ref{para-perp-v-gen}) simplify to,
\begin{equation}
{\vec{V}}^{\,0}_{s1} (\vec{q}) = A^d_F\,{\vec{V}}^{\,0,\parallel}_{s1} (\vec{q}_0)
\, ; \hspace{0.50cm}
{\vec{V}}^{\,\Delta}_{s1} (\vec{q})
= A^d_F\,{\vec{V}}^{\Delta,\parallel}_{s1} (\vec{q}_0) \, .
\label{para-perp-v}
\end{equation}

The $s1$ velocities of Eq. (\ref{g-parallel-velocities-def}) and the $c$
fermion velocity have Cartesian components given by,
\begin{eqnarray}
{\vec{V}}^{\Delta,\parallel}_{s1} (\vec{q}_0) & = & -{\rm sgn}\,\{
e_{s1} (q_{0x_1}) - e_{s1} (q_{0x_2})\} 
{\vert\Delta\vert\over W_{s1}}\,[v_{s1} (q_{0x_1}),- v_{s1} (q_{0x_2})]  \, ,
\nonumber \\
{\vec{V}}^{\, 0,\parallel}_{s1} (\vec{q}_0) & = &
[v_{s1} (q_{0x_1}),v_{s1} (q_{0x_2})] 
\, ; \hspace{0.35cm}
{\vec{V}}_{c} (\vec{q}^{\,h}) =
[v_{c} (q^{h}_{x_1}),v_{c} (q^{h}_{x_2})] \, ,
\label{g-velocities}
\end{eqnarray}
where $e_{\gamma} (q)$ and 
$v_{\gamma} (q)$ with $\gamma =c,s1$ are the
elementary functions and elementary velocities, respectively, 
of Eq. (\ref{pairf}), $\vert\Delta\vert$ one half the energy parameter defined in
Eq. (\ref{D-D-HM-x<<1}), and $W_{s1}$ the 
energy bandwidth of the $s1$ fermion auxiliary dispersion provided in Eq. (\ref{general-epsilon}).

The following velocities vanish: ${\vec{V}}_c (\vec{\pi}) = {\vec{V}}_{c} (0)
= {\vec{V}}^0_{s1} (0) = 0$. The velocities ${\vec{V}}^{\Delta}_{s1} (\vec{q})$ and
${\vec{V}}^{\Delta,\parallel}_{s1} (\vec{q}_0)$ have the same absolute value
$V^{\Delta}_{s1} =V^{\Delta}_{s1} ({\vec{q}})$, which can be written as,
\begin{equation}
V^{\Delta}_{s1} ({\vec{q}}) = {\vert\Delta\vert\over\sqrt{2}}\,G_{s1}({\vec{q}}) \, ; \hspace{0.35cm}
G_{s1}({\vec{q}}) = G^{\parallel}_{s1}({\vec{q}}_0) = {2\over W_{s1}}\sqrt{v_{s1} (q_{0x_1})^2+v_{s1} (q_{0x_2})^2\over 2} \, ,
\label{v-D-Fs-G}
\end{equation}
where the function $G^{\parallel}_{s1}({\vec{q}}_0)$ is such that, 
\begin{equation}
G^{\parallel}_{s1}({\vec{q}}_0) \in (0,1) \, ; \hspace{0.25cm}
G^{\parallel}_{s1} ({\vec{q}}_{Bs1}) =
\vert\sin 2\phi\vert
\, ; \hspace{0.25cm}
G^{\parallel}_{s1} ({\vec{q}}^{\,N}_{Bs1}) = 1
\, ; \hspace{0.25cm}
G^{\parallel}_{s1} ({\vec{q}}^{\,AN}_{Bs1}) = 0 \, .
\label{G-s-limits}
\end{equation}
Here the expression $G^{\parallel}_{s1} ({\vec{q}}_{Bs1}) =
\vert\sin 2\phi\vert$ is exact for $0<x\ll 1$ and is expected to be a
good approximation for $0<x<x_*$ and the auxiliary momenta
${\vec{q}}^{\,N}_{Bs1}$ and ${\vec{q}}^{\,AN}_{Bs1}$
are particular cases of the general auxiliary momentum
${\vec{q}}_{Bs1}$ defined in Eq. (\ref{qFc-qBs1}) and
point in the nodal and anti-nodal directions, respectively.
The corresponding momenta ${\vec{q}}^{\,d\,N}_{Bs1}$ and
${\vec{q}}^{\,d\,AN}_{Bs1}$ are particular cases of
the general $s1$ boundary-line momentum ${\vec{q}}^{\,d}_{Bs1}$
of Eq. (\ref{q-Bs1}). If as discussed in Section III 
one assumes that the hole concentration $x_h$ above which the Fermi line is
particle like obeys the inequality $x_{c2}\leq x_h<x_*$ (rather than $x_h\geq x_*$),
for the hole-concentration range $x_h<x<x_*$ for which the Fermi line is particle like that the range of
$\phi$ that the expression $G^{\parallel}_{s1} ({\vec{q}}_{Bs1}) =
\vert\sin 2\phi\vert$ refers to is given in Eq. (\ref{kF}) so that
the minimum magnitude of $G^{\parallel}_{s1} ({\vec{q}}_{Bs1})$ rather than 
vanishing reads $G^{\parallel}_{s1} ({\vec{q}}_{Bs1}) \approx 2\phi_{AN}$
where $\phi_{AN}(x)$ is small. Such a minimum magnitude is reached at $\phi=\phi_{AN}$ and $\phi=\pi/2-\phi_{AN}$ 
instead of at $\phi =0$ and $\phi =\pi/2$, respectively.

At the $s1$ boundary line the velocity ${\vec{V}}_{s1} (\vec{q})$ reads,
\begin{equation}
{\vec{V}}_{s1} ({\vec{q}}^{\,d}_{Bs1}) = {\vec{V}}^{\Delta}_{s1} ({\vec{q}}^{\,d}_{Bs1}) 
\, ; \hspace{0.35cm}  {\vec{V}}^{\Delta}_{s1} ({\vec{q}}^{\,d}_{Bs1}) = A_F^d\,
{\vec{V}}^{\Delta,\parallel}_{s1} ({\vec{q}}_{Bs1}) \, ,
\label{BL-Vs1}
\end{equation}
where the second expression is valid for $x_{c1}<x<x_{c2}$. 
To reach such an expression we used the relations (\ref{para-perp-v})
and that $A^d_{s1}=A_F^d$ for momenta at the $s1$ boundary line and $x_{c1}<x<x_{c2}$.
The equality ${\vec{V}}_{s1} ({\vec{q}}^{\,d}_{Bs1}) = {\vec{V}}^{\Delta}_{s1} ({\vec{q}}^{\,d}_{Bs1})$
implies that the velocity ${\vec{V}}^{\Delta}_{s1} ({\vec{q}}^{\,d}_{Bs1})$ plays
an important role in the quantum liquid physics. The corresponding auxiliary velocity 
${\vec{V}}^{\Delta,\parallel}_{s1} ({\vec{q}}_{Bs1})$ and
the absolute value $V^{\Delta}_{s1} ({\vec{q}}^{\,d}_{Bs1})$
of both ${\vec{V}}^{\Delta}_{s1} ({\vec{q}}^{\,d}_{Bs1})$ and
${\vec{V}}^{\Delta,\parallel}_{s1} ({\vec{q}}_{Bs1})$
are given by,
\begin{equation}
{\vec{V}}^{\Delta,\parallel}_{s1} ({\vec{q}}_{Bs1}) = -{\rm sgn}\,\{
e_{s1} (q_{Bs1x_1}) - e_{s1} (q_{Bs1x_2})\} 
{\Delta\over W_{s1}}\,[v_{s1} (q_{Bs1x_1}),- v_{s1} (q_{Bs1x_2})] \, ; \hspace{0.35cm}
V^{\Delta}_{s1} ({\vec{q}}^{\,d}_{Bs1}) = {\vert\Delta\vert\over\sqrt{2}}\,G^{\parallel}_{s1}({\vec{q}}_{Bs1})  \, ,
\label{v-D-Fs}
\end{equation}
respectively.

For the $s1$ band nodal directions such that $q_{0x_1}=\pm q_{0x_2}$ 
the velocity ${\vec{V}}_{s1} (\vec{q}) = {\vec{\nabla}}_{\vec{q}}\,\epsilon_{s1} (\vec{q})$
simplifies to ${\vec{V}}_{s1} ({\vec{q}}) = 
{\vec{V}}^{0}_{s1}({\vec{q}})$ so that it is not
well defined for the four momenta
of the $s1$ boundary line pointing in the
nodal directions. Hence in that case we consider
instead the velocity at the $s1$ boundary-line momenta
${\vec{q}}^{\,d\,N}_{Bs1}$ whose auxiliary
momenta ${\vec{q}}^{\,N}_{Bs1}$ given in Eq. (\ref{qFc-qBs1}) 
point in the infinitesimal vicinity of the nodal 
directions and hence for the quadrant where $q_{0x_1}=q_{0x_2}<0$
have Cartesian components ${\vec{q}}^{\,N}_{Bs1}=[-q^N_{Bs1x_1}(1+\delta q),-q^N_{Bs1x_1}(1+\delta q)]$
where $\delta q\rightarrow 0$.
(According to Eq. (\ref{q-N-Fs}) one has for small values of
$x$ and $U/4t\geq u_0$ that $q^N_{Bs1x_1}=q^N_{Bs1x_2}=-[\pi/2](1-x)$.)
For such nodal momentum values the corresponding auxiliary velocity reads, 
\begin{equation}
{\vec{V}}^{\Delta,\parallel}_{s1} ({\vec{q}}^{\,N}_{Bs1}) 
= -{\rm sgn}\{\delta q\} {\vert\Delta\vert\over W_{s1}} 
\,v_{s1} (q^N_{Bs1x_1})
\left[\begin{array}{c}
+1 \\
-1
\end{array} \right] .
\label{Vs-N-D}
\end{equation}

Finally, the use of the energy-dispersion expression given in 
Eq. (\ref{c-band-x-small}) in the $c$ fermion
velocity expression provided in Eq. (\ref{g-velocities})
leads to,
\begin{equation}
{\vec{V}}_{c} (\vec{q}^{\,h}) = 
- {{\vec{q}}^{\,h}\over m^*_{c}} \, ,
\label{VSx}
\end{equation}
for $x\ll 1$ and $U/4t\geq u_0$.

\subsection{The angle $\phi$, Fermi-line one-electron spectrum, and 
corresponding Fermi velocity}

Complementarily to the results on the Fermi line momentum values
reported in Subsection III-C, here we study several one-electron physical quantities. 
Unfortunately, except for the expression $e_{c}(q)=-U/2+2t\cos q$ valid for $U/4t\gg 1$ 
and that of $e_{s1} (q)$ given in Eq. (\ref{e-s1-q}) of Appendix C
for small values of the hole concentration and $U/4t\geq u_0$, one
does not know accurately the form of the elementary functions $e_c (q)$ and $e_{s1} (q)$
that control the dependence on the momentum Cartesian components
of the $c$ and $s1$ fermion energy dispersions given in Eqs. 
(\ref{general-epsilon}) and (\ref{c-band}) 
and corresponding velocities of Eq. (\ref{g-velocities-def}). However, one
is aware of some properties and symmetries of such functions as for instance those behind the
limiting values reported in Eqs. (\ref{F-s-limits}) and (\ref{G-s-limits}) for the related 
functions $F_{s1} ({\vec{q}}_{Bs1})$ and $G_{s1} ({\vec{q}}_{Bs1})$
of Eqs. (\ref{bands-bipt}) and (\ref{v-D-Fs-G}), respectively. 
Those imply that the following 
expression is exact for small hole concentration values,
$m=0$, and $U/4t\geq u_0$ and is expected to be a good 
approximation for finite hole concentration values $0 <x<x_*$ provided that $u_0\leq U/4t\leq u_{\pi}$,
\begin{equation}
2\phi = {\rm sgn}\left({k^h_{Fx_2}\over 
k^h_{Fx_1}}\right)\arctan\left({G^{\parallel}_{s1} ({\vec{q}}_{Bs1})
\over F^{\parallel}_{s1} ({\vec{q}}_{Bs1})}\right) 
= {\rm sgn}\left({k^h_{Fx_2}\over 
k^h_{Fx_1}}\right)\arctan\left({\sqrt{2}\,V^{\Delta}_{s1} ({\vec{q}}^{\,d}_{Bs1})\over
\vert\Delta_{s1} ({\vec{q}}^{\,d}_{Bs1})\vert}\right) .
\label{2phi-F}
\end{equation}
Here $\phi$ is the angle of Eq. (\ref{kF}), which defines the direction of the hole Fermi momentum 
whose expression is given in that equation and in Eq. (\ref{kF-qFc-qBs1}). 

The dependence on $\phi$ of the angles $\phi^{d}_{s1}=\phi^{d}_{s1} (\phi)$ and 
$\phi^{d}_c=\phi^{d}_c  (\phi)$ provided in Eqs. (\ref{phiF-s1}) and (\ref{phiF-c}), respectively,
is valid only for the isotropic Fermi-velocity range $x_{c1}<x<x_{c2}$. Such angles 
define the relative directions of the $c$ fermion hole momentum and $s1$ fermion
momentum on the right-rand side of Eq. (\ref{kF-qFc-qBs1}) for the hole Fermi momentum, which
for that range are perpendicular to each other.
In turn, expression (\ref{2phi-F}) for $2\phi$ does not involve any relation between $c$
fermion and $s1$ fermion quantities. Indeed, it involves the ratio
$V^{\Delta}_{s1} ({\vec{q}}^{\,d}_{Bs1})/\vert\Delta_{s1} ({\vec{q}}^{\,d}_{Bs1})\vert$
that refers only to $s1$ fermion quantities. This is why Eq. (\ref{2phi-F})
is a good approximation for the whole range $0<x<x_*$. 
If as discussed in Section III the hole concentration $x_h$ above which the Fermi line is particle like belongs
to the range $x_{c2}\leq x_h\leq x_*$, then for $x_h<x<x_*$ the angle $\phi$ of Eq.
(\ref{2phi-F}) belongs to the ranges given in Eq. (\ref{kF}). (The angle $\phi_{AN}$
appearing in such ranges vanishes for $x\leq x_h$ and is small for $x\in (x_h,x_*)$.)

Consistently with Eq. (\ref{2phi-F}), 
the absolute value $V^{\Delta}_{Bs1}\equiv V^{\Delta}_{s1} ({\vec{q}}^{\,d}_{Bs1})$ of 
both the velocities ${\vec{V}}^{\Delta}_{s1} ({\vec{q}}^{\,d}_{Bs1})$ and
${\vec{V}}^{\Delta,\parallel}_{s1} ({\vec{q}}_{Bs1})$ of 
Eq. (\ref{BL-Vs1}) reads,
\begin{equation}
V^{\Delta}_{Bs1} \equiv V^{\Delta}_{s1} ({\vec{q}}^{\,d}_{Bs1}) = {\vert\Delta\vert\over\sqrt{2}}\vert\sin 2\phi\vert \, .
\label{pairing-en-v-Delta}
\end{equation}
Since the angle $\phi_{AN}$ is small, for the hole-concentration range $x_h<x<x_*$ we consider 
corrections up to first order in $\phi_{AN}$, so that ${\rm max}\,\{\delta E_F\}\approx \vert\Delta\vert$.
Moreover,
the minimum magnitude of the velocity (\ref{pairing-en-v-Delta}) reads
${\rm min}\,\{V^{\Delta}_{Bs1}\}\approx [\phi_{AN}\,\sqrt{2}]\vert\Delta\vert$. In turn,
for $0<x<x_h$ its minimum magnitude is zero for momenta ${\vec{q}}^{\,d\,AN}_{Bs1}$
whose auxiliary momenta point in the anti-nodal directions.

Hence the above two velocities can be written as,
\begin{equation}
{\vec{V}}^{\Delta}_{Bs1} = V^{\Delta}_{Bs1}\,{\vec{e}}_{\phi^{\Delta,d}_{Bs1}} \, ; \hspace{0.35cm}
{\vec{V}}^{\Delta,\parallel}_{s1} ({\vec{q}}_{Bs1}) =
V^{\Delta}_{Bs1}\,{\vec{e}}_{\phi^{\Delta,\parallel}_{Bs1}}  \, .
\label{v-D-Fs-simplified}
\end{equation}
Here the angles $\phi^{\Delta,d}_{Bs1}\equiv \phi^{\Delta}_{s1} ({\vec{q}}^{\,d}_{Bs1})$
and $\phi^{\Delta,\parallel}_{Bs1}\equiv \phi^{\Delta,\parallel}_{s1} ({\vec{q}}_{Bs1})$ read,
\begin{equation}
\phi^{\Delta,d}_{Bs1} = \phi^{\Delta,\parallel}_{Bs1} + \phi_F^d
\, ; \hspace{0.35cm}
\phi^{\Delta,\parallel}_{Bs1} = -\arctan\left({v_{s1} (q_{Bs1x_2})\over
v_{s1} (q_{Bs1x_1})}\right) \, ,
\label{phi-Delta}
\end{equation}
where the expression for $\phi^{\Delta,d}_{Bs1}$ is valid for $x_{c1}<x<x_{c2}$ and
follows from Eq. (\ref{BL-Vs1}).

For $0<x<x_h$ the velocity (\ref{pairing-en-v-Delta}) is a function
of the angle $\phi$ with the following limiting behaviors,
\begin{equation}
V^{\Delta}_{Bs1}  = {\vert\Delta\vert\over\sqrt{2}} \, , \hspace{0.35cm}
\phi=\pi/4 \, ; \hspace{0.50cm}
V^{\Delta}_{Bs1}  = 0 \, , \hspace{0.35cm}
\phi=0,\pi/2 \, .
\label{limiting-pairing-en-v-Delta}
\end{equation}
For $x_h<x<x_*$ the maximum magnitude remains the
same whereas $V^{\Delta}_{Bs1}\approx [\phi_{AN}\,\sqrt{2}]\vert\Delta\vert$
for $\phi=\phi_{AN}$ and $\phi=[\pi/2-\phi_{AN}]$.
That for $0<x<x_h$ the velocity $V^{\Delta}_{Bs1}$ vanishes for
auxiliary momenta pointing in the anti-nodal 
directions together with the $s1$ band remaining full for $x> 0$ and $m=0$ 
ground states has important physical consequences.
Indeed, finite-energy excitations involving creation of
$s1$ fermion holes whose auxiliary momenta point in the anti-nodal directions
refer to real-space states. The energy of such excitations
equals the pseudogap energy scale. As further discussed in Ref.
\cite{cuprates}, such pseudogap real-space states are observed
in experiments on hole-doped superconductors \cite{k-r-spaces}.

In turn, the velocity ${\vec{V}}_{c} (\vec{q}^{\,h})$ of Eq. (\ref{VSx})
is for $\vec{q}^{\,h}=\vec{q}^{\,h\,d}_{Fc}$ given by,
\begin{equation}
\vec{V}_{c} (\vec{q}^{\,h\,d}_{Fc}) =
-V_{Fc}\,{\vec{e}}_{\phi^d_{c}}
\, ; \hspace{0.50cm} V_{Fc} =
{q^h_{Fc} (\phi)\over m_c^*} \approx {\sqrt{x\pi}\,2\over m_c^*} \, ,
\label{Vc-FS}
\end{equation}
where the angle $\phi^{d}_c$ is for $x\in (x_{c1},x_{c2})$ provided in 
Eq. (\ref{phiF-c}) and the mass $m_c^*=1/2r_c t$ is that of the
$c$ fermion energy dispersion given in Eq. (\ref{bands}). It is associated with the charge mass ratio $r_c$ of 
Eq. (\ref{m*c/mc-UL}). 

According to the $c$ Fermi-line definition of Eq. (\ref{g-FS}) 
and following the form of the general 
$c$ fermion energy dispersion expression provided in Eq. (\ref{c-band}), 
the latter energy vanishes for $c$ band momentum values
belonging to the $c$ Fermi line. The use of the $c$ and $s1$ fermion energy
dispersions given in Eqs. (\ref{general-epsilon}) and (\ref{bands}) and related expressions provided
in Eqs. (\ref{bands-bipt})-(\ref{D-x}) in the general energy 
functional of Eq. (\ref{DE}) leads for $u_0\leq U/4t\leq u_{\pi}$ 
and hole concentrations $0 <x<x_*$ to the following general expression
for the one-electron energy spectrum 
at the Fermi line whose hole Fermi momenta are given in
Eq. (\ref{kF-qFc-qBs1}),
\begin{equation}
E_F = \mu + \delta E_F (\phi) \, ; \hspace{0.35cm}
\delta E_F (\phi) = - \epsilon_{s1} ({\vec{q}}^{\,d}_{Bs1}) = \vert\Delta_{s1} ({\vec{q}}^{\,d}_{Bs1})\vert = 
\vert\Delta\vert\vert\cos 2\phi\vert \, .
\label{E-F}
\end{equation}
Here the zero-temperature chemical potential $\mu$ is given in Eqs.
(\ref{mux}) and (\ref{mu-Hm}), it and $\delta E_F (\phi)$ are
the isotropic and anisotropic Fermi-energy terms, respectively, and
the equality $\vert\Delta_{s1} ({\vec{q}}^{\,d}_{Bs1})\vert = 
\vert\Delta\vert\vert\cos 2\phi\vert$ follows from Eqs. (\ref{2phi-F}) and (\ref{v-D-Fs-simplified}).
The Fermi-line anisotropic and gapped one-electron spectrum 
of excitation momentum and energy ${\vec{k}}^h_{F}$ and
$E_F $, respectively, has a $d$-wave-symmetry like structure. 
Its relation to the physics observed in the hole-doped 
cuprates \cite{two-gaps,k-r-spaces,duality,2D-MIT,Basov,ARPES-review,Tsuei,pseudogap-review}
is discussed in Ref. \cite{cuprates}. 

The value of the momentum ${\vec{q}}^{\,d}_{Bs1}$ appearing
in the energy spectrum (\ref{E-F}) refers according to Eq. (\ref{kF-qFc-qBs1}) to exactly one
value of the Fermi hole momentum ${\vec{k}}^h_{F}$. The 
inverse relation is two-valued, with each
Fermi hole momentum ${\vec{k}}^h_{F}$ corresponding to the momenta 
${\vec{q}}^{\,+1}_{Bs1}$ and ${\vec{q}}^{\,-1}_{Bs1}$
associated with two alternative and degenerate one-electron excited states 
with the same excitation momentum ${\vec{k}}^h_{F}$ and 
energy $E_F$ given in Eqs. (\ref{qFc-qBs1}) and (\ref{E-F}), respectively. 
Let us confirm that such excited states refer to electrons 
at the same point of the Fermi line but with different Fermi velocities.
This is equivalent to show that the electronic velocity at
the Fermi hole momentum ${\vec{k}}^h_{F}$ depends on
the doublicity $d=\pm 1$, unlike the corresponding energy
$E_F$ and momentum ${\vec{k}}^h_{F}$.

Since following the relation between quantum overlaps
and the transformation laws of the $s1$ fermion
operators discussed in Ref. \cite{companion} and according to the numbers
and number deviations of Table \ref{tableIV} of Appendix A
creation of one electron involves creation of one $c$ fermion 
and one $s1$ fermion hole, the Fermi velocity is straightforwardly 
given by,
\begin{equation}
\vec{V}^{\,d}_F = [\vec{V}_{c} (\vec{q}^{\,h\,d}_{Fc})
- {\vec{V}}^{\Delta}_{s1} ({\vec{q}}^{\,d}_{Bs1})] = V^{\,d}_{F}\,{\vec{e}}_{\phi^{d}_{V_F}} \, ,
\label{VF-pm1}
\end{equation}
where
\begin{equation}
V^{d}_F = V_{Fc}\,\sqrt{1 - 2r_{\Delta}\cos (\phi_c^d-\phi^{\Delta,d}_{Bs1})+r_{\Delta}^2}  
\, ; \hspace{0.35cm}
\phi^{d}_{V_F} = \arctan\left({\sin \phi^{d}_c - r_{\Delta}\,\sin \phi^{\Delta,d}_{Bs1}
\over \cos \phi^{d}_c - r_{\Delta}\,\cos \phi^{\Delta,d}_{Bs1}}\right) \, .
\label{VF-pm1-modulus}
\end{equation}
Here $r_{\Delta}$ is the velocity ratio of Eq. (\ref{x-h}) and
$\phi^{\Delta}_{Bs1}$ and $\phi_c^d$ the angles that
define the direction of the $s1$ and $c$ fermion velocities of Eqs.
(\ref{v-D-Fs-simplified}) and (\ref{Vc-FS}), respectively. 

The form of the expressions given in Eq. (\ref{VF-pm1-modulus})
confirms that both the absolute value and
angle of the Fermi velocity depend indeed on the doublicity
$d=\pm 1$. Note however that $r_{\Delta} =V^{\Delta}_{Bs1}/V_{Fc}=\eta_{\Delta}\vert\sin 2\phi\vert\ll 1$ 
for $x>x_{c1}$ so that the velocity $V^{d}_F$ of Eq. (\ref{VF-pm1-modulus}) becomes independent
of both $d$ and $\phi$ and given approximately by $V_F^d\approx V_F=V_{Fc}$.

\subsection{Charge and spin excitations}

The Fermi velocity of Eqs. (\ref{VF-pm1})-(\ref{VF-pm1-modulus})
involves both the charge $c$ fermion velocity $\vec{V}_{c} (\vec{q}^{\,h})$ 
and spin $s1$ fermion velocity ${\vec{V}}^{\Delta}_{s1} ({\vec{q}})$. In contrast,
the velocities corresponding to the charge and spin 
excitations involve only the velocities $\vec{V}_{c} (\vec{q}^{\,h})$
and ${\vec{V}}_{s1} ({\vec{q}})$, respectively. For two-electron
charge and spin excitations the matrices $A_{s1}^{d}$ and $A_{F}^{d}$ are the 
$2\times 2$ unit matrix so that the $s1$ boundary line is not deformed. Therefore,
${\vec{q}}^{\,d}_{Bs1}={\vec{q}}_{Bs1}$ for the $s1$ boundary line of such states,
so that the $s1$ fermion velocity ${\vec{V}}^{\Delta}_{s1} ({\vec{q}}_{Bs1})$ and
$c$ fermion velocity $\vec{V}_{c} (\vec{q}^{\,h}_{Fc})$ are independent and read,
\begin{equation}
{\vec{V}}^{\Delta}_{s1} ({\vec{q}}_{Bs1}) =
V^{\Delta}_{Bs1}\,{\vec{e}}_{\phi^{\Delta}_{s1}} 
\, ; \hspace{0.50cm} \vec{V}_{c} (\vec{q}^{\,h}_{Fc}) =
-V_{Fc}\,{\vec{e}}_{\phi_{c}} \, ,
\label{Vs1-Vc-no-par}
\end{equation}
respectively. Here the velocity absolute values $V^{\Delta}_{Bs1}$ 
and $V_{Fc}$ are those provided in Eqs. (\ref{pairing-en-v-Delta})
and (\ref{Vc-FS}), respectively,
the angle $\phi^{\Delta}_{s1}$ equals that given in 
Eq. (\ref{phi-Delta}) and thus reads,
\begin{equation}
\phi^{\Delta}_{s1} = -\arctan\left({v_{s1} (q_{Bs1x_2})\over
v_{s1} (q_{Bs1x_1})}\right) \, ,
\label{phi-Delta-no-par}
\end{equation}
and $\phi_{c} = \phi$.

The $c$ fermions describe the charge degrees of freedom of
the rotated electrons that singly occupy sites and thus
carry the electronic charge $-e$, which remains invariant
under the electron - rotated-electron unitary transformation.
Hence they couple to external charge potentials and carry the
elementary charge currents. Since in contrast to the $s1$ 
fermions the $c$ fermions do not 
emerge from a Jordan-Wigner transformation, their direct
$c$ - $c$ fermion interactions vanish or are 
very weak and the $c$ fermion elementary charge current reads,
\begin{equation}
{\vec{j}}_c ({\vec{q}}^{\,h}) = -e\,\alpha_U\,{\vec{V}}_{c} ({\vec{q}}^{\,h}) \, ; \hspace{0.35cm}
{\vec{j}}_c ({\vec{q}}^{\,h}_{Fc}) = -e\,{q^{h}_{Fc}\over m_c^{\rho}}\,{\vec{e}}_{\phi_{c}+\pi}
\approx-e\,{\sqrt{x\pi}\,2\over m_c^{\rho}}\,{\vec{e}}_{\phi_{c}+\pi} \, ; \hspace{0.35cm}
\alpha_U  \equiv {m_c^*\over m_c^{\rho}} \, .
\label{jc}
\end{equation}
Here $m_c^{\rho}$ is a renormalized transport mass and 
$q^{h}_{Fc}\approx \sqrt{x\pi}\,2$ is a good approximation
for hole concentrations $x\in (0,x_{c1})$ and is expected to be a reasonably 
good approximation for $x\in (x_{c1},x_{c2})$ provided that $U/4t\geq u_0$. Since for finite values
of $U/4t$ the model (\ref{H}) does not commute with the charge current operator,
it follows that $m_c^{\rho}\leq m_c^{\infty}=1/2t$ for $U/4t\geq u_0$ where the
equality refers to the limit $U/4t\rightarrow\infty$. Indeed, the Drude peak exhausts
the conductivity sum-rule both for $U/4t\rightarrow 0$ and
$U/4t\rightarrow\infty$ and otherwise its spectral weight is smaller
than that associated with such a sum-rule. For $U/4t\geq u_0$ one 
then estimates,
\begin{equation}
{m_c^{\rho}\over m_c^{\infty}} = 2t\,m_c^{\rho} \approx r_c
\, ; \hspace{0.5cm} \alpha_U = {m_c^*\over m_c^{\rho}} \approx
{m_c^*\over m_c^{\infty}}{1\over r_c} =
{1\over r_c^2} \, , \hspace{0.35cm} U/4t\geq u_0 \, ,
\label{ratios}
\end{equation}
where $r_c$ is the charge mass ratio given in Eq. (\ref{m*c/mc-UL}).
For instance, for $U/4t\approx u_*=1.525$ one finds
$m_c^{\rho}/m_c^{*}=1/\alpha_U\approx r_c^2
= (\pi\,0.27)^2\approx 0.72$, whereas $m_c^{\rho}/m_c^{*}=1$
for $U/4t\rightarrow\infty$.

For the Hubbard model on the 1D lattice the
general description of Ref. \cite{companion} also applies
and since for that model there is only $c$ and $s1$ fermion zero-momentum 
forward-scattering, the transport
mass $m_c^{\rho}$ can be evaluated explicitly by
suitable use of the 1D exact solution and is 
given in Eq. (139) of Ref. \cite{mass}.

Finally, we consider the general spin spectrum. According to the results of Ref. \cite{companion}
and numbers and number deviations of Table \ref{tableIV} of Appendix A, 
both the spin-singlet and spin-triplet 
excitations relative to the $m=0$ ground state involve creation of two $s1$ fermion
holes. The use of Eqs. (\ref{DE}) and (\ref{DP}) then leads
to the following general spin spectrum,
\begin{equation} 
\delta E_{spin} =-\epsilon_{s1} ({\vec{q}})-\epsilon_{s1} ({\vec{q}}\,')
\, ; \hspace{0.25cm} 
\delta \vec{P}=\delta {\vec{q}}_{c}^{\,0}-{\vec{q}}-{\vec{q}}\,' 
\, ; \hspace{0.25cm} 
\delta {\vec{q}}_{c}^{\,0} = \vec{\pi} \, , \hspace{0.05cm} x=0
\, ; \hspace{0.25cm} 
\delta {\vec{q}}_{c}^{\,0} = 0 \, , \hspace{0.05cm} x>0 \, ,
\label{DE-spin}
\end{equation}
where the momentum shift $\delta {\vec{q}}_{c}^{\,0}$
refers to the momentum ${\vec{q}}_{\gamma}^{\,0}$ 
of Eq. (\ref{q-j-f-Q-c-0-s1-2D}) for $\gamma =c$. For spin excitations of
the square-lattice quantum liquid one has that $\delta {\vec{q}}_{c}^{\,0}=\vec{\pi}=[\pi,\pi]$
at $x=0$ and $\delta {\vec{q}}_{c}^{\,0}= 0$ for $x>0$ for initial $m=0$ ground states. 
Derivative of the excitation energy $\delta E_{spin}$ of
Eq. (\ref{DE-spin}) relative to the corresponding
momentum $\delta \vec{P}$ confirms that the spin
velocity is $[{\vec{V}}_{s1} ({\vec{q}})+{\vec{V}}_{s1} ({\vec{q}}\,')]$
and hence involves only the $s1$ fermion velocity.

\section{Relation to the exact solution of the 1D model,
agreement of the square-lattice quantum liquid 
with results obtained by the standard formalism of many-body physics,
and relation to other schemes}

The general quantum liquid of $c$ and $s1$ fermions introduced in Ref.
\cite{companion} and further studied in this paper 
refers to the Hubbard model in the one- and two-electron subspace 
on a 1D or square lattice. For both such lattices that
quantum problem is non-perturbative in terms of electron operators so 
that rewriting the theory in terms of the standard formalism of many-electron 
physics is an extremely complex problem. 

In this section we discuss the use of the $c$ and $s1$ fermion description to
construct a dynamical theory for the 1D Hubbard model that provides 
correlation-function expressions both at low and finite energy and recovers 
the well-known low-enery behavior of that 1D quantum liquid. Furthermore,
we show that the predictions of the square-lattice quantum liquid approach
concerning the spin spectrum at half filling agree both with experiments 
on LCO and results obtained by the standard formalism of many-body 
physics. Finally, we discuss the relation of the square-lattice
quantum liquid of $c$ and $s1$ fermions to other schemes.

\subsection{The $c$ and $s1$ fermion description of the 1D Hubbard model}

For the 1D model there is an exact solution \cite{Lieb,Takahashi,Martins}.
However, it does not provide correlation-function expressions 
at finite energy. The relation to that solution of the general $c$ and $s1$ fermion description 
and its underlying theory has been clarified \cite{companion,1D} and one can profit from
such a relation to further develop and test the theory. That allows
going beyond the range of the exact solution alone and derive 
one- and two-electron spectral-weights and corresponding spectral-
and correlation-function expressions for both low and finite energy \cite{V,TTF}.
Fortunately and as discussed in the following, in the low-energy limit such a
general theory recovers the so called Tomonaga-Luttinger-liquid (TTL)
behavior and anomalous scaling dimension of spin and charge correlations \cite{LE}.

The general rotated-electron description introduced in Ref. \cite{companion}
for the Hubbard model on the square lattice has already been 
considered in Ref. \cite{1D} for the same model on the 1D lattice.
Within the notation of the latter reference, the $\eta$-spinons, spinons, $c$ fermions,
$s1$ fermions, independent $\eta$-spinons, and independent spinons 
considered here are called in Ref. \cite{companion} holons, spinons, $c$ pseudoparticles, 
$s1$ pseudoparticles, Yang holons, and HL spinons, respectively. 
(HL stands for Heilmann and Lieb \cite{1D}.) Moreover, the $2\nu$-$\eta$-spinon composite 
$\eta\nu$ fermions and $2\nu$-spinon $s\nu$ composite fermions considered in Ref. \cite{companion}
are called in Ref. \cite{1D} $c\nu$ pseudoparticles and $s\nu$ pseudoparticles, respectively.
The relation to the rotated-electron occupancy configurations
of all such objects that generate the energy eigenstates of the 1D
model are in Refs. \cite{companion} and \cite{1D} 
the same. (The studies of the former reference
extended the description to the Hubbard model on the
square lattice.)

The $c$ and $s1$ fermion scheme refers
to the limit $N_a\gg 1$ where the thermodynamic Bethe-ansatz 
equations of Takahashi apply \cite{Takahashi}. Exploring the
relation of such a scheme to the exact solution has allowed the construction
of a 1D $c$ and $s1$ fermion dynamical theory, which provides general
expressions for the finite-energy correlation and
spectral dynamical functions \cite{V}, beyond those provided
by bosonization and conformal-field theory.
Such a dynamical theory was recently 
successfully used to study the finite-energy singular features in 
photoemission of the organic compound tetrathiafulvalene-tetracyanoquinodimethane 
(TTF-TCNQ) metallic phase \cite{TTF}. The $c$ and $s1$ fermion
dynamical theory refers to the 1D Hubbard model.
More recently, other methods for the study of finite-energy spectral and 
dynamical functions of 1D correlated systems were
introduced \cite{Affleck,Affleck-09,Glazman,Glazman-09}. Both the finite-energy spectral-weight
distributions studied by the $c$ and $s1$ fermion dynamical theory 
for the 1D Hubbard model and the methods of
Refs. \cite{Affleck,Affleck-09,Glazman,Glazman-09} for other 1D correlated problems include 
power-law singularities near well-defined finite-energy branch lines with exponents depending on the interaction
strength and the excitation momentum. 

As discussed in Ref. \cite{Affleck-09}, the study of the singularities of spectral
functions using models such as the 1D Hubbard model is not limited
to low energies or to weak interactions and the
idea of extracting exponents of finite-energy spectral functions from
the Bethe-ansatz exact solution by combining it with the present $c$ and $s1$ fermion
description has appeared in the pseudofermion
dynamical theory for the 1D Hubbard model of Ref. \cite{V}. In turn, according
to the discussions of Ref. \cite{Glazman-09}, our method relies in part on the integrability
of the 1D Hubbard model, whereas the phenomenology developed in that reference
does not require any special property of the underlying microscopic interaction. 
The solvability of the 1D Hubbard model is in the $N_a\gg 1$ limit 
the $c$ and $s1$ fermion theory refers to associated with the 
occurrence of an infinite set of conservation laws \cite{Martins,Prosen}.
As discussed in Ref. \cite{companion}, within our description such laws follow from the commutation
of the Hamiltonian with each of the infinite $\alpha\nu$ translation generators ${\hat{\vec{q}}}_{\alpha\nu}$
in the presence of the fictitious magnetic fields ${\vec{B}}_{\alpha\nu} ({\vec{r}}_j)$
considered in that reference whose eigenvalues are the components of the microscopic 
momenta of the $\alpha\nu$ fermions. Such microscopic momenta are for the 1D model good quantum numbers
for the whole Hilbert space. For the model on the square lattice and $x>0$ the $c$ and $s1$
fermions undergo inelastic collisions \cite{cuprates0,cuprates}. In contrast, due 
to the occurrence of the above conservation laws in 1D they
have only zero-momentum forward scattering, associated with
two-fermion phase shifts \cite{V}. Importantly, the studies of Ref. \cite{LE} 
reveal that when both momenta in the argument of such phase shifts lie at or
near the $c$ Fermi line and/or $s1$ boundary line, their square in units of $2\pi$ fully determines the
anomalous dimensions of the primary fields of conformal-field
theory. The low-energy physics of correlated 1D problems 
has some universal properties described by the 
TLL \cite{Voit}. The investigations of Ref. \cite{LE} confirm that the 
1D $c$ and $s1$ fermion dynamical theory
gives a correct description of the TLL behavior and anomalous scaling dimensions of spin 
and charge correlations. Such a consistency provides further evidence of the validity 
of our general method. Indeed, the finite-energy $c$ and $s1$ fermion dynamical theory 
is shown in that reference to recover in the limit of low energy the usual low-energy TLL results. 
It is an improved 1D version of the $c$ and $s1$ fermion scheme used in our studies for the model on the
square lattice, which also refers to the general rotated-electron
description of Refs. \cite{1D,companion}. 

The construction of the $c$ and $s1$ fermion dynamical theory of Ref. \cite{V} 
profits from the relation of the quantum
objects of our description to the quantum numbers of the Bethe-ansatz solution.
The derivation of the general correlation-function and dynamical-function
expressions by that dynamical theory takes into account implicitly that
the in the case of the spin-neutral two-spinon $s1$ fermion branch such quantum numbers
are the eigenvalues of the $s1$ translation generator ${\hat{\vec{q}}}_{s1}$
in the presence of the fictitious magnetic field ${\vec{B}}_{s1} ({\vec{r}}_j)$
of Eq. (\ref{A-j-s1-3D}). The consistency of our description concerning the
isomorphism between the $c$ and $\alpha\nu$ fermion microscopic momentum
values and the quantum numbers of the exact solution is addressed in Ref. \cite{companion}.

\subsection{Agreement of the square-lattice quantum liquid 
with experiments and results obtained by the standard formalism 
of many-body physics}

It is desirable that the results of the square-lattice quantum liquid studied
in this paper are compared with those of the standard formalism of 
many-body physics. Unfortunately, such a quantum liquid is non-perturbative
in terms of electron operators so that, in contrast to a 3D isotropic
Fermi liquid \cite{Landau,Pines}, rewriting the theory in terms of it is an extremely complex problem. 

In spite of the lack of an exact solution for the model on a square lattice and the
non-perturbative character of the quantum problem in terms of electrons, in this
subsection results achieved by the square-lattice quantum liquid description
are compared with those obtained by methods relying on the standard formalism 
of many-body physics. Unfortunately, there are not many controlled results
for the Hubbard model on the square lattice from approximations relying on
that formalism. Here we consider the interesting problem of the spin-excitation 
spectrum of the half-filling Hubbard model on the square lattice. 

Within the present description and as discussed below, since the $c$ and $s1$ fermion
momentum values are for the model in the one- and two-electron
subspace good quantum numbers, that problem refers to an 
effectively non-interacting limit whereas in terms of electrons it is an
involved many-body problem. Fortunately, there are reliable results on
that particular problem obtained by controlled approximations of the standard formalism of many-electron 
physics: within such approximations its solution requires 
summing up an infinite set of ladder diagrams, to find the spin-wave dispersion of 
the half-filled Hubbard model on the square lattice in a spin-density-wave-broken 
symmetry ground state \cite{LCO-Hubbard-NuMi}. 
In turn, within our description the spin spectrum is
that of Eq. (\ref{DE-spin}), which involves the creation of two holes in the $s1$ band
whose energy dispersion $\epsilon_{s1} ({\vec{q}})$ is given in Eq. (\ref{Esx0}). 

Agreement between the two methods is
both a further checking of the validity of our description
and a confirmation that the $c$ and $s1$ fermion
interactions are indeed residual and their momentum
values good quantum numbers for the model on the square lattice in the
one- and two-electron subspace.
(The results of Refs. \cite{companion,1D} confirm
that for 1D these momentum values are good
quantum numbers for the whole Hilbert space.) 
As a side result, in this subsection we find the magnitudes
for $U/4t\approx 1.525$
of the energy scales $\mu^0$ and $2\Delta_0$ whose limiting values are 
given in Eqs. (\ref{38}) of Appendix A and below in Table III, respectively, and $W^0_{s1}$ 
of Eq. (\ref{Wg}). That value of $U/4t$ corresponds to the critical hole concentration
$x_*\approx 0.27$ and is that appropriate to the description 
by the theory of the properties of the 
parent compound LCO. In reference \cite{cuprates} strong evidence is 
provided that such a $U/4t$ value is also that appropriate to
the hole-doped cuprates with superconducting zero-temperature
critical hole concentrations $x\approx 0.05$ and $x\approx 0.27$.

The spin-triplet excitations relative to the $x=0$ and
$m=0$ absolute ground state involve creation
of two holes in the $s1$ band along with a shift
$\vec{\pi}/N_a^2$ of all discrete momentum 
values of the full $c$ band so that the
general spin spectrum (\ref{DE-spin}) reads, 
\begin{equation}
\omega (\vec{k})= [-\epsilon_{s1} ({\vec{q}}) -\epsilon_{s1} ({\vec{q}}\,')]
\, ; \hspace{0.25cm} \vec{k} = [\vec{\pi} - {\vec{q}} - {\vec{q}}\,'] \, ,
\label{DE-spin-x0}
\end{equation}
where $\vec{\pi}=\pm [\pi,\pm\pi]$ and the $s1$ fermion energy dispersion 
$\epsilon_{s1} ({\vec{q}})$ is given in Eq. (\ref{Esx0}). 
For $x=0$ and $m=0$ both the $c$ and $s1$ bands are full for 
the initial ground state and since the $c$ band remains full for
the excited states one can ignore the $s1$ - $s1$ and $s1$ - $c$
fermion interactions. Indeed, then the residual fermion interactions
studied in Ref. \cite{cuprates0} have little effect on the occupancy 
configurations of the two holes created in the $s1$ band upon
the two-electron spin-triplet excitations. This is consistent 
with the lack of a $c$ Fermi line for the initial ground 
state and the lack of $s1$ band holes other than the two holes created 
upon the spin-triplet excitation so that in spite of the $s1$ - $s1$ fermion
long-range interactions associated with the effective vector potential
of Eq. (\ref{A-j-s1-3D}) the exclusion principle, phase-space restrictions,
and momentum and energy conservation drastically limit 
the number of available momentum occupancy configurations of the
final excited states. 

The excitation spectrum (\ref{DE-spin-x0}) refers both to coherent and
incoherent spin spectral weight. In contrast to the 1D case where
as mentioned in the previous subsection a suitable $c$ and $s1$
fermion dynamical theory is available \cite{V,TTF}, for the square-lattice
quantum liquid there are within the present status of the theory  
no suitable tools to calculate matrix elements between the ground state
and one- and two-electron excited states. Hence, one cannot
calculate explicitly spin-spin correlation functions. Within the
1D $c$ and $s1$ fermion dynamical theory, the sharp features of the spin
two-electron spectral weight distributions result from processes
where one of the two $s1$ fermion holes is created at the
$s1$ boundary line.  

The coherent spin spectral weight is here 
associated with a Goldstone-mode-like gapless spin-wave spectrum.
It consists of sharp $\delta$-peaks having as background the incoherent
spectral-weight distribution. 
From comparison with the results of Ref. \cite{LCO-Hubbard-NuMi},
we have confirmed that such spectral weight is generated by 
processes corresponding to well-defined values of the
momenta ${\vec{q}}$ and ${\vec{q}}\,'$ of the general spectrum 
(\ref{DE-spin-x0})
such that one hole is created at a momentum pointing in the nodal
directions of the $s1$ band and the other hole at a momentum
belonging the $s1$ band boundary line, as expected from analogy with
the 1D spectral-weight distributions. 
The incoherent part corresponds to the remaining values of ${\vec{q}}$ and ${\vec{q}}\,'$
of the excitation spectrum (\ref{DE-spin-x0}). 
The occurrence of the Goldstone-mode-like gapless spin-wave spectrum 
follows from the long-range antiferromagnetic order of the initial $x=0$ and
$m=0$ ground state. In turn, for the $x>0$ short-range spin ordered phase 
the spin weight distribution associated with the general spin spectrum provided in 
Eq. (\ref{DE-spin}) has no coherent part. In 1D it has not coherent part both
for $x=0$ and $x>0$, due to the lack of a ground-state long-range antiferromagnetic
order. 

As mentioned above, for the original electrons the problem
is highly correlated and involves an infinite
set of ladder diagrams and no simple analytical 
expression was found for the spin-wave energy 
spectrum \cite{LCO-Hubbard-NuMi}. In contrast, 
for the $s1$ fermion description it is effectively non interacting and
described by simple analytical expressions. 
Let us profit from symmetry 
and limit our analysis to the sector $k_x\in (0,\pi)$ and 
$k_y\in (0,k_x)$ of the $(\vec{k},\omega)$ space.
Within the description of the quantum problem
used here, for $1/k_B\,T\rightarrow\infty$ the coherent 
spin-spectral-weight distribution derived in Ref. \cite{LCO-Hubbard-NuMi}
corresponds to a surface of energy and momentum given by,
\begin{equation}
\omega (\vec{k}) =
{\mu^0\over 2}\left\vert\sin\left({k_x+k_y\over 2}\right)\right\vert
+ W^0_{s1}\left\vert\sin\left({k_x-k_y\over 2}\right)\right\vert \, ; 
\hspace{0.25cm} \vec{k} = \vec{\pi}- {\vec{q}}
- {\vec{q}}\,'  \, .
\label{om-SW}
\end{equation} 
This is a particular case of the general spin spectrum given in Eq.
(\ref{DE-spin-x0}), which corresponds to the above-mentioned specific processes 
associated with the following choices for $\vec{\pi}$, ${\vec{q}}$, 
and ${\vec{q}}\,'$,
\begin{eqnarray}
\vec{\pi} & = & [\pi,-\pi] \, ,
\nonumber \\
{\vec{q}} & = & \left[{\pi\over 2}-{(k_x+k_y)\over 2},-{\pi\over 2}-{(k_x+k_y)\over 2}\right] \, ,
\nonumber \\
{\vec{q}}\,' & = & \left[{\pi\over 2}-{(k_x-k_y)\over 2},-{\pi\over 2}+{(k_x-k_y)\over 2}\right] \, ,
\label{k-SW-I}
\end{eqnarray} 
for the sub-sector such that $k_x\in (0,\pi)$, $k_y\in (0,k_x)$ for $k_x\leq\pi/2$,
and $k_y\in (0,\pi-k_x)$ for $k_x\geq\pi/2$ and,
\begin{eqnarray}
\vec{\pi} & = & [\pi,\pi] \, ,
\nonumber \\
{\vec{q}} & = & \left[{\pi\over 2}-{(k_x+k_y)\over 2},{3\pi\over 2}-{(k_x+k_y)\over 2}\right] \, ,
\nonumber \\
{\vec{q}}\,' & = & \left[{\pi\over 2}-{(k_x-k_y)\over 2},-{\pi\over 2}+{(k_x-k_y)\over 2}\right]
\, ,
\label{k-SW-II}
\end{eqnarray} 
for the sub-sector such that $k_y\in (0,\pi)$, $k_x\in (\pi-k_y,\pi)$ for $k_y\leq\pi/2$,
and $k_x\in (k_y,\pi)$ for $k_y\geq\pi/2$, respectively.
Note that as mentioned above, the components of the $s1$ band momenta ${\vec{q}}$ appearing 
in Eqs. (\ref{k-SW-I}) and (\ref{k-SW-II}) are such that
$q_{x_1}-q_{x_2}=-\pi$ and thus belong to the half-filling $s1$ boundary line defined by Eq. (\ref{g-FS-x-0}), whereas 
those of the momenta 
${\vec{q}}\,'$ in the same equations obey the relation
${q'}_{x_1}=-{q'}_{x_2}$ so that point in the nodal directions. 
\begin{figure}
\includegraphics[width=6cm,height=6cm]{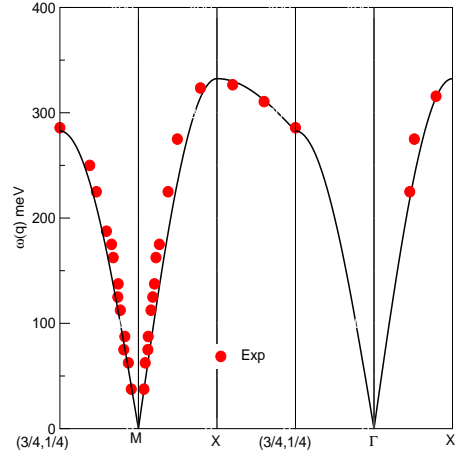}
\caption{\label{fig1} The theoretical spin spectra (\ref{direc-GO})-(\ref{direc-XO}) 
(solid lines) plotted in the second Brillouin zone for 
$\mu^0= 565.6$ meV and $W^0_{s1}= 49.6$ meV and the 
experimental data of Ref. \cite{LCO-neutr-scatt} (circles) in meV. Such
theoretical magnitudes correspond to $t\approx 0.295$ eV and 
$U\approx 1.800$ eV, so that $U/4t\approx 1.525$. The 
momentum is given in units of $2\pi$. The solid lines plotted here 
correspond to the simple analytical expressions provided in 
Eqs. (\ref{direc-GO})-(\ref{direc-XO}). Those plotted in Fig. 5 of Ref. \cite{LCO-Hubbard-NuMi}
are very similar yet are obtained within the standard formalism of 
many-body physics by summing up an infinite number
of ladder diagrams. Experimental points from Ref. \cite{LCO-neutr-scatt}.}
\end{figure}

Next let us consider the high symmetry directions in the  
Brillouin zone. Indeed, the curves plotted in Fig. 5 of Ref. \cite{LCO-Hubbard-NuMi}
refer only to such directions and the use of our above general expressions
leads for $U/4t=1.525$ to an excellent agreement with such curves. These directions 
correspond also to those measured by high-resolution inelastic 
neutron scattering in LCO, as plotted in Fig. 3 (A) of Ref. \cite{LCO-neutr-scatt}.
We denote such symmetry directions by $MO$, $\Gamma O$, $XM$, $\Gamma X$, 
and $XO$. They connect the momentum-space points $M=[\pi,\pi]$, $O=[\pi/2,\pi/2]$, 
$\Gamma =[0,0]$, and $X=[\pi,0]$ of the general spin-wave spectrum 
provided in Eq. (\ref{om-SW}). The use of that equation reveals that the spin-wave 
excitation spectrum is in such symmetry directions given by, 
\begin{eqnarray}
\omega_{\Gamma O} (\vec{k}) & = & {\mu^0\over 2}\sin (k_i) \, ,
\nonumber \\
\vec{k} & = & [\pi,-\pi] - [\pi/2 -k_i,-\pi/2 -k_i] - [\pi/2,-\pi/2]
\nonumber \\
& = & [k_i,k_i] \, ; \hspace{0.50cm} k_i = k_x = k_y \in (0,\pi/2) \, ,
\label{direc-GO}
\end{eqnarray}
\begin{eqnarray}
\omega_{MO} (\vec{k}) & = & {\mu^0\over 2}\sin (k_i) \, ,
\nonumber \\
\vec{k} & = & [\pi,\pi] - [\pi/2 -k_i,3\pi/2 -k_i] - [\pi/2,-\pi/2]
\nonumber \\
& = & [k_i,k_i] \, ; \hspace{0.50cm} k_i = k_x = k_y \in (\pi/2,\pi) \, ,
\label{direc-MO}
\end{eqnarray}
\begin{eqnarray}
\omega_{\Gamma X} (\vec{k}) & = &
\left[{\mu^0\over 2} +W^0_{s1}\right]\sin (k_x/2) \, ,
\nonumber \\
\vec{k} & = & [\pi,-\pi] - [\pi/2-k_x/2,-\pi/2 -k_x/2] 
\nonumber \\
& - & [\pi/2 -k_x/2,-\pi/2+k_x/2]
\nonumber \\
& = & [k_x,0]   \, ; \hspace{0.50cm} k_x \in (0,\pi) \, ,
\label{direc-GX}
\end{eqnarray}
\begin{eqnarray}
\omega_{XM} (\vec{k}) & = &
\left[{\mu^0\over 2} +W^0_{s1}\right]\cos (k_y/2) \, ,
\nonumber \\
\vec{k} & = & [\pi,\pi] - [-k_y/2,\pi -k_y/2] - [k_y/2,-k_y/2]
\nonumber \\
& = & [\pi,k_y]  \, ; \hspace{0.50cm} k_y \in (0,\pi) \, ,
\label{direc-XM}
\end{eqnarray}
\begin{eqnarray}
\omega_{XO} (\vec{k}) & = &
{\mu^0\over 2} -W^0_{s1} \cos (k_x) \nonumber \\
& = & {\mu^0\over 2} +W^0_{s1} \cos (k_y) \, ,
\nonumber \\
\vec{k}  & = & [\pi,-\pi] - [0,-\pi] - [\pi -k_x,-\pi +k_x]
\nonumber \\
& = & [\pi,\pi] - [0,\pi] - [k_y,-k_y]
\nonumber \\
& = &  [k_x,\pi -k_x] 
\, ; \hspace{0.25cm} k_x \in (\pi/2,\pi)
\nonumber \\
& = &  [\pi -k_y,k_y] 
\, ; \hspace{0.25cm} k_y \in (0,\pi/2) \, .
\label{direc-XO}
\end{eqnarray}

The theoretical spin excitation spectra (\ref{direc-GO})-(\ref{direc-XO})
are plotted in Fig. 1 (solid line) for $\mu^0= 565.6$ meV and 
$W^0_{s1}= 49.6$ meV together with the experimental results (circles) 
for $T=10$ K. This gives a Mott-Hubbard gap $2\mu^0= 1131.2$ meV.
The spin-spectrum expressions provided in Eqs. (\ref{direc-GO})-(\ref{direc-XO})  
refer to the first Brillouin zone. In Fig. 1 we plot them in the second
Brillouin zone, alike in Fig. 3 (A) of Ref. \cite{LCO-neutr-scatt}
and Fig. 5 of Ref. \cite{LCO-Hubbard-NuMi}.
An excellent quantitative agreement is reached for these magnitudes of the involved
energy scales, which according to the complementary results of Ref.
\cite{LCO-Hubbard-NuMi} correspond to $U/4t\approx 1.525$
and $t\approx 295$ meV. Moreover, from the use of the relation 
$\Delta_0 \approx r_s\,4W_{s1}^0$ of Eq. (\ref{Delta-0-gen})
we find that $\Delta_0\approx 84$ meV for $U/4t\approx u_*=1.525$. 
The magnitudes of the energy scales $\Delta_0$, $4W^0_{s1}$, and $\mu^0/2$ found 
here for $U/4t\approx 1.525$ are given in Table III together with those for
$U/4t\ll 1$, $U/4t =u_0$, and $U/4t\gg 19$. The magnitude of energy parameter 
$\Delta_0$ interpolates between that of $\mu^0/2$ for
$U/4t\ll 1$ and $4W^0_{s1}$ for $U/4t\gg 1$. It reaches a
maximum magnitude approximately given by
$\Delta_0 \approx t/\pi \approx 0.318\,t$ at $U/4t=u_0\approx 1.302$.
\begin{table}
\begin{tabular}{|c|c|c|c|} \hline
$U/4t$ & $\Delta_0$ & $4W^0_{s1}$ & $\mu^0/2$ \\
\hline
$\ll 1$ & $16t\,e^{-\pi\sqrt{4t/U}}$ & $16t$ & $16t\,e^{-\pi\sqrt{4t/U}}$ \\
\hline
$u_0$ & $t/\pi\approx 0.318\,t$ & $e^{1}t/\pi\approx 0.865\,t$ & $e^{1}t/\pi\approx 0.865\,t$ \\
\hline
$u_*$ & $0.285\,t$ & $0.673\,t$ & $0.959\,t$ \\
\hline
$\gg 19$ & $\pi\,[4t]^2/U$ & $\pi\,[4t]^2/U$ & $[U/4-2t]$ \\
\hline
\end{tabular}
\caption{Approximate magnitudes of the energy scales $2\Delta_0$, $4W^0_{s1}$,
and $\mu^0/2$ for different values of $U/4t$.
The energy parameter $\Delta_0$ interpolates between $\mu^0/2\approx 16t\,e^{-\pi\sqrt{4t/U}}$ for
$U/4t\ll 1$ and $4W^0_{s1}\approx \pi\,[4t]^2/U$ for $U/4t\gg 19$ and 
goes through a maximum magnitude ${\rm max}\,\{\Delta_0\} \approx t/\pi$ at the $U/4t$ value
$U/4t=u_0\approx 1.302$ at which $\mu^0/2\approx 4W^0_{s1}$. The magnitudes of 
$\Delta_0$, $4W^0_{s1}$, and $\mu^0/2$ found here for $U/4t=u_*=1.525$ are also
given.} 
\label{tableIII}
\end{table}

For the intermediate value $U/4t\approx u_*=1.525\in (u_0,u_1)$ corresponding to Fig. 1 
our investigations find the relation $W_{s1}^0 \approx 0.168\,t$, which for a constant value
of $t$ provides a $W_{s1}^0$ magnitude about twelve times smaller than that
found by the use of the limiting expression $W^0_{s1} \approx J \approx 4\pi\,t^2/U\approx 2.060\,t$ 
for that $U/4t$ value. Hence for the model on the square lattice the usual energy scale 
$J \approx 4\pi\,t^2/U$ controls the physics for a smaller $U/4t$ range than
in 1D, which corresponds to very large $U/4t\gg 19$ values. 
As discussed in Ref. \cite{companion}, the value $U/4t\approx 19$
is that at which the energy scale $J \approx 4\pi\,t^2/U$ 
reads $J =0.168\, t$ and thus has the magnitude 
that $W_{s1}^0$ reaches at $U/4t\approx 1.525$. This is why the large-$U/4t$ limiting expressions
$W^0_{s1} \approx J$ and $\Delta_0\approx 4J$ are valid for $U/4t\gg 19$.
It follows that the intermediate-$U/4t$ range plays a major role in the
physics of the square-lattice quantum liquid studied in this paper and in
Refs. \cite{cuprates0,cuprates}. 

The above results are consistent with the momentum
values of the $c$ and $s1$ fermions being good
quantum numbers. Furthermore, they confirm that the predictions of the 
square-lattice quantum liquid theory concerning the spin spectrum at half filling 
agree both with experiments on the parent compound LCO and results 
obtained by the standard formalism of many-body physics. 

\subsection{Relation to other schemes where the fermions emerge
from individual spin-$1/2$ spins or spinons}

For most previous studies on the large-$U$ Hubbard
model and $t-J$ model on a square lattice involving for instance
the slave particle formalism \cite{2D-MIT,Fazekas,Xiao-Gang} or Jordan-Wigner 
transformations \cite{Feng} the spinless fermions arise from individual 
spin-$1/2$ spins or spinons. In contrast, within our $s1$ fermion 
extended Jordan-Wigner transformation the $s1$ fermions emerge from spin-neutral
two-spinon composite $s1$ bond particles.
Indeed, here the hard-core bosons refer to spin-neutral two-spinon $s1$ 
bond operators. Often in previous related schemes involving
spin-liquid mean-field theories, which refer in general to individual
spin-$1/2$ spinons or spinless fermions that emerge from them, such objects
are coupled to a gauge field \cite{2D-MIT,Xiao-Gang}.
Since in 2D gauge theory is confining, in
some of these liquids individual spin-$1/2$
spinons are not observable, as in for instance in the
simple dimer state \cite{Xiao-Gang}. 

Here the original hard-core particles themselves are spin-neutral
two-spinon objects. Furthermore, since the spin-$1/2$ spinons 
have been constructed to inherently referring here to the rotated-electron
singly occupied sites, the one-rotated-electron-per-site
constraint is naturally fulfilled for $U/4t>0$ so that the fluctuations 
needed to enforce such a constraint are for the square-lattice model
incorporated in the electron - rotated-electron unitary operator $\hat{V}$. 
Related resonating-valence-bond pictures for spin-singlet 
occupancy configurations of ground states 
were considered long ago \cite{Fazekas,Pauling}, yet
often problems arise in the construction of energy
eigenstates due to bond states being overcomplete 
and non-orthogonal.  

Motivated by previous studies where the spinless
fermions arise from individual spin-$1/2$ spins or spinons, 
we have also performed a Jordan-Wigner transformation 
for the model both on the square and 1D lattices 
whose spinless fermions emerge from the spin-$1/2$ spinon
operators $s^{\pm}_{\vec{r}_j}$ of Eq. (\ref{sir-pir}).
According to Eqs. (\ref{albegra-s-p-m}) and (\ref{albegra-s-com}) such operators have
a hard-core character and the corresponding
spinons are well-defined for $U/4t>0$.
One then finds out that for 1D the momentum values of
the emerging spinless fermions do not coincide with the quantum numbers
of the exact solution. In contrast, the discrete momentum
values of the $s1$ fermions emerging from spin-neutral 
two-spinon $s1$ bond particles correspond to
such quantum numbers, as confirmed in Ref. \cite{companion}.  

For the model on the square lattice in the one- and two-electron
subspace considered in this paper and of interest for the further investigations 
on real materials of Refs. \cite{cuprates0,cuprates}, our description occupancy
configurations of the $c$ fermions and $s1$ fermions
generate according to the analysis of Ref. \cite{companion} energy eigenstates.
This is in contrast to the alternative scheme in
terms of spinless fermions emerging from the spin-$1/2$ spinon
operators $s^{\pm}_{\vec{r}_j}$ of Eq. (\ref{sir-pir}), which generates
a complete set of states in that subspace, which however are
not in general energy eigenstates. The same applies to the 1D model in the whole Hilbert
space. Following the results of
Ref. \cite{companion}, the momentum values of the $c$ and $s1$
fermions are good quantum numbers for the model on the square
lattice in the one- and two-electron subspace. Our scheme uses several approximations 
to access the shape of the $s1$ boundary
line, and the $c$ and $s1$ fermion energy dispersions. Our corresponding
results are not exact, yet they rely on a general subspace structure 
and correct symmetries, which are compatible with the unknown
exact expressions of the quantities under consideration. We expect that our approximations
provide a good description of the physics for intermediate
values of $U/4t$, consistently with the results
of the previous subsection concerning the half-filling spin spectrum.
That spin spectrum is found above to lead 
for the model on the square lattice to analytical expressions 
for the spin-wave spectrum that agree with the corresponding numerical curves obtained 
by a different controlled approximation involving the sum of an infinite set of electronic 
ladder diagrams \cite{LCO-Hubbard-NuMi}.

For the model on the square lattice the above alternative scheme leads to
a vector potential of the same general form as that given
in Eq. (\ref{A-j-s1}), but with the operator $n_{\vec{r}_j,s1}$ 
of Eq. (\ref{n-j-s1}) replaced by $[1/2+s^{x_3}_{\vec{r}_j}]$
and the real-space variable $\vec{r}_j$ referring to the spin
effective lattice rather than to the $s1$ effective lattice. Except
that such a scheme refers here to the spins of the
rotated electrons that singly occupy sites and is valid for $U/4t>0$, it is identical to the spin
problem and corresponding Jordan-Wigner transformation
considered in Ref. \cite{Wang}.

A qualitative difference, which has impact on the physics,
is that a $x\geq 0$ and $m=0$ ground state is described here
by a full $s1$ momentum band whereas for the spinless fermions
emerging from the spin-$1/2$ spinon operators 
$s^{\pm}_{\vec{r}_j}$ of Eq. (\ref{sir-pir}) it is
described by a half-filled spin band. This is alike the
{\it Wigner-Jordan (WJ) fermions} 
associated with the WJ representation of
quantum spins with antiferromagnetic interaction of Ref. 
\cite{Wang}, which at zero temperature 
fill exactly half of the total states. In turn, within our
$c$ and $s1$ fermion description the $s1$ momentum
band is full for the initial $x\geq 0$ and $m=0$ ground state and 
addition or removal of one electron leads to the emergence 
of a single $s1$ fermion hole in that band. Indeed, following
the transformation laws under the electron - rotated-electron
unitary transformation of the objects whose occupancy
configurations generate the energy eigenstates of the square-lattice
model in the one- and two-electron subspace,
excited states with three or a larger odd number 
of $s1$ fermion holes have nearly no overlap with
one-electron excitations \cite{companion}.
Furthermore, symmetry selection rules 
\cite{companion} impose that such an overlap 
vanishes exactly for excited states with an even
number of $s1$ fermion holes.

The emergence of a single hole in the $s1$ band upon 
creation or annihilation of one electron plays a major role 
in the strong phase-space, exclusion-principle, and energy 
and momentum conservation restrictions of the $s1$ - $s1$ fermion
and $s1$ - $c$ fermion scattering. Within the emergence of a single $s1$ fermion hole
there are obviously no available final channels associated with $s1$ - $s1$ fermion scattering.
The above restrictions, together with the isotropic and anisotropic
character of the $c$ Fermi line and $s1$ boundary line,
respectively, leads for $x>0$ to unusual one-electron
scattering properties \cite{cuprates}. This includes an anisotropic one-electron inverse 
lifetime and scattering rate linear in the 
excitation energy $\omega$, in excellent quantitative agreement with 
that observed in recent angle-resolved photoemission studies  
of the optimally doped high-temperature superconductor 
LSCO \cite{Kam,PSI-ANI-07}.  
Such an inverse lifetime is different from that characteristic of an 
isotropic conventional 2D electron liquid, which is
proportional to $(\omega/E_F)^2\ln (E_F/\omega)$ where $E_F$ 
denotes the Fermi energy \cite{2D-1/tau-1}. 

In turn, for the spinless fermions emerging from the spin-$1/2$ spinon
operators $s^{\pm}_{\vec{r}_j}$ given in Eq. (\ref{sir-pir}),
that their momentum band is half filled implies that the ground-state 
configurations contain as many spinless fermions as spinless 
fermion holes. It follows that in contrast to the $s1$ fermions, upon one-electron excitations there
are available final channels for both spinless fermion - spinless fermion and spinless fermion - $c$ fermion inelastic
collisions. In addition, since the momenta carried by
such spinless fermions are not good quantum numbers, their
interactions are not residual so that in contrast to
the $c$ and $s1$ fermion description one cannot use
Fermi's golden rule in terms of a collision integral as that associated with the 
$c$ - $s1$ fermion interactions to calculate the 
one-electron lifetime \cite{cuprates0}. In
principle the final expressions should be the same independently
of the representation. However, the one-electron scattering problem is much more complex
when expressed in terms of the interactions of the spinless fermions emerging 
from the spin-$1/2$ spinon operators $s^{\pm}_{\vec{r}_j}$ and so far
we could not solve it. 

Finally, only within our description do the $s1$ fermion occupancy configurations 
that generate the spin degrees of freedom of $x>0$ and $m=0$ ground states
of the 1D model become in the $U/4t\rightarrow\infty$ limit 
those of the spins of the well-known spin-charge factorized wave function introduced 
independently by Woynarovich \cite{Woy} and by Ogata and Shiba \cite{Ogata},
respectively. Moreover, only within that description become such ground-state
configurations of the model on a square lattice in the $U/4t\rightarrow\infty$ limit 
and within the suitable mean-field approximation (\ref{ficti-B}) for the fictitious magnetic field 
${\vec{B}}_{s1}$ of Eq. (\ref{A-j-s1-3D}) those of a $\nu_{s1}=1$ full lowest Landau level with 
$N_{a_{s1}}^2=N/2$ one-$s1$-fermion degenerate states of the 2D QHE.

\section{Concluding remarks}

The operator description introduced in Ref. \cite{companion} for the Hubbard model on the square 
lattice in the one- and two-electron subspace as defined in that rteference has been constructed to
inherently the discrete momentum values of the $c$ and $s1$ fermions beeing good quantum numbers. 
Complementarily, in this paper we derive approximate results and expressions for the $s1$ band boundary line, 
$c$ and $s1$ fermion energy dispersions, and corresponding $c$ and $s1$ fermion velocities 
of the square-lattice quantum liquid introduced in that reference.

That the $s1$ band remains full alike for for $x> 0$ and $m=0$ ground states alike for
$x=0$ is related to the degree of localization in real space of the $s1$ fermion holes,
which upon one- and two-electron excitations emerge near the $s1$ boundary line at momenta pointing in or near the 
anti-nodal directions. Indeed, the $s1$ 
fermion velocity studied in Section V vanishes for $s1$ boundary-line momenta 
pointing in such directions. The corresponding finite-energy real-space excitations 
occur near the pseudogap energy and break
translational symmetry locally, due to  
the square $s1$ effective lattice spacing being given 
by $a_{s1}=\sqrt{2/(1-x)}\,a$ rather than by the original
square-lattice spacing $a$. In turn, for $x>0$ the $c$ band
has for the ground state an isotropic $c$ Fermi line 
associated with a finite $c$ fermion velocity so that the 
low-energy one- and two-electron excitations involving creation of 
$c$ fermions or $c$ fermion holes near the $c$ Fermi
line and $s1$ fermion holes with momenta pointing in
and near the nodal directions have a delocalized character. 
Consistently, in contrast to the isotropic $c$ Fermi line it is found
in this paper that the $s1$ boundary line is rather anisotropic,
the corresponding $s1$ fermion velocity vanishing and
reaching its maximum magnitude for momenta pointing
in the anti-nodal and nodal directions, respectively. 

Such properties of the $c$ Fermi line and $s1$ boundary line are behind the
anisotropy of the Fermi velocity, which is larger for smaller $x$. They remain
valid for the square-lattice quantum liquid weakly perturbed by the effects of 
3D anisotropy investigated in Refs. \cite{cuprates0,cuprates}. The results of
that reference are fully consistent
with the two-gap scenario observed in the hole-doped
cuprates superconductors \cite{two-gaps,k-r-spaces}.
For instance, there is a direct connection to the recent experimental 
studies of Ref. \cite{k-r-spaces}. Indeed, the excitations involving
coherent virtual-electron pairs whose spinless charge-$-2e$ zero-momentum $c$ 
fermion pairs are assisted by spin-singlet two-spinon composite $s1$ fermions of momenta 
pointing in or near the nodal directions correspond to the 
superconducting superfluid of delocalized Cooper pairs
in momentum space observed in such experiments. In turn,
the finite-energy excitations involving $s1$ fermions whose momenta point in and near the
anti-nodal directions refer to the locally translational-breaking 
pseudogap states in real space observed in the same experiments.

Furthermore, the investigations of Refs. \cite{cuprates0,cuprates}
provide evidence that accounting for the interplay of the electronic
correlations described by the square-lattice quantum liquid of 
$c$ and $s1$ fermions studied in this paper with the weak effects of 3D
anisotropy and intrinsic disorder leads to a successful theoretical description of 
the unusual properties of the hole-doped cuprates. 
Indeed, the corresponding scheme introduced in Ref. \cite{cuprates}, leads to quantitative
agreement with the universal properties of the hole-doped
cuprates with superconducting zero-temperature critical 
hole concentrations $x_c\approx 0.05$ and
$x_*\approx 0.27$. For instance, the isotropic
character of the $c$ Fermi line and the anisotropy of the
$s1$ boundary line found in this paper combined with the $s1$ band 
being full and having a single hole for $x>0$ and $m=0$ 
ground states and their one-electron excited states, respectively, leads to unusual one-electron 
scattering properties controlled by the $c$ - $s1$ fermion inelastic collisions.
The investigations of that reference profit from such a $c$ and
$s1$ fermion description of the one-electron scattering problem and find a strongly 
anisotropic one-electron inverse lifetime and
scattering rate in excellent quantitative agreement with 
that observed in recent angle-resolved photoemission studies of the optimally
doped high-temperature superconductor LSCO \cite{Kam,PSI-ANI-07}. The studies of Ref. \cite{cuprates} also
successfully address the normal-state linear-$T$ resistivity
in that and other hole doped cuprates \cite{LSCO-resistivity,Y-resistivity}, 
neutron resonance mode observed in families of hole doped
cuprates with critical temperature $T_c\approx 95$ K \cite{two-gaps},
hole-concentration $x$ dependence of several physical quantities such as
$T_c$ and the superfluid density, and the LSCO neutron-scattering 
low-energy incommensurate peaks \cite{2D-MIT,pseudogap-review}.

In conclusion, the results of Refs. \cite{cuprates0,cuprates} confirm
that addition to the square-lattice quantum liquid of $c$ and $s1$
fermions studied in this paper of small 3D anisotropy and intrinsic disorder 
effects leads to a successful description of the unusual properties observed
in the hole-doped cuprates.

\begin{acknowledgments}
I thank Miguel A. N. Ara\'ujo, Daniel Arovas, Miguel A. Cazalilla, Karlo Penc, 
Nuno M. R. Peres, Pedro D. Sacramento, and Maria J. Sampaio 
for discussions and the support of the ESF Science Program INSTANS and grant
PTDC/FIS/64926/2006.
\end{acknowledgments}
\appendix

\section{The one- and two-electron subspace}

The square-lattice quantum liquid of $c$ and $s1$ fermions
studied in this paper refers to the one- and two-electron subspace of the Hubbard model
on the square lattice as defined in Ref. \cite{companion}.
Here we provide some basic information about such a
subspace needed for the studies of this paper.
The results summarized in the following are derived and discussed
in that paper and in Ref. \cite{s1-bonds}.

For hole concentrations $0\leq x\leq 1$ and subspaces spanned by energy eigenstates
with constant eigenvalue $S_c$ of the generator of the
global $U(1)$ symmetry of Ref. \cite{bipartite}
there is a vacuum $\vert 0_{\eta s}\rangle$, which remains
invariant under the electron - rotated-electron unitary 
transformation. For $x=1$ such a vacuum coincides with the
electronic vacuum whereas for $0\leq x<1$ it refers to the
$m=(1-x)$ fully polarized ground state.  
For the 1D model one can derive a closed form
analytical expression for the critical magnetic field $H_c$ for
fully polarized ferromagnetism as function of the electronic density $n=(1-x)$ and
$u=U/4t$. It is given in Eq. (2) of Ref. \cite{renor}
and plotted in Fig. 1 of that reference. Alike in 1D, it is expected
that for the model on the square lattice the magnitude $H_c$ of the in-plane critical field
for fully polarized ferromagnetism is for all values of $U/4t$ 
a decreasing function of the hole concentration $x$, which vanishes
for $x\rightarrow 1$. It is expected as well that for $x<1$ it is a decreasing function of $U/4t$,
which vanishes for $U/4t\rightarrow\infty$. Such a vacuum reads,
\begin{equation}
\vert 0_{\eta s}\rangle = \vert 0_{\eta};N_{a_{\eta}}^D\rangle\times\vert 0_{s};N_{a_{s}}^D\rangle
\times\vert GS_c;2S_c\rangle \, ,
\label{23}
\end{equation}
where the $\eta$-spin $SU(2)$ vacuum $\vert 0_{\eta};N_{a_{\eta}}^D\rangle$ 
associated with $N_{a_{\eta}}^D$ independent $+1/2$
$\eta$-spinons, the spin $SU(2)$ vacuum $\vert 0_{s};N_{a_{s}}^D\rangle$ 
with $N_{a_{s}}^D$ independent $+1/2$ spinons, and the $c$ $U(1)$
vacuum $\vert GS_c;2S_c\rangle$ with $N_c=2S_c$ $c$ fermions
remain invariant under the electron - rotated-electron unitary transformation. 
For such a fully polarized problem the on-site electronic repulsion has no
effects so that for $m\rightarrow (1-x)$ and $H\rightarrow H_c$ 
the rotated electrons become electrons. In that limit the $N_c=2S_c=N$ $c$ fermions are 
the non-interacting spinless fermions, which describe the charge degrees of freedom of the
$N$ electrons. The $c$ $U(1)$ vacuum $\vert GS_c;2S_c\rangle$ can be written as,   
\begin{equation}
\vert GS_c;2S_c\rangle = \prod_{j=1}^{2S_c}f^{\dag}_{{\vec{q}}_{j},c}\vert GS_c;0\rangle \, .
\label{GS-c}
\end{equation}
Here $\vert GS_c;0\rangle$ refers to the $x=1$ and $N=0$ electronic vacuum,
\begin{equation}
\vert 0_{elec}\rangle = \vert 0_{\eta};N_{a}^D\rangle\times\vert 0_{s};0\rangle
\times\vert GS_c;0\rangle \, .
\label{electron-vacuum}
\end{equation}
The state (\ref{GS-c}) is a particular case of the state of the same
form appearing in Eq. (\ref{LWS-1-2-el}). In turn, the vacuum (\ref{electron-vacuum})
corresponds to a limiting case of the vacuum (\ref{23}) associated with
the values $N_{a_{\eta}}^D=N_{a}^D$, $N_{a_{s}}^D=0$, and $2S_c=0$. 

The $c$ fermion, $\alpha\nu$ fermion, and independent $\eta$-spinon and independent spinon
description of Ref. \cite{companion} refers to a complete set of $4^{N_a^D}$
momentum eigenstates $\vert \Phi_{U/4t}\rangle={\hat{V}}^{\dag}\vert \Phi_{\infty}\rangle$ 
generated from application onto the corresponding $U/4t\rightarrow\infty$ states $\vert \Phi_{\infty}\rangle$ 
of the electron - rotated-electron unitary operator ${\hat{V}}^{\dag}$. Such states can
be generated from their LWS as,
\begin{equation}
\vert \Phi_{U/4t}\rangle = \prod_{\alpha
=\eta,\,s}\frac{({\hat{S}}^{\dag}_{\alpha})^{L_{\alpha,\,-1/2}}}{
\sqrt{{\cal{C}}_{\alpha}}}\vert \Phi_{LWS;U/4t}\rangle
\, ; \hspace{0.35cm} {\cal{C}}_{\alpha} = \delta_{L_{\alpha,\,-1/2},\,0} +
\prod_{l=1}^{L_{\alpha,\,-1/2}}l\,[\,L_{\alpha}+1-l\,] \, , 
\label{non-LWS}
\end{equation}
where $L_{\alpha,\,-1/2}$ is the number of $\eta$-spin-projection $-1/2$ independent $\eta$-spinons $(\alpha =\eta$) or
spin-projection $-1/2$ independent spinons $(\alpha =s$). The corresponding LWS
$\vert \Phi_{LWS;U/4t}\rangle={\hat{V}}^{\dag}\vert \Phi_{LWS;\infty}\rangle$ and $U/4t\rightarrow\infty$
LWS belonging to the same $V$ tower read,
\begin{equation}
\vert \Phi_{LWS;U/4t}\rangle =
[\prod_{\alpha}\prod_{\nu}\prod_{{\vec{q}}\,'}f^{\dag}_{{\vec{q}}\,',\alpha\nu}\vert 0_{\alpha};N_{a_{\alpha}^D}\rangle]
[\prod_{{\vec{q}}}f^{\dag}_{{\vec{q}},c}\vert GS_c;0\rangle] \, ; \hspace{0.25cm} f^{\dag}_{{\vec{q}}\,',\alpha\nu} =
{\hat{V}}^{\dag}\,{\mathcal{F}}^{\dag}_{{\vec{q}}\,',\alpha\nu}\,{\hat{V}} \, ; \hspace{0.25cm} f^{\dag}_{{\vec{q}},c} =
{\hat{V}}^{\dag}\,{\mathcal{F}}^{\dag}_{{\vec{q}},c}\,{\hat{V}} \, ,
\label{LWS-full-el}
\end{equation}
and
\begin{equation}
\vert \Phi_{LWS;\infty}\rangle =
[\prod_{\alpha}\prod_{\nu}\prod_{{\vec{q}}\,'}{\mathcal{F}}^{\dag}_{{\vec{q}}\,',\alpha\nu}\vert 0_{\alpha};N_{a_{\alpha}^D}\rangle]
[\prod_{{\vec{q}}}{\mathcal{F}}^{\dag}_{{\vec{q}},c}\vert GS_c;0\rangle] \, ,
\label{LWS-full-el-infty}
\end{equation}
respectively. Here ${\mathcal{F}}^{\dag}_{{\vec{q}},c}$ and ${\mathcal{F}}^{\dag}_{{\vec{q}}\,',\alpha\nu}$
are the $c$ fermion and $\alpha\nu$ fermion creation operators for $U/4t\rightarrow\infty$, respectively.

For the 1D Hubbard model the states (\ref{non-LWS}) and (\ref{LWS-full-el}) are both
energy and momentum eigenstates. In turn, since for the model on the square lattice the
set of $\alpha\nu$ translation generators ${\hat{\vec{q}}}_{\alpha\nu}$ in the presence of the fictitious magnetic fields 
${\vec{B}}_{\alpha\nu}$ considered in Ref. \cite{companion} whose eigenvalues are
the components of the $\alpha\nu$ band momenta do not in general commute with
the Hamiltonian and the corresponding set of $\alpha\nu$ fermion numbers $\{N_{\alpha\nu}\}$ are not in general conserved,
such states are not energy eigenstates and the microscopic momenta carried by the
$\alpha\nu$ fermions are not in general good quantum numbers. However, such states have
been constructed to inherently being $S_{\eta}$, $S_{\eta}^z$, $S_s$, $S_s^z$, $S_c$, and momentum eigenstates 
and each energy eigenstate $\vert \Psi_{U/4t}\rangle={\hat{V}}^{\dag}\vert \Psi_{\infty}\rangle$ can 
be expressed as a suitable superposition of a well-defined set of 
$\vert \Phi_{U/4t}\rangle={\hat{V}}^{\dag}\vert \Phi_{\infty}\rangle$ momentum eigenstates (\ref{non-LWS})
with the same momentum eigenvalue and values of
$S_{\eta}$, $S_{\eta}^z$, $S_s$, $S_s^z$, $S_c$, $C_{\eta}=\sum_{\nu}\nu\,N_{\eta\nu}$, 
and $C_{s}=\sum_{\nu}\nu\,N_{s\nu}$ and the same $c$ fermion momentum distribution 
function $N_c ({\vec{q}})$ \cite{companion}. As discussed in this paper, for the model
on the square lattice in the one- and two-electron subspace the form of the states
(\ref{non-LWS}) and (\ref{LWS-full-el}) simplifies to that provided in Eq. (\ref{LWS-1-2-el})
and as justified in Ref. \cite{companion} such states are energy eigenstates so that the microscopic momenta carried by the
$c$ and $s1$ fermions are good quantum numbers

For spin densities $m=0$ and $m=(1-x)$ and $U/4t\rightarrow\infty$ and 
$U/4t\geq 0$, respectively, the $c$ fermions are for $0<x<1$ non-interacting and for the model
on the square lattice have an energy dispersion given by, 
\begin{equation}
\epsilon^0_{c} (\vec{q})= -2t\,[\cos (q_{x_1})+\cos (q_{x_2})-\cos (k_{Fx_1})
-\cos (k_{Fx_2})] \, .
\label{EcGen-H-c}
\end{equation}
Here $k_{Fx_1}$ and $k_{Fx_2}$ are the Cartesian coordinates of the
Fermi momentum $\vec{k}_F$,
\begin{equation}
\vec{k}_F = k_F (\phi) \,{\vec{e}}_{\phi} \, ;
\, ; \hspace{0.50cm}
\phi = \arctan \left({k_{Fx_2}\over 
k_{Fx_1}}\right) \, ; \hspace{0.35cm}
\phi \in (0,2\pi) \, .
\label{k-F-m-max}
\end{equation}
For such values of $m$ and $U/4t$ the corresponding Fermi line 
encloses a momentum area,
\begin{equation}
\int_{0}^{2\pi}{d\phi\over 2\pi}\,
\pi\left[k_{F} (\phi)\right]^2 = (1+x)\,4\pi^2 \, .
\label{KF-sum-rule-m-max}
\end{equation} 
It is twice the area enclosed by the Fermi line at $m=0$ and $U/4t=0$. 

Two energy scales that play a key role in the square-lattice quantum
liquid are the chemical potential $\mu$ and the maximum $s1$ 
fermion spinon-pairing energy $2\vert\Delta\vert$ of Eq. (\ref{D-D-HM-x<<1}),
which in the limit $x\rightarrow 0$ equals the energy scale $2\Delta_0$ whose
magnitudes are provided in Table III. Within the LWS representation, the convention that $\mu$ has the same sign 
as the hole concentration $x$ is used. For $0<x<1$ and 
$m=0$ the chemical potential is an increasing 
function of the hole concentration $x$ whose values
belong to the range,
\begin{equation}
\mu^0\leq\mu (x)\leq\mu^1 \, ; \hspace{0.50cm}  0<x<1
\, , \hspace{0.35cm} m= 0 \, ,
\label{36}
\end{equation}
where $\mu^1\equiv \lim_{x\rightarrow 1}\mu$.
$\mu^1$ reads,
\begin{equation}
\mu^1 =  [U/2 + 2Dt] \, ;
\hspace{0.50cm} D = 1,2 \, .
\label{37}
\end{equation}
Moreover, $\mu^0$ has the following approximate limiting behaviors \cite{companion},
\begin{eqnarray}
\mu^0 \approx {U\over 2\pi^2}\left({[8\pi]^2 t\over U}\right)^{D/2}e^{-2\pi \left({t\over U}\right)^{1/D}} 
\, , \hspace{0.25cm} U/4t\ll 1 \, ;
\hspace{0.5cm}
\mu^0 \approx  [U/2 - 2Dt] \, ,
\hspace{0.25cm}  U/4t\gg 1  \, ,
\hspace{0.25cm} D = 1,2 \, ,
\label{38}
\end{eqnarray}
so that $\mu^0\rightarrow 0$ as
$U/4t\rightarrow 0$ whereas $\mu^0\rightarrow\infty$ for
$U/4t\gg 1$ for both the model on the 1D and square
lattices. As discussed in previous sections, $2\mu^0$ is the 
half-filling Mott-Hubbard gap. 

Another energy scale that plays an important role
is the energy $\epsilon_{c}$ for addition onto a
$x\geq 0$ and $m=0$ ground state of 
one $c$ fermion and the corresponding energy energy $-\epsilon_{c}$ 
for removal from that state of one $c$ fermion, which
obey the inequalities,
\begin{equation}
0 \leq \epsilon_{c}\leq [4Dt - W^p_c]   
\, ; \hspace{0.50cm}
0 \leq -\epsilon_{c}\leq W^p_c  
\, ; \hspace{0.50cm} D = 1,2 \, ,
\label{42}
\end{equation}
respectively. In turn, the energy $-\epsilon_{s1}$ 
for addition to a $x\geq 0$ and $m=0$ ground state of one $s1$ fermion hole obeys the inequality,
\begin{equation}
0 \leq -\epsilon_{s1} \leq {\rm max}\,\{W_{s1},(D-1)\vert\Delta\vert\} 
\, , \hspace{0.35cm} D = 1,2 \, .
\label{43}
\end{equation}
In such inequalities $W^p_c$ and $W^h_c=[4Dt - W^p_c]$ are 
the ground-state $c$ fermion and $c$ fermion-hole 
energy-dispersion bandwidths, respectively, $W^h_c=[4Dt-W^p_c]\in (0,4Dt)$ 
increases monotonously upon increasing the hole concentration within the range 
$x\in (0,1)$, $W_{s1}$ is the $s1$ fermion auxiliary energy-dispersion bandwidth of
Eq. (\ref{D-D-HM-x<<1}), and the energy parameter $\vert\Delta\vert$ 
corresponds for the model on the square lattice to the 
maximum $s1$ fermion pairing energy per spinon of Eq. (\ref{D-x}).
The energy $W_{s1}^0=\lim_{x\rightarrow 0}W_{s1}$
has the following approximate limiting behaviors \cite{companion},
\begin{eqnarray}
W_{s1}^0 & = & 2Dt  \, ;
\hspace{0.50cm}  U/4t = 0 \, ,
\nonumber \\
& \approx & 2D\pi\,t^2/U \, ;
\hspace{0.50cm} U/4t\gg 1 \, .
\label{51}
\end{eqnarray}
For the model on the square lattice the large-$U/4t$
expression is expected to be a good approximation for very large values
$U/4t\gg 19$, as justified in this paper and Ref. \cite{companion}.

The one- and two-electron subspace is spanned by a given 
initial $x\geq 0$ and $m=0$ ground state and a well-defined 
set of its excited energy eigenstates of excitation energy $\omega<2\mu$
for $x>0$ and $\omega<\mu^0$ for $x=0$. Such states
have no $-1/2$ $\eta$-spinons, $2\nu$-$\eta$-spinon
composite $\eta\nu$ fermions, and $2\nu'$-spinon composite $s\nu'$ fermions with $\nu'\geq 3$ 
spinon pairs so that $N_{\eta\nu}=0$ and $N_{s\nu'}=0$ for $\nu'>1$, whereas
according to the studies of Ref. \cite{companion} the numbers 
of independent $\pm 1/2$ spinons and 
that of $s2$ fermions are restricted to the following ranges,
\begin{equation}
L_{s,\,\pm 1/2} = 0,1  \, ;
\hspace{0.50cm} N_{s2} = 0 \, ,
\hspace{0.35cm} {\cal{N}} = 1 
\, ; \hspace{0.50cm}
2S_s + 2N_{s2} =0,1,2  \, ,
\hspace{0.35cm} {\cal{N}} = 2 \, .
\label{46}
\end{equation}
Here ${\cal{N}}=1,2$ refers to the ${\cal{N}}$-electron
operators whose application onto the ground
state leads to the excited states under consideration.
Furthermore, the numbers of $c$ and $s1$ fermions
read $N_c =N=(1-x)\,N_a^D$ and
$N_{s1} = [N/2-2N_{s2}-S_s]=(1-x)\,N_a^D/2 -
[2N_{s2}+S_s]$, respectively.

From analysis of the ground-state occupancies found in Ref. \cite{companion}, 
for the one- and two-electron subspace there occur the fourteen   
classes of elementary excited states relative to an initial $x\geq 0$ and
$m=0$ ground state whose numbers and number deviations are
given in Table \ref{tableIV}. The table provides the values of the deviations $\delta N_c^h$ 
and numbers $N_{s1}^h$ of such excited states, 
corresponding electron number deviations $\delta N_{\uparrow}$ 
and $\delta N_{\downarrow}$, and independent-spinon numbers $L_{s,\,+1/2}$ and 
$L_{s,\,-1/2}$ and $s2$ fermion numbers 
$N_{s2}$ restricted to the value ranges of that subspace.
The spin $S_s$ and deviations $\delta S_c$, $\delta N_{s1}=[\delta S_c-S_s-2N_{s2}]$, and $\delta N_{a_{s1}}=[\delta S_c +S_s]$ 
of each excitation are also given.

In reference \cite{companion} it is found from analysis
of the transformation laws under the electron - rotated-electron
unitary transformation of the objects whose occupancy
configurations generate the energy eigenstates of
the Hubbard model that
nearly the whole one- and two-electron spectral weight is 
contained in the one- and two-electron subspace whose 
numbers are given in Table IV. If in addition we restrict our
considerations to the LWS-subspace of the one- and
two-electron subspace, then $L_{s,\,-1/2}=0$ 
in Eq. (\ref{46}), whereas
the values $L_{s,\,+1/2} = 0,1$ for $N_{s2} = 0$ and
${\cal{N}} = 1$ remain valid and in $2S_s + 2N_{s2} =0,1,2$
one has that $2S_s=L_{s,\,+1/2}$ for ${\cal{N}} = 2$ in that equation.

For the one- and two-electron subspace of the 
Hamiltonian (\ref{H}) only the $c$ fermions and
$s1$ fermions play an active role.
The numbers $N_a^D$ and $N_{a_{s1}}^D$ of sites of the 
$c$ and $s1$ effective lattices, respectively, equal the number of
corresponding $c$ and $s1$ band discrete momentum
values. The number $N_{a_{s1}}^D$, 
that of $s1$ fermions $N_{s1}$, and that of both unoccupied 
sites of the $s1$ effective lattice and $s1$ band holes 
$N^h_{s 1}$ are for the one- and two-electron subspace given by,
\begin{eqnarray}
N_{a_{s1}}^D & = & N_{s1} + N^h_{s1} = 
N_{a_s}^D/2 + S_s = S_c + S_s \, ; \hspace{0.50cm}
N_{a_{s}}^D = (1-x)\,N_a^D \, ,
\nonumber \\
N_{s1} & = & N_{a_s}^D/2 - S_s +2N_{s2}
= S_c - S_s - 2N_{s2} 
\, ; \hspace{0.50cm}
N^h_{s 1} = 2S_s +2N_{s2} =0,1,2 
\, ,
\label{68}
\end{eqnarray}
respectively, where $N_{a_{s}}^D=(1-x) N_a^D$ is the number of sites of the
related spin effective lattice. 
\begin{table}
\begin{tabular}{|c|c|c|c|c|c|c|c|c|c|c|c|c|c|c|} 
\hline
numbers & charge & +1$\uparrow$el. & -1$\downarrow$el. & +1$\downarrow$el. & -1$\uparrow$el. & singl.spin & 
tripl.spin & tripl.spin & tripl.spin & $\pm$2$\uparrow\downarrow$el. & +2$\uparrow$el. & -2$\downarrow$el. & +2$\downarrow$el. & -2$\uparrow$el. \\
\hline
$\delta N_c^h$ & $0$ & $-1$ & $1$ & $-1$ & $1$ & $0$ & $0$ & $0$ & $0$ & $\mp 2$ & $-2$ & $2$ & $-2$ & $2$ \\
\hline
$N_{s1}^h$ & $0$ & $1$ & $1$ & $1$ & $1$ & $2$ & $2$ & $2$ & $2$ & $0$ & $2$ & $2$ & $2$ & $2$ \\
\hline
$\delta N_{\uparrow}$ & $0$ & $1$ & $0$ & $0$ & $-1$ & $0$ & $1$ & $-1$ & $0$ & $\pm 1$ & $2$ & $0$ & $0$ & $-2$ \\
\hline
$\delta N_{\downarrow}$ & $0$ & $0$ & $-1$ & $1$ & $0$ & $0$ & $-1$ & $1$ & $0$ & $\pm 1$ & $0$ & $-2$ & $2$ & $0$ \\
\hline
$L_{s,\,+1/2}$ & $0$ & $1$ & $1$ & $0$ & $0$ & $0$ & $2$ & $0$ & $1$ & $0$ & $2$ & $2$ & $0$ & $0$ \\
\hline
$L_{s,\,-1/2}$ & $0$ & $0$ & $0$ & $1$ & $1$ & $0$ & $0$ & $2$ & $1$ & $0$ & $0$ & $0$ & $2$ & $2$ \\
\hline
$N_{s2}$ & $0$ & $0$ & $0$ & $0$ & $0$ & $1$ & $0$ & $0$ & $0$ & $0$ & $0$ & $0$ & $0$ & $0$ \\
\hline
$S_s$ & $0$ & $1/2$ & $1/2$ & $1/2$ & $1/2$ & $0$ & $1$ & $1$ & $1$ & $0$ & $1$ & $1$ & $1$ & $1$ \\
\hline
$\delta S_c$ & $0$ & $1/2$ & $-1/2$ & $1/2$ & $-1/2$ & $0$ & $0$ & $0$ & $0$ & $\pm 1$ & $1$ & $-1$ & $1$ & $-1$ \\
\hline
$\delta N_{s1}$ & $0$ & $0$ & $-1$ & $0$ & $-1$ & $-2$ & $-1$ & $-1$ & $-1$ & $\pm 1$ & $0$ & $-2$ & $0$ & $-2$ \\
\hline
$\delta N_{a_{s1}}$ & $0$ & $1$ & $0$ & $1$ & $0$ & $0$ & $1$ & $1$ & $1$ & $\pm 1$ & $2$ & $0$ & $2$ & $0$ \\
\hline
\end{tabular}
\caption{The deviations $\delta N_c^h=-2\delta S_c$ and numbers $N_{s1}^h=[2S_s+2N_{s2}]$ for the 
fourteen classes of one- and two-electron excited states of the $x>0$ and $m=0$ ground state that span the one- and two-electron 
subspace as defined in Ref. \cite{companion}, corresponding electron number deviations 
$\delta N_{\uparrow}$ and $\delta N_{\downarrow}$, and independent-spinon numbers $L_{s,\,+1/2}$ and 
$L_{s,\,-1/2}$ and $s2$ fermion numbers $N_{s2}$ restricted to value ranges of that subspace.
The spin $S_s$ and deviations $\delta S_c$, $\delta N_{s1}=[\delta S_c-S_s-2N_{s2}]$, and $\delta N_{a_{s1}}=[\delta S_c +S_s]$ 
of each excitation are also provided.}
\label{tableIV}
\end{table} 

It is found in Ref. \cite{s1-bonds} that for the one- and 
two-electron subspace and $N_a^D\gg 1$, 
the operators $g_{{\vec{r}}_{j},s1}$ (and  $g^{\dag}_{{\vec{r}}_{j},s1}$)
related to the annihilation and creation of local $s1$ fermion operators
through Eq. (\ref{JW-f+}), which annihilate (and create) a 
$s1$ bond particle at a site of the $s1$ effective lattice
of real-space coordinate ${\vec{r}}_{j}$, have the following 
general form both for the model on the $1D$ and 
square lattices,
\begin{equation}
g_{\vec{r}_{j},s1} = \sum_{g=0}^{N_{s1}/2D-1} h_{g}\, a_{\vec{r}_{j},s1,g}  
\, ; \hspace{0.35cm} g_{\vec{r}_{j},s1}^{\dag} = 
\left(g_{{\vec{r}}_{j},s1}\right)^{\dag} 
\, ; \hspace{0.50cm}
a_{\vec{r}_{j},s1,g} =
\sum_{d=1}^{D}\sum_{l=\pm1}
\, b_{\vec{r}_{j}+{\vec{r}_{d,l}}^{\,0},s1,d,l,g} \, .
\label{46-b}
\end{equation}
Hence the expression of the operator $g_{\vec{r}_{j},s1}^{\dag}$ 
involves the operators,
\begin{equation}
a_{\vec{r}_{j},s1,g}^{\dag} = \left(a_{{\vec{r}}_{j},s1,g}\right)^{\dag}
\, ; \hspace{0.50cm}
b_{\vec{r},s1,d,l,g}^{\dag} = \left(b_{\vec{r},s1,d,l,g}\right)^{\dag}
\, .
\label{47-b}
\end{equation}
The operators $a_{\vec{r}_{j},s1,g}^{\dag}$ and $a_{\vec{r}_{j},s1,g}$ 
create and annihilate, respectively, a superposition of $2D=2,4$ 
two-site bonds of the type studied in Ref. \cite{s1-bonds} 
and $b_{\vec{r},s1,d,l,g}^{\dag}$ and $b_{\vec{r},s1,d,l,g}$ are
two-site one-bond operators whose expression is given in that reference. 

\section{Confirmation of the gauge structure of the
underlying $s1$ effective lattice}

The $s1$ bond-particle description of Ref. \cite{s1-bonds} involves a change of gauge structure 
\cite{s1-bonds,Xiao-Gang} so that the real-space coordinates of the sites of the 
$s1$ effective lattice correspond to one of the two sub-lattices of the square spin effective lattice. 
Here it is confirmed that for the limit $N_a^D\gg 1$ that the description used in our studies refers to and for hole concentrations $x$ such that
the density $n=(1-x)$ is finite the two choices of $s1$ effective lattice lead to the same description.

The initial ground states of the one- and two-electron
subspace are $x\geq 0$, $m=0$, and $N_{s1}^h=0$ states.
Here $N_{s1}^h=[N_{a_{s1}}^D-N_{s1}]$ is the
number of unoccupied sites of the $s1$ effective
lattice. Its expression is for the one- and two-electron
subspace given in Eq. (\ref{68}) of Appendix A
along with that of the number $N_{s1}$ of $s1$ fermions.
We emphasize that for the present limit $N_{a_{s1}}^D \gg1$ the transformation
of Eq. (\ref{fs1-q-x}) leads precisely to the same operators
$f_{\vec{q}_j,s1}^{\dag}$ and $f_{\vec{q}_j,s1}$ independently
on whether the real-space coordinates $\vec{r}_{j'}$ of the 
corresponding operators $f_{\vec{r}_{j'},s1}^{\dag}$ and
$f_{\vec{r}_{j'},s1}$, respectively, are those of any of the two
sub-lattices of the spin effective lattice. Indeed, the two 
choices of real-space variables of the $s1$ effective lattice
considered in Ref. \cite{s1-bonds} are for $N_{s1}^h=0$ 
states related both by the transformations
$\vec{r}_{j'}\rightarrow\vec{r}_{j'}+a_{s1}\,{\vec{e}}_{x_1}$
and $\vec{r}_{j'}\rightarrow\vec{r}_{j'}+a_{s1}\,{\vec{e}}_{x_2}$.
If in the expressions of Eq. (\ref{fs1-q-x}) we replace
$\vec{r}_{j'}$ by $\vec{r}_{j'}+a_{s1}\,{\vec{e}}_{x_1}$ or
$\vec{r}_{j'}+a_{s1}\,{\vec{e}}_{x_2}$ and transform the
summations in integrals, for the periodic boundary
conditions considered in Ref. \cite{s1-bonds} we
obtain within the limit $N_{a_{s1}}^D\rightarrow\infty$
exactly the same operators $f_{\vec{q}_j,s1}^{\dag}$ and 
$f_{\vec{q}_j,s1}$.

This confirms that the change of gauge structure 
\cite{s1-bonds,Xiao-Gang} used in the construction of the $s1$
effective lattice, which consists in choosing one of the two sub-lattices
of the square spin effective lattice to play the role of
$s1$ effective lattice, does not affect the momentum values 
of the $s1$ fermions: One reaches the same momentum 
values and thus the same physics for the two possible choices of $s1$ effective-lattice 
real-space coordinates. 

That change of gauge structure was 
made in Ref. \cite{s1-bonds} for the $N_{s1}^h=0$
configuration state that generates the spin
degrees of freedom of the $x\geq 0$ and $m=0$ ground states. 
As discussed in that paper, all $N_{s1}^h$-finite 
configuration states associated with the excited
states generated by application of one-
and two-electron operators onto such ground
states can be straightforwardly constructed 
from the $N_{s1}^h=0$ configuration state. The point
is that the momentum $s1$ fermion operators obtained
from the transformation of Eq. (\ref{fs1-q-x}) are
precisely the same for any finite number $N_{s1}^h$
of $s1$ band holes, independently on whether the corresponding $N_{s1}^h$-finite configuration 
states are generated from the $N_{s1}^h=0$ configuration 
state when described by occupancy of one or the other 
sublattice sites. This confirms that
for the limit $N_{a_{s1}}^D \gg1$ the two alternative
choices of real-space coordinates of the $s1$ effective
lattice refer indeed to a gauge structure,
as stated in Ref. \cite{Xiao-Gang}. 

That such a change of gauge structure 
leads to the same $s1$ fermions independently
of the choice of the real-space coordinates of the $s1$ effective
lattice is a result that applies both to the model on the square and 1D lattices. 
                          
\section{The $x$ dependence of the elementary function $e_{s1} (q)$ for $0<x\ll 1$ and $U/4t\geq u_0$}

The goal of this Appendix is the evaluation up to first order in $x$ of the elementary function 
$e_{s1} (q)$ that for $0<x\ll 1$ and $U/4t\geq u_0$ leads to the expressions
$q^{N}_{Bs1}\approx \sqrt{2}[\pi/2](1-x)$ and
$q^{AN}_{Bs1}\approx \pi -\sqrt{x\,2\pi}$ for the absolute values of the auxiliary 
nodal and anti-nodal momenta of Eq. (\ref{q-N-Fs}), respectively, energy
scale $2\vert\Delta\vert=2\Delta_0(1-x/x_*^0)$ and parameter $x_*^0=2r_s/\pi$ 
given in Eq. (\ref{D-D-HM-x<<1}). Specifically,
such expressions are reached provided that $e_{s1} (q)\approx -[W_{s1}^0/2]\,\cos q$ 
remains a good approximation for $0<x\ll 1$ and $U/4t\geq u_0$ except for $q$ values near both
zero and $q^{AN}_{Bs1}$ and is given by,
\begin{eqnarray}
e_{s1} (q) & \approx &  -{W_{s1}^0\over 2}\cos q_U
\, ; \hspace{0.35cm} q \in (0,q_U) \, ,
\nonumber \\ 
& \approx &  -{W_{s1}^0\over 2}\cos q
\, ; \hspace{0.35cm} q \in (q_U,\pi -q_U) \, ,
\nonumber \\
& \approx &  -{W_{s1}^0\over 2}\cos \left(\pi -\sqrt{[\pi -q^{AN}_{Bs1}]^2+[q_U]^2}\right)
\, ; \hspace{0.35cm} q \in (\pi -q_U,q^{AN}_{Bs1}) \, ,
\nonumber \\
q_U & = & \sqrt{{x\,\pi\over r_s}(1 - r_s)}
\, ; \hspace{0.35cm} 0<x\ll 1 \, , \hspace{0.35cm} U/4t\geq u_0 \, .
\label{e-s1-q}
\end{eqnarray}
The above expressions are then consistently obtained
for $0<x\ll 1$ and $U/4t\geq u_0$ by expanding up to first order in $x$ the solution
of the equation,
\begin{equation}
\epsilon^{0,\parallel}_{s1} ({\vec{q}}_{Bs1}^{\,AN}) = -{W_{s1}^0\over 2}\left[\cos q_U  +
\cos \left(\pi -\sqrt{[\pi -q^{AN}_{Bs1}]^2+[q_U]^2}\right)-x\,\pi\right] = 0 \, ;
\hspace{0.25cm} 0<x\ll 1 \, , \hspace{0.15cm} U/4t\geq u_0 \, .
\label{q-AN-Bs1}
\end{equation}

In turn, use in Eq. (\ref{Delta-x-small}) of the expressions provided in Eq. (\ref{e-s1-q}) for 
the $s1$ elementary function leads to the following $x$ dependence for
the energy scale $2\vert\Delta\vert$,                      
\begin{equation}
2\vert\Delta\vert =
\Delta_0 \left[\cos q_U -
\cos \left(\pi -\sqrt{[\pi -q^{AN}_{Bs1}]^2+[q_U]^2}\right)\right] \, ,
\hspace{0.35cm} 0<x\ll 1 \, , \hspace{0.15cm} U/4t\geq u_0 \, .
\label{D-D-HM-pre}
\end{equation}
Taking into account that $[\pi -q^{AN}_{Bs1}]^2\approx  x\,2\pi$ and
expanding within the $0<x\ll 1$ limit the right-hand side of
Eq. (\ref{D-D-HM-pre}) up to first order in $x$ then leads to the $x$ dependence
$2\vert\Delta\vert=(1-x/x_*^0)2\Delta_0$ for $U/4t\geq u_0$
provided in Eq. (\ref{D-D-HM-x<<1}) where $x_*^0=2r_s/\pi$.

The expression given in Eq. (\ref{e-s1-q})
has only physical meaning up to first order in $x$ and up to that order
the function $[e_{s1} (0) +e_{s1} (q_{Bs1}^{AN})]/W_{s1}^0=-x\,\pi/2$ is independent
of $U/4t$ so that the solution of Eq. (\ref{q-AN-Bs1}) is given
by $[\pi -q^{AN}_{Bs1}]^2\approx  x\,2\pi$. In turn,
$[e_{s1} (0) -e_{s1} (q_{Bs1}^{AN})]/W_{s1}^0$ depends on $U/4t$
and through Eq. (\ref{D-D-HM-pre}) is behind the $U/4t$ dependence of the 
parameter $x_c^0 = 2r_s/\pi$ of Eq. (\ref{D-D-HM-x<<1}) for the
range $U/4t\geq u_0$, which the present derivation refers to. 


\end{document}